\documentclass[twocolumn,times,tighten]{aastex631}
\usepackage{CJK}
\usepackage{caption}

\newcommand{\at}{\textit{@}}
\newcommand{\PM}{$\pm$}
\newcommand{\udm}{cm$^{-3}$pc}  
\newcommand{\du}{$^\circ$} 

\begin{document}
\title{FAST Pulsar Database: II. Scattering Profiles of 149 Pulsars}

\author[0000-0002-1056-5895]{W.~C. Jing}
\affiliation{National Astronomical Observatories, Chinese Academy of Sciences, Jia-20 Datun Road, ChaoYang District, Beijing 100012, China}
\affiliation{School of Astronomy and Space Science, University of Chinese Academy of Sciences, Beijing 100049, China}

\author[0000-0002-9274-3092]{J.~L. Han}\thanks{E-mail: hjl@nao.cas.cn}
\affiliation{National Astronomical Observatories, Chinese Academy of Sciences, Jia-20 Datun Road, ChaoYang District, Beijing 100012, China}
\affiliation{School of Astronomy and Space Science, University of Chinese Academy of Sciences, Beijing 100049, China}
\affiliation{Key Laboratory of Radio Astronomy and Technology,  Chinese Academy of Sciences, Beijing 100101, China }

\author[0009-0004-3433-2027]{C. Wang}
\affiliation{National Astronomical Observatories, Chinese Academy of Sciences, Jia-20 Datun Road, ChaoYang District, Beijing 100012, China}
\affiliation{School of Astronomy and Space Science, University of Chinese Academy of Sciences, Beijing 100049, China}
\affiliation{Key Laboratory of Radio Astronomy and Technology,  Chinese Academy of Sciences, Beijing 100101, China }
  
\author[0000-0002-6437-0487]{P.~F. Wang}
\affiliation{National Astronomical Observatories, Chinese Academy of Sciences, Jia-20 Datun Road, ChaoYang District, Beijing 100012, China}
\affiliation{School of Astronomy and Space Science, University of Chinese Academy of Sciences, Beijing 100049, China}
\affiliation{Key Laboratory of Radio Astronomy and Technology,  Chinese Academy of Sciences, Beijing 100101, China }

\author[0000-0002-4704-5340]{T. Wang}
\affiliation{National Astronomical Observatories, Chinese Academy of Sciences, Jia-20 Datun Road, ChaoYang District, Beijing 100012, China}

\author[0000-0002-5915-5539]{N.~N. Cai}
\affiliation{National Astronomical Observatories, Chinese Academy of Sciences, Jia-20 Datun Road, ChaoYang District, Beijing 100012, China}

\author[0000-0003-1778-5580]{J. Xu}
\affiliation{National Astronomical Observatories, Chinese Academy of Sciences, Jia-20 Datun Road, ChaoYang District, Beijing 100012, China}
\affiliation{Key Laboratory of Radio Astronomy and Technology,  Chinese Academy of Sciences, Beijing 100101, China }

\author[0009-0009-6590-1540]{Z.~L. Yang}
\affiliation{National Astronomical Observatories, Chinese Academy of Sciences, Jia-20 Datun Road, ChaoYang District, Beijing 100012, China}
\affiliation{School of Astronomy and Space Science, University of Chinese Academy of Sciences, Beijing 100049, China}

\author[0000-0002-6423-6106]{D.~J. Zhou}
\affiliation{National Astronomical Observatories, Chinese Academy of Sciences, Jia-20 Datun Road, ChaoYang District, Beijing 100012, China}

\author[0009-0008-1612-9948]{Yi Yan}
\affiliation{National Astronomical Observatories, Chinese Academy of Sciences, Jia-20 Datun Road, ChaoYang District, Beijing 100012, China}

\author[0009-0003-2212-4792]{W.~Q. Su}
\affiliation{National Astronomical Observatories, Chinese Academy of Sciences, Jia-20 Datun Road, ChaoYang District, Beijing 100012, China}
\affiliation{School of Astronomy and Space Science, University of Chinese Academy of Sciences, Beijing 100049, China}

\author[0000-0002-6730-4987]{X.~Y. Gao}
\affiliation{National Astronomical Observatories, Chinese Academy of Sciences, Jia-20 Datun Road, ChaoYang District, Beijing 100012, China}

\author[0000-0003-1946-086X]{L. Xie}
\affiliation{National Astronomical Observatories, Chinese Academy of Sciences, Jia-20 Datun Road, ChaoYang District, Beijing 100012, China}

\begin{abstract}
The turbulent ionized interstellar medium diffracts radio waves and makes them propagate in multiple paths. The pulse-broadening observed at low frequencies results from the scattering effect of interstellar clouds of ionized gas. 
During the Galactic Plane Pulsar Snapshot (GPPS) survey and other projects by using the Five-hundred-meter Aperture Spherical radio Telescope (FAST), we detect the pulse-broadening for 149 pulsars in the radio frequency band between 1.0 and 1.5~GHz, including 68 newly discovered pulsars in the GPPS survey and 81 previously known pulsars. 
We find that a more accurate dispersion measure can be obtained from aligning the front edge of the scattered subband pulses at the 1/4 or 1/2 peak level for most pulsars with one dominant component in the intrinsic profile, and the best DM values from aligning the intrinsic profile components from the model-fitting. From the pulse profiles at a few subbands we derive the pulse-broadening timescale and the scattering spectral index. These scattering parameters are measured for the first time for 113 pulsars. For 36 pulsars with previously detected scattering features, our measurements of the pulse-broadening timescale are consistent with results in the literature. We find that pulsars behind spiral arms show a stronger scattering effect due to greater density fluctuations in the arm regions. With a properly derived dispersion measure and careful calibration, we also present polarization profiles for 82 pulsars in three subbands of FAST observations.
\end{abstract}


\section{Introduction} \label{sec:intro}
 
Ionized gas in the interstellar medium can cause two kinds of effects when radio pulses from pulsars pass through. One is the dispersion effect on the pulses, which causes a longer delay of pulses at lower frequencies. 
The more electrons accumulated along the path, the larger the pulse delay. The dispersion measure (DM) for pulses is thus defined as $DM= \int_{\rm psr}^{\rm us} n_e(l) {\rm d} l,$ in units of pc~cm$^{-3}$. Here $n_{\rm e}(l)$ is the electron density at a distance $l$ in units of cm$^{-3}$, and ${\rm d}l$ is the unit length along the line of sight in pc. The distribution of electron density in the Milky Way can be described by the NE2001 model \citep{cl02} and YMW16 model \citep{ymw17}. For most pulsars without distance measurements, their pulsar distances can be estimated from observed DM values by using the models. 
The second effect is the scattering caused by the irregular distribution of ionized gas clouds, which leads to multipath propagation of radio waves \citep{Scheuer68} and is shown by the angular broadening of the point source, the intensity scintillation in the frequency and time domain, and the pulse broadening \citep[see][and references herein]{Rickett77_review, Rickett90_review}. 

The scattering effect broadens the intrinsic narrow pulses asymmetrically, so that the scattered profile can be expressed by the convolution of the intrinsic profile with a pulse broadening function (PBF) \citep{ss70}:
\begin{equation} \label{eq:pbf}
    {\rm PBF}(t) = \tau_{\rm s}^{-1} {\rm exp} (-t/\tau_{\rm s}) U(t)
\end{equation}
where $\tau_{\rm s}$ represents the pulse-broadening timescale, and $U(t)$ is the Heaviside step function \citep{bcc04}.
This PBF assumes a thin screen in the middle between the pulsar and the observer \citep{Cronyn70}. The thin-screen model is generally used to describe the interstellar scattering \citep[e.g.][]{cwb85, okp21}.
The pulse-broadening timescale  $\tau_{\rm s}$ depends on the observing frequency $\nu$, and is typically described by a power-law relation \citep[e.g.][]{bkc+24}: 
\begin{equation}
\tau_{\rm s} (\nu_{\rm GHz}) = \tau_{\rm 1GHz} \cdot \nu_{\rm GHz}^{-\alpha} 
\end{equation}
where $\alpha$ is the scattering spectral index.

Previously, \citet{lkm01} and \citet{lh03} used the unscattered profile at a high frequency as the intrinsic profile to measure or model the pulse-broadening at lower frequencies. \citet{lrk15} allows the possible evolution of the profile components with frequency and then fits data for $\tau_{\rm s}$. \citet{nab13} simultaneously fit the intrinsic profile, pulse-broadening timescale at 1~GHz and DM by assuming $\alpha=4$. \citet{fpb+24} tried to fit pulse profiles in the 2-D frequency-time waterfall plot for fast radio bursts and radio pulsars.
The pulse-broadening timescale $\tau_{\rm s}$ at 1 GHz,  $\tau_{\rm 1GHz}$ and the scattering spectral index $\alpha$ are two conventional characteristic parameters to describe the scattering effect. For the Kolmogorov turbulence, one should get $\alpha=4.4$  \citep{Rickett90_review}. For the Gaussian turbulence, $\alpha=4.0$ \citep[e.g.][]{lj75a,lj75b}. The $\alpha$ derived from observations typically has a value around 4.0 with a deviation of about 0.6 \citep{okp21}.

The pulse-broadening timescale $\tau_{\rm s}$ is related to observed DMs of pulsars. It varies with $\rm DM^{2.2}$ for the turbulent density distribution with a Kolmogorov spectrum, or $\rm DM^{2.0}$ for the medium with a Gaussian spectrum \citep{Rickett77_review}. 
The DM dependence is not linear, and the observed $\tau_{\rm s}-{\rm DM}$ relation shows a steeper slope at high DM values than $\rm DM^{2.2}$ \citep[e.g.][]{Sutton1971,bwh+92,rmd97}. Considering the frequency dependence and DM dependence together, one can get a joint empirical formula, e.g.  
\begin{equation}
    \log(\tau_{\rm s}) = a + b (\log DM) + c (\log DM)^2 - \alpha \log\nu_{\rm GHz}
    \label{eq:tau2dm}
\end{equation}
by \citet{bcc04}, where $a=-6.46$, $b=0.154$, $c=1.07$ and $\alpha=3.86\pm0.16$, %
or 
\begin{equation}
 \tau_{\rm 1GHz} = 1.9\times10^{-7}{\rm ms} \times DM^{1.5}\times(1+3.55\times10^{-5}DM^{3})
 \end{equation}
by \citet{coc22}. The data are scattered from this relation by 0.76 dex.
The slope against DM becomes slightly steeper at higher DM due to the large density fluctuations in the inner Galaxy \citep{xz17,ddb17,cwb85,cws16}. Observed values of $\tau$ exhibit a spread of about two orders of magnitude around the empirical curves \citep{coc22}. This is because of various fluctuations \citep{oc26}. On the one hand, there are large-scale electron density fluctuations distribution in the Galactic disk, halo and spiral arms. On the other hand, there are the small-scale density fluctuations caused by supernova shocks, stellar winds, HII regions \citep{tc93,cl97,cl01,cl02,cl03,occ21,hsl25,pmf+26}.

The scattering effect not only broadens the pulse profile but also redistributes the polarized emission into profile tails. The polarization profiles can be explained as the original profiles convolved with the PBF \citep{kha72}. As a result, the polarization position angle (PA) curve tends to be flattened for the exponential tail of the profile \citep{lh03}. This has been further verified by more observations \citep[e.g.][]{Karastergiou09}, including the data in this paper. 

Currently, the Five-hundred-meter Aperture Spherical radio Telescope (FAST) \citep{nan2006,nlj+11} is the most sensitive telescope in the world. Since 2019 
it has been working for many pulsar projects, including the Galactic Plane Pulsar Snapshot (GPPS) survey \citep{hww21}. Up to now, more than 851 new pulsars have been discovered by this survey \citep{hww21,hzw+24}. FAST has also been used for many related pulsar studies, such as the polarization measurements of 682 pulsars \citep{whx+23}, the HI absorption measurements of a pulsar \citep{jhh+23}, detection of dwarf pulses \citep{cyh+23, yhz+24}, observations for mode-changing of pulsars \citep{ywh23} and pulsar timing \citep{shw+23}. 

This paper presents the database of scattering profiles for 149 pulsars observed by the FAST during the FAST GPPS survey \citep{hww21,hzw+24} and other projects by the co-authors \citep[e.g.][]{xhw+22}. Among them, 68 pulsars were discovered by the GPPS survey. Section \ref{sec:obs} describes the FAST observations and data processing, including a method to get better DM values from the scattered signals than those from conventional routines. Section \ref{sec:result} presents the results and relevant discussion. Conclusions are given in Section \ref{sec:summary}.

\section{FAST Observations and Data Processing} \label{sec:obs}

All observations have been carried out by using the FAST L-band 19-beam receiver \citep{jth20}, which works in the radio band of 1.0 -- 1.5 GHz. The digital bands can record the 4 ($XX$, $YY$, Re[$X^{*}Y$] and Im[$X^{*}Y$]) or 2 ($XX$, $YY$) polarization channel data for all 2048 frequency channels covering the band with a sampling time of 49.152~{\textmu}s. Polarimetric calibration is achieved using the amplitude-modulated noise signals with a period of 2.01326592~s and a duty cycle of 50\% \citep{whx+23}. 

\begin{figure}
  \centering
  \includegraphics[width=0.47\textwidth]{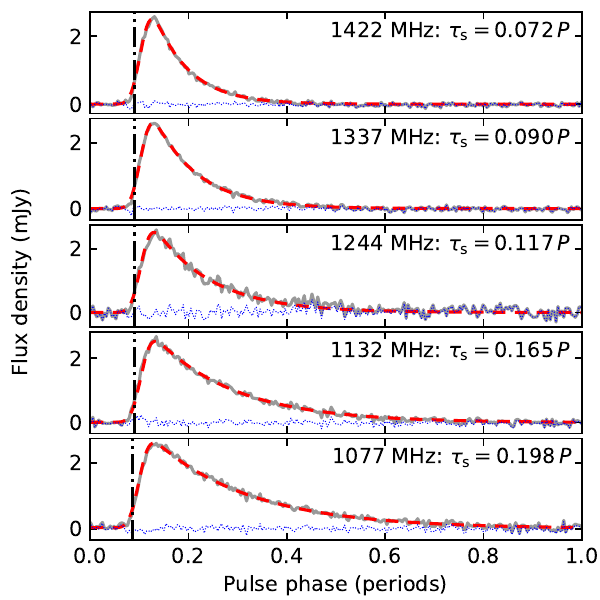}
  \caption{An example for fitting scattered features for sub-band profiles of FAST observations. The observed profiles (solid grey line) for PSR J2052+4421g are presented for 4 subbands. The fitted (dashed) profiles are obtained by a Gaussian function convolved with the standard PBF. The central frequencies and pulse-broadening timescales expressed as a fraction of the pulsar period, are marked in each subpanel for a sub-band profile. 
  }
  \label{fig:2052profile}
\end{figure}

Folding pulsar data and polarimetric calibration were processed in the following steps. Data from frequency channels at the lower edge covering a band of 45 MHz and at the higher edge covering a band of 30 MHz have been discarded because of the low gains \citep{jth20}. 
The data are first folded into a pulse stack by using {\it DSPSR} \citep{sb11}, based on the pulsar ephemeris from the Pulsar Catalogue \citep{mht05}. For pulsars without timing solutions, we use {\it PDMP} from \textsc{psrchive}\footnote{\url{http://psrchive.sourceforge.net}} \citep{hvm04}
to get the best pulsar period. Radio frequency interference (RFI) in some channels is mitigated manually by using {\it PAZ} \citep{sdo12} or automatically by using the new tool $2\sigma CRF$ \citep{chs+23}. 

When a pulsar is strongly scattered, the pulse is much more broadened at lower frequencies. Ideally, profiles with high signal-to-noise ratios from very fine frequency channels over a wide frequency range are needed to fit the scattering model. In practice, a compromise has to be made on the bandwidth and signal-to-noise ratio. In practice, we measure the scattering parameters using 3 to 5 subband profiles of the FAST pulsar observations, depending on signal strength and RFI situation, and then use the 32 subband data to verify the results.

\begin{figure}
  \centering
  \includegraphics[width=0.47\textwidth]{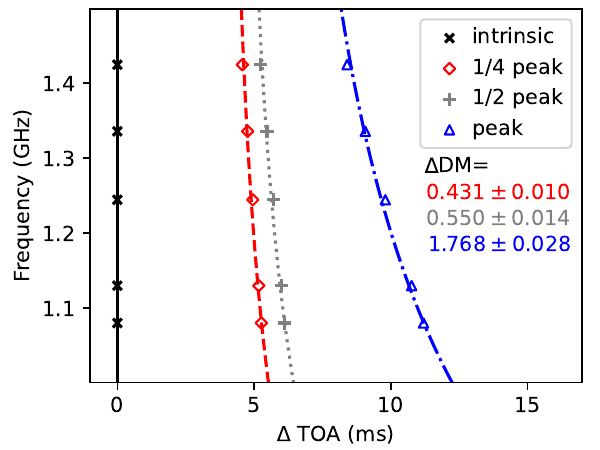}
  \caption{TOA delays for subband profiles for PSR J2052 +4421g measured by using the DM values from aligning the peak, the lead edge at the half peak level (1/2) and at the quarter peak level (1/4), compared to TOA from the exactly de-convolved intrinsic profile component alignment without scattering delay. There is a small DM offset caused by different alignments, depending on the pulsar period, as indicated inside a plot.}  
  \label{fig:TOA}
\end{figure}

\begin{figure}
  \centering
  \includegraphics[width=0.47\textwidth]{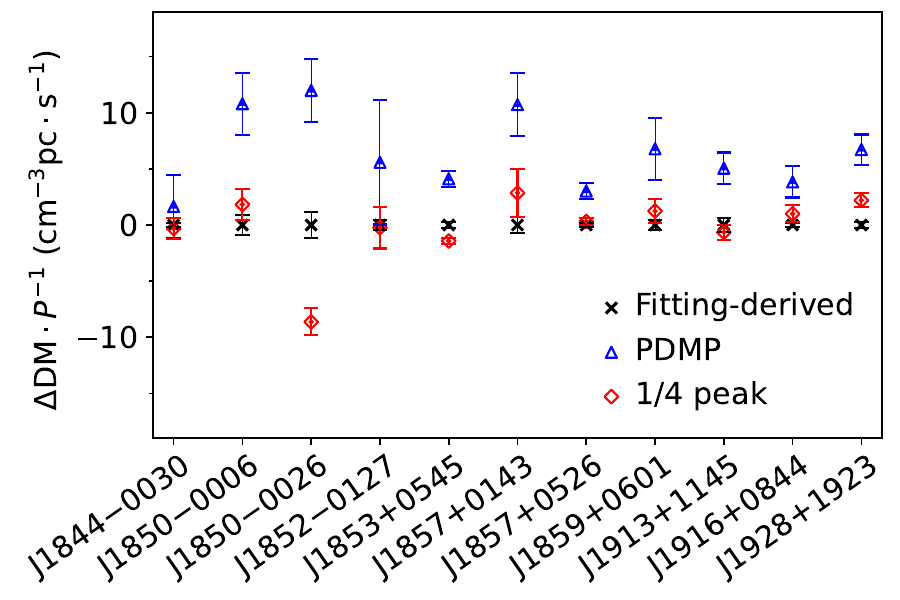}
  \caption{Comparison of DM estimates derived by using three alignments: DM values in the 2-D fitting \citep{okp21}, the DM value directly measured by the {\it PDMP} tool (based on the peak alignment), and the DM value from the front-edge alignment at the 1/4 peak level.}
\label{fig:dm2dm}
\end{figure}

\subsection{Preliminary determination of DM from scattering profiles} \label{sec:dm}

Obtaining unbiased DM measurements can be rather difficult for highly scattered pulsars. An incorrect DM value can lead to the misalignment of scattering profiles. The conventional method for measuring pulsar DM is to align the pulse peaks of all subband profiles. For broadened pulses with shifted peaks at lower frequencies \citep{okp21}, such a method causes a large systematic deviation of a real DM value. 

To minimize the DM bias, some new methods have been proposed, either through fitting profiles based on specific assumptions \citep{gkk17,afm+23} or by using data-driven methods that avoid excessive assumptions about PBFs and intrinsic profiles \citep[e.g.][]{hss+19}. 
For pulsars with a single dominant component scattered with the pulse broadening function of an exponential function given by the thin screen discussed above, we can get a DM by straight-forwardly aligning the leading edge at a half (1/2) or a quarter (1/4) level of the peak, as demonstrated in Fig.~\ref{fig:2052profile}. This method was first presented by \citet{akh70}. We evaluate the possible DM offsets by different alignments of the scattered profile of PSR J2046+4253g, a newly-discovered pulsar by the FAST GPPS survey. To remove white noise from subband profiles, we simultaneously fit the pulse-broadening timescale at 1 GHz $\tau_{\rm1GHz}$ and the scattering spectral index $\alpha$, assuming the same intrinsic profile across the observing band, similar to \citet{nab13}. 

The observed and fitted subband profiles are aligned well in Figure \ref{fig:2052profile}. Using these scattering parameters, we generate a noise-free waterfall plot that can reveal the frequency-dependent times of arrival (TOAs) for the half and quarter-level alignment of the front edges and the peak positions for all sub-band profiles, with an example shown in Figure \ref{fig:TOA}, which in fact cause different systematic DM deviation.
Taking the DM values from the careful 2-D fitting method \citep{okp21} as the reference, we compare the DM values obtained from the normal {\it PDMP} in the \textsc{psrchive} package and the 1/4 peak level alignment, as shown in Fig.~\ref{fig:dm2dm}, and find that the DM values obtained by the 1/4 peak alignment are much better than those obtained by the {\it PDMP} and the offset is almost ignorable if the intrinsic profiles and PBFs are properly assumed.

\subsection{Jointly fitting subband profiles for DM and scattering parameters} \label{sec:method}

The DM of a scattered pulsar preliminarily obtained by aligning the leading edge of subband profiles at the half or quarter peak power level can be used for the further joint fitting of DM-corrected subband profiles for the pulse-broadening timescale and new DM iteratively.

There are four methods in literature to get the pulse-broadening timescale from the scattering profiles: (1) directly fitting the observed profiles using the convolution of the best-trial intrinsic profile and the PBF \citep[e.g.][]{cwb85, kl07}; (2) the deconvolution of the scattering profiles by using the ``CLEAN'' method \citep[e.g.][]{bcc03, yl24, bkj26}; (3) the deconvolution of the profiles in the Fourier domain \citep{ki93}; (4) the coherent cyclic spectral analysis to resolve the frequency-dependent time delay due to multipath propagation from base-band data \citep[e.g.][]{Demorest11}. 
The last method works on the base-band data, which is not applicable here. The deconvolution method and the clean method can work well on subband profiles with high signal-to-noise ratios (S/Ns), while our subband profiles do not always have high S/Ns. Because our subband profiles have a wide range of signal-to-noise ratios,  we use the direct fitting method and acknowledge the possible parameter bias related to the assumed intrinsic profile. Fortunately, most pulsars have one dominant profile component or two or three components that can be easily figured out from the highest-frequency subband profile. The thin screen model can describe the scattering medium for most pulsars. 

We fit subband profiles for the typical pulse-broadening timescale at 1~GHz, the scattering spectral index $\alpha$ and the model DM iteratively. The details are outlined below.

\begin{figure}
\centering
\includegraphics[width=0.98\columnwidth]{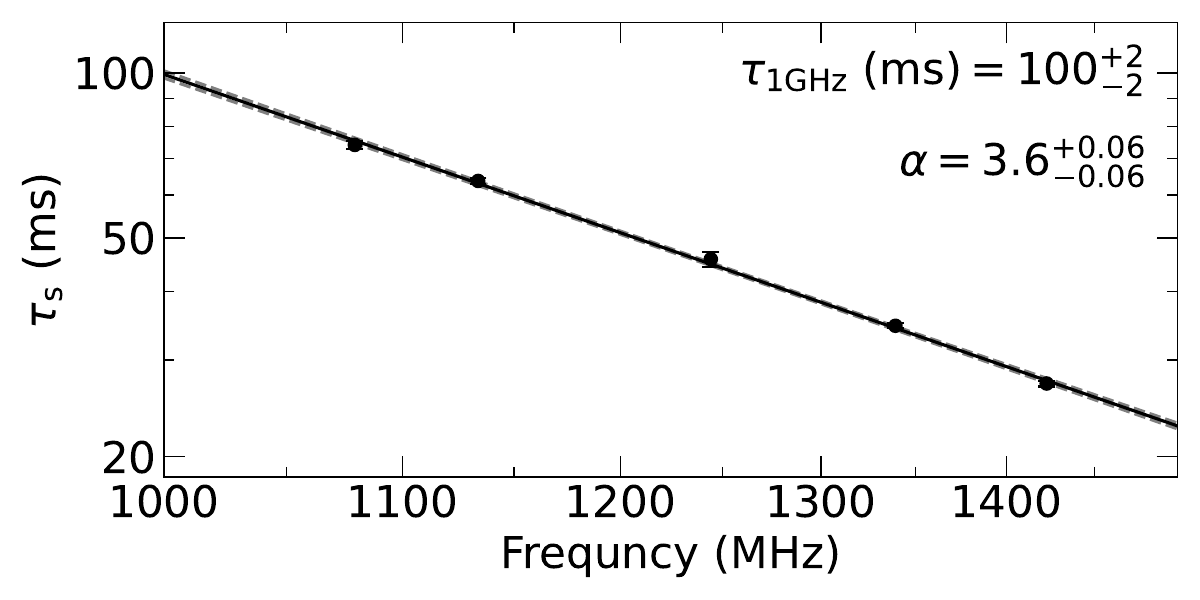} 
\caption{Fitting the pulse-broadening timescale $\tau_{\rm s}$ to get the scattering spectral index $\alpha$. The pulse-broadening timescales are obtained from Fig.~\ref{fig:2052profile} using the technique described in Section 2.2.1. From such a power-law fitting, one can get the frequency dependence of $\tau_{\rm s}$ and $\alpha$ (with the $\pm 1\sigma$ uncertainties).}
\label{fig:fitJ2052}
\end{figure}

\begin{table*}[htpb]
    \centering
    \caption{Fitted scattering parameters for 149 pulsars derived from FAST observations}
    \label{tab:spar-2d}
   \setlength{\tabcolsep}{0.25cm}
\footnotesize
\renewcommand{\arraystretch}{0.85}  
\begin{tabular}{lrrrcrrrrl}
        \hline\noalign{\smallskip}
\multicolumn{1}{c}{PSR Name} & \multicolumn{1}{c}{$l$} & \multicolumn{1}{c}{$b$} & \multicolumn{1}{c}{$P$} & \multicolumn{1}{c}{FAST Obs. Date} & \multicolumn{1}{c}{$T_{\rm obs}$} & \multicolumn{1}{c}{DM$\rm_{1/4}$} & \multicolumn{1}{c}{DM} & \multicolumn{1}{c}{$\tau_{\rm1GHz}$} & \multicolumn{1}{c}{$\alpha$} \\
\multicolumn{1}{c}{ } & \multicolumn{1}{c}{(\du)} & \multicolumn{1}{c}{(\du)} & \multicolumn{1}{c}{(ms)} & \multicolumn{1}{c}{yyyy-mm-dd} & \multicolumn{1}{c}{(s)} & \multicolumn{1}{c}{(\udm)} & \multicolumn{1}{c}{(\udm)} & \multicolumn{1}{c}{(ms)} & \multicolumn{1}{c}{ } \\
\multicolumn{1}{c}{(1)} & \multicolumn{1}{c}{(2)} & \multicolumn{1}{c}{(3)} & \multicolumn{1}{c}{(4)} & \multicolumn{1}{c}{(5)} & \multicolumn{1}{c}{(6)} & \multicolumn{1}{c}{(7)} & \multicolumn{1}{c}{(8)} & \multicolumn{1}{c}{(9)} & \multicolumn{1}{c}{(10)} \\
\hline\noalign{\smallskip}
J0248$+$6021    &136.902&   0.697& 217.12&2023-08-14             &1100&$ 369.7 ( 2)$&$ 369.6 ( 2)$&$ 34.7 ( 3)$&$3.91 ( 3)$\\ 
J1822$-$1252    & 18.218&   0.394&2070.94&2026-04-28             & 600&$ 885   ( 9)$&$ 891   ( 4)$&$185   ( 3)$&$3.8  ( 3)$\\ 
J1822$-$1400    & 17.252&$-$0.176& 214.75&2026-04-24             & 900&$ 649.5 ( 8)$&$ 649.3 ( 9)$&$  6.7 ( 1)$&$3.27 ( 8)$\\ 
J1823$-$1344g   & 17.563&$-$0.200& 467.66&2026-03-29             & 900&$1101   (26)$&$1099   ( 2)$&$ 97   ( 8)$&$4.1  ( 4)$\\ 
J1824$-$1118    & 19.809&   0.741& 435.79&2025-08-18             & 900&$ 601.5 ( 2)$&$ 601.3 ( 2)$&$ 28.20( 7)$&$3.890( 9)$\\ 
J1829$-$1132g   & 20.202&$-$0.496& 244.78&2026-04-19             & 900&$ 849   ( 3)$&$ 848.5 ( 7)$&$ 35   ( 4)$&$4.3  ( 4)$\\ 
J1831$-$1127g   & 20.468&$-$0.843&1402.86&2026-04-27             & 600&$ 524   ( 4)$&$ 525.0 (11)$&$ 36   ( 4)$&$2.9  ( 4)$\\ 
J1833$-$0204g   & 29.022&   3.050&   4.43&2021-10-18             & 900&$ 312.1 ( 1)$&$ 312.08( 4)$&$  0.8 ( 1)$&$3.2   (7)$\\ 
J1833$-$0556    & 25.623&   1.232&1521.63&2025-08-07             & 900&$ 450   ( 4)$&$ 453   ( 4)$&$167   ( 5)$&$2.98 (10)$\\ 
J1833$-$0559    & 25.514&   1.321& 483.50&2025-08-07             & 900&$ 346   ( 1)$&$ 347   ( 1)$&$169   ( 6)$&$3.5  ( 2)$\\ 
J1834$-$0602    & 25.640&   0.965& 487.96&2025-10-23             & 900&$ 442.5 ( 6)$&$ 440.7 ( 2)$&$ 22.8 ( 4)$&$5.65 (10)$\\ 
J1834$-$0812    & 23.709&   0.001& 491.19&2025-10-21             & 900&$1125   (13)$&$1136   ( 7)$&$267   (40)$&$1.8  ( 5)$\\ 
J1837$-$0604    & 25.960&   0.265&  96.31&2020-10-10             & 600&$ 454.4 ( 5)$&$ 455.5 ( 3)$&$ 50   ( 2)$&$5.4  ( 1)$\\ 
J1838$-$0453    & 27.070&   0.708& 380.96&2025-09-11             & 600&$ 615.6 ( 4)$&$ 615.6 ( 3)$&$ 25.9 ( 5)$&$4.60 ( 8)$\\ 
J1838$-$0508g   & 26.920&   0.464& 263.05&2026-01-03             & 900&$ 786   ( 3)$&$ 785.8 (10)$&$ 43   ( 3)$&$3.8  ( 3)$\\ 
J1840$-$0445    & 27.486&   0.203& 422.35&2024-08-30             &1700&$ 364   ( 8)$&$ 365.6 (15)$&$288   (25)$&$4.2  ( 4)$\\ 
J1840$-$0643    & 25.662&$-$0.567&  35.58&2020-08-13             & 600&$ 493   ( 1)$&$ 494.0 ( 6)$&$ 26   ( 2)$&$3.8  ( 2)$\\ 
J1841$-$0157    & 30.099&   1.216& 663.38&2021-09-01             & 900&$ 473.4 ( 3)$&$ 473.5 ( 4)$&$ 14.44( 6)$&$3.50  (2)$\\ 
J1841$-$0353g   & 28.375&   0.359& 778.64&2025-03-04             & 900&$ 948   ( 4)$&$ 949   ( 2)$&$104.5 ( 9)$&$2.99 ( 9)$\\ 
J1842$-$0153    & 30.283&   1.022&1054.32&2021-09-03             & 900&$ 424.3 ( 6)$&$ 423.8 ( 7)$&$ 31.4 ( 2)$&$3.38  (3)$\\
J1842$-$0258g   & 29.202&   0.738& 596.30&2023-10-26             & 900&$ 355.7 ( 9)$&$ 355.7 ( 7)$&$ 20.7 ( 5)$&$3.7   (1)$\\ 
J1842$-$0309    & 29.077&   0.584& 404.95&2023-10-26             & 900&$ 943.6 (17)$&$ 948   ( 1)$&$ 82   ( 3)$&$3.8   (2)$\\
J1843$-$0050    & 31.295&   1.360& 782.66&2021-10-09             & 900&$ 511.6 ( 7)$&$ 511.6 ( 4)$&$ 15.5 ( 3)$&$3.61  (6)$\\ 
J1843$-$0137    & 30.543&   1.087& 669.88&2021-09-23             & 900&$ 483.3 ( 6)$&$ 483.4 ( 8)$&$ 25.5 ( 3)$&$3.53  (4)$\\ 
J1843$-$0157g   & 30.294&   0.847& 508.90&2021-08-22; 2021-09-03 &1200&$ 751   ( 2)$&$ 751   ( 2)$&$110   ( 5)$&$3.4   (2)$\\ 
J1843$-$0310g   & 29.163&   0.381& 285.15&2024-06-16             & 900&$ 1281  ( 6)$&$1283   ( 2)$&$222   (19)$&$5.5  ( 3)$\\ 
J1843$-$0355    & 28.484&   0.056& 132.31&2021-05-28             & 300&$ 794   ( 2)$&$ 794.06( 9)$&$ 38.4 ( 3)$&$5.16 ( 9)$\\ 
J1844$-$0030    & 31.710&   1.271& 641.15&2020-04-21             & 900&$ 603.1 ( 7)$&$ 603.0 ( 7)$&$ 11.0 ( 2)$&$3.19  (5)$\\
J1844$-$0136g   & 30.685&   0.843& 417.27&2024-04-21             & 900&$ 901   ( 4)$&$ 903   ( 2)$&$ 73   ( 7)$&$4.2   (3)$\\ 
J1844$-$0142g   & 30.587&   0.847& 348.74&2021-09-23; 2021-10-18;&2100&$ 794.5 (15)$&$ 794.4 ( 9)$&$ 28   ( 2)$&$4.9   (2)$\\ 
                &       &        &       &2021-11-11             \\
J1844$-$0202g   & 30.269&   0.720& 761.45&2023-02-18             & 900&$ 628   ( 6)$&$ 628   ( 6)$&$ 64   ( 7)$&$4.5   (5)$\\ 
J1844$-$0240    & 29.804&   0.254& 581.55&2022-11-20             & 300&$ 315   ( 5)$&$ 315   ( 3)$&$ 64   ( 5)$&$3.3   (3)$\\ 
J1844$-$0244    & 29.727&   0.235& 507.72&2022-11-20             & 300&$ 422.0 ( 6)$&$ 422.0 ( 6)$&$ 20.3 ( 3)$&$3.48  (5)$\\
J1844$-$0256    & 29.574&   0.119& 273.00&2022-11-20             & 300&$ 819   ( 2)$&$ 820.0 ( 7)$&$ 98   ( 5)$&$3.8   (2)$\\ 
J1844$-$0302    & 29.395&   0.243&1198.63&2022-11-20             & 300&$ 542   ( 2)$&$ 542   ( 2)$&$ 29   ( 2)$&$4.4   (3)$\\
J1844$-$0310    & 29.343&   0.036& 525.02&2021-02-21             & 600&$ 827   ( 2)$&$ 827.8 ( 7)$&$73(4)^\dag$&$4.1(2)^\dag$\\ 
J1844$-$0538    & 27.073&$-$0.941& 255.73&2025-09-26             & 600&$ 410.9 ( 3)$&$ 410.7 ( 2)$&$ 11.92(13)$&$4.06 ( 6)$\\ 
J1845$-$0103g   & 31.272&   0.847&  99.07&2021-10-22; 2021-11-11 &1200&$ 750.1 ( 6)$&$ 750.6 ( 2)$&$ 34   ( 1)$&$4.46 (10)$\\ 
J1845$-$0142g   & 30.734&   0.593& 126.19&2022-01-20; 2022-01-30 &1200&$ 564   ( 4)$&$ 563.1 ( 6)$&$ 52   (12)$&$3.1  ( 7)$\\ 
J1845$-$0144g   & 30.709&   0.466& 593.94&2021-12-25; 2022-01-13;&2100&$ 937.7 (14)$&$ 937.7 (14)$&$216   ( 5)$&$4.14 ( 8)$\\ 
                &       &        &       &2022-01-30             \\
J1845$-$0229Ag  & 30.024&   0.212& 657.72&2022-11-27             & 900&$ 832   ( 5)$&$ 842   ( 4)$&$238   (22)$&$4.6  ( 3)$\\ 
J1845$-$0229Bg  & 30.098&   0.085& 463.05&2023-05-30             & 900&$ 846   (31)$&$ 876   ( 7)$&$234   (65)$&$5.1  ( 9)$\\ 
J1845$-$0243g   & 29.828&   0.066& 345.26&2023-05-26             & 900&$ 747   ( 4)$&$ 750   ( 2)$&$191   (24)$&$5.1  ( 5)$\\ 
J1845$-$0254g   & 29.677&$-$0.012& 492.65&2023-07-06             & 900&$ 735   (35)$&$ 769   ( 7)$&$516   (74)$&$4.7  ( 5)$\\ 
J1845$-$0316    & 29.391&$-$0.255& 207.64&2025-04-11             & 300&$ 495.1 ( 4)$&$ 495.1 ( 3)$&$ 21   ( 1)$&$5.1  ( 2)$\\
J1846$-$0211g   & 30.391&   0.166& 788.17&2022-09-02             & 900&$ 838   ( 3)$&$ 838   ( 4)$&$114   ( 8)$&$5.1  ( 3)$\\ 
J1846$-$0513    & 27.682&$-$1.207&  23.38&2025-09-21             &1280&$ 310.32( 7)$&$ 310.33( 4)$&$  2.07( 6)$&$4.8  ( 2)$\\ 
J1848$-$0055    & 31.797&   0.171& 274.58&2021-10-24             & 300&$1155   ( 2)$&$1157.8 (14)$&$119   ( 9)$&$4.0  ( 3)$\\ 
J1849$-$0013    & 32.501&   0.338& 491.75&2020-08-15             & 900&$ 343.6 ( 9)$&$ 345.1 ( 6)$&$ 71   ( 3)$&$4.2  ( 2)$\\ 
J1849$-$0040    & 32.075&   0.198& 672.54&2021-01-30; 2021-11-11 &1200&$1265   ( 4)$&$1265   ( 4)$&$278   (13)$&$4.9  ( 2)$\\  
J1849$-$0200g   & 31.000&   0.600& 326.76&2022-06-11             & 900&$ 883   ( 4)$&$ 887   ( 3)$&$160   (33)$&$2.3  ( 6)$\\ 
J1850$-$0002    & 32.823&   0.356& 893.35&2022-11-06             &3400&$ 537   ( 4)$&$ 537   ( 2)$&$ 42   ( 1)$&$3.4  ( 1)$\\ 
J1850$-$0006    & 32.764&   0.093&2191.30&2020-02-23; 2020-08-21 &1200&$ 632   ( 2)$&$ 629   ( 3)$&$229   ( 3)$&$3.70 ( 4)$\\ 
J1850$-$0020    & 33.081&   0.453&1574.70&2021-01-30; 2020-04-01;&3900&$ 598   ( 2)$&$ 598   ( 2)$&$ 34   ( 1)$&$3.6  ( 2)$\\ 
                &       &        &       &2020-04-02; 2020-08-01;\\
                &       &        &       &2020-08-19             \\
J1850$-$0026    & 32.407&   0.066& 166.64&2020-04-02; 2020-08-01;&2700&$ 947.0 ( 2)$&$ 947.5 ( 2)$&$35.6(4)^\dag$&$4.19(4)^\dag$\\ 
                &       &        &       &2020-08-19             \\
J1850$-$0031    & 32.369&   0.041& 734.15&2020-04-02; 2020-08-01;&2700&$ 895.0 (16)$&$ 895   ( 2)$&$105   ( 2)$&$4.68 ( 5)$\\
                &       &        &       &2020-08-19             \\
\multicolumn{10}{r}{ ... to be continued.} \\ 
\hline
\end{tabular}
\end{table*}
\addtocounter{table}{-1}
\begin{table*}[htpb]
    \centering
    \caption{-- {\it continued}}
   \setlength{\tabcolsep}{0.25cm}
\footnotesize
\renewcommand{\arraystretch}{0.85}  
\begin{tabular}{lrrrcrrrrl}
        \hline\noalign{\smallskip}
\multicolumn{1}{c}{PSR Name} & \multicolumn{1}{c}{$l$} & \multicolumn{1}{c}{$b$} & \multicolumn{1}{c}{$P$} & \multicolumn{1}{c}{FAST Obs. Date} & \multicolumn{1}{c}{$T_{\rm obs}$} & \multicolumn{1}{c}{DM$\rm_{1/4}$} & \multicolumn{1}{c}{DM} & \multicolumn{1}{c}{$\tau_{\rm1GHz}$} & \multicolumn{1}{c}{$\alpha$} \\
\multicolumn{1}{c}{ } & \multicolumn{1}{c}{(\du)} & \multicolumn{1}{c}{(\du)} & \multicolumn{1}{c}{(ms)} & \multicolumn{1}{c}{yyyy-mm-dd} & \multicolumn{1}{c}{(s)} & \multicolumn{1}{c}{(\udm)} & \multicolumn{1}{c}{(\udm)} & \multicolumn{1}{c}{(ms)} & \multicolumn{1}{c}{ } \\
\multicolumn{1}{c}{(1)} & \multicolumn{1}{c}{(2)} & \multicolumn{1}{c}{(3)} & \multicolumn{1}{c}{(4)} & \multicolumn{1}{c}{(5)} & \multicolumn{1}{c}{(6)} & \multicolumn{1}{c}{(7)} & \multicolumn{1}{c}{(8)} & \multicolumn{1}{c}{(9)} & \multicolumn{1}{c}{(10)} \\
\hline\noalign{\smallskip}
J1850$-$0050g   & 32.103&$-$0.170& 221.54&2021-09-03; 2021-10-09 &1800&$1060   ( 4)$&$1064   ( 2)$&$157   ( 9)$&$4.3  ( 2)$\\ 
J1850$+$0242    & 35.265&   1.403&   4.48&2020-12-22             & 300&$ 539.7 ( 8)$&$ 539.7 ( 2)$&$  5.9 ( 7)$&$4.8  ( 4)$\\ 
J1851$-$0029    & 32.542&   0.335& 518.68&2021-02-11; 2021-03-06 & 600&$ 519.5 ( 3)$&$ 519.5 ( 3)$&$ 13.3 ( 2)$&$3.42 ( 5)$\\  
J1851$-$0108g   & 31.883&$-$0.466&  87.09&2021-09-23; 2021-10-05 &1200&$ 669.6 ( 6)$&$ 669.6 ( 4)$&$ 25   ( 2)$&$4.6  ( 3)$\\ 
J1851$-$0241    & 30.515&   1.186& 435.19&2022-11-22             & 300&$ 511   ( 3)$&$ 511   ( 2)$&$105   ( 9)$&$4.1  ( 3)$\\ 
J1851$+$0233    & 35.181&   1.232& 344.02&2020-05-28             & 300&$ 606   ( 8)$&$ 610   ( 3)$&$ 84   (18)$&$4.1  ( 7)$\\ 
J1852$-$0127    & 31.706&$-$0.802& 429.01&2021-09-01             & 300&$ 427.7 ( 7)$&$ 427.9 ( 9)$&$ 66.2 ( 8)$&$3.89 ( 4)$\\
J1852$+$0013    & 33.282&   0.174& 957.75&2020-04-17; 2020-08-11;&3600&$ 542.7 ( 3)$&$ 542.7 ( 3)$&$ 21.3 ( 2)$&$3.63 ( 4)$\\ 
                &       &        &       &2020-08-17; 2020-09-20 \\
J1852$+$0018    & 33.326&$-$0.085& 318.77&2019-08-25; 2020-04-17;&2100&$ 452.9 ( 7)$&$ 452.9 ( 8)$&$ 20   ( 2)$&$3.7  ( 3)$\\ 
                &       &        &       &2020-09-15             \\
J1852$+$0031    & 33.523&   0.017&2180.47&2020-08-16; 2021-11-14 &1800&$ 752.0 (15)$&$ 745.6 (13)$&$803   ( 2)$&$3.720( 8)$\\ 
J1852$+$0056g   & 33.864&   0.254&1177.91&2020-04-04; 2020-08-19;&6300&$ 893.4 (12)$&$ 893.6 (15)$&$118   ( 3)$&$4.82 ( 8)$\\ 
                &       &        &       &2020-08-20; 2020-09-11;\\
                &       &        &       &2020-11-06; 2021-02-07;\\
                &       &        &       &2022-02-07             \\
J1852$+$0309g   & 35.821&   1.271&   5.58&2021-03-06             & 900&$ 358.0 ( 4)$&$ 358.18( 6)$&$  3.6 ( 6)$&$4.3  ( 7)$\\ 
J1853$-$0049g   & 32.389&$-$0.731&  16.78&2022-05-09             & 900&$ 405.4 ( 4)$&$ 405.6 ( 2)$&$  5.4 (12)$&$3.8  ( 8)$\\ 
J1853$-$0054g   & 32.421&$-$0.974& 308.35&2021-12-21; 2022-01-13 &1200&$ 441.6 ( 6)$&$ 441.6 ( 8)$&$ 17   ( 1)$&$4.0  ( 2)$\\ 
J1853$+$0237g   & 35.552&   0.635& 427.36&2020-04-19; 2021-03-20 &1200&$ 715   ( 7)$&$ 718   ( 2)$&$ 78   ( 9)$&$4.5  ( 5)$\\ 
J1853$+$0505    & 37.650&   1.956& 905.06&2021-04-22             & 300&$ 263.0 ( 5)$&$ 260.4 ( 6)$&$209   ( 1)$&$3.60 ( 2)$\\
J1853$+$0545    & 38.353&   2.064& 126.41&2020-08-01             & 300&$ 197.72( 3)$&$ 197.73 (4)$&$ 20.9 ( 1)$&$3.89 ( 2)$\\ 
J1854$-$00      & 33.849&$-$0.394& 716.87&2020-04-20             & 300&$ 492   ( 2)$&$ 491.7 (17)$&$ 18.0 ( 5)$&$2.8  ( 1)$\\   
J1854$+$0131g   & 34.647&   0.000&2043.73&2020-02-17; 2020-08-17;&3000&$ 451   ( 8)$&$ 450    (8)$&$607   (27)$&$3.0  ( 2)$\\ 
                &       &        &       &2020-09-10; 2020-10-05 \\
J1855$-$0221g   & 31.345&$-$2.076&   2.77&2023-03-05             & 900&$ 223.25( 7)$&$ 223.32( 4)$&$ 0.8  ( 2)$&$4.4  ( 7)$\\ 
J1855$+$0205    & 35.281&   0.007& 246.83&2019-09-13             & 300&$ 866.1 ( 8)$&$ 866.1 ( 6)$&$ 17.8 ( 3)$&$3.4  ( 2)$\\ 
J1855$+$0422    & 37.314&   1.052&1678.11&2020-12-04             & 300&$ 453.9 ( 9)$&$ 449   ( 1)$&$ 50.5 ( 9)$&$4.60 ( 8)$\\    
J1855$+$0455g   & 37.753&   1.398& 101.01&2020-08-13             & 900&$ 372   ( 2)$&$ 372.0 ( 5)$&$ 21   ( 2)$&$3.8  ( 4)$\\ 
J1855$+$0511g   & 37.985&   1.534&1421.47&2024-02-17             & 900&$ 285   ( 2)$&$ 284.7 (18)$&$160   ( 4)$&$3.72 (10)$\\ 
J1855$+$0527    & 38.227&   1.642&1393.47&2020-04-19             & 300&$ 353   ( 3)$&$ 353   ( 3)$&$ 99   ( 4)$&$3.9  ( 2)$\\    
J1856$+$0245    & 36.007&   0.057&  80.91&2020-11-04             & 600&$ 621   ( 2)$&$ 619.1 ( 6)$&$ 29   ( 4)$&$4.9  ( 6)$\\
J1857$+$0143    & 35.168&$-$0.571& 139.79&2019-09-17             & 300&$ 248.5 ( 4)$&$ 248.3 ( 2)$&$ 41.5 ( 9)$&$3.39 ( 8)$\\
J1857$+$0210    & 35.586&$-$0.393& 630.97&2021-11-07             & 300&$ 778.8 ( 6)$&$ 778.9 ( 8)$&$ 32.0 ( 9)$&$4.6  ( 2)$\\   
J1857$+$0214    & 35.577&$-$0.241& 333.92&2020-02-12; 2020-08-22 &1800&$ 972   (12)$&$ 982   ( 7)$&$196   (43)$&$2.7  ( 7)$\\ 
J1857$+$0300    & 36.274&   0.072& 772.68&2020-05-20             & 300&$ 690   ( 2)$&$ 690   ( 2)$&$ 18   ( 2)$&$4.9  ( 4)$\\  
J1857$+$0526    & 38.438&   1.187& 349.96&2020-05-14; 2021-08-26 &1800&$ 464.85(10)$&$ 465.1 ( 1)$&$ 24.0 ( 3)$&$3.82 ( 4)$\\
J1858$+$0215    & 35.725&$-$0.493& 745.76&2020-04-19             & 300&$ 697   ( 1)$&$ 697   ( 1)$&$ 42   ( 2)$&$3.1  ( 2)$\\ 
J1858$+$0244g   & 36.122&$-$0.212&   2.61&2021-12-10             & 900&$ 282.45( 4)$&$ 282.57( 3)$&$  1.6 ( 2)$&$5.3  ( 5)$\\ 
J1859$+$0430    & 37.819&   0.343& 336.32&2020-04-02             & 900&$ 780   ( 4)$&$ 780   ( 2)$&$ 24   ( 3)$&$4.0  ( 6)$\\ 
J1859$+$0601    & 39.245&   0.903&1044.24&2020-03-04; 2020-08-21 &1500&$ 273.5 (11)$&$ 273.8 (13)$&$ 68   ( 3)$&$2.9  ( 2)$\\
J1900$+$0438    & 38.067&   0.169& 312.29&2020-04-04; 2020-08-15 &1800&$ 622   ( 4)$&$ 623   ( 3)$&$193   (17)$&$4.8  ( 3)$\\   
J1901$+$0300    & 36.811&$-$0.974&   7.80&2020-10-28             &1670&$ 253.70( 2)$&$ 253.69( 2)$&$  0.52( 1)$&$3.52 ( 8)$\\ 
J1901$+$0459    & 38.490&   0.087& 877.07&2020-12-19             & 600&$1100   ( 6)$&$1100   ( 5)$&$121   (12)$&$4.8  ( 4)$\\     
J1902$-$0107g   & 33.211&$-$2.996&   6.12&2022-07-21             & 900&$ 290.2 ( 2)$&$ 290.02( 8)$&$  2.6 ( 3)$&$3.4  ( 5)$\\ 
J1903$+$0327    & 37.336&$-$1.014&   2.15&2020-05-23             & 900&$297.526( 9)$&$297.533( 4)$&$  0.45( 3)$&$4.3  ( 5)$\\ 
J1905$+$0600    & 39.838&$-$0.277& 441.25&2020-02-19             & 300&$ 727.1 ( 5)$&$ 727.1 ( 5)$&$ 19.7 ( 2)$&$3.80 ( 4)$\\ 
J1907$+$0534    & 39.717&$-$0.988&1138.32&2020-05-14             & 300&$ 522.7 ( 7)$&$ 522.7 ( 7)$&$ 20.5 ( 4)$&$3.41 ( 9)$\\     
J1908$+$0833    & 42.469&   0.171& 512.15&2020-10-26             & 300&$ 698   ( 2)$&$ 698   ( 2)$&$ 35   ( 2)$&$4.6  ( 3)$\\
J1908$+$0839    & 42.560&   0.229& 185.41&2020-10-26             & 600&$ 513.3 ( 2)$&$ 513.3 ( 3)$&$  8.9 ( 1)$&$2.75 ( 5)$\\   
J1908$+$0909    & 42.972&   0.494& 336.55&2019-12-26             & 300&$ 465.3 ( 2)$&$ 465.3 ( 2)$&$  7.4 ( 2)$&$2.36 ( 6)$\\  
J1910$+$0534    & 40.056&$-$1.668& 452.85&2021-06-21             & 300&$ 478.2 ( 5)$&$ 478.2 ( 6)$&$ 14.4 ( 2)$&$2.97 ( 5)$\\
J1910$+$1026    & 44.438&   0.504& 531.59&2020-09-01; 2020-11-23 &1200&$ 714.0 (12)$&$ 713.9 (13)$&$ 23   ( 2)$&$3.3  ( 2)$\\   
J1911$+$0101A   & 36.111&$-$3.918&   3.62&2021-11-19             & 750&$ 202.54( 2)$&$ 202.56( 1)$&$  0.42( 1)$&$3.38 (10)$\\ 
J1911$+$0925    & 43.675&$-$0.220& 323.87&2020-08-06             & 900&$ 481   ( 3)$&$ 481   ( 2)$&$ 48   ( 7)$&$4.0  ( 5)$\\ 
J1913$+$1000    & 44.285&$-$0.194& 837.16&2021-01-13             & 900&$ 420.9 ( 7)$&$ 420.3 ( 6)$&$ 17.0 ( 3)$&$2.90 ( 6)$\\ 
J1913$+$1102    & 45.250&   0.194&  27.27&2021-11-18             & 600&$ 339.0 ( 8)$&$ 338.7 ( 2)$&$ 12   ( 2)$&$3.6  ( 6)$\\ 
J1913$+$11025   & 45.286&   0.127& 923.90&2025-08-23             & 900&$ 620   ( 2)$&$ 620   ( 2)$&$ 47   ( 2)$&$2.5  ( 2)$\\ 
J1913$+$1145    & 45.920&   0.476& 306.05&2020-05-23             & 300&$ 641.9 ( 2)$&$ 641.9 ( 2)$&$ 15.9 ( 2)$&$4.19 ( 5)$\\
J1914$+$1054g   & 45.213&$-$0.000& 138.87&2020-12-28             & 900&$ 417.6 ( 6)$&$ 417.6 ( 7)$&$ 12   ( 2)$&$4.0  ( 6)$\\ 
J1916$+$0844    & 43.538&$-$1.493& 440.02&2020-12-15             & 300&$ 338.6 ( 2)$&$ 338.5 ( 3)$&$ 12.27( 8)$&$3.34 ( 3)$\\
\multicolumn{10}{r}{ ... to be continued.} \\ 
\hline
\end{tabular}
\end{table*}
\addtocounter{table}{-1}
\begin{table*}[htpb]
    \centering
    \caption{-- {\it continued}}
   \setlength{\tabcolsep}{0.25cm}
\footnotesize
\renewcommand{\arraystretch}{0.85}  
\begin{tabular}{lrrrcrrrrl}
        \hline\noalign{\smallskip}
\multicolumn{1}{c}{PSR Name} & \multicolumn{1}{c}{$l$} & \multicolumn{1}{c}{$b$} & \multicolumn{1}{c}{$P$} & \multicolumn{1}{c}{FAST Obs. Date} & \multicolumn{1}{c}{$T_{\rm obs}$} & \multicolumn{1}{c}{DM$\rm_{1/4}$} & \multicolumn{1}{c}{DM} & \multicolumn{1}{c}{$\tau_{\rm1GHz}$} & \multicolumn{1}{c}{$\alpha$} \\
\multicolumn{1}{c}{ } & \multicolumn{1}{c}{(\du)} & \multicolumn{1}{c}{(\du)} & \multicolumn{1}{c}{(ms)} & \multicolumn{1}{c}{yyyy-mm-dd} & \multicolumn{1}{c}{(s)} & \multicolumn{1}{c}{(\udm)} & \multicolumn{1}{c}{(\udm)} & \multicolumn{1}{c}{(ms)} & \multicolumn{1}{c}{ } \\
\multicolumn{1}{c}{(1)} & \multicolumn{1}{c}{(2)} & \multicolumn{1}{c}{(3)} & \multicolumn{1}{c}{(4)} & \multicolumn{1}{c}{(5)} & \multicolumn{1}{c}{(6)} & \multicolumn{1}{c}{(7)} & \multicolumn{1}{c}{(8)} & \multicolumn{1}{c}{(9)} & \multicolumn{1}{c}{(10)} \\
\hline\noalign{\smallskip}
J1918$+$1340g   & 48.197&   0.254& 232.98&2020-08-07             & 900&$ 572.6 ( 4)$&$ 572.6 ( 3)$&$ 26.3 ( 7)$&$4.6  ( 2)$\\ 
J1919$+$1314    & 47.895&$-$0.088& 571.45&2021-10-12             & 300&$ 612.4 ( 6)$&$ 612.4 ( 7)$&$ 17.2 ( 6)$&$2.89 (10)$\\
J1919$+$1341    & 48.251&   0.168&  11.66&2020-03-03; 2020-08-07;& 27000&$ 394 ( 1)$&$ 394.0 ( 3)$&$  2.1 ( 1)$&$4.7  ( 3)$\\ 
                &       &        &       &2021-05-13; 2021-06-29;\\
                &       &        &       &2021-06-29; 2021-07-15;\\
                &       &        &       &2021-11-14; 2021-12-26;\\
                &       &        &       &2022-01-05; 2022-05-29;\\
                &       &        &       &2022-06-25; 2022-11-05;\\
                &       &        &       &2023-01-24; 2023-02-09;\\
                &       &        &       &2023-03-08; 2023-08-08;\\
                &       &        &       &2023-08-09; 2023-08-10;\\
                &       &        &       &2023-08-15; 2023-08-19;\\
                &       &        &       &2023-08-23; 2023-08-26;\\
                &       &        &       &2023-08-28; 2023-10-14;\\
                &       &        &       &2023-11-10; 2024-01-19;\\
                &       &        &       &2024-01-30             \\
J1920$+$1110    & 46.152&$-$1.199& 509.92&2020-05-09             & 300&$ 186.7 ( 5)$&$ 186.7 ( 6)$&$ 20.7 ( 3)$&$3.58 ( 5)$\\
J1920$+$1340g   & 48.334&$-$0.005&1525.92&2023-10-14             & 900&$1053   (24)$&$1053   ( 7)$&$352   (38)$&$2.0  ( 5)$\\ 
J1921$+$1216g   & 47.214&$-$0.866&   3.00&2023-11-26             & 900&$ 256.09( 2)$&$ 256.12( 1)$&$  0.97( 7)$&$3.03 ( 5)$\\ 
J1921$+$1259g   & 47.873&$-$0.583& 573.16&2024-02-17             & 900&$ 357   ( 3)$&$ 356.6 (16)$&$ 38   ( 2)$&$4.2  ( 3)$\\ 
J1921$+$1340g   & 48.515&$-$0.297&4603.04&2023-02-14             &4590&$ 723   ( 5)$&$ 723   ( 5)$&$160   ( 5)$&$3.69 (10)$\\ 
J1921$+$1505g   & 49.738&   0.381& 611.90&2020-08-12             & 900&$ 513   ( 2)$&$ 510.6 (16)$&$ 25   ( 3)$&$4.3  ( 5)$\\ 
J1922$+$1512g   & 49.933&   0.212&2357.21&2022-05-12             & 900&$ 395   ( 2)$&$ 395   ( 3)$&$ 78   ( 2)$&$3.16 ( 7)$\\ 
J1924$+$1713    & 51.988&   0.720& 758.43&2020-08-08             & 900&$ 539.0 (13)$&$ 539   ( 2)$&$ 39   ( 2)$&$3.7  ( 2)$\\ 
J1928$+$1245    & 48.534&$-$2.290&   3.02&2020-09-29             & 600&$ 179.19( 2)$&$179.192( 5)$&$ 0.274( 8)$&$3.0  ( 2)$\\ 
J1928$+$1923    & 54.281&   1.016& 817.40&2020-08-03             & 300&$ 478.2 ( 5)$&$ 476.4 ( 5)$&$ 48.2 ( 5)$&$3.78 ( 4)$\\
J1929$+$19      & 54.173&   0.568& 339.24&2020-11-21             & 900&$ 524.2 ( 9)$&$ 524.5 ( 9)$&$ 32.1 ( 3)$&$2.85 ( 3)$\\ 
J1946$+$2433g   & 60.915&$-$0.254&   8.57&2022-02-02             & 900&$ 275.2 ( 7)$&$ 275.20( 9)$&$  4.6 (12)$&$4.8  (10)$\\ 
J1946$+$2757g   & 63.847&   1.487& 742.99&2021-07-07; 2021-07-22 &1800&$ 458   ( 3)$&$ 450   ( 2)$&$125   ( 5)$&$3.1  ( 2)$\\ 
J1950$+$2414    & 61.111&$-$1.186&   4.30&2020-05-13             & 300&$ 142.04( 2)$&$ 142.08( 2)$&$  0.88( 4)$&$3.2  ( 3)$\\  
J1959$+$3141g   & 68.424&   1.083& 514.53&2023-08-25             & 900&$ 340.6 (10)$&$ 340.7 ( 2)$&$ 15.3 ( 6)$&$3.8  ( 2)$\\ 
J2004$+$3429    & 71.425&   1.571& 240.99&2020-04-19             & 600&$ 354.1 ( 2)$&$ 354.1 ( 3)$&$ 11.5 ( 1)$&$4.16 ( 5)$\\   
J2005$+$3411g   & 71.261&   1.229& 651.04&2020-08-08; 2021-11-14 &1800&$ 483.0 ( 4)$&$ 483.0 ( 4)$&$ 42.6 (10)$&$3.83 ( 6)$\\ 
J2015$+$3404g   & 72.255&$-$0.466&   4.28&2023-01-06             & 900&$ 207.98( 4)$&$ 208.05( 4)$&$  1.6 ( 2)$&$2.8  ( 3)$\\ 
J2019$+$3718g   & 75.379&   0.690& 505.23&2023-10-17             & 900&$ 461.6 ( 5)$&$ 461.6 ( 7)$&$ 24.0 ( 3)$&$4.28 ( 5)$\\ 
J2020$+$3806g   & 76.145&   0.990& 525.29&2025-02-28             & 900&$ 496.8 ( 4)$&$ 496.2 ( 3)$&$ 10.2 ( 3)$&$4.9  ( 2)$\\ 
J2021$+$3651    & 75.222&   0.111& 103.77&2020-03-03; 2020-04-23 & 900&$ 369.1 ( 5)$&$ 369.2 ( 5)$&$ 14.5 ( 6)$&$2.9  ( 2)$\\  
J2021$+$4024g   & 78.176&   2.114& 370.56&2020-08-12; 2020-10-13 &1800&$ 678.8 (10)$&$ 678.8 ( 5)$&$ 82.1 (14)$&$3.3  ( 3)$\\ 
J2022$+$3845g   & 76.911&   1.017&1008.87&2020-08-08; 2020-10-13 &1800&$ 483   ( 2)$&$ 483   ( 3)$&$258   ( 8)$&$4.3  ( 2)$\\ 
J2022$+$3910g   & 77.273&   1.237&2036.25&2025-08-13             & 900&$ 640   ( 4)$&$ 641   ( 4)$&$173   ( 5)$&$4.7  ( 2)$\\ 
J2030$+$3818g   & 77.449&$-$0.508& 133.72&2021-03-17; 2021-08-11 &1800&$ 597.8 ( 7)$&$ 596.8 ( 3)$&$  7.8 ( 9)$&$2.0  ( 5)$\\ 
J2030$+$3944g   & 78.599&   0.297& 306.19&2020-08-10             & 900&$ 934.7 (10)$&$ 934.8 ( 8)$&$114   ( 4)$&$3.4  ( 2)$\\ 
J2032$+$4055g   & 79.838&   0.646&  48.74&2023-10-15             & 900&$ 372.4 ( 7)$&$ 373.7 ( 5)$&$ 38   ( 4)$&$3.1  ( 4)$\\ 
J2041$+$3934g   & 79.745&$-$1.470& 379.97&2024-11-18             & 900&$ 408   ( 1)$&$ 408.2 (13)$&$234   (12)$&$4.8  ( 2)$\\ 
J2041$+$4551    & 84.766&   2.334&1160.00&2022-05-01             & 300&$ 306   ( 1)$&$ 306.2 (16)$&$ 28.4 ( 5)$&$3.50 ( 2)$\\ 
J2045$+$4431g   & 84.126&   0.975& 141.82&2024-11-18             & 900&$ 428   ( 5)$&$ 428.5 (15)$&$ 16.6 ( 9)$&$3.1  ( 2)$\\ 
J2046$+$4236g   & 82.756&$-$0.381& 523.84&2023-10-15             & 900&$ 466   (10)$&$ 466   ( 4)$&$273   (62)$&$3.7  ( 6)$\\ 
J2046$+$4253g   & 83.016&$-$0.259& 331.15&2022-09-07             & 900&$ 622.1 ( 8)$&$ 622.2 ( 9)$&$127   ( 4)$&$3.87 ( 9)$\\ 
J2052$+$4421g   & 84.836&$-$0.169& 375.34&2020-06-07; 2020-08-12;&2700&$ 543.8 ( 7)$&$ 543.8 ( 9)$&$ 95   ( 1)$&$3.46 ( 4)$\\ 
&&&&2021-11-14&&&\\
J2057$+$4557g   & 86.597&   0.152& 225.38&2023-11-23             & 300&$ 457.0 (16)$&$ 456.9 ( 3)$&$ 48   ( 1)$&$3.8  ( 1)$\\ 
J2205$+$6012    &103.686&   3.696&   2.42&2023-10-18             & 240&$ 157.55( 1)$&$157.604( 6)$&$  0.29( 1)$&$3.5  ( 2)$\\ 
\hline
    \end{tabular}
\tablecomments{ Columns (1) pulsar name, (2) Galactic longitude, (3) Galactic latitude, (4) spin period, (5) FAST observation dates. For some pulsars data from a few dates are added together to get the scattering parameters; (6) the total length of observations in seconds;  (7) dispersion measure from the alignment of the front edge at the 1/4 peak level, (8) dispersion measure from the joint fitting, 
(9) typical pulse-broadening timescale at 1~GHz $\tau_{\rm1GHz}$, (10) scattering spectral index $\alpha$.  $^\dag$ The scattering parameters of two pulsars are obtained from the model fitting of subband profiles, allowing the frequency evolution of profile components.} 
\end{table*}


%





\begin{figure}
\centering
\includegraphics[width=0.99\columnwidth]{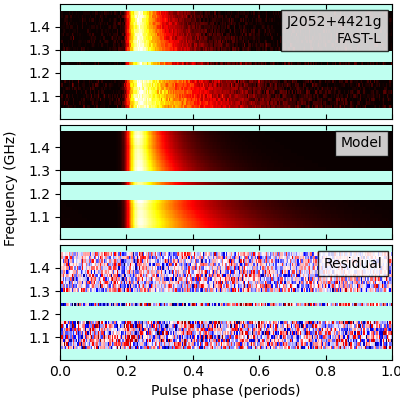} 
\caption{Comparison with the observed 32-subband data by the FAST (top panel) with the modeled profiles (middle panel) produced by the fitted scattering parameters. The scattering parameters should have been reasonably obtained if the residual image (bottom panel) is random and has no obvious structure. Here is only an example for PSR J2025+4421g, and some channels with RFI have been discarded. Comparisons for all pulsars are presented in Fig.~A\ref{fig:2d-all}.}
\label{fig:2dfitJ2052}
\end{figure}

\subsubsection{Fitting the subband profiles independently} \label{sec:individual_fit}

The pulse-broadening timescale $\tau_{\rm s}$ in each subband can be directly obtained from the fitting on each subband profile \citep{gkk17,okp21}. Then, one could get the scattering spectral index $\alpha$ from these fitted timescales for these subbands.

The observed subband profiles are modeled as the convolution of the intrinsic profile, the PBF and also the instrumental smearing function. 
The intrinsic profile is assumed to be a sum of multiple Gaussian components, $\Sigma G(t)$ \citep{Kramer94}, where $t$ is the pulse phase. Each Gaussian component has three free parameters: amplitude $A$, center phase $\mu$, and the characteristic width $\sigma$ \citep{okp21}. The number of Gaussian components is initially obtained from the subband profile at the highest frequency suffering the least scattering effect, and we limit the number to no more than 3 components to avoid over-fitting \citep{bts19}.
The PBF is modeled as an exponential decay as Equation \ref{eq:pbf} representing the interstellar medium scattering effect \citep{lkm01, lmg04}. 
Instrumental smearing effects \citep{khl02} are described by the rectangular functions representing intra-channel dispersion $D(t)$ and sampling time $S(t)$.
Thus, the scattered profile of each subband is then modeled by
\begin{equation}
    I(t) = S(t) \otimes D(t) \otimes {\rm PBF}(t) \otimes \Sigma G(t),
    \label{eq:convolution}
\end{equation}
here $\otimes$ denotes the convolution. A flat baseline is included for cases where the scattering tail is sufficiently long to spread energy across the entire phase of pulsar rotation \citep{gk16}. 
The subband profiles from different subbands are fitted independently using the different intrinsic profiles, as done by \citet{ldk13, lkk15, lrk15}.

In the practical fitting calculations, the least squares method {\it curve\_fit} in the Python package {\it scipy} is used which provides an initial estimate of the fitting parameters for each subband profile. Subsequently, the Monte Carlo Markov Chain (MCMC) method {\it emcee} is implemented to refine the parameters and assess the fitting errors. The best-fit parameters and their errors are determined using the 50th, 16th, and 84th percentiles from the MCMC sampling \citep[see][]{okp21}. The convergence of the MCMC fitting has been visually inspected. 

The pulse broadening timescales $\tau_{\rm s}$ measured in this manner of 3 to 5 subbands are obtained, which are listed in Table~A\ref{tab:taus} and then used for a further fitting with a power-law function by the MCMC method to get the scattering spectral index $\alpha$ and also the typical pulse-broadening timescale $\tau_{\rm 1GHz}$, as shown in Figure~\ref{fig:fitJ2052}.

\begin{figure}[htbp]
  \centering
  \includegraphics[width=0.43\textwidth]{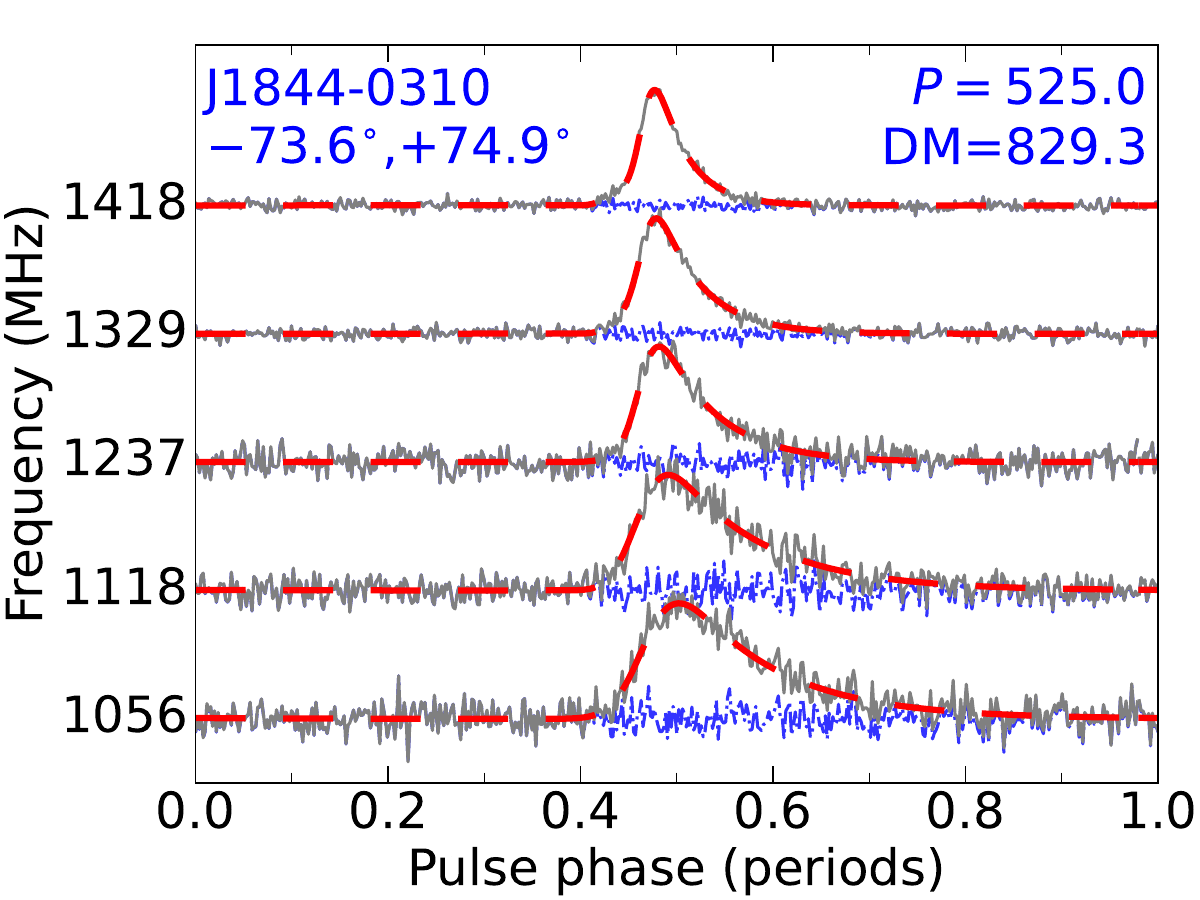}\\
  \includegraphics[width=0.43\textwidth]{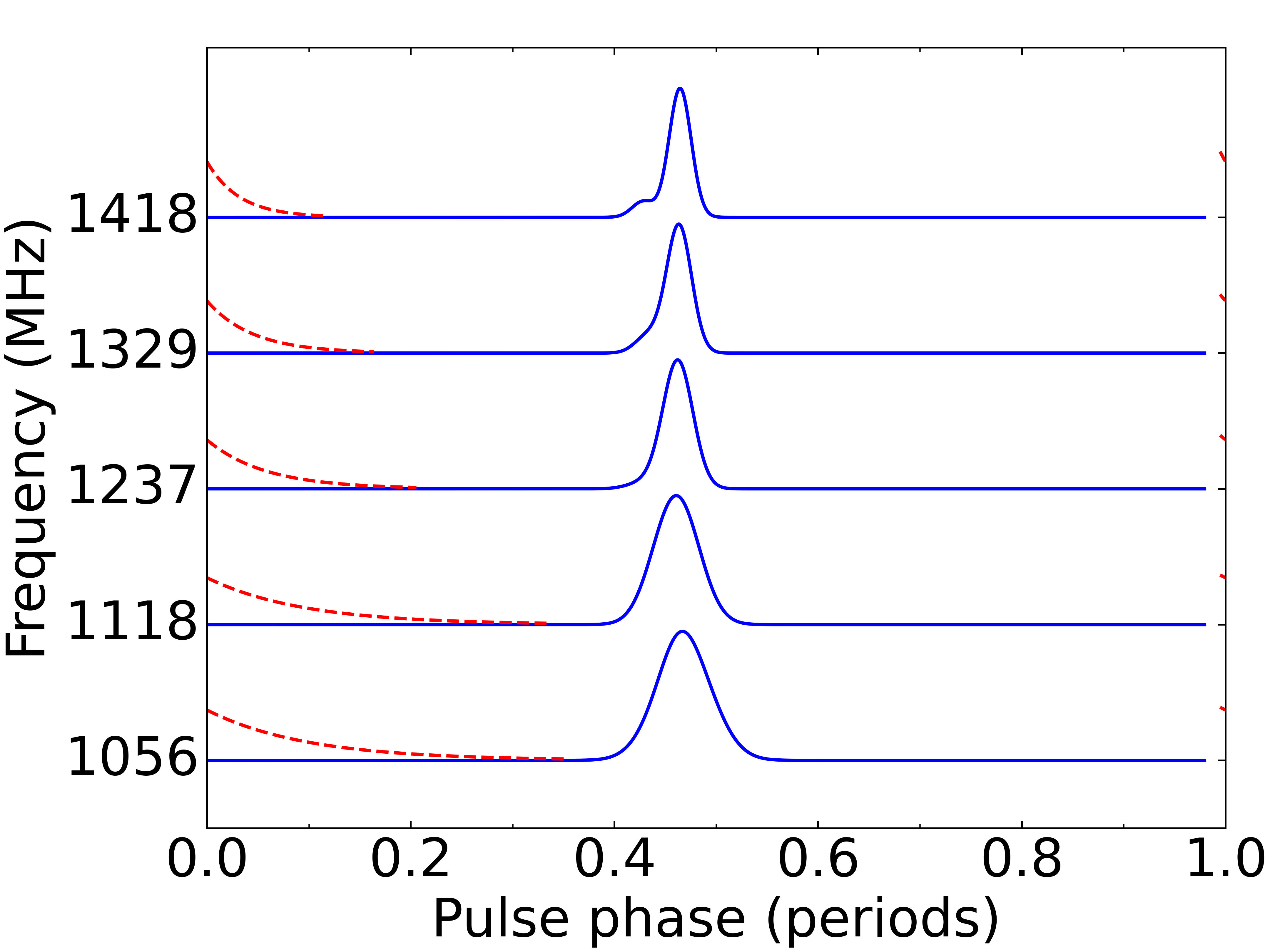}\\
  \includegraphics[width=0.40\textwidth]{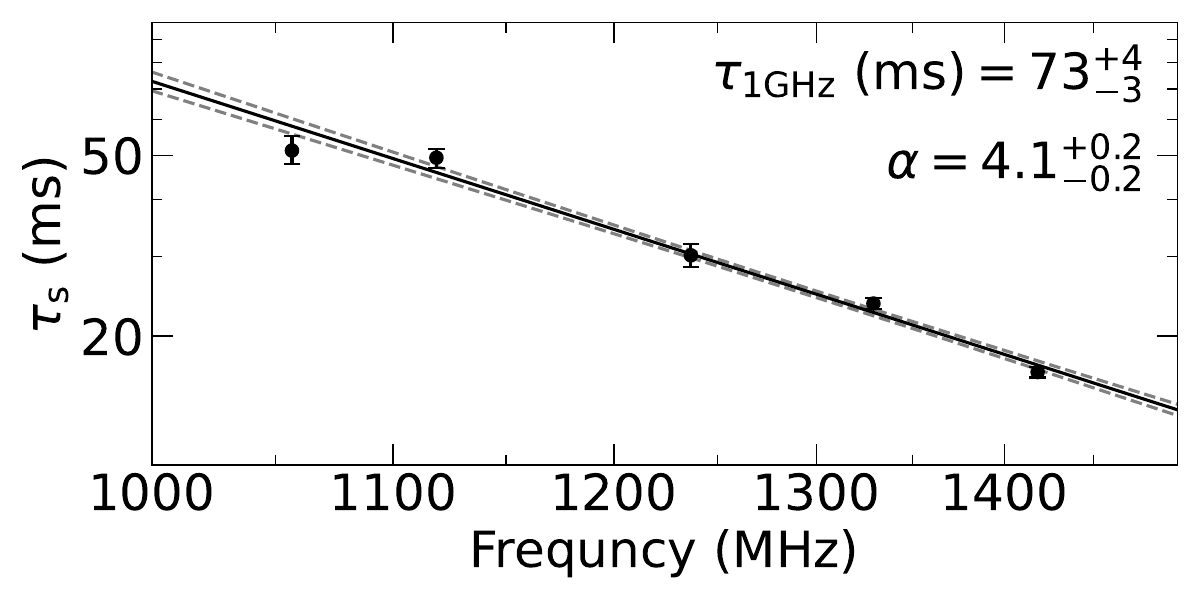}
  \caption{The subband profiles from the FAST observations for PSR J1844$-$0310 and the results for scattering parameter fittings. {\it Top panel:} profiles in solid lines and fittings in dashed lines for 5 subbands, together with residuals in short-dashed lines; {\it Middle panel:} the intrinsic subband profiles obtained from fittings in solid lines together with the PBFs in short-dashed lines; {\it Bottom panel:} the power-law fitting to the pulse-broadening timescales for the scattering spectral index $\alpha$.
  }
\label{fig:spec1844}
\end{figure}
\begin{figure}[htbp]
  \centering
  \includegraphics[width=0.43\textwidth]{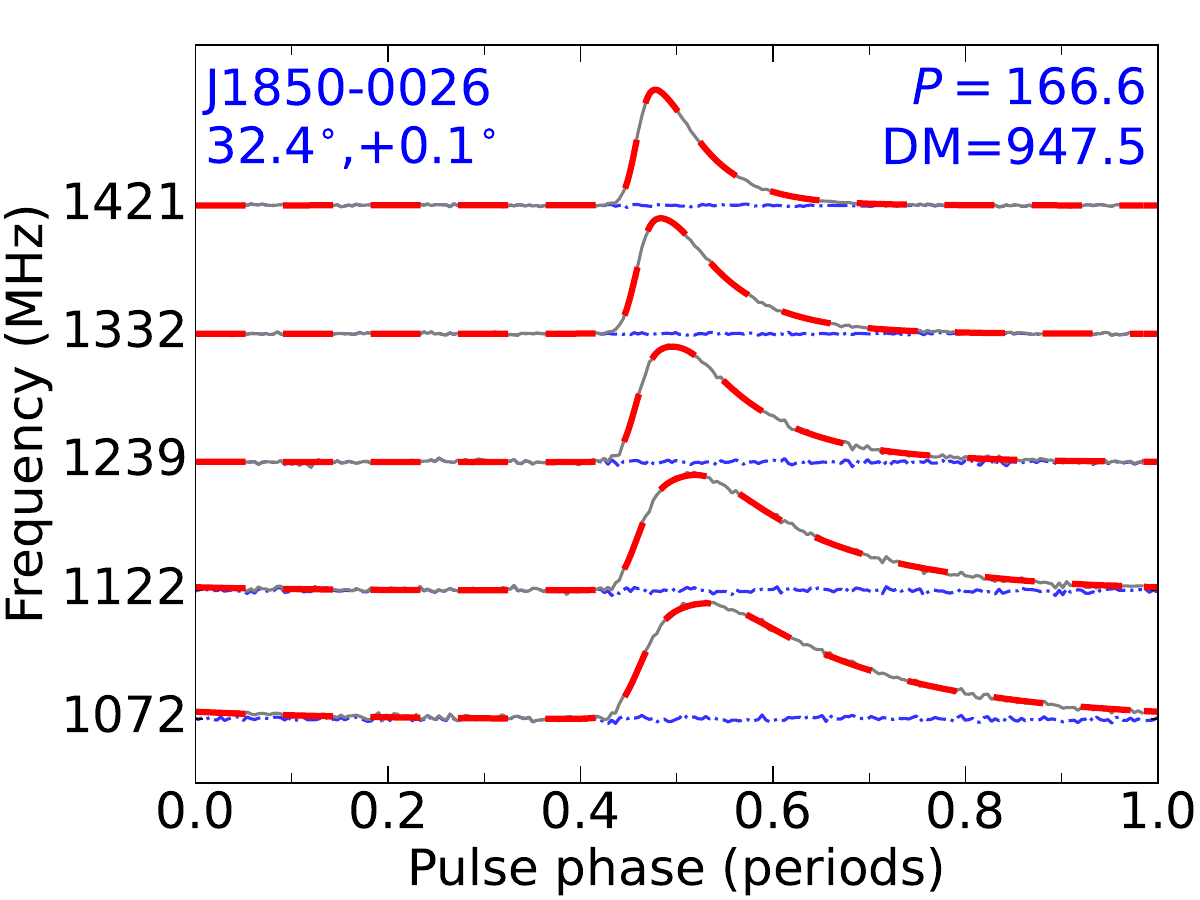}\\
  \includegraphics[width=0.43\textwidth]{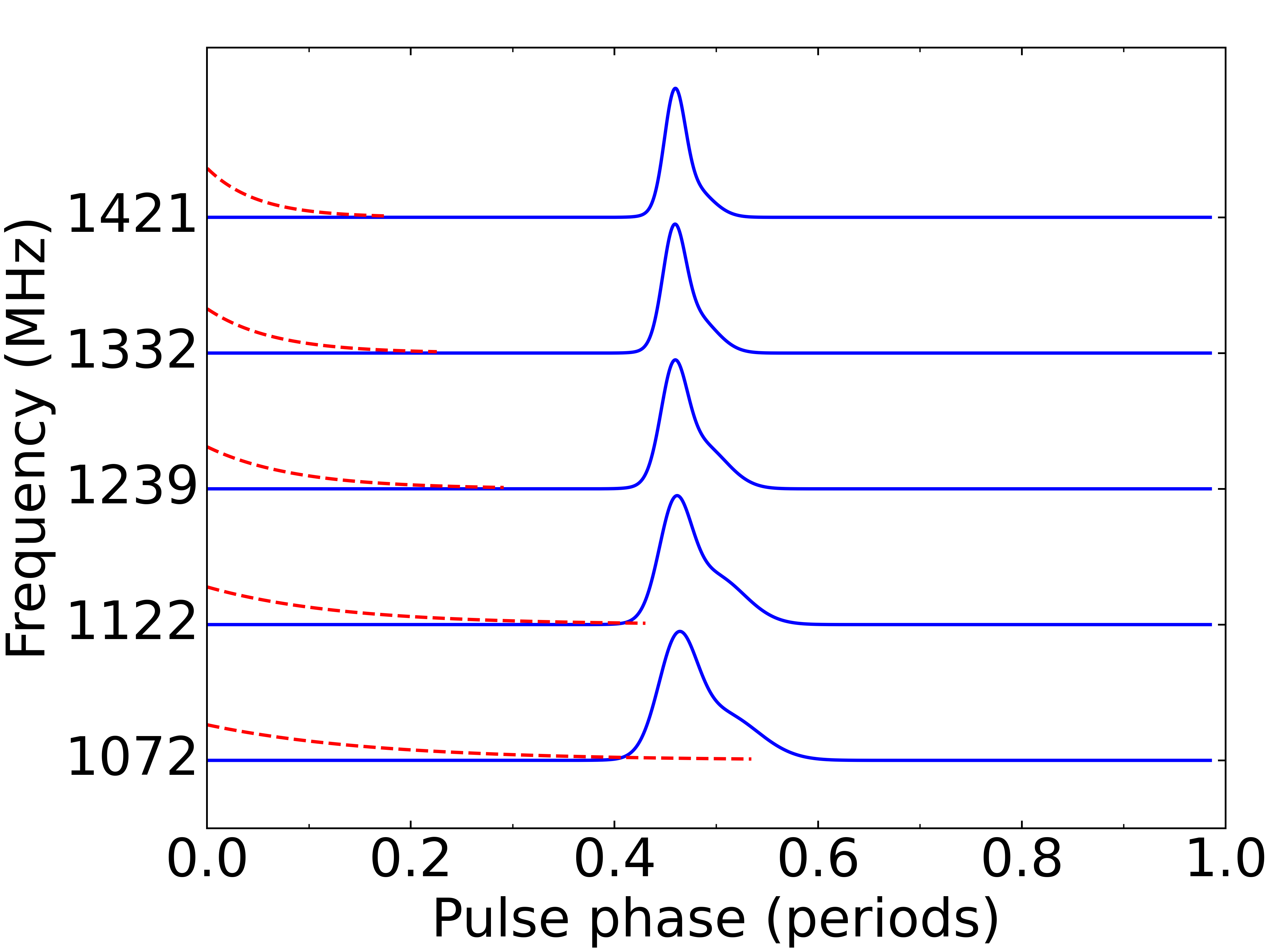}\\
  \includegraphics[width=0.40\textwidth]{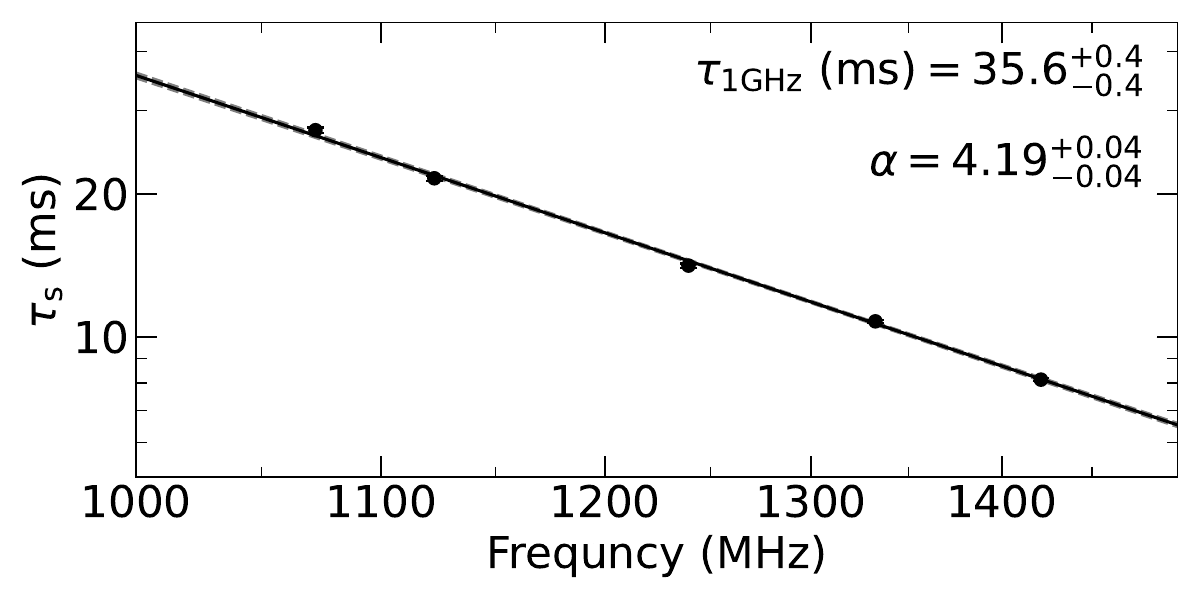}\\
  \caption{The same as Figure~\ref{fig:spec1844} but for PSR J1850$-$0026.}
\label{fig:spec1850}
\end{figure}

\subsubsection{Joint fitting subband profiles}

If one can restore the unscattered profiles at multiple frequencies, the DM of a scattered pulsar can be optimized \citep[e.g.][]{sjk+24}. If the length of scattered profiles is larger than the pulsar periods, the baseline of the scattered profile \citep[e.g.][]{gkk17} is hard to measure. We do our best to compensate for the baseline by a freely fitted porfile offset or simply skip the data from such subbands or channels. Here, we model the frequency dependence of scattered profiles by assuming the same intrinsic profile in the observing band between 1.0 and 1.5~GHz. Then, we perform the best fit for DM and scattering parameters to the subband profiles.

The intrinsic profile is first modeled from the subband profiles at the highest frequency, as done in the last sub-section. 
The modeled profile of each subband is the convolution of the PBF and the intrinsic profile. Next, when we fit subband profiles together, both ISM effects and intrinsic profiles are optimized simultaneously. 
The real DM is obtained from the alignment of the main peak of the modeled intrinsic profiles from all subbands.

The Monte Carlo Markov Chain (MCMC) method {\it emcee} is implemented to fit good subband profiles together. We get the model parameters and determine the uncertainties using the 50\%, 16\%, and 84\% percentiles from the MCMC sampling \citep[e.g.][]{okp21}. The convergence of the MCMC has been visually inspected. A new DM is obtained by aligning the main peak of the modeled intrinsic profile components. The above steps are iterated until the DM value is consistent with the previous one in 1$\sigma$ uncertainty. The final results for pulsar DM and scattering parameters are in the last three columns in Table~\ref{tab:spar-2d}.

\subsubsection{Validation of the fitting results by using the 2-D phase-frequency waterfall plot}

The 2-D phase-frequency waterfall plot with 32 subbands in the FAST observation band of 1.0 -- 1.5~GHz is used to evaluate the goodness of fitted scattering parameters. Based on the fitted parameters obtained above, we generate the model subband profiles and then scale them to match the observed subband profiles. For some channels or subbands, we have to adjust the baselines of the profiles for several strongly scattered pulsars. After matching and subtracting, the residual waterfall plot should be random and have no structure as shown in Fig.~\ref{fig:2dfitJ2052} if the assumed intrinsic profiles with one, two or three components and the pulse-broadening functions are reasonable. For almost all cases, the joint fittings have produced excellent matching to the 2-D phase-frequency waterfall plots, as shown in Figure~A\ref{fig:2d-all}. There are only two exceptional cases in Figure~A2, see details below.

\begin{table*}[ht]
    \centering
    \caption{Comparison of pulse-broadening timescale and scattering spectral index from literature and obtained in this paper by FAST observations}
    \label{tab:compare}
\footnotesize
\renewcommand{\arraystretch}{0.85}  
\setlength{\tabcolsep}{3mm}
    \begin{tabular}{lrrrrrrrrr}
        \hline\noalign{\smallskip}
\multicolumn{1}{c}{PSR Name} & \multicolumn{1}{c}{Ref} & \multicolumn{1}{c}{$\nu_{\rm Ref}$} & \multicolumn{1}{c}{$\tau_{\rm Ref}$} & \multicolumn{1}{c}{$\alpha_{\rm Ref}$} & \multicolumn{1}{c}{$\tau_{\rm FAST}$} & \multicolumn{1}{c}{$\alpha_{\rm FAST}$} \\
\multicolumn{1}{c}{ } & \multicolumn{1}{c}{ } & \multicolumn{1}{c}{(MHz)} & \multicolumn{1}{c}{(ms)} & \multicolumn{1}{c}{ } & \multicolumn{1}{c}{(ms)} & \multicolumn{1}{c}{ } \\
\multicolumn{1}{c}{(1)} & \multicolumn{1}{c}{(2)} & \multicolumn{1}{c}{(3)} & \multicolumn{1}{c}{(4)} & \multicolumn{1}{c}{(5)} & \multicolumn{1}{c}{(6)} & \multicolumn{1}{c}{(7)} \\
\hline\noalign{\smallskip}                                                                                         
J1822$-$1400 &[1]& 1000&$  6.81( 9)$&$3.31 ( 8)$&$  6.7 ( 1)$&$3.27 ( 8)$\\
J1824$-$1118 &[1]& 1000&$ 26.9 ( 2)$&$3.77 ( 2)$&$ 28.20( 7)$&$3.890( 9)$\\
J1833$-$0559 &[1]& 1000&$180   ( 6)$&$3.78 ( 9)$&$169   ( 6)$&$3.5  ( 2)$\\
J1837$-$0604 &[1]& 1000&$ 62   ( 6)$&$5.01 ( 2)$&$ 50   ( 2)$&$5.4  ( 1)$\\
J1842$-$0153 &[1]& 1000&$ 32.7 ( 3)$&$3.92 ( 5)$&$ 31.4 ( 2)$&$3.38 ( 2)$\\
J1844$-$0030 &[1]& 1000&$ 12.3 ( 3)$&$4.1  ( 2)$&$ 11.0 ( 2)$&$3.19 ( 5)$\\
J1844$-$0244 &[1]& 1000&$ 20.0 ( 4)$&$3.31 ( 9)$&$ 20.3 ( 3)$&$3.48 ( 5)$\\
J1844$-$0538 &[1]& 1000&$ 11.73( 7)$&$4.07 ( 3)$&$ 11.92(13)$&$4.06 ( 6)$\\
J1850$-$0006 &[1]& 1000&$260   ( 7)$&$4.2  ( 1)$&$229   ( 3)$&$3.70 ( 4)$\\
J1850$-$0026 &[1]& 1000&$ 46.5 ( 6)$&$4.946( 2)$&$ 35.6 ( 4)$&$4.19 ( 4)$\\
J1852$-$0127 &[1]& 1000&$ 57   ( 3)$&$3.6  ( 1)$&$ 66.2 ( 8)$&$3.89 ( 4)$\\
J1852$+$0031 &[3]& 1175&$495   (25)$&          -&$443.5 (13)$&$3.720( 8)$\\
             &[3]& 1475&$225   (14)$&          -&$190.3 ( 9)$&          -\\
             &[2]& 2700&$ 36   (16)$&$2.8  (10)$&           -&          -\\
J1853$+$0505 &[3]& 1175&$124   (14)$&          -&$117.0 ( 7)$&$3.60 ( 2)$\\
             &[3]& 1475&$ 54   ( 3)$&          -&$ 51.6 ( 5)$&          -\\
J1853$+$0545 &[1]& 1000&$ 20.0 ( 3)$&$3.13 ( 4)$&$ 20.9 ( 1)$&$3.89 ( 2)$\\
             &[3]& 1175&$ 13.6 (20)$&          -&$ 11.16( 7)$&          -\\
             &[3]& 1475&$  7.1 ( 9)$&          -&$  4.61( 5)$&          -\\
             &[3]& 2380&$  1.5 ( 2)$&          -&           -&          -\\
J1855$+$0205 &[4]& 1000&$ 16.3 ( 9)$&          -&$ 17.3 ( 5)$&$3.4  ( 2)$\\
J1855$+$0422 &[3]& 1175&$ 27   ( 3)$&          -&$ 24.1 ( 6)$&$4.60 ( 8)$\\
J1856$+$0245 &[4]& 1000&$ 18.0 ( 6)$&          -&$ 29   ( 4)$&$4.9  ( 6)$\\
J1857$+$0143 &[1]& 1000&$ 50   ( 2)$&$4.0  ( 1)$&$ 41.5 ( 9)$&$3.39 ( 8)$\\
             &[5]& 1170&$ 24   ( 5)$&          -&$ 24.4 ( 7)$&          -\\
             &[5]& 2600&$  0.05( 1)$&          -&           -&          -\\
             &[5]& 4850&$  0.03( 2)$&          -&           -&          -\\
J1857$+$0210 &[3]& 1175&$ 13.4 (36)$&          -&$ 15.2 ( 7)$&$4.6  ( 2)$\\
J1857$+$0526 &[1]& 1000&$ 24.0 ( 6)$&$3.78 ( 7)$&$ 24.0 ( 3)$&$3.82 ( 4)$\\
             &[3]& 1175&$ 14.5 (17)$&          -&$ 13.0 ( 2)$&          -\\
             &[3]& 1475&$  6.2 (13)$&          -&$  5.44(10)$&          -\\
J1858$+$0215 &[3]& 1175&$ 38   ( 3)$&          -&$ 25.5 (15)$&$3.1  ( 2)$\\
             &[3]& 1475&$ 18.4 (47)$&          -&$ 12.6 (12)$&          -\\
J1859$+$0601 &[1]& 1000&$125   (26)$&$4.4  ( 7)$&$ 68   ( 3)$&$2.9  ( 2)$\\
J1903$+$0327 &[6]& 1400&$  0.126(1)$&          -&$  0.12( 3)$&$4.3  ( 5)$\\
J1908$+$0839 &[3]& 1175&$  5.6 (13)$&          -&$  4.4 ( 2)$&$2.75 ( 5)$\\
             &[3]& 1475&$  2.6 ( 6)$&          -&$  2.23( 7)$&          -\\
J1908$+$0909 &[3]& 1175&$  4.9 ( 9)$&          -&$  5.1 ( 2)$&$2.36 ( 6)$\\
J1910$+$0534 &[3]& 1175&$ 12.5 ( 8)$&          -&$  8.9 ( 2)$&$2.97 ( 5)$\\
J1911$+$0925 &[7]& 1000&$ 38   ( 8)$&$4.3  ( 7)$&$ 48   ( 7)$&$4.0  ( 5)$\\
J1913$+$1000 &[3]& 1175&$ 11.1 (40)$&          -&$ 10.6 ( 3)$&$2.90 ( 6)$\\
J1913$+$11025&[7]& 1000&$ 51   ( 4)$&$1.6  ( 2)$&$ 71   ( 8)$&$2.6  ( 3)$\\
J1913$+$1145 &[1]& 1000&$ 15.2 ( 7)$&$3.9  ( 2)$&$ 15.9 ( 2)$&$4.19 ( 5)$\\
             &[3]& 1175&$  9.2 (10)$&          -&$  8.1 ( 2)$&          -\\
             &[3]& 1475&$  4.3 ( 5)$&          -&$  3.12( 8)$&          -\\
J1916$+$0844 &[1]& 1000&$ 12.8 ( 1)$&$4.14 ( 7)$&$ 12.27( 8)$&$3.34 ( 3)$\\
             &[3]& 1175&$  7.7 (10)$&          -&$  7.16( 6)$&          -\\
             &[3]& 1475&$  3.6 ( 9)$&          -&$  3.35( 5)$&          -\\
J1919$+$1314 &[4]& 1000&$ 18.7 ( 9)$&          -&$ 18.1 ( 5)$&$2.89 (10)$\\
J1920$+$1110 &[3]& 1175&$ 14.0 (33)$&          -&$ 11.6 ( 2)$&$3.58 ( 5)$\\
             &[3]& 1475&$  6.3 (21)$&          -&$  5.1 ( 2)$&          -\\
J1924$+$1713 &[7]& 1000&$ 39   (18)$&$4.5  (15)$&$ 39   ( 2)$&$3.7  ( 2)$\\
J1928$+$1923 &[1]& 1000&$ 48.6 ( 5)$&$3.94 ( 5)$&$ 48.2 ( 5)$&$3.78 ( 4)$\\
J2205$+$6012 &[8]& 1000&$  0.32( 1)$&$3.71 ( 6)$&$  0.29( 1)$&$3.5  ( 2)$\\
\hline
    \end{tabular}
\tablecomments{ Columns (1) pulsar name, (2) reference for public scattering parameters, (3) central frequency of the pulse broadening timescale of the reference, (4) pulse-broadening timescale at the given frequency, (5) public scattering spectral index, (6) pulse-broadening timescale at the reference frequency interpolated by scattering parameters in this paper, (9) scattering spectral index measured by FAST. The references are marked as: 
 [1] \citet{okp21}; 
 [2] \citet{lkm01}; 
 [3] \citet{bcc04}; 
 [4] \citet{nab13}; 
 [5] \citet{ldk13}; 
 [6] \citet{crl08}; 
 [7] \citet{psf22}; 
 [8] \citet{dcs22}.
 }
\end{table*}

\subsubsection{Two exceptional cases}

The two exceptional cases are PSR J1844$-$0310 and J1850$-$0026. Through simple joint fitting using the same intrinsic profiles, we got the scattering spectral indexes $5.7\pm0.2$ for PSR J1844$-$0310 and $5.76\pm0.02$ for PSR J1850$-$0026. 
Both of them show remarkable residuals in the 2-D phase-frequency waterfall plot, and the $\alpha$ value for PSR J1850$-$0026 is greater than that given by \citet{okp21}, as listed in Table~\ref{tab:compare}.
The extremely low and seemingly implausible error values observed for the two pulsars are likely the result of the fitting process converging to a local optimum.
However, when we check the fitting to the profiles of individual subbands and the 2-D residual plots, we understand that there is a weak frequency-evolving component on the leading edge of PSR J1844$-$0310 and a weak component on the trailing edge of J1850$-$0026.

As shown in Figure~\ref{fig:spec1844}, the weak leading component of PSR J1844$-$0310 becomes more prominent and is mixed with the main component at lower frequencies, making the main component broader at lower frequencies. By taking the frequency-evolved intrinsic profiles, we fit the observed subband profiles using the method in Section \ref{sec:individual_fit}, we get the scattering parameters of $\tau_{\rm1GHz}=73\pm4$~ms and $\alpha=4.2\pm0.2$.

The extra component on the trailing side of PSR J1850$-$0026 also has frequency evolution and causes more broadening of the observed profiles at the lower frequencies, not only the scattering tails, as shown in Figure~\ref{fig:spec1850}. As done above, by taking the frequency-evolved intrinsic profiles, we fit the observed subband profiles using the method in Section \ref{sec:individual_fit}, we get the broadening timescales for all subband profiles, and then fit the scattering parameters. Finally, we obtain  $\tau_{\rm1GHz}=35.6\pm0.4$~ms and $\alpha=4.19\pm0.04$.  

We, therefore, understand that in some cases, only when the reasonable frequency evolution of profile components is considered in the model-fitting, can the scattering parameters be properly estimated.

\section{Results and discussion} 
\label{sec:result}

In total, we have measured the scattering parameters for 122 pulsars using data from GPPS and other FAST projects, as indicated by the lengthening tails at lower frequencies in Figure~\ref{fig:2052profile} and Figure~\ref{fig:2dfitJ2052}. Plots for all 122 pulsars are presented in Figure ~A\ref{fig:2d-all}.
The fitted scattering parameters for all 122 pulsars are listed in Table~\ref{tab:spar-2d}, including two DM values respectively obtained by the front-edge alignment and joint fitting, $\tau_{\rm1GHz}$, and $\alpha$. It can be observed that DM values derived from the two approaches are consistent within the 3$\sigma$ error.

\begin{figure}[htbp]
  \centering
  \includegraphics[width=0.9\columnwidth]{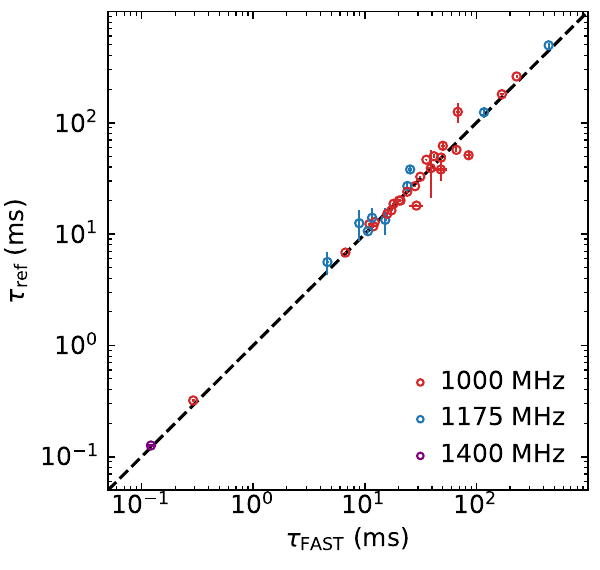}
  \caption{Comparison of pulse-broadening timescales $\tau$ measured by FAST and the values in the literature.}
\label{fig:comp_refs}
\end{figure}

\subsection{Pulsar scattering parameters in literature and comparison with FAST measurements} \label{subsec:comparison}

To investigate the DM and direction dependence of scattering effects and to compare FAST-measured scattering parameters with the literature values, we have to collect all measurements of scattering parameters previously published. We noticed that \citet{hs24} collected scattering times for 473 pulsars from 54 papers. In our full collection, 497 pulsars have the pulse-broadening timescale measured in the reference. We only use measurements of $\tau_{\rm s}$ with a relative uncertainty better than 10 percent. Most of them are the reliable measurements from wide-band observations \citep[][]{okp21} with prominent scattering features in profiles. 

Many pulsars have their scattering parameters measured by 
\citet{bcc04,nab13,zvk13,ldk13,lkk15,lrk15,gkk17,kmn15,kjm17,kmj19,okp21,kzu22} where observations were made at multiple frequencies. Some pulsars at specific Galactic positions also show scattering profiles  \citep{jkl06,dcl09,bts19,kbm19,sjm+19,grf22,kzu22,psf22,sbb+23}. In addition, scattering parameters have been observed for some specific pulsars, such as PSRs J0026+6320 and J2217+5733 \citep{sjm+19}, J0534+2200 (Crab) \citep{kkk02,kll07,bwk07,btk08,oob15,ecc13,esd16,mtb17}, J0540-6919 via giant pulse observations \citep{jr03,gsa21}, J0835-4510 (Vela) \citep{akh70}, J1208-5936 \citep{bbb23}, J1227-6208 \citep{bkc+24}, J1410-6132 \citep{bjk08}, J1550-5418 \citep{crh07}, J1638-4713 \citep{lfd+23}, J1717-4054 \citep{khs14}, J1744-2946 \citep{ldj+24}, J1745-2900 \citep{sle14,ppe15}, J1746-2829 \citep{wcb+23}, J1747-2809 towards the Galactic Center \citep{crg09}, J1751-2737 at the Galactic Center \citep{bdf17}, J1752-2806 \citep{xot19}, J1813-1749 near the Galactic Center \citep{crh21}, J1823-3021A from giant-pulse observations \citep{hbfa25}, J1841-0050 \citep{crc12,wwh20}, J1939+2134 \citep{cwd+90}, J1949+3426 \citep{acd19}, J1959+2048 from giant-burst observations \citep{mkp17}, J2113+4644 \citep{ss10}, and J2205+6012 \citep{dcs22}.

After collecting all the data and discarding some uncertain values, we get the scattering parameters of 243 pulsars.

Among the 149 pulsars we measured, the scattering parameters for 113 pulsars are determined for the first time, including 68 newly discovered pulsars by the FAST GPPS survey. In contrast, the scattering parameters of the other 36 pulsars have been measured previously in the literature, and they are also listed in Table~\ref{tab:compare}.

Based on our measurements of $\tau_s$ in different subbands in Table~A\ref{tab:taus} and the scattering spectral index $\alpha$ in Table~\ref{tab:spar-2d}, we interpolate the pulse-broadening timescale $\tau_{\rm FAST}$ at the frequency specified in the literature where $\tau_{\rm ref}$ was obtained. Then we compare the $\tau_{\rm FAST}$ with the $\tau_{\rm ref}$, and find that they are in good agreement with each other, as shown in Figure~\ref{fig:comp_refs}.

\begin{figure}
  \centering
  \includegraphics[width=1\columnwidth]{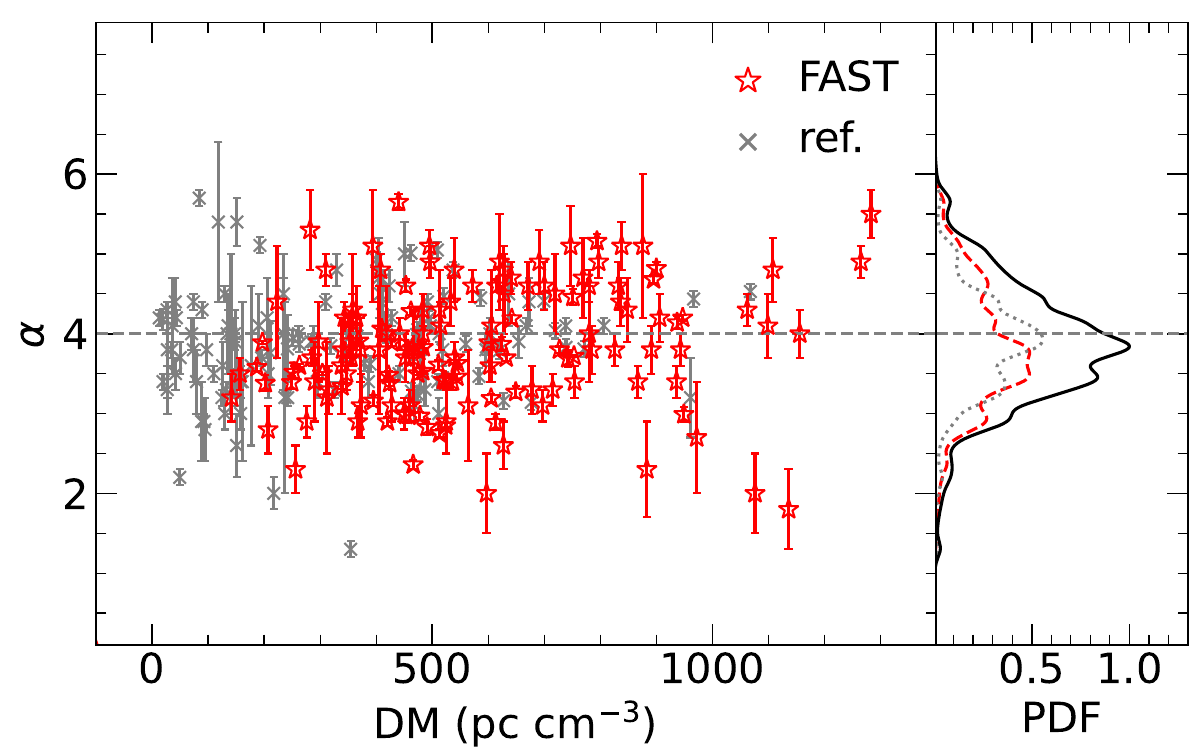}
  \caption{Distribution of scattering spectral indexes $\alpha$ for pulsars measured by FAST and pulsars in literature, peaked around $\alpha=4.0$. The probability distribution function (PDF) in the right subpanel is composited by summing the Gaussian functions defined by the $\alpha$ and uncertainty.
  }
\label{fig:alpha-dm}
\end{figure}

Among 149 pulsars, we find that the scattering spectral indexes for 23 pulsars (including 4 pulsars with the error-bar of $\alpha$ greater than 0.5) have been determined previously, as listed in Table~\ref{tab:compare}. The $\alpha$ value of PSR J1913+11025 obtained by \citet{psf22} is $1.6\pm0.2$, which is very low compared with the others. We noticed that there is a possible weak component on the trailing tail in the profile of \citet{psf22} and also our subband profiles by the FAST. Then, we fit the profiles with an intrinsic profile with two Gaussian components and get $\alpha=2.5\pm0.2$.

As mentioned in the introduction, the scattering spectral indexes could be different for different turbulent mediums, for example, $\alpha=4.0$ for the Gaussian distribution of fluctuations, or $\alpha=4.4$ for the general medium with a Kolmogorov spectrum \citep{cl91}. One can find that the $\alpha$ values in Figure~\ref{fig:alpha-dm} have a wide distribution around 4.0, and are consistent with the results given by e.g. \citet{lrk15} and \citet{okp21}.
Physically, several reasons can cause the deviations of the $\alpha$ values: (1) the inner cutoff of the wavenumber spectrum, reducing $\alpha$ as low as 2 when the scattering angle is less than a critical cross-over angle \citep{cl03,lkk15}; (2) a low $\alpha$ down to 2.6 in the highly supersonic turbulence with a short-wave-dominated density spectrum \citep{xz17}; (3) a non-Kolmogorov form for the density spectrum \citep{bcc04}; (4) a finite size of screens such as filamentary structures or clumps \citep{cl01}; (5) the anisotropic scattering media \citep{bmg10,gk16}.

\subsection{\texorpdfstring{Dependence of $\tau_{\rm s}$ with DM}{Dependence of tau with DM}}

\begin{figure}
  \centering
  \includegraphics[width=0.98\columnwidth]{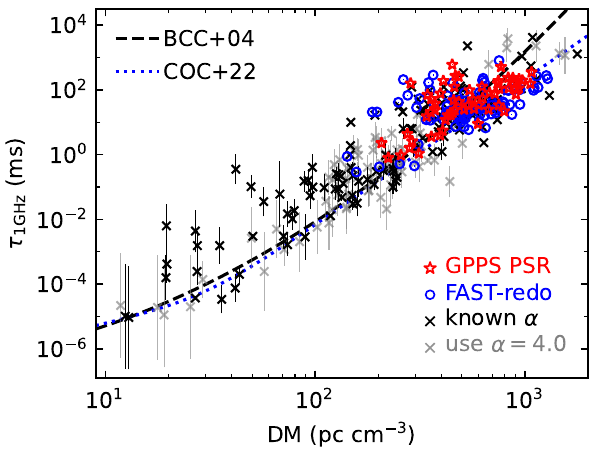}
  \caption{The distribution of the pulse-broadening timescale $\tau_{\rm s}$ at 1~GHz versus pulsar DM. The values of newly discovered pulsars in the GPPS survey and the known pulsars re-detected by the FAST are specially marked by stars and circles. The other values are extrapolated from the measurements at other frequencies either by using the previously estimated $\alpha$ or simply assuming $\alpha=4.0$. Two empirical relations given by \citet{bcc04} and \citet{coc22} are outlined by a dashed line and a dotted line.}
\label{fig:tau-dm}
\end{figure}

\begin{figure}
  \centering
  \includegraphics[width=1\columnwidth]{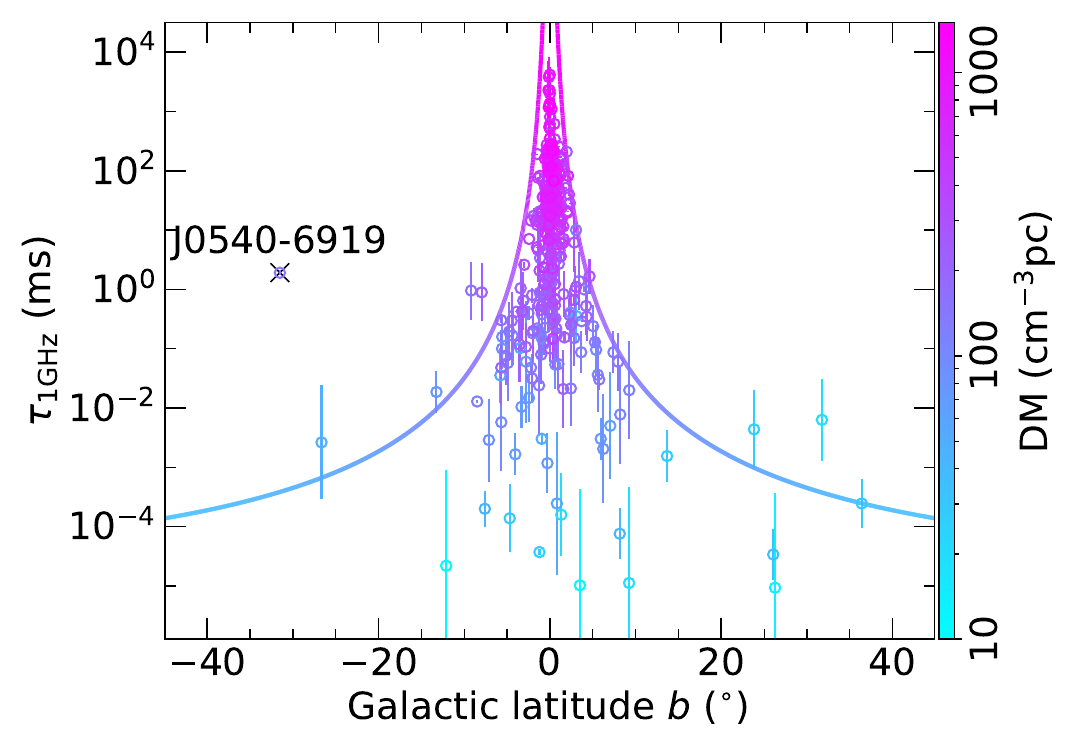}
  \caption{Pulse-broadening timescale  $\tau_{\rm1GHz}$ of pulsars at various Galactic latitudes, together with the upper limit curve calculated by the DM upper limit.}
\label{fig:tau-gb}
\end{figure}

\begin{figure*}
  \centering
  \includegraphics[width=0.7\textwidth]{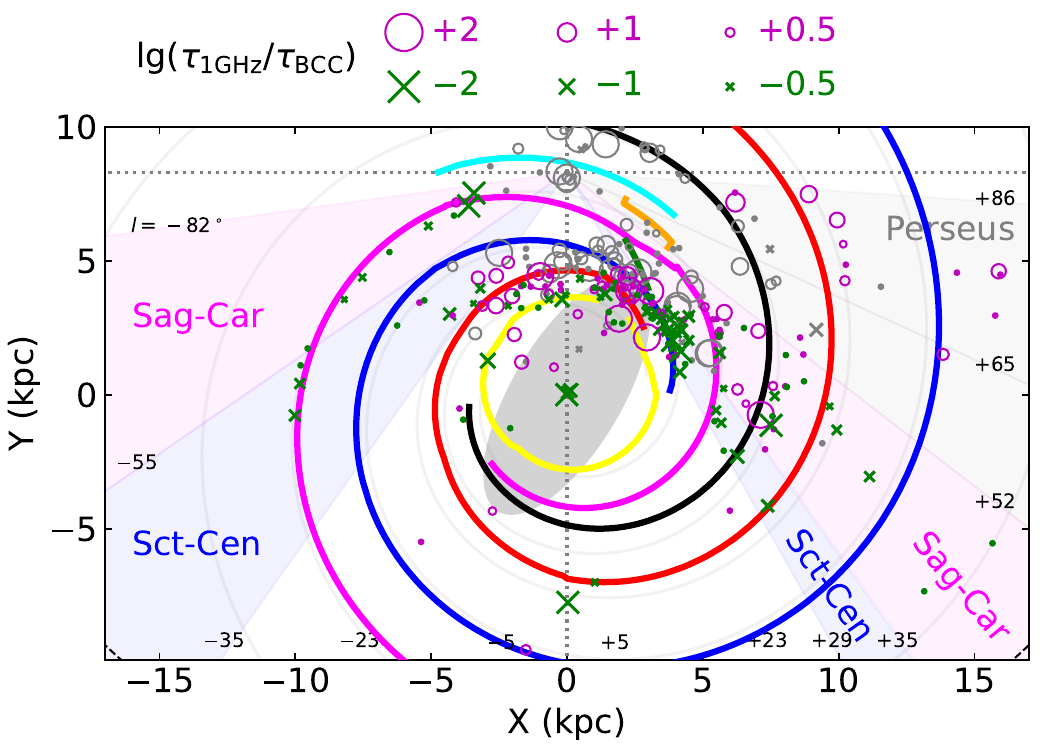}
  \caption{Distribution of scattered pulsars with $|b|<5^\circ$ in various spiral arm regions outlined by \citet{hh14} and \citet{rmb+19}. Crosses and circles indicate the relative $\tau_{\rm1GHz}$ deviation from the empirical prediction given by Equation \ref{eq:tau2dm}. Pulsars with $\rm DM<300~pc~cm^{-3}$ are plotted in gray. The Galactic disk is divided into and labeled with dominated spiral arms, namely the Perseus arm, the Sagittarius-Carina (Sag-Car) arm, and the Scutum-Centaurus (Scu-Cen) arm. 
  }
\label{fig:tau-plane}
\end{figure*}

\begin{figure*}
  \centering
\includegraphics[width=0.8\textwidth]{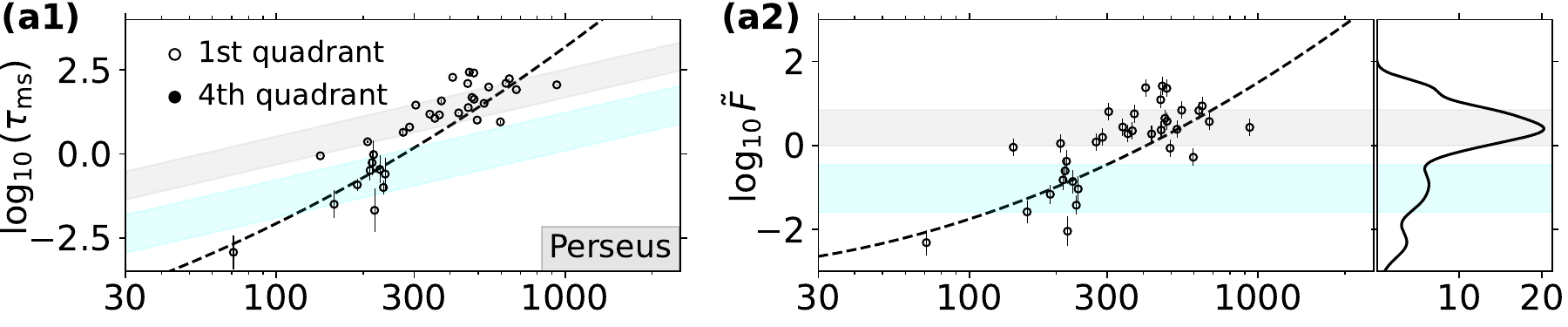}
\includegraphics[width=0.8\textwidth]{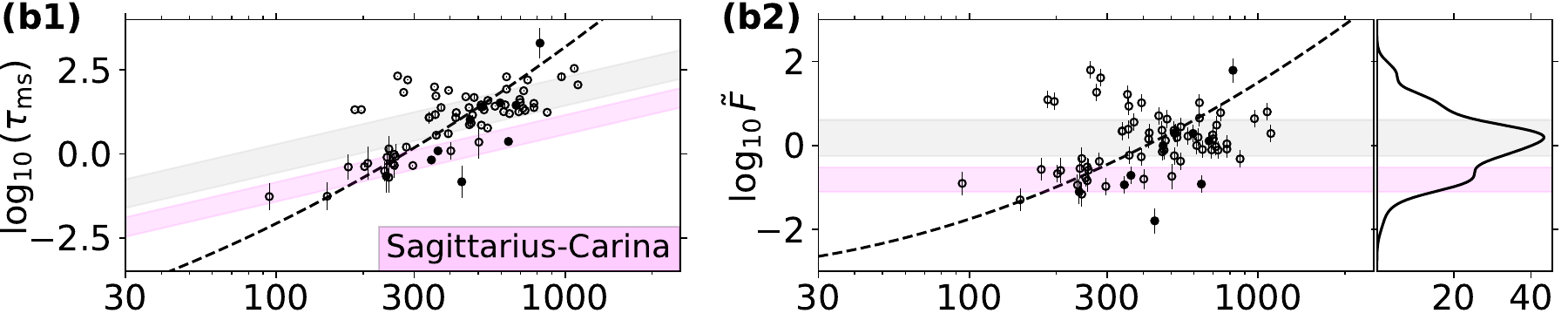}
\includegraphics[width=0.8\textwidth]{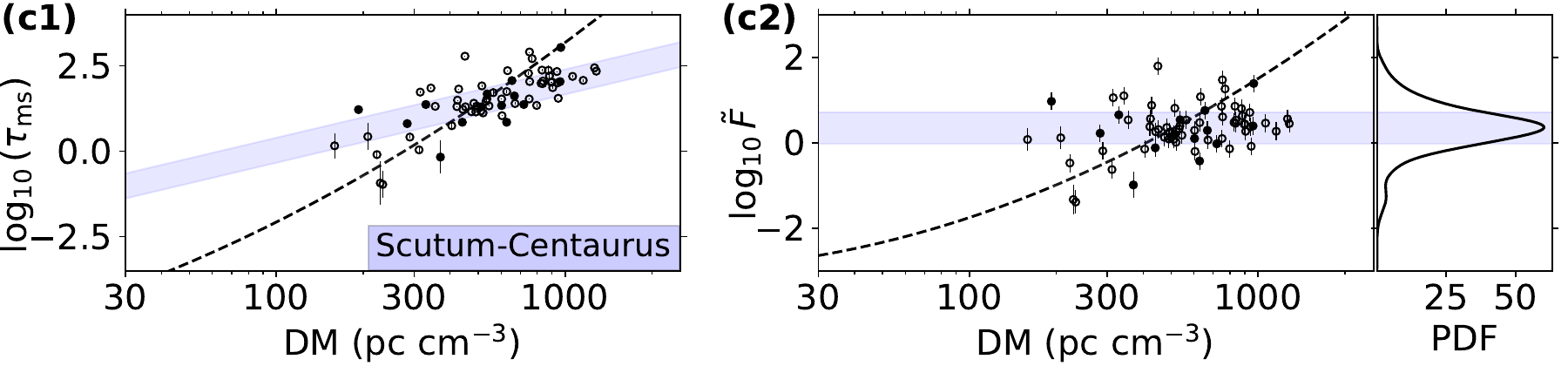}
  \caption{The pulse-broadening timescales $\tau_{\rm1GHz}$ and the fluctuation parameter $\tilde{F}$ for scattered pulsars in different regions dominated by different spiral arms as marked in Figure~\ref{fig:tau-plane}. The dashed line is drawn according to the empirical $\tau_{\rm s}$-DM formula in Equation \ref{eq:tau2dm}. The probability distribution function (PDF) of $\tilde{F}$ for pulsars in the highlighted area is expressed with the Gaussian functions in the left sub-panels. The open and filled circles label pulsars in the first and fourth Galactic quadrants.}
\label{fig:fluc}
\end{figure*}

For analyzing the DM and direction dependence of scattering effect, all measured pulse-broadening timescales $\tau_{\rm s}$ are scaled to $\tau_{\rm 1GHz}$ by using the estimated scattering spectral index $\alpha$. If no observed $\alpha$ is available, we take $\alpha = 4.0$ for scaling $\tau_{\rm s}$. The accuracy of such extrapolated pulse-broadening timescales rests on the uncertainty of $\alpha$ \citep{sdo80}, and we take $\Delta \alpha=0.6$ that is the standard deviation of $\alpha$ given by \citet{okp21}.

The pulse-broadening timescales $\tau_{\rm1GHz}$ are related to the dispersion measure by an empirical formula expressed in Eq.(\ref{eq:tau2dm}) \citep{bcc04,coc22}, as shown in Figure~\ref{fig:tau-dm}. Our new FAST measurements contribute new data at the high DMs, and are consistent with the trend.

\subsection{Dependence of $\tau_{\rm s}$ on the spatial distribution of the interstellar medium}

The scattering effect is caused by ionized gas clouds between a pulsar and us. Therefore the pulse-broadening timescale  $\tau_{\rm1GHz}$ and the scattering spectral index $\alpha$ depend on the foreground clouds \citep[e.g.][]{lkk15}. 
Some low-DM pulsars show varying scattering effects \citep{bts19} because they are behind HII regions or supernova remnants \citep{kmj19}. Here we investigate the dependence of $\tau_{\rm s}$ on the spatial distribution of the interstellar medium. 

First of all, the scattering effect varies with the Galactic latitude \citep{cad84,occ21,kpp22,hs24}. The upper limit of the pulse-broadening timescale, as shown in Figure \ref{fig:tau-gb}, can be derived by combining the $\tau_{\rm s}$-DM relation (Equation \ref{eq:tau2dm}) with the upper limit of the Galactic DM of ${\rm DM}_{\rm max} = (23.5\pm 2.5) {\rm~pc~cm^{-3}} / \sin(b)$ \citep{occ20}.
Notably, the large pulse-broadening timescale of PSR J0540$-$6919 should be attributed to the medium in the Large Magellanic Cloud \citep{cws16}.

\begin{figure}
\centering
\includegraphics[width=0.4\textwidth]{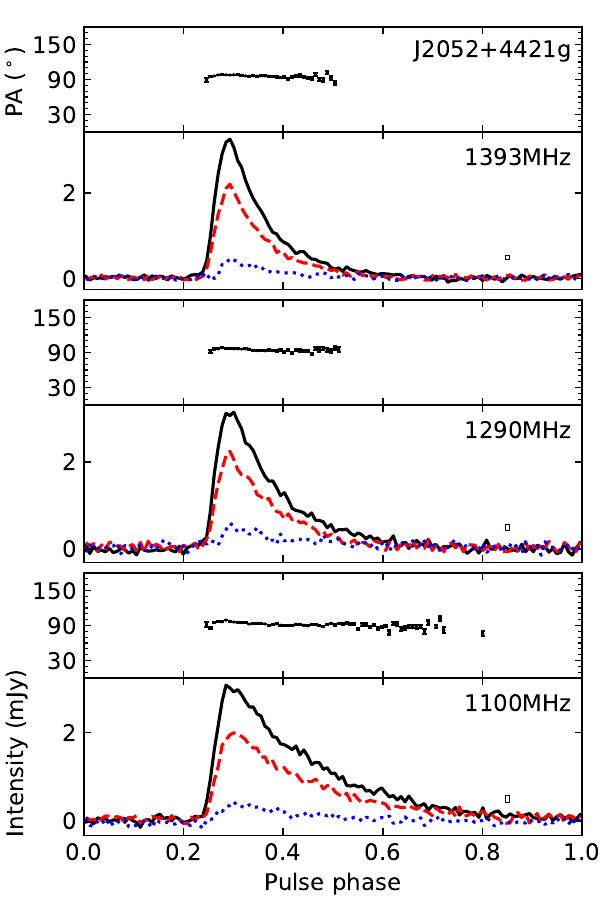}
\caption{Polarization profiles of PSR J2052+4421g at three subbands of FAST observations. For each subband, the total intensity profile is drawn by the solid line, the linear polarization profile by the dashed line, and the circular polarization profile by the dotted line in the lower sub-panel. The box indicates the phase bin size and $\pm1\sigma$. The polarization position angles (PAs) with error bars are shown in the top sub-panel. The polarization profiles of another 40 well-calibrated pulsars are presented in Fig.~A\ref{fig:pol}.
}
\label{fig:ex_gpps0019pol}
\end{figure}

Secondly, one may find that pulsars behind different spiral arms or HII regions show different dependence of $\tau_{\rm s}$ with DM. 
By using the scattering parameters measured in this paper and compiled data from literature, we aim to explore the dependence of $\tau_{\rm s}$ on different density fluctuations in different spiral arms \citep{cr98,occ22} 
%
\begin{equation}
    \tilde{F} = (\tau_{\rm 1GHz} / {\rm\,0.48~ms}) (DM/100\,{\rm~cm^{-3}pc)}^{-2}
\end{equation}
in different Galactic disk regions as divided in Figure \ref{fig:tau-plane}. 
Here $\tilde{F}$ is a fluctuation parameter, characterizing density fluctuations in turbulence \citep{cws16,occ21}.
Our approach is similar to the work by \citet{abs86} and \citet{ps08} and the investigations of the Sun-local region, the inner disk, and the outer disk conducted by \citet{kmn15} and \citet{hs24}. 
The Galactic disk region is divided into different sections based on the Galactic longitudes \citep[e.g.][]{pbk+26}. Specifically, regions dominated by the Perseus arm, the Sagittarius-Carina (Sag-Car) arm, and the Scutum-Centaurus (Scu-Cen) arm are separated, with each arm corresponding to certain ranges of Galactic longitudes. 

The pulse-broadening timescales $\tau_{\rm1GHz}$ and fluctuation parameter $\tilde{F}$ for pulsars in these radial sectors are plotted in Figure~\ref{fig:fluc}(a1) to (c1) and (a2) to (c2). The fluctuation parameter $\tilde{F}$ of different arm regions is shown in Figure~\ref{fig:fluc}(a2) to (c2).
It is clearly shown in the figure that the data are clustered for different DM values and different values of $\tilde{F}$, which implies that the interstellar medium in different spiral arms has different fluctuation properties. Their probability density functions (PDF) of fluctuation parameter $\tilde{F}$ for scattered pulsars can be expressed by multiple Gaussian functions, for example, for these pulsars in the highlighted regions corresponding to spiral arms at different distances. In the Perseus Arm region there are two groups of pulsars; In the Sagittarius-Carina region there are also two groups of pulsars with different DMs and fluctuation parameters; In contrast, for Scutum-Centaurus region, only one pulsar group shows the scattering effect. 

The grouping of pulsars with high fluctuations and high DMs implies that they are behind a given identifiable spiral arm. Therefore, the distances of these pulsars can be constrained as proposed in 
\citet{fc89,cr98,lcb04}. For example, the two fluctuation groups in the Perseus arm region (see Figure~\ref{fig:fluc} a1-a2) should correspond to the pulsars in or behind the local arm or the pulsars in or behind the Perseus arm.

\subsection{Polarization profiles}

Using the DMs from the joint fitting and calibrating the FAST data, we can get the subband polarization profiles for the pulsars with strong scattering features. Polarimetric calibration of the pulse profiles has been conducted by using \textsc{psrchive} \citep{vanStraten04} following the detailed procedures given in  \citet{whx+23} \citep{vanStraten04}. 
The Faraday rotation measure (RM) is obtained from the {\it RMFIT} in \textsc{psrchive} software routine. After Faraday rotation is fitted and corrected for the channel data, we get the final observed rotation measures (RM) and polarization profiles from the observation data. 

Figure~\ref{fig:ex_gpps0019pol} is one example, showing the polarization profiles of PSR J2052+4421g at three subbands of FAST observations. The profiles are more extended at lower frequencies. Polarization angle curves flatten in the scattered profile tails, as discussed by \citet{lh03}. The subband polarization profiles for another 40 pulsars observed by FAST are shown in Figure A\ref{fig:pol}, and their polarization parameters are listed in Table A\ref{tab:pol}. 

\section{Conclusions} \label{sec:summary}

We present the pulse-broadening timescales and scattering spectral indexes for 149 pulsars derived from FAST observations at the L-band of 1.0 -- 1.5 GHz. The measurements of 113 pulsars are obtained for the first time. For 36 pulsars with the scattering parameters reported in the literature, our measurements are generally consistent with previously reported values, though there are a few exceptions.

We find that aligning the front edge of the subband profiles at the half or quarter of the profile peak can yield better DM values for most pulsars than those obtainable using {\it PDMP}. However, the best DM value can be determined by aligning the main peak of the modeled intrinsic profiles obtained from joint fitting the scattered subband profiles. 

We find that in the regions with spiral arms, the pulsars could be grouped by their DM values and the relative fluctuations, which indicate the density irregularities in the spiral arm regions. On the other hand, the grouping of pulsars can probably be a rough indication of pulsar locations. 

From the FAST observations, the polarized profiles have been obtained for 82 pulsars with scattered profiles at three subbands, which constantly show the flattened polarization angle curves in the scattered profile tails, as suggested by \citet{lh03}.

\normalem

\section*{Acknowledgements}  
We thank the referee, Prof. Ramesh Bhat, for very careful reading and helpful suggestions. The authors are supported by the Natural Science Foundation of China: No. 11988101 and 11833009, the National SKA Program of China (2020SKA0120100), the National SKA Program of China (grant No. 2022SKA0120103) and the Chinese Academy of Sciences via the project JZHKYPT-2021-06.
FAST is a Chinese national mega-science facility, built and operated by the National Astronomical Observatories, Chinese Academy of Sciences. We all appreciate the excellent performance of the FAST and the operation team. 

\bibliographystyle{aasjournal}
\bibliography{ms2024-0320}

\section*{Appendix}

\captionsetup[table]{name={\bf Table A}}
\captionsetup[figure]{name={\bf Figure A}}

\setcounter{figure}{0}
\setcounter{table}{0}

This appendix presents big databases, including two long tables and figures.

The FAST observations were carried out in the band of 1.0-1.5 GHz, with various RFI in different frequency channels for different pulsars at different times. Considering these various RFI situations, we measured the pulse-broadening timescales at these sub-bands, and published them directly in Table~A\ref{tab:taus}.

\begin{figure*}
\centering
\includegraphics[width=0.24\textwidth,height=0.24\textwidth]{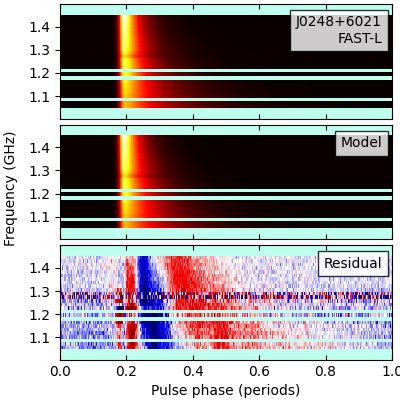}
\includegraphics[width=0.24\textwidth,height=0.24\textwidth]{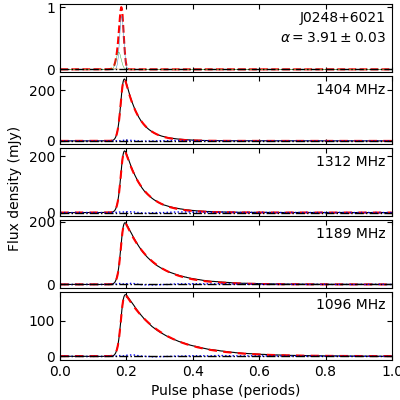}
\includegraphics[width=0.24\textwidth,height=0.24\textwidth]{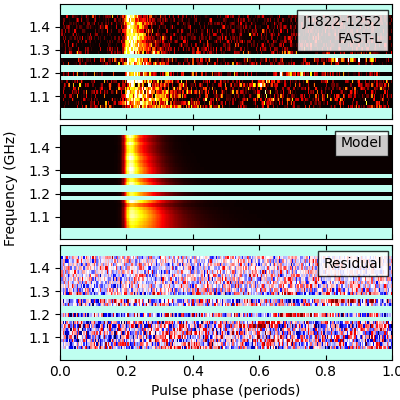}
\includegraphics[width=0.24\textwidth,height=0.24\textwidth]{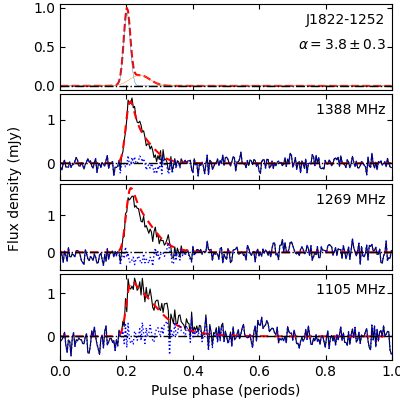}
\includegraphics[width=0.24\textwidth,height=0.24\textwidth]{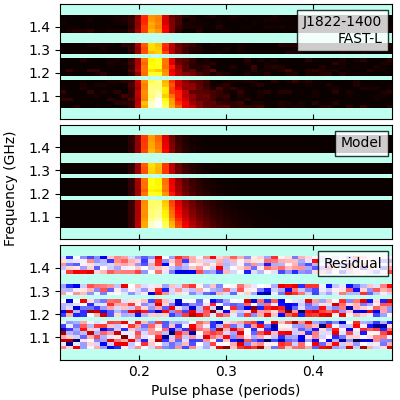}
\includegraphics[width=0.24\textwidth,height=0.24\textwidth]{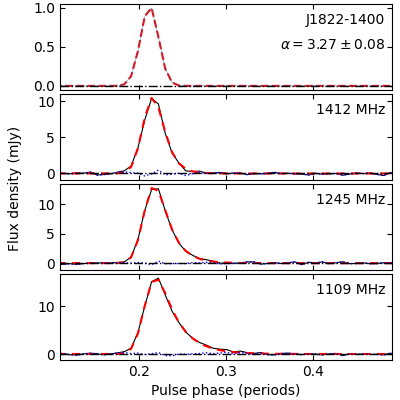}
\includegraphics[width=0.24\textwidth,height=0.24\textwidth]{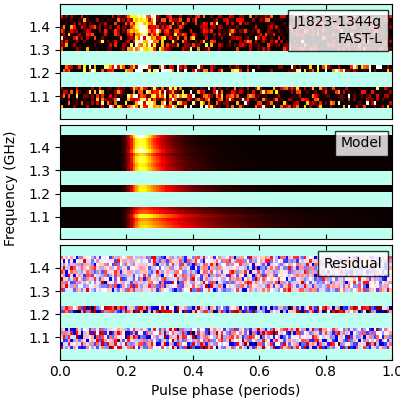}
\includegraphics[width=0.24\textwidth,height=0.24\textwidth]{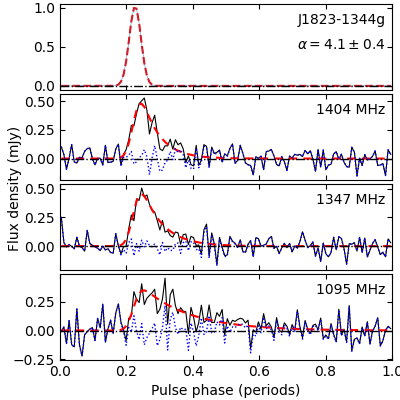}
\includegraphics[width=0.24\textwidth,height=0.24\textwidth]{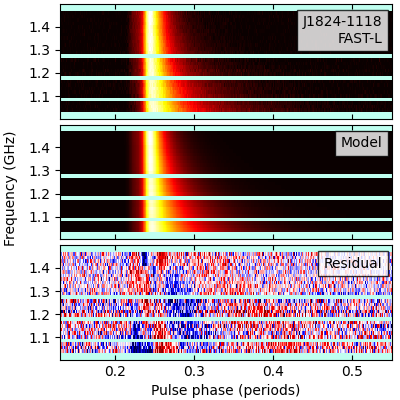}
\includegraphics[width=0.24\textwidth,height=0.24\textwidth]{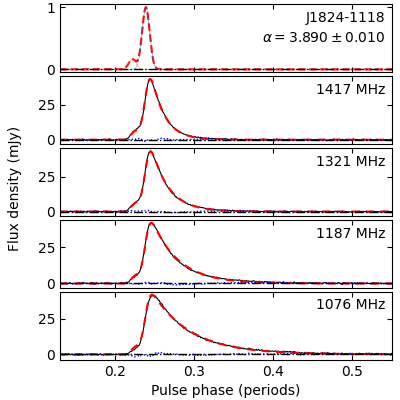}
\includegraphics[width=0.24\textwidth,height=0.24\textwidth]{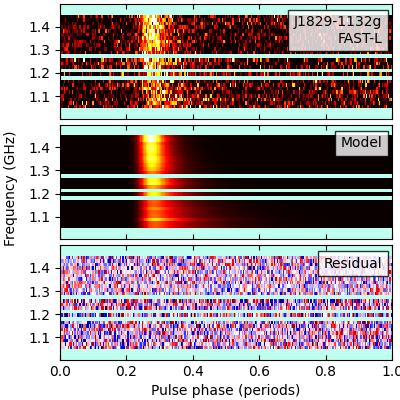}
\includegraphics[width=0.24\textwidth,height=0.24\textwidth]{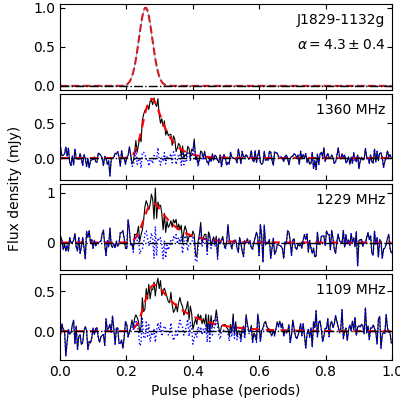}
\includegraphics[width=0.24\textwidth,height=0.24\textwidth]{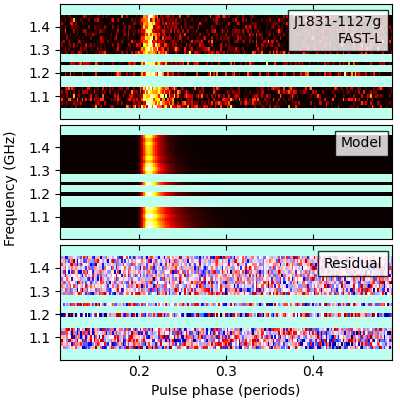}
\includegraphics[width=0.24\textwidth,height=0.24\textwidth]{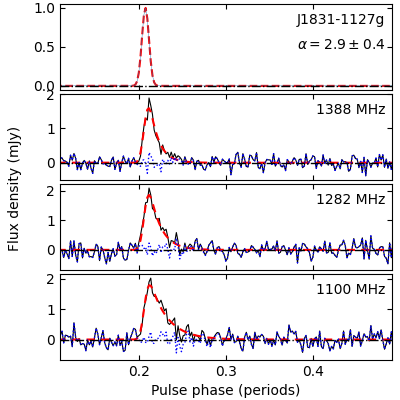}
\includegraphics[width=0.24\textwidth,height=0.24\textwidth]{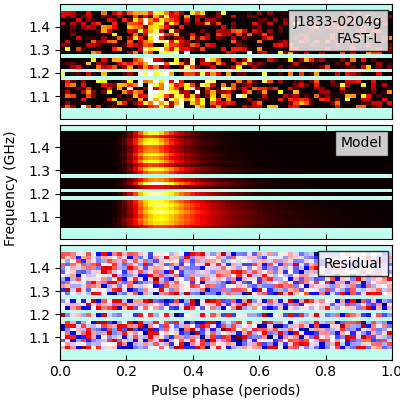}
\includegraphics[width=0.24\textwidth,height=0.24\textwidth]{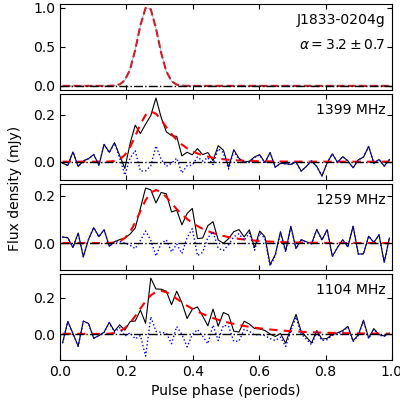}
\includegraphics[width=0.24\textwidth,height=0.24\textwidth]{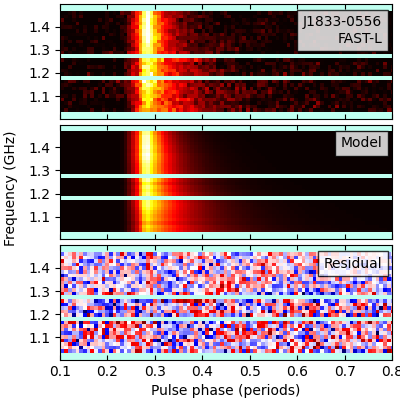}
\includegraphics[width=0.24\textwidth,height=0.24\textwidth]{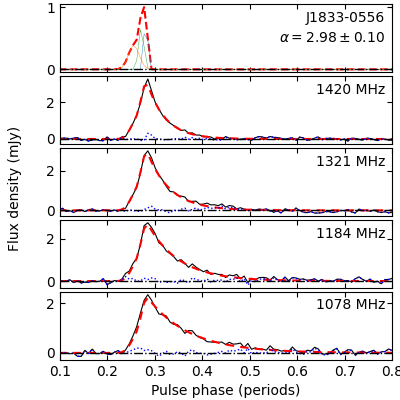}
\includegraphics[width=0.24\textwidth,height=0.24\textwidth]{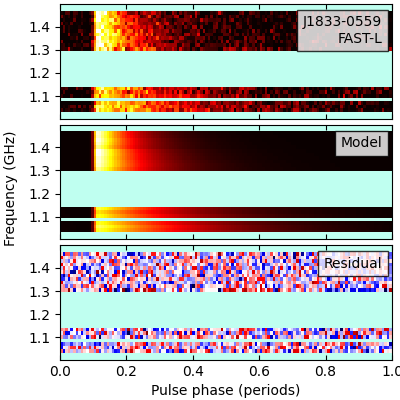}
\includegraphics[width=0.24\textwidth,height=0.24\textwidth]{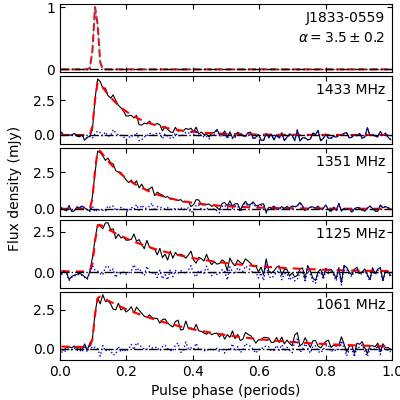}
\caption{The observed and modeled scattering profiles for 147 pulsars. Comparing the phase-time waterfall 2D intensity plot for the 32 channels in the FAST observation band (left top) with the modeled scattered profiles (left middle), one can get the residual plot (left bottom). The intrinsic profiles and the modeled subband profiles are given on the right.}
\label{fig:2d-all}
\end{figure*}

Figure~A\ref{fig:2d-all} presents the scattering profiles in 3-5 subbands for 118 pulsars which have been well-fitted in the joint model-fitting to subband profiles, except for PSR J1844-0310 and J1850-0026, which probably have frequency-evolution for intrinsic profile components and are presented in Figure~A\ref{fig:2psr-bad}.

The polarized subband profiles for 82 scattered pulsars are presented in Figure~A\ref{fig:pol}. The polarization parameters are given in Table~A\ref{tab:pol}.

\begin{table*}
    \centering
    \caption{Pulse-broadening timescales $\tau_{\rm s}$ measured for subband profiles from FAST observations. The FAST observation band is divided into 3 -- 5 subbands according to the RFI situation so that the central frequencies of subbands differ from pulsar to pulsar and are given in the table.}
    \label{tab:taus}
\footnotesize
\renewcommand{\arraystretch}{0.85}  
    \begin{tabular}{lrrrrr}
        \hline\noalign{\smallskip}
\multicolumn{1}{c}{PSR Name}&\multicolumn{1}{c}{$\tau_{\rm s}$\at$\nu_1$ }&\multicolumn{1}{c}{$\tau_{\rm s}$\at$\nu_2$}&\multicolumn{1}{c}{$\tau_{\rm s}$\at$\nu_3$}&\multicolumn{1}{c}{$\tau_{\rm s}$\at$\nu_4$}&\multicolumn{1}{c}{$\tau_{\rm s}$\at$\nu_5$}\\
\multicolumn{1}{c}{ }&\multicolumn{1}{c}{(ms{\at}MHz)}&\multicolumn{1}{c}{(ms{\at}MHz)}&\multicolumn{1}{c}{(ms{\at}MHz)}&\multicolumn{1}{c}{(ms{\at}MHz)}&\multicolumn{1}{c}{(ms{\at}MHz)}\\
\hline\noalign{\smallskip}
J0248$+$6021 & 23.45\PM0.01\at1096& 17.87\PM0.01\at1178& 11.34\PM0.00\at1320&  9.12\PM0.00\at1404&                    \\ 
J1822$-$1252 &   228\PM  19\at1106&   107\PM   6\at1269&    72\PM   9\at1380&                    &                    \\ 
J1822$-$1400 &  4.94\PM0.08\at1111&   3.2\PM 0.1\at1245&   2.0\PM 0.1\at1402&                    &                    \\ 
J1823$-$1344g&84$_{-19}^{+28}$\at1095&    25\PM   4\at1351&    23\PM   4\at1404&                    &                 \\ 
J1824$-$1118 & 20.90\PM0.06\at1076& 14.30\PM0.04\at1186&  9.48\PM0.02\at1321&  7.47\PM0.03\at1417&                    \\ 
J1829$-$1132g&    22\PM   2\at1111&    14\PM   1\at1229&    10\PM   0\at1358&                    &                    \\ 
J1831$-$1127g&    30\PM   2\at1102&    25\PM   2\at1273&    15\PM   1\at1381&                    &                    \\ 
J1833$-$0204g&   0.5\PM 0.1\at1105&   0.4\PM 0.1\at1259&   0.2\PM 0.1\at1399&                    &                    \\
J1833$-$0556 &   139\PM   4\at1078&   102\PM   3\at1183&    73\PM   2\at1322&    56\PM   3\at1420&                    \\ 
J1833$-$0559 &   160\PM  10\at1061&   126\PM  10\at1125&    61\PM   1\at1351&    47\PM   2\at1433&                    \\ 
J1834$-$0602 &  14.2\PM 0.3\at1094&  11.1\PM 0.2\at1175&   5.3\PM 0.1\at1326&     6\PM   1\at1404&                    \\ 
J1834$-$0812 &   347\PM  43\at1328&   230\PM  16\at1374&   245\PM  15\at1420&                    &                    \\ 
J1837$-$0604 &    38\PM   3\at1096&  22.1\PM 0.8\at1187&  13.7\PM 0.3\at1314&  10.2\PM 0.2\at1404&                    \\ 
J1838$-$0453 &  16.7\PM 0.4\at1095&  11.8\PM 0.5\at1171&   7.5\PM 0.1\at1327&   6.1\PM 0.2\at1404&                    \\ 
J1838$-$0508g&    31\PM   4\at1107&    21\PM   2\at1262&    12\PM   1\at1380&                    &                    \\ 
J1840$-$0445 &475$_{-192}^{+717}$\at1108&   113\PM  10\at1267&    81\PM   5\at1380&                    &              \\ 
J1840$-$0643 &  11.3\PM 0.4\at1328&   9.4\PM 0.3\at1375&   8.6\PM 0.3\at1420&                    &                    \\ 
J1841$-$0157 & 10.29\PM0.09\at1085&   7.4\PM 0.1\at1182&  4.83\PM0.08\at1337&   4.0\PM 0.1\at1420&                    \\
J1841$-$0353g&   105\PM   6\at1084&    76\PM   4\at1177&    47\PM   2\at1325&    31\PM   1\at1411&                    \\ 
J1842$-$0153 &  23.7\PM 0.2\at1092&  17.8\PM 0.1\at1183& 11.70\PM0.09\at1319&   8.5\PM 0.1\at1420&                    \\
J1842$-$0258g&  17.2\PM 0.7\at1085&  12.1\PM 0.5\at1190&   7.6\PM 0.3\at1319&   4.6\PM 0.4\at1421&                    \\
J1842$-$0309 &    53\PM   3\at1084&    45\PM   2\at1187&    31\PM   1\at1320&    21\PM   0\at1421&                    \\
J1843$-$0050 &  15.0\PM 0.3\at1072&  12.7\PM 0.3\at1121&   6.9\PM 0.2\at1351&   5.5\PM 0.3\at1432&                    \\
J1843$-$0137 &  20.2\PM 0.3\at1084&  14.5\PM 0.3\at1174&   9.1\PM 0.1\at1329&   6.9\PM 0.1\at1418&                    \\
J1843$-$0157g&    85\PM   7\at1093&    62\PM   5\at1174&    44\PM   1\at1323&    32\PM   1\at1420&                    \\
J1843$-$0310g&   893\PM 409\at1102&    82\PM  12\at1258&    47\PM   3\at1389&                    &                    \\ 
J1843$-$0355 &22$_{-2}^{+3}$\at1108&  11.8\PM 0.6\at1261&   8.3\PM 0.2\at1380&                    &                   \\ 
J1844$-$0030 &   8.5\PM 0.2\at1086&   6.5\PM 0.2\at1176&   4.3\PM 0.2\at1334&   3.6\PM 0.2\at1420&                    \\
J1844$-$0136g&    46\PM   3\at1085&    33\PM   2\at1187&    22\PM   1\at1317&    14\PM   1\at1421&                    \\
J1844$-$0142g&  16.3\PM 0.8\at1085&    13\PM   0\at1172&   7.8\PM 0.4\at1323&   5.6\PM 0.4\at1419&                    \\
J1844$-$0202g&    33\PM   5\at1095&    26\PM   4\at1288&    12\PM   1\at1398&                    &                    \\
J1844$-$0240 &    46\PM   7\at1096&    35\PM   3\at1331&    19\PM   2\at1421&                    &                    \\
J1844$-$0244 &  14.3\PM 0.3\at1087&  11.0\PM 0.3\at1180&   8.1\PM 0.2\at1321&   6.6\PM 0.2\at1420&                    \\
J1844$-$0256 &    67\PM  12\at1085&    51\PM   5\at1179&    35\PM   2\at1326&    23\PM   1\at1418&                    \\
J1844$-$0302 &  17.4\PM 0.8\at1077&  11.2\PM 0.8\at1188&   7.6\PM 0.5\at1328&   5.9\PM 0.5\at1420&                    \\
J1844$-$0310 &    52\PM   2\at1056&    48\PM   2\at1119&    29\PM   1\at1237&  23.8\PM 0.6\at1328&  16.8\PM 0.4\at1418\\
J1844$-$0538 &   7.9\PM 0.1\at1095&   6.1\PM 0.1\at1174&  3.87\PM0.09\at1320&   3.0\PM 0.1\at1404&                    \\ 
J1845$-$0103g&    25\PM   2\at1085&    17\PM   1\at1179&  10.0\PM 0.5\at1323&   7.1\PM 0.4\at1420&                    \\
J1845$-$0142g&   307\PM 194\at1077&    34\PM   4\at1187&    26\PM   2\at1318&    16\PM   1\at1421&                    \\
J1845$-$0144g&   158\PM   5\at1086&   105\PM   4\at1172&    66\PM   1\at1326&    53\PM   1\at1420&                    \\
J1845$-$0229Ag&   118\PM   9\at1094&    85\PM   3\at1334&    35\PM   1\at1420&                    &                    \\
J1845$-$0229Bg&    61\PM   8\at1278&    68\PM  13\at1324&    43\PM   4\at1375&                    &                    \\
J1845$-$0243g&284$_{-119}^{+484}$\at1106&    69\PM  10\at1258&    45\PM   4\at1400&                    &                    \\
J1845$-$0254g&168$_{-33}^{+52}$\at1342&   153\PM  29\at1385&   184\PM  36\at1430&                    &                    \\
J1845$-$0316 &  15.7\PM 0.7\at1077&  10.4\PM 0.5\at1183&   6.6\PM 0.4\at1326&   4.6\PM 0.3\at1420&                    \\
J1846$-$0211g&    84\PM   7\at1086&    43\PM   3\at1191&    25\PM   2\at1319&    20\PM   2\at1421&                    \\
J1846$-$0513 &  1.18\PM0.05\at1110&  0.70\PM0.04\at1266&  0.47\PM0.03\at1391&                    &                    \\ 
J1848$-$0055 &109$_{-22}^{+37}$\at1086&    69\PM   9\at1171&    37\PM   2\at1317&    32\PM   1\at1420&                    \\
J1849$-$0013 &    55\PM   1\at1085&  38.0\PM 0.8\at1181&  27.0\PM 0.4\at1325&  21.5\PM 0.3\at1420&                    \\
J1849$-$0040 &   152\PM  21\at1083&    99\PM  10\at1198&    72\PM   4\at1320&    51\PM   1\at1420&                    \\
J1849$-$0200g&   962\PM 498\at1086&245$_{-65}^{+177}$\at1174&   117\PM   9\at1331&    87\PM   4\at1421&                    \\
J1850$-$0002 &    31\PM   1\at1077&    24\PM   1\at1177&  13.8\PM 0.8\at1330&  11.5\PM 0.8\at1419&                    \\
J1850$-$0006 &   176\PM   3\at1085&   132\PM   2\at1180&    86\PM   1\at1315&    61\PM   1\at1420&                    \\
J1850$-$0020 &    23\PM   0\at1085&    14\PM   0\at1183&  10.4\PM 0.9\at1324&   9.7\PM 0.9\at1420&                    \\
J1850$-$0026 &  27.3\PM 0.4\at1072&  21.6\PM 0.2\at1122&  14.1\PM 0.1\at1239& 10.81\PM0.08\at1332&  8.15\PM0.06\at1421\\
J1850$-$0031 &    67\PM   1\at1084&  50.2\PM 0.9\at1184&  32.5\PM 0.4\at1323&  22.7\PM 0.3\at1421&                    \\
J1850$-$0050g&    73\PM   9\at1241&    54\PM   2\at1316&    43\PM   1\at1370&    36\PM   1\at1432&                    \\
J1850$+$0242 &    14\PM   5\at1347&    12\PM   6\at1377&7$_{-3}^{+9}$\at1408&                    &                    \\
J1851$-$0029 &   9.7\PM 0.1\at1085&   7.1\PM 0.1\at1186&   4.6\PM 0.1\at1330&  3.86\PM0.08\at1420&                    \\
\multicolumn{6}{r}{ ... to be continued. } \\
\hline
\end{tabular}
\end{table*}

\addtocounter{table}{-1}
\begin{table*}
    \centering
    \caption{-- {\it continued}}
\footnotesize
\renewcommand{\arraystretch}{0.85}  
    \begin{tabular}{lrrrrr}
        \hline\noalign{\smallskip}
\multicolumn{1}{c}{PSR Name}&\multicolumn{1}{c}{$\tau_{\rm s}$\at$\nu_1$ }&\multicolumn{1}{c}{$\tau_{\rm s}$\at$\nu_2$}&\multicolumn{1}{c}{$\tau_{\rm s}$\at$\nu_3$}&\multicolumn{1}{c}{$\tau_{\rm s}$\at$\nu_4$}&\multicolumn{1}{c}{$\tau_{\rm s}$\at$\nu_5$}\\
\multicolumn{1}{c}{ }&\multicolumn{1}{c}{(ms{\at}MHz)}&\multicolumn{1}{c}{(ms{\at}MHz)}&\multicolumn{1}{c}{(ms{\at}MHz)}&\multicolumn{1}{c}{(ms{\at}MHz)}&\multicolumn{1}{c}{(ms{\at}MHz)}\\
\hline\noalign{\smallskip}
J1851$-$0108g&    15\PM   1\at1086&     9\PM   1\at1140&   6.4\PM 0.4\at1334&                    &                    \\
J1851$-$0241 &    80\PM  10\at1078&    50\PM   5\at1186&    29\PM   2\at1318&    28\PM   1\at1420&                    \\
J1851$+$0233 &198$_{-88}^{+203}$\at1095&    29\PM   5\at1257&    21\PM   2\at1401&                    &                    \\
J1852$-$0127 &    49\PM   1\at1072&    43\PM   1\at1120&    25\PM   0\at1273&  19.7\PM 0.3\at1349&                    \\
J1852$+$0013 &  16.6\PM 0.2\at1085&  12.4\PM 0.2\at1182&   9.1\PM 0.1\at1322&   7.9\PM 0.1\at1421&                    \\
J1852$+$0018 &  14.4\PM 0.7\at1086&  11.8\PM 0.6\at1184&   6.7\PM 0.4\at1321&   5.5\PM 0.3\at1420&                    \\
J1852$+$0031 &   635\PM   2\at1085&   462\PM   1\at1175& 286.4\PM 0.5\at1324& 215.8\PM 0.5\at1420&                    \\
J1852$+$0056g&    74\PM   3\at1086&    49\PM   1\at1183&  31.8\PM 0.8\at1321&  22.3\PM 0.7\at1420&                    \\
J1852$+$0309g&2$_{-1}^{+5}$\at1117&   1.1\PM 0.2\at1223&   0.9\PM 0.1\at1356&                    &                    \\
J1853$-$0049g&4$_{-1}^{+3}$\at1092&   2.6\PM 0.4\at1247&   1.8\PM 0.2\at1360&                    &                    \\
J1853$-$0054g&    11\PM   0\at1086&   9.0\PM 0.6\at1177&   6.2\PM 0.4\at1325&   3.3\PM 0.3\at1420&                    \\
J1853$+$0237g&    54\PM   7\at1099&    25\PM   2\at1274&    18\PM   1\at1399&                    &                    \\
J1853$+$0505 &   145\PM   0\at1088& 120.5\PM 0.8\at1169&  76.6\PM 0.5\at1319&  56.3\PM 0.5\at1420&                    \\
J1853$+$0545 &  14.7\PM 0.1\at1087& 11.62\PM0.08\at1157&  7.13\PM0.03\at1319&  5.44\PM0.03\at1420&                    \\
J1854$-$00   &  14.0\PM 0.5\at1077&   9.9\PM 0.5\at1190&   7.1\PM 0.5\at1322&   5.5\PM 0.4\at1421&                    \\
J1854$+$0131g&   496\PM  39\at1077&   542\PM  42\at1129&   237\PM  10\at1336&   199\PM   7\at1423&                    \\
J1855$-$0221g&  0.58\PM0.06\at1098&  0.29\PM0.03\at1259&  0.16\PM0.01\at1399&                    &                    \\
J1855$+$0205 &    17\PM   1\at1086&   8.7\PM 0.8\at1186&   7.1\PM 0.5\at1315&   5.2\PM 0.4\at1420&                    \\
J1855$+$0422 &  33.3\PM 0.7\at1084&  21.9\PM 0.6\at1187&  13.8\PM 0.4\at1333&  11.2\PM 0.4\at1419&                    \\
J1855$+$0455g&    16\PM   2\at1076&    11\PM   1\at1181&   6.7\PM 0.7\at1315&     5\PM   1\at1421&                    \\
J1855$+$0511g&   116\PM   4\at1079&    77\PM   2\at1186&    59\PM   1\at1318&    42\PM   2\at1419&                    \\
J1855$+$0527 &    73\PM   3\at1086&    55\PM   2\at1178&    32\PM   1\at1320&    24\PM   1\at1420&                    \\
J1856$+$0245 &17$_{-3}^{+5}$\at1084&14$_{-2}^{+10}$\at1187&     6\PM   1\at1317&     1\PM   1\at1420&                    \\
J1857$+$0143 &    32\PM   1\at1086&  23.5\PM 0.7\at1187&  16.9\PM 0.4\at1313&  12.6\PM 0.3\at1420&                    \\
J1857$+$0210 &    20\PM   1\at1082&    14\PM   0\at1170&   7.1\PM 0.6\at1322&   5.8\PM 0.5\at1419&                    \\
J1857$+$0214 &   119\PM  22\at1216&    80\PM   6\at1379&    80\PM   6\at1406&                    &                    \\
J1857$+$0300 &    18\PM   3\at1070&    13\PM   1\at1103&     2\PM   1\at1391&                    &                    \\
J1857$+$0526 &  17.2\PM 0.2\at1087&  12.4\PM 0.1\at1183&  8.36\PM0.07\at1317&  6.37\PM0.06\at1418&                    \\
J1858$+$0215 &    32\PM   1\at1085&    25\PM   2\at1183&    17\PM   1\at1319&     9\PM   1\at1421&                    \\
J1858$+$0244g&   1.2\PM 0.2\at1098&  0.46\PM0.04\at1262&  0.30\PM0.02\at1399&                    &                    \\
J1859$+$0430 &    16\PM   2\at1087&    12\PM   1\at1184&   9.3\PM 0.8\at1317&     5\PM   1\at1419&                    \\
J1859$+$0601 &    70\PM   3\at1086&    63\PM   2\at1187&    39\PM   1\at1320&    30\PM   1\at1421&                    \\
J1900$+$0438 &196$_{-58}^{+187}$\at1108&    68\PM   5\at1265&    37\PM   2\at1399&                    &                    \\
J1901$+$0300 &  0.35\PM0.01\at1099&  0.22\PM0.01\at1291&  0.18\PM0.01\at1381&                    &                    \\ 
J1901$+$0459 &    72\PM   7\at1087&    51\PM   6\at1173&    28\PM   2\at1336&    26\PM   2\at1420&                    \\
J1902$-$0107g&   1.3\PM 0.3\at1103&   1.0\PM 0.2\at1263&   0.5\PM 0.1\at1397&                    &                    \\
J1903$+$0327 &  0.43\PM0.02\at1104&  0.16\PM0.01\at1341&  0.12\PM0.01\at1422&                    &                    \\
J1905$+$0600 &  14.7\PM 0.3\at1086&  10.0\PM 0.2\at1184&   6.5\PM 0.1\at1312&   5.1\PM 0.2\at1421&                    \\
J1907$+$0534 &  14.9\PM 0.4\at1087&  11.0\PM 0.4\at1187&   8.4\PM 0.3\at1319&   6.4\PM 0.3\at1420&                    \\
J1908$+$0833 &    17\PM   1\at1104&    13\PM   1\at1269&   8.0\PM 0.7\at1399&                    &                    \\
J1908$+$0839 &   6.9\PM 0.1\at1087&   5.7\PM 0.1\at1183&  4.24\PM0.08\at1320&  3.59\PM0.07\at1419&                    \\
J1908$+$0909 &  6.11\PM0.09\at1086&  4.96\PM0.07\at1185&  4.15\PM0.05\at1312&  3.69\PM0.06\at1421&                    \\
J1910$+$0534 &  11.7\PM 0.3\at1085&   9.1\PM 0.4\at1170&   5.8\PM 0.3\at1321&   4.0\PM 0.4\at1422&                    \\
J1910$+$1026 &    17\PM   1\at1080&    12\PM   1\at1188&   9.4\PM 0.6\at1318&   7.0\PM 0.5\at1420&                    \\
J1911$+$0101A&  0.36\PM0.02\at1084&  0.29\PM0.02\at1188&  0.28\PM0.02\at1320&  0.25\PM0.02\at1420&                    \\
J1911$+$0925 &    32\PM   6\at1087&    22\PM   3\at1180&    14\PM   1\at1314&    12\PM   2\at1419&                    \\
J1913$+$1000 &  12.9\PM 0.2\at1096&  10.6\PM 0.3\at1160&2.3$_{-0.9}^{+0.7}$\at1334&1$_{-1}^{+1}$\at1404&              \\ 
J1913$+$1102 &   8.4\PM 0.3\at1078&   5.4\PM 0.1\at1185&  3.61\PM0.06\at1317&  2.58\PM0.03\at1421&                    \\
J1913$+$11025&    66\PM   4\at1069&    71\PM   5\at1119&    35\PM   2\at1348&    37\PM   2\at1426&                    \\
J1913$+$1145 &  10.0\PM 0.2\at1087&   7.7\PM 0.1\at1181&  5.12\PM0.09\at1330&  4.05\PM0.07\at1420&                    \\
J1914$+$1054g&     7\PM   1\at1085&     6\PM   1\at1168&   3.6\PM 0.6\at1320&   3.5\PM 0.5\at1420&                    \\
J1916$+$0844 &  9.50\PM0.08\at1085&  7.34\PM0.08\at1174&  4.60\PM0.07\at1321&  3.37\PM0.06\at1420&                    \\
J1918$+$1340g&    20\PM   1\at1085&    12\PM   1\at1167&   7.4\PM 0.4\at1320&   6.3\PM 0.5\at1420&                    \\
J1919$+$1314 &  15.3\PM 0.5\at1087&  11.0\PM 0.6\at1166&   7.2\PM 0.3\at1328&   6.4\PM 0.3\at1420&                    \\
J1919$+$1341 &   1.2\PM 0.1\at1085&  0.77\PM0.08\at1185&  0.63\PM0.05\at1319&  0.45\PM0.07\at1421&                    \\
J1920$+$1110 &  14.9\PM 0.2\at1087&  10.7\PM 0.2\at1187&   7.1\PM 0.1\at1317&   5.3\PM 0.2\at1420&                    \\
J1920$+$1340g&   270\PM  63\at1088&   236\PM  23\at1275&   230\PM  21\at1383&                    &                    \\
J1921$+$1216g&  0.59\PM0.05\at1098&  0.41\PM0.03\at1262&  0.30\PM0.01\at1400&                    &                    \\
J1921$+$1259g&    41\PM   2\at1077&    16\PM   2\at1169&    13\PM   1\at1321&    10\PM   1\at1422&                    \\
J1921$+$1340g&   103\PM   3\at1061&    53\PM   2\at1235&  42.2\PM 0.8\at1328&  32.5\PM 0.7\at1421&                    \\
J1921$+$1505g&    14\PM   1\at1086&     8\PM   1\at1188&     3\PM   1\at1314&     3\PM   1\at1420&                    \\
J1922$+$1512g&    59\PM   1\at1085&    45\PM   1\at1182&  33.1\PM 0.8\at1318&    24\PM   0\at1421&                    \\
J1924$+$1713 &    30\PM   1\at1086&    21\PM   1\at1182&  13.7\PM 0.7\at1330&  10.3\PM 0.6\at1422&                    \\
\hline\noalign{\smallskip}
    \end{tabular}
\end{table*}

\addtocounter{table}{-1}
\begin{table*}
    \centering
    \caption{-- {\it continued}}
\footnotesize
\renewcommand{\arraystretch}{0.85}  
    \begin{tabular}{lrrrrr}
        \hline\noalign{\smallskip}
\multicolumn{1}{c}{PSR Name}&\multicolumn{1}{c}{$\tau_{\rm s}$\at$\nu_1$ }&\multicolumn{1}{c}{$\tau_{\rm s}$\at$\nu_2$}&\multicolumn{1}{c}{$\tau_{\rm s}$\at$\nu_3$}&\multicolumn{1}{c}{$\tau_{\rm s}$\at$\nu_4$}&\multicolumn{1}{c}{$\tau_{\rm s}$\at$\nu_5$}\\
\multicolumn{1}{c}{ }&\multicolumn{1}{c}{(ms{\at}MHz)}&\multicolumn{1}{c}{(ms{\at}MHz)}&\multicolumn{1}{c}{(ms{\at}MHz)}&\multicolumn{1}{c}{(ms{\at}MHz)}&\multicolumn{1}{c}{(ms{\at}MHz)}\\
\hline\noalign{\smallskip}
J1928$+$1923 &  35.4\PM 0.4\at1084&  27.3\PM 0.4\at1172&  16.4\PM 0.3\at1333&  11.7\PM 0.2\at1420&                    \\
J1928$+$1245 &  0.18\PM0.01\at1094&  0.15\PM0.01\at1186&  0.12\PM0.01\at1316&  0.10\PM0.01\at1404&                    \\ 
J1929$+$19   &  27.3\PM 0.3\at1093&  20.8\PM 0.3\at1175&  13.0\PM 0.2\at1327&  10.9\PM 0.2\at1405&                    \\ 
J1946$+$2433g&3$_{-1}^{+7}$\at1124&   1.8\PM 0.4\at1267&   1.2\PM 0.3\at1375&                    &                    \\
J1946$+$2757g&    89\PM   5\at1086&    68\PM   7\at1159&    40\PM   2\at1336&    34\PM   2\at1421&                    \\
J1950$+$2414 &  0.57\PM0.05\at1086&  0.44\PM0.06\at1160&  0.29\PM0.03\at1333&  0.25\PM0.02\at1419&                    \\
J1959$+$3141g&  13.7\PM 0.8\at1085&   8.4\PM 0.5\at1186&   5.3\PM 0.5\at1318&   4.7\PM 0.5\at1421&                    \\
J2004$+$3429 &   7.7\PM 0.1\at1084&   5.5\PM 0.1\at1181&  3.50\PM0.06\at1329&  2.74\PM0.04\at1421&                    \\
J2005$+$3411g&  31.1\PM 0.6\at1085&  22.0\PM 0.6\at1173&  14.3\PM 0.3\at1328&  10.6\PM 0.3\at1421&                    \\
J2015$+$3404g&   1.4\PM 0.2\at1106&  0.82\PM0.07\at1263&  0.65\PM0.04\at1400&                    &                    \\
J2019$+$3718g&  16.3\PM 0.4\at1072&  13.6\PM 0.3\at1149&   9.6\PM 0.2\at1248&   7.9\PM 0.1\at1348&   6.3\PM 0.1\at1433\\
J2020$+$3806g&   7.1\PM 0.1\at1083&   5.6\PM 0.1\at1177&   3.9\PM 0.1\at1321&   3.4\PM 0.1\at1412&                    \\ 
J2021$+$3651 &    13\PM   1\at1085&   8.0\PM 0.8\at1171&   6.4\PM 0.5\at1327&   5.0\PM 0.4\at1421&                    \\
J2021$+$4024g&    59\PM   2\at1086&    48\PM   1\at1171&    29\PM   0\at1328&  24.0\PM 0.6\at1421&                    \\
J2022$+$3845g&   171\PM   7\at1085&   133\PM   6\at1171&    65\PM   2\at1333&    56\PM   1\at1422&                    \\
J2022$+$3910g&   173\PM   7\at1077&    60\PM   3\at1168&    35\PM   1\at1373&                    &                    \\ 
J2030$+$3818g&     7\PM   1\at1076&     6\PM   1\at1122&     5\PM   2\at1301&     3\PM   0\at1352&                    \\
J2030$+$3944g&    98\PM   6\at1087&    63\PM   3\at1180&    47\PM   1\at1318&    32\PM   0\at1422&                    \\
J2032$+$4055g&71$_{-32}^{+96}$\at1109&    21\PM   2\at1252&  12.8\PM 0.8\at1397&                    &                    \\
J2041$+$3934g&   126\PM  13\at1073&   160\PM  18\at1146&    84\PM  15\at1198&    66\PM   3\at1353&                    \\
J2041$+$4551 &  22.9\PM 0.1\at1085& 16.95\PM0.09\at1173& 11.61\PM0.05\at1318&  9.36\PM0.05\at1420&                    \\
J2045$+$4431g&    21\PM   4\at1096&26$_{-7}^{+12}$\at1284&    17\PM   2\at1399&                    &                    \\
J2046$+$4236g&311$_{-113}^{+468}$\at1146&    87\PM   7\at1343&    67\PM   5\at1421&                    &                    \\
J2046$+$4253g&    98\PM   5\at1086&    61\PM   2\at1183&  38.8\PM 0.9\at1321&  30.6\PM 0.6\at1422&                    \\
J2052$+$4421g&    71\PM   1\at1087&  56.7\PM 0.7\at1174&  35.4\PM 0.4\at1330&  27.4\PM 0.3\at1419&                    \\
J2057$+$4557g&    35\PM   0\at1085&    28\PM   0\at1174&  15.9\PM 0.3\at1320&  13.5\PM 0.3\at1413&                    \\
J2205$+$6012 &  0.24\PM0.02\at1087&  0.21\PM0.02\at1121&  0.10\PM0.00\at1349&  0.08\PM0.00\at1412&                    \\ 
\multicolumn{6}{r}{ ... ended. } \\
\hline\noalign{\smallskip}
    \end{tabular}
\end{table*}


\newpage

\addtocounter{figure}{-1}
\begin{figure*}
\centering
\includegraphics[width=0.24\textwidth,height=0.24\textwidth]{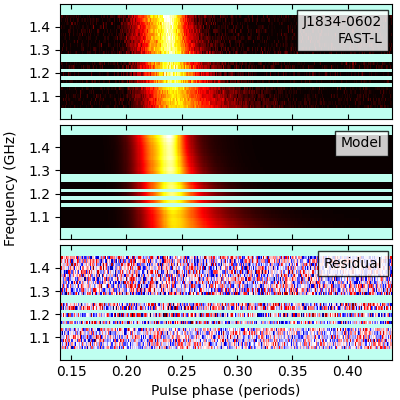}
\includegraphics[width=0.24\textwidth,height=0.24\textwidth]{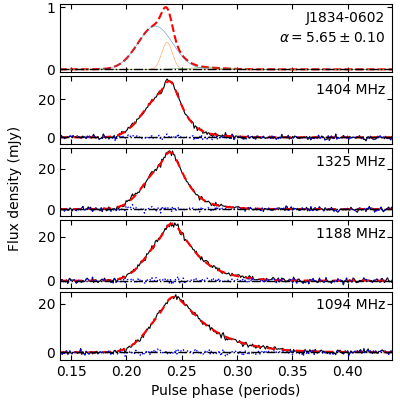}
\includegraphics[width=0.24\textwidth,height=0.24\textwidth]{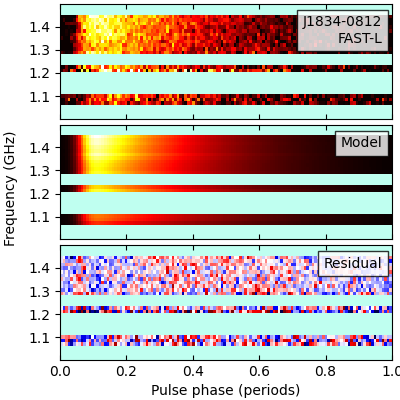}
\includegraphics[width=0.24\textwidth,height=0.24\textwidth]{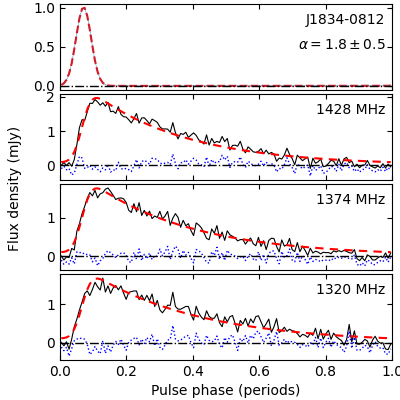}
\includegraphics[width=0.24\textwidth,height=0.24\textwidth]{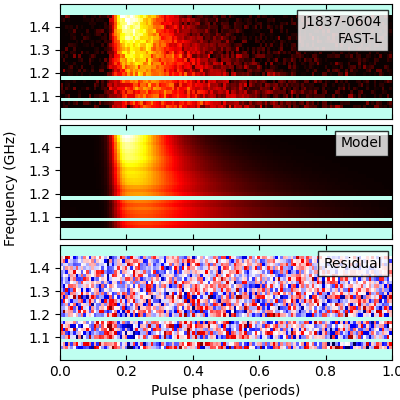}
\includegraphics[width=0.24\textwidth,height=0.24\textwidth]{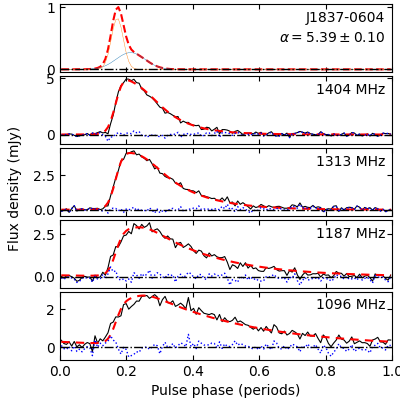}
\includegraphics[width=0.24\textwidth,height=0.24\textwidth]{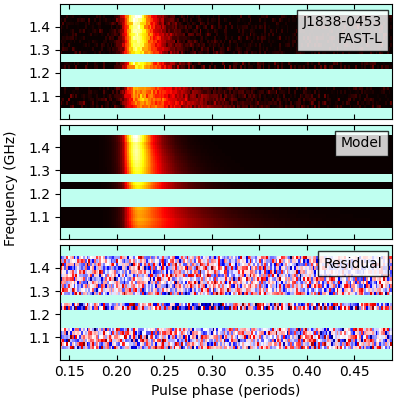}
\includegraphics[width=0.24\textwidth,height=0.24\textwidth]{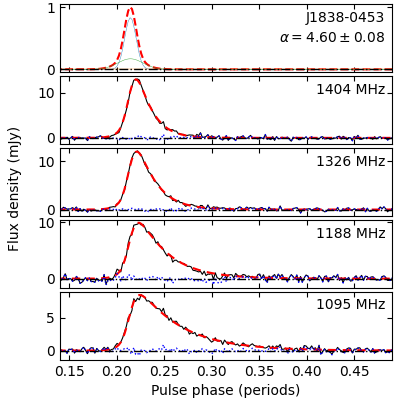}
\includegraphics[width=0.24\textwidth,height=0.24\textwidth]{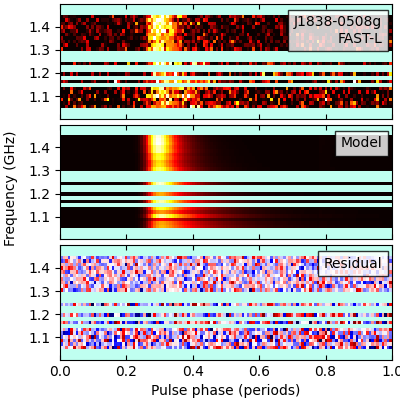}
\includegraphics[width=0.24\textwidth,height=0.24\textwidth]{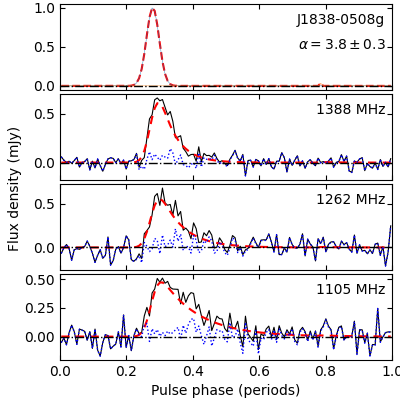}
\includegraphics[width=0.24\textwidth,height=0.24\textwidth]{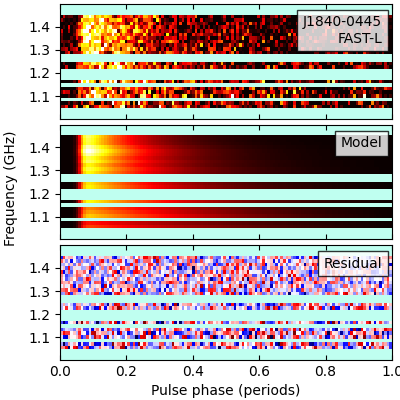}
\includegraphics[width=0.24\textwidth,height=0.24\textwidth]{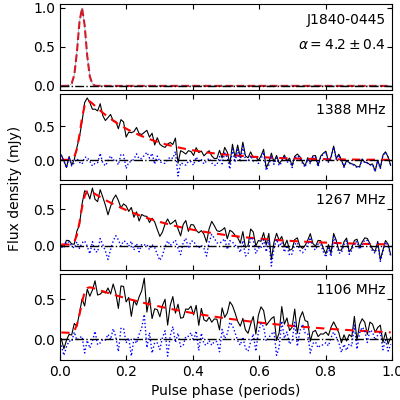}
\includegraphics[width=0.24\textwidth,height=0.24\textwidth]{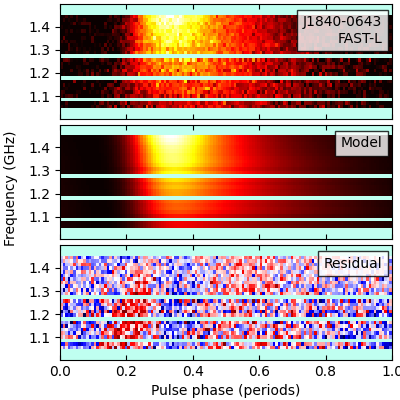}
\includegraphics[width=0.24\textwidth,height=0.24\textwidth]{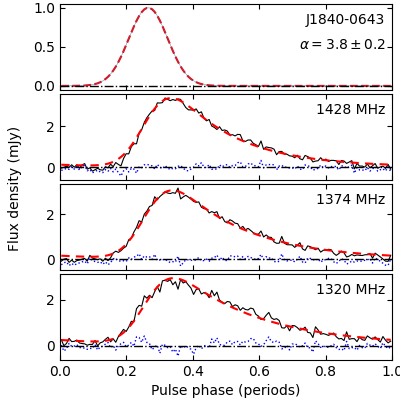}
\includegraphics[width=0.24\textwidth,height=0.24\textwidth]{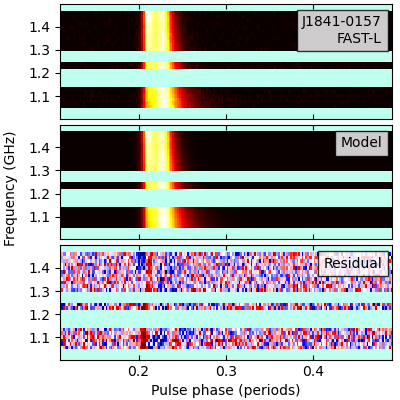}
\includegraphics[width=0.24\textwidth,height=0.24\textwidth]{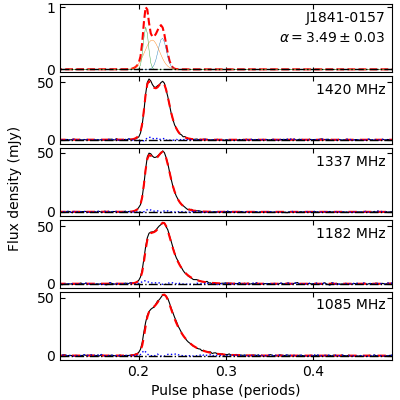}
\includegraphics[width=0.24\textwidth,height=0.24\textwidth]{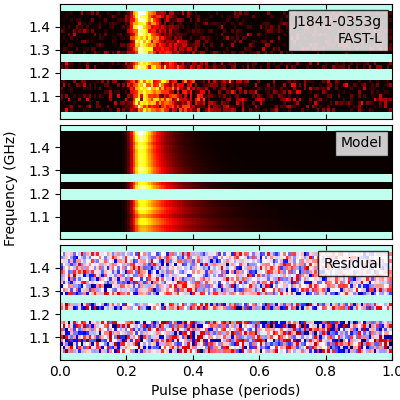}
\includegraphics[width=0.24\textwidth,height=0.24\textwidth]{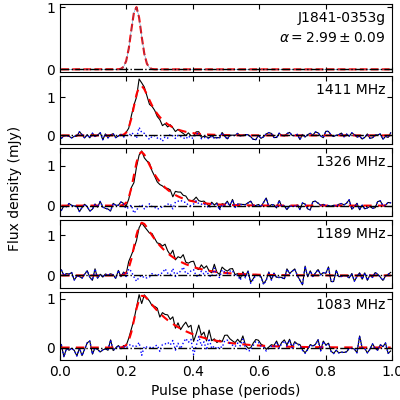}
\includegraphics[width=0.24\textwidth,height=0.24\textwidth]{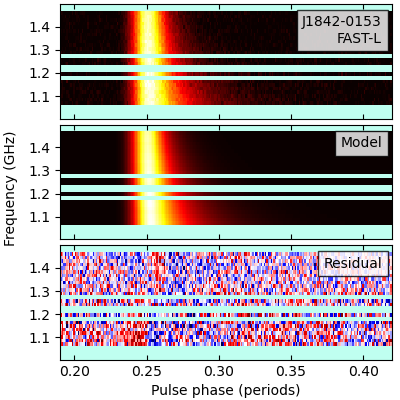}
\includegraphics[width=0.24\textwidth,height=0.24\textwidth]{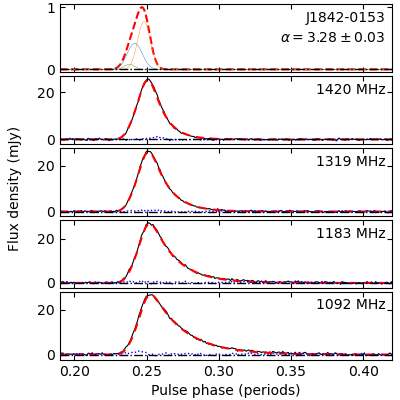}
\caption{--- {\it to be continued.}}
\end{figure*}

\addtocounter{figure}{-1}
\begin{figure*}
\centering
\includegraphics[width=0.24\textwidth,height=0.24\textwidth]{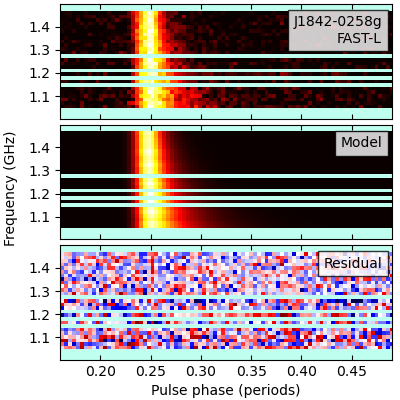}
\includegraphics[width=0.24\textwidth,height=0.24\textwidth]{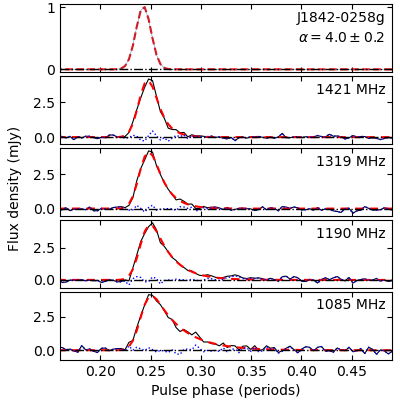}
\includegraphics[width=0.24\textwidth,height=0.24\textwidth]{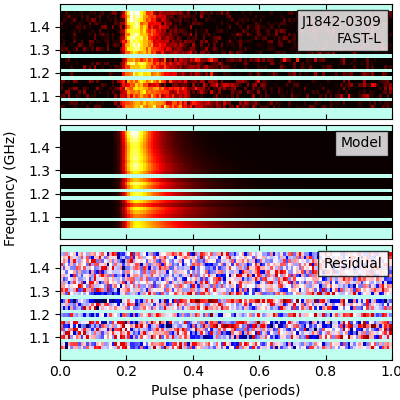}
\includegraphics[width=0.24\textwidth,height=0.24\textwidth]{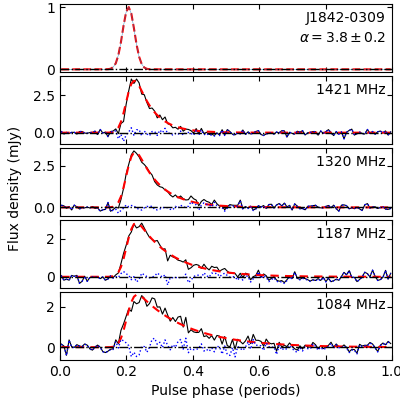}
\includegraphics[width=0.24\textwidth,height=0.24\textwidth]{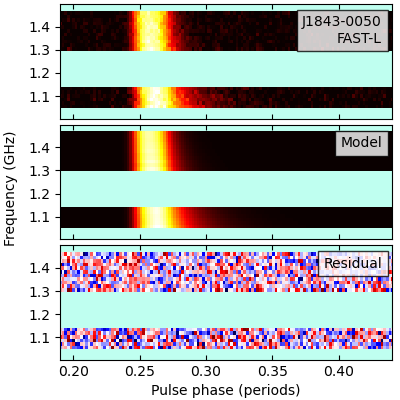}
\includegraphics[width=0.24\textwidth,height=0.24\textwidth]{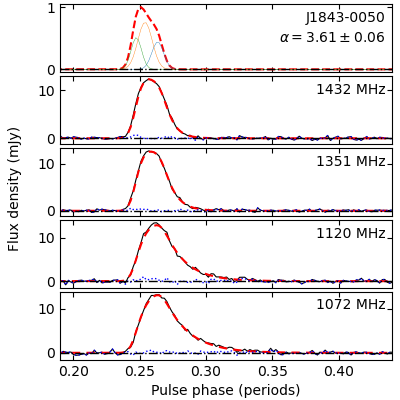}
\includegraphics[width=0.24\textwidth,height=0.24\textwidth]{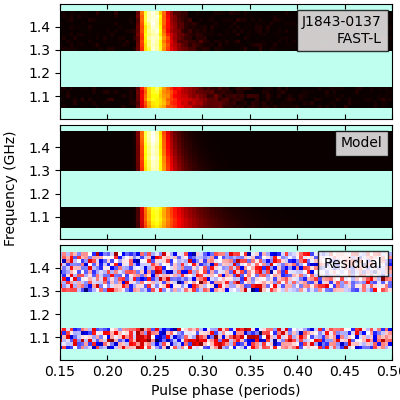}
\includegraphics[width=0.24\textwidth,height=0.24\textwidth]{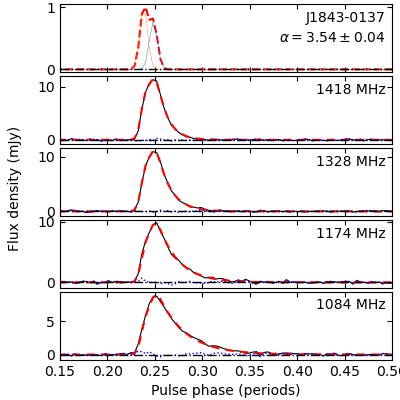}
\includegraphics[width=0.24\textwidth,height=0.24\textwidth]{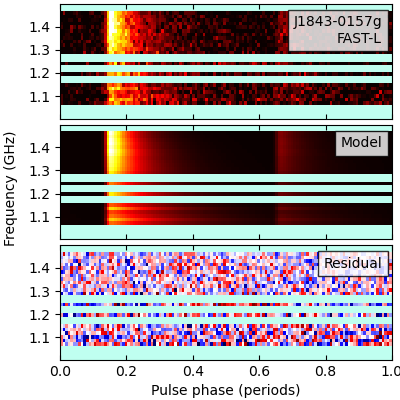}
\includegraphics[width=0.24\textwidth,height=0.24\textwidth]{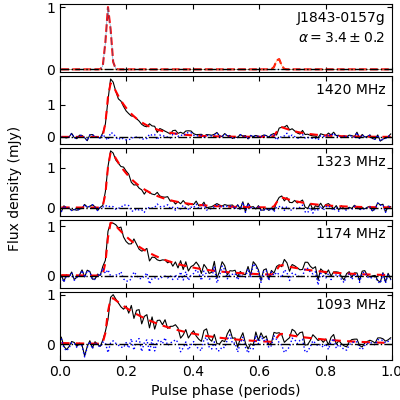}
\includegraphics[width=0.24\textwidth,height=0.24\textwidth]{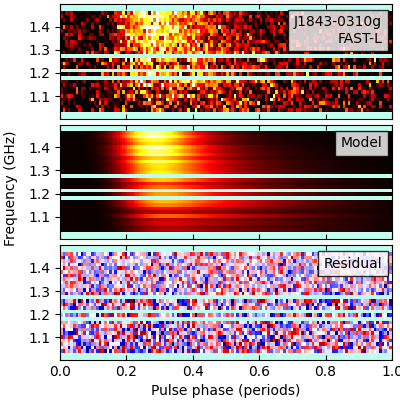}
\includegraphics[width=0.24\textwidth,height=0.24\textwidth]{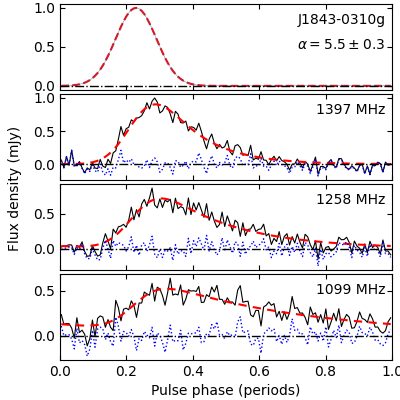}
\includegraphics[width=0.24\textwidth,height=0.24\textwidth]{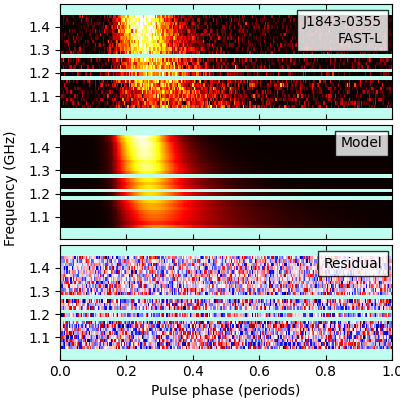}
\includegraphics[width=0.24\textwidth,height=0.24\textwidth]{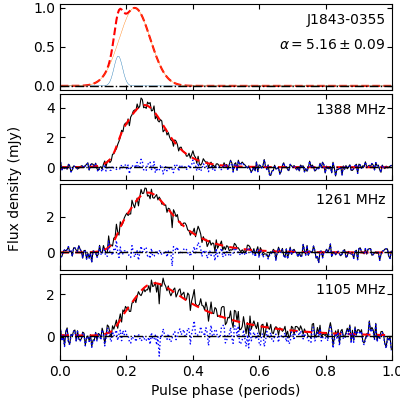}
\includegraphics[width=0.24\textwidth,height=0.24\textwidth]{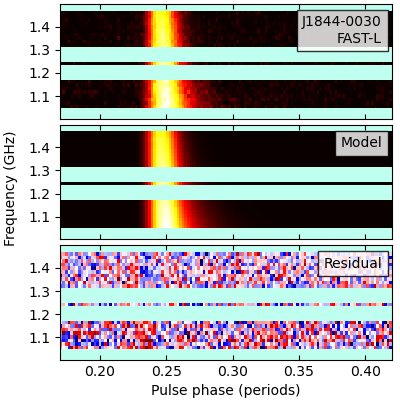}
\includegraphics[width=0.24\textwidth,height=0.24\textwidth]{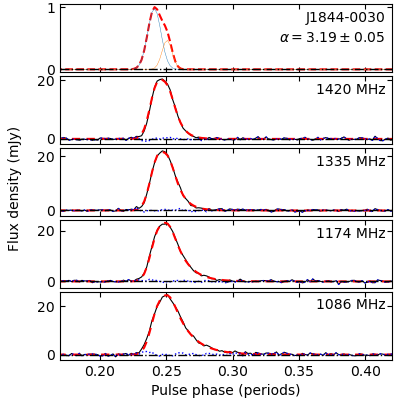}
\includegraphics[width=0.24\textwidth,height=0.24\textwidth]{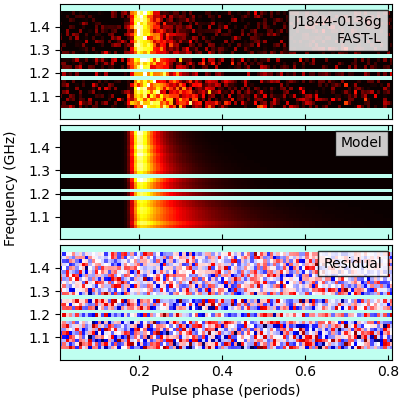}
\includegraphics[width=0.24\textwidth,height=0.24\textwidth]{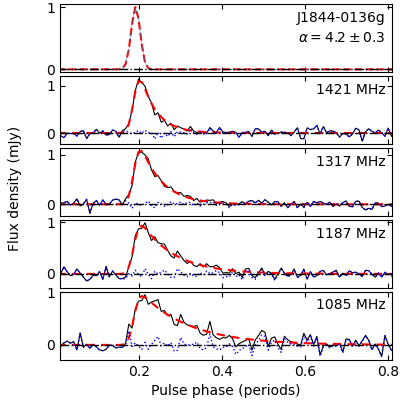}
\includegraphics[width=0.24\textwidth,height=0.24\textwidth]{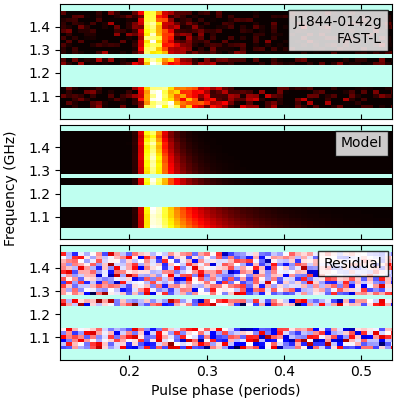}
\includegraphics[width=0.24\textwidth,height=0.24\textwidth]{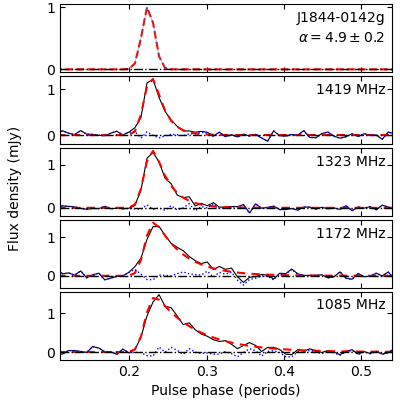}
  \caption{--- {\it to be continued.}}
\end{figure*}

\addtocounter{figure}{-1}
\begin{figure*}
\centering
\includegraphics[width=0.24\textwidth,height=0.24\textwidth]{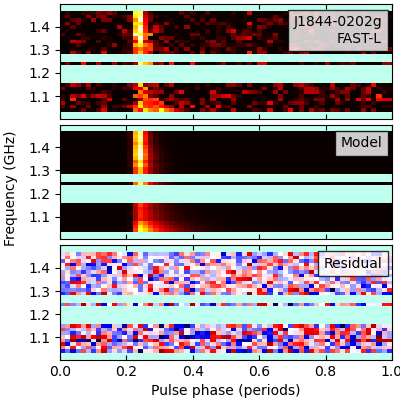}
\includegraphics[width=0.24\textwidth,height=0.24\textwidth]{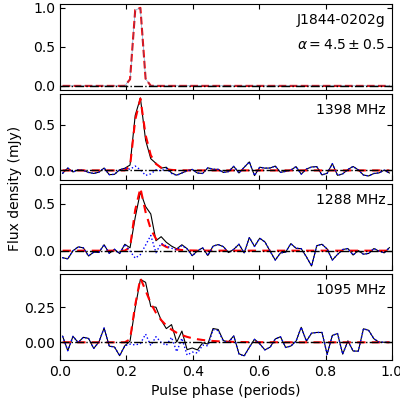}
\includegraphics[width=0.24\textwidth,height=0.24\textwidth]{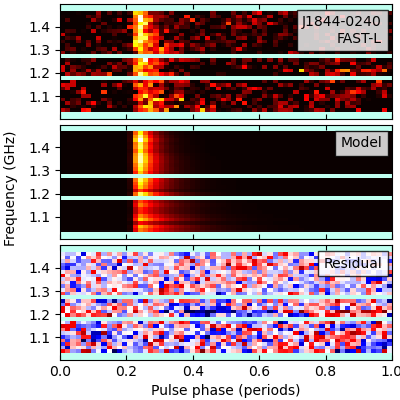}
\includegraphics[width=0.24\textwidth,height=0.24\textwidth]{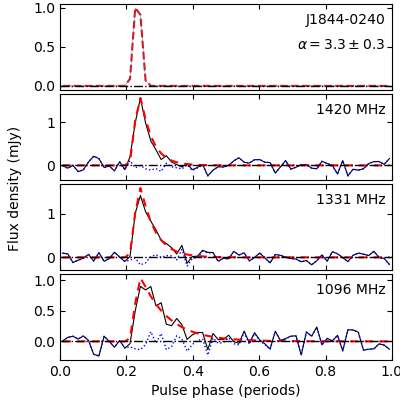}
\includegraphics[width=0.24\textwidth,height=0.24\textwidth]{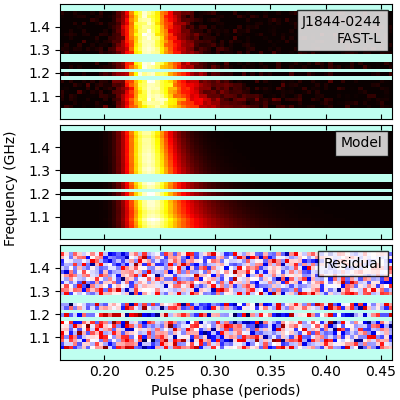}
\includegraphics[width=0.24\textwidth,height=0.24\textwidth]{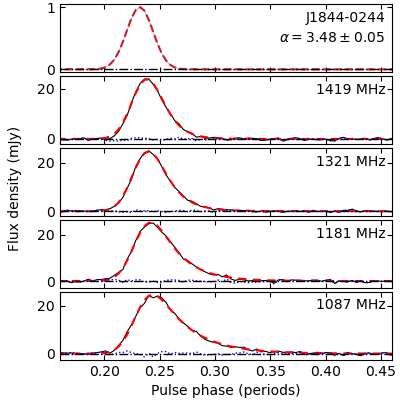}
\includegraphics[width=0.24\textwidth,height=0.24\textwidth]{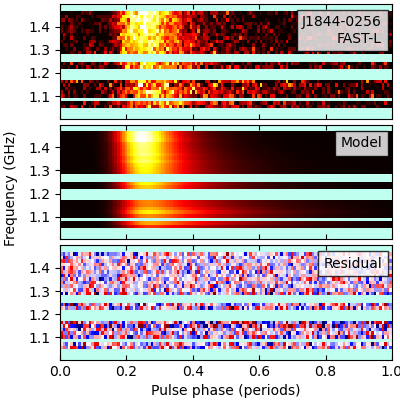}
\includegraphics[width=0.24\textwidth,height=0.24\textwidth]{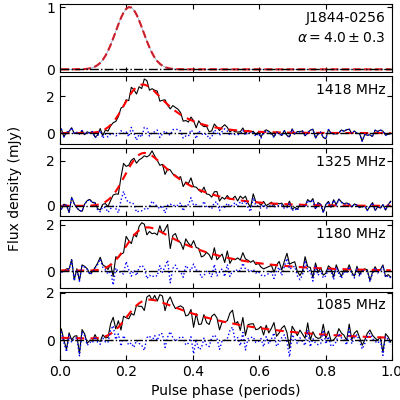}
\includegraphics[width=0.24\textwidth,height=0.24\textwidth]{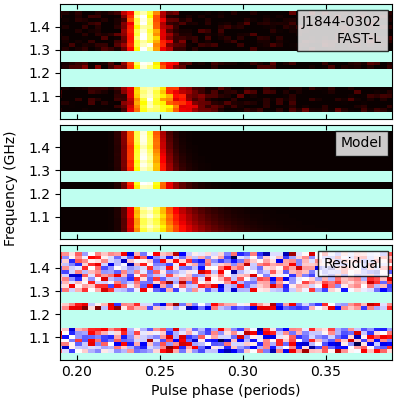}
\includegraphics[width=0.24\textwidth,height=0.24\textwidth]{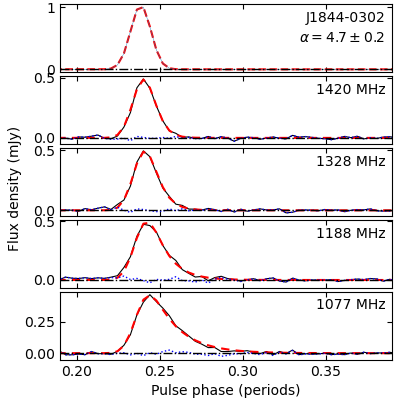}
\includegraphics[width=0.24\textwidth,height=0.24\textwidth]{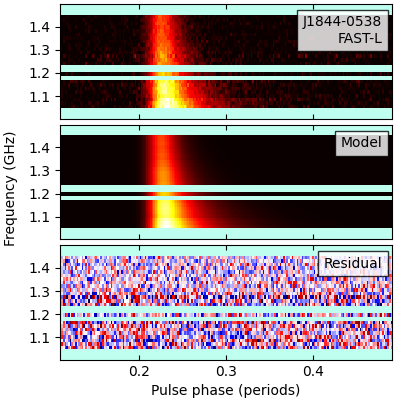}
\includegraphics[width=0.24\textwidth,height=0.24\textwidth]{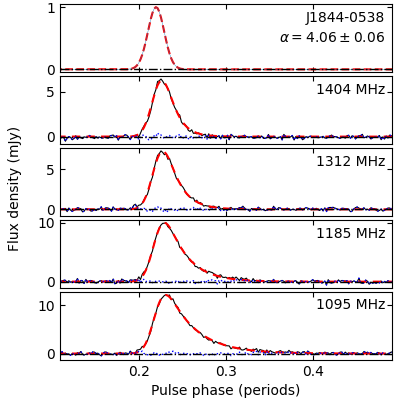}
\includegraphics[width=0.24\textwidth,height=0.24\textwidth]{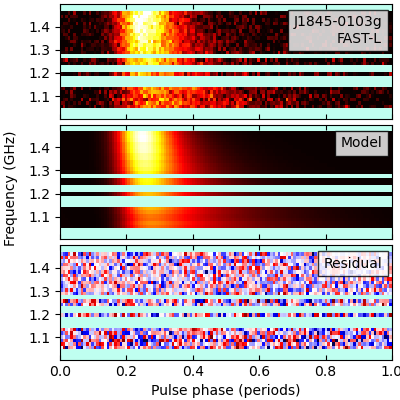}
\includegraphics[width=0.24\textwidth,height=0.24\textwidth]{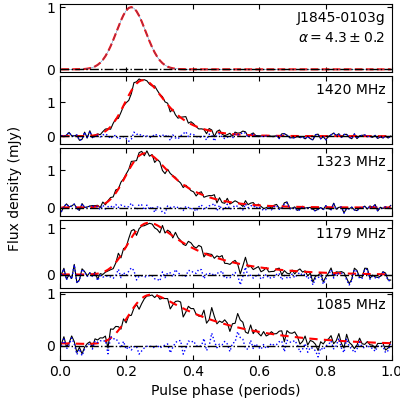}
\includegraphics[width=0.24\textwidth,height=0.24\textwidth]{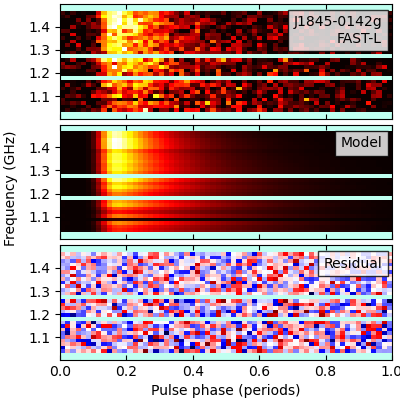}
\includegraphics[width=0.24\textwidth,height=0.24\textwidth]{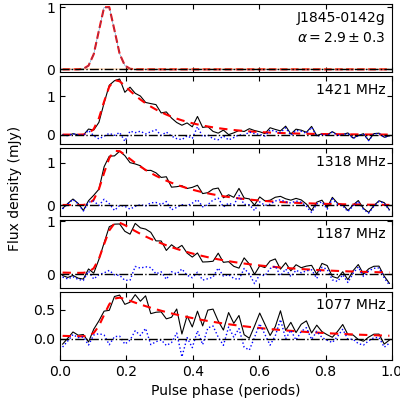}
\includegraphics[width=0.24\textwidth,height=0.24\textwidth]{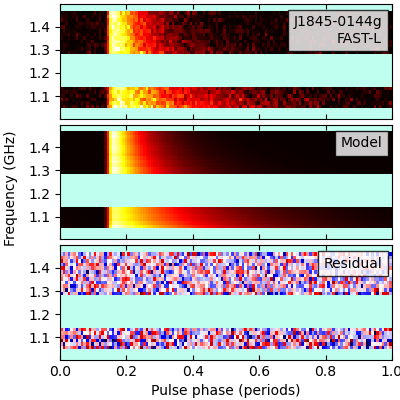}
\includegraphics[width=0.24\textwidth,height=0.24\textwidth]{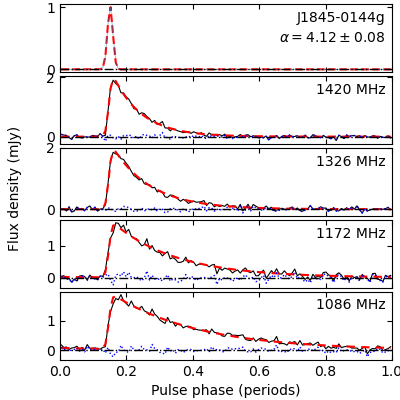}
\includegraphics[width=0.24\textwidth,height=0.24\textwidth]{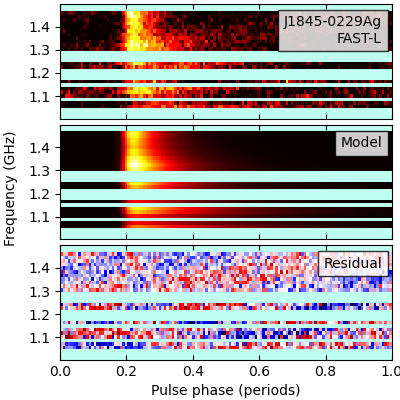}
\includegraphics[width=0.24\textwidth,height=0.24\textwidth]{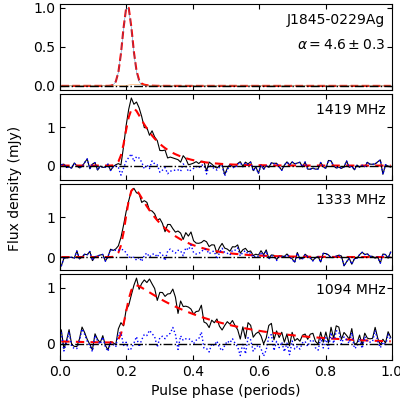}
\caption{--- {\it to be continued.}}
\end{figure*}

\addtocounter{figure}{-1}
\begin{figure*}
\centering
\includegraphics[width=0.24\textwidth,height=0.24\textwidth]{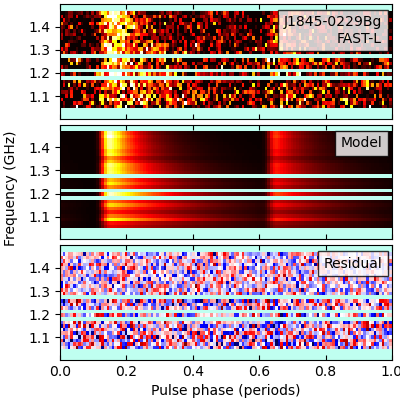}
\includegraphics[width=0.24\textwidth,height=0.24\textwidth]{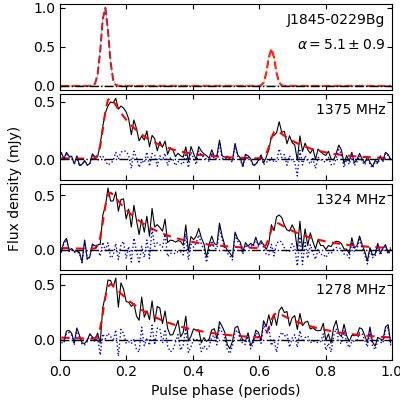}
\includegraphics[width=0.24\textwidth,height=0.24\textwidth]{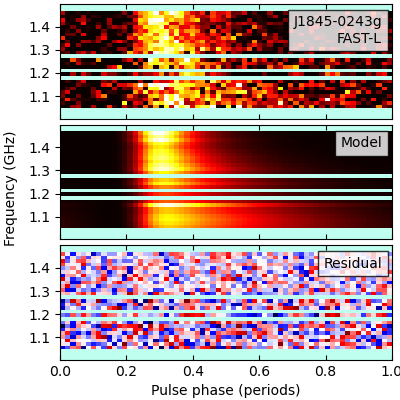}
\includegraphics[width=0.24\textwidth,height=0.24\textwidth]{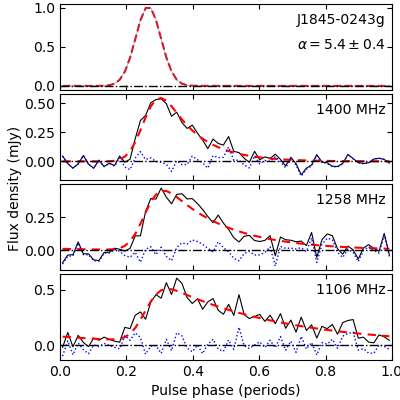}
\includegraphics[width=0.24\textwidth,height=0.24\textwidth]{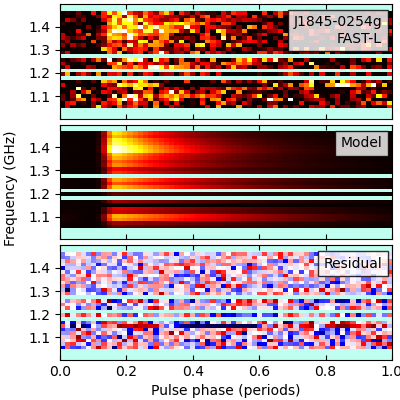}
\includegraphics[width=0.24\textwidth,height=0.24\textwidth]{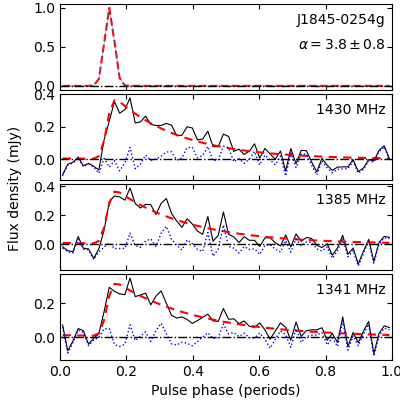}
\includegraphics[width=0.24\textwidth,height=0.24\textwidth]{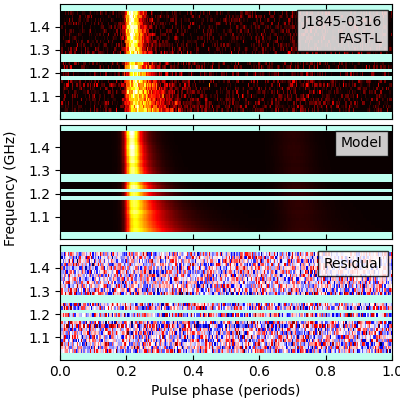}
\includegraphics[width=0.24\textwidth,height=0.24\textwidth]{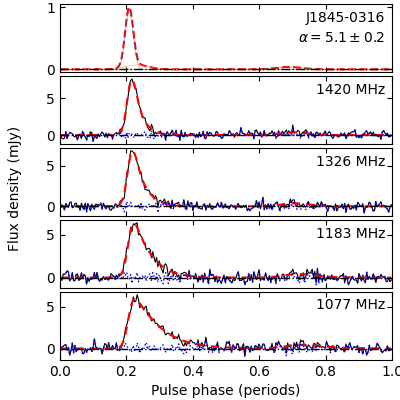}
\includegraphics[width=0.24\textwidth,height=0.24\textwidth]{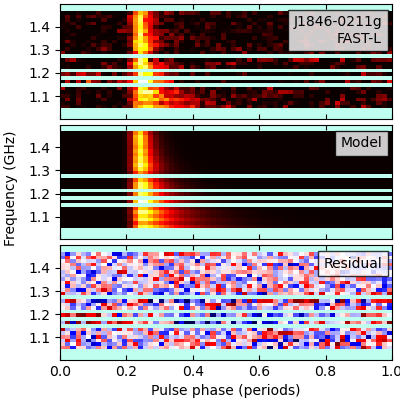}
\includegraphics[width=0.24\textwidth,height=0.24\textwidth]{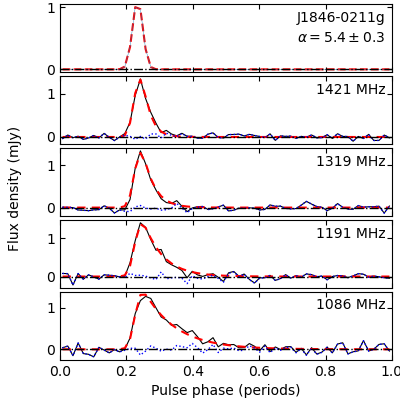}
\includegraphics[width=0.24\textwidth,height=0.24\textwidth]{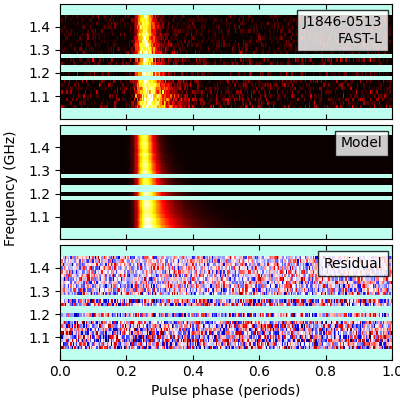}
\includegraphics[width=0.24\textwidth,height=0.24\textwidth]{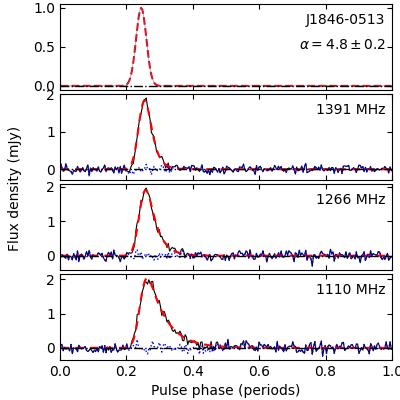}
\includegraphics[width=0.24\textwidth,height=0.24\textwidth]{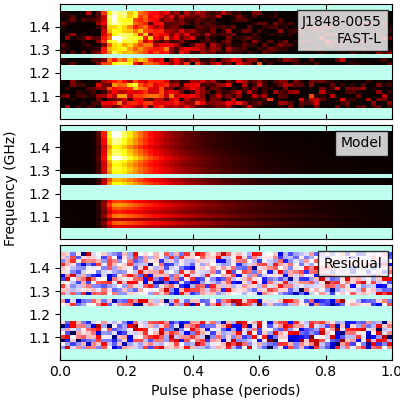}
\includegraphics[width=0.24\textwidth,height=0.24\textwidth]{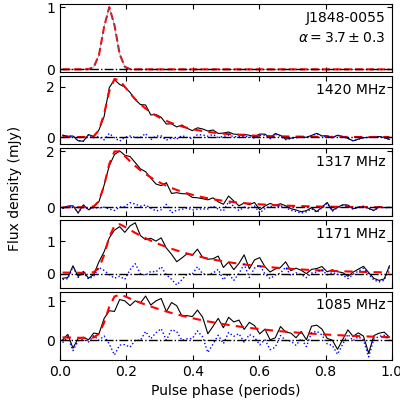}
\includegraphics[width=0.24\textwidth,height=0.24\textwidth]{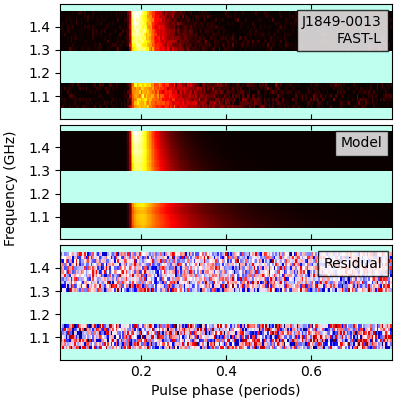}
\includegraphics[width=0.24\textwidth,height=0.24\textwidth]{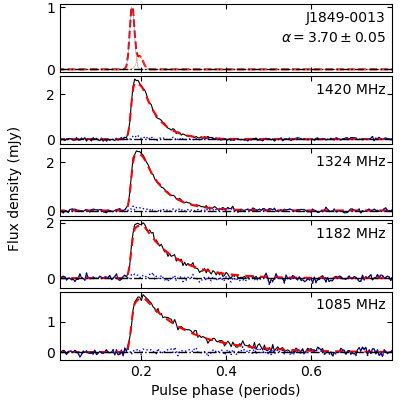}
\includegraphics[width=0.24\textwidth,height=0.24\textwidth]{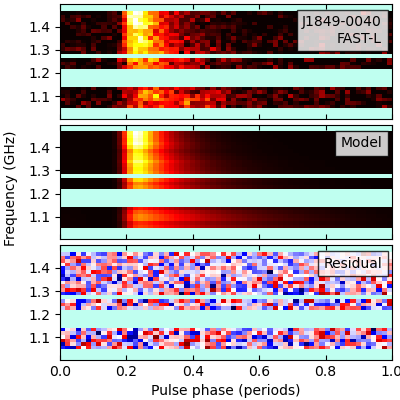}
\includegraphics[width=0.24\textwidth,height=0.24\textwidth]{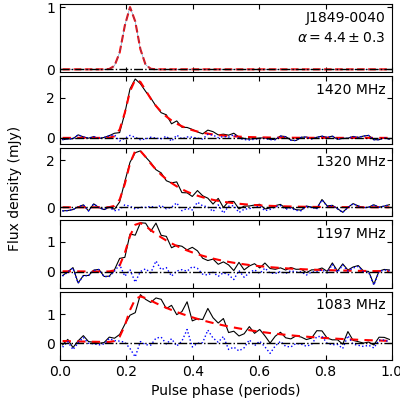}
\includegraphics[width=0.24\textwidth,height=0.24\textwidth]{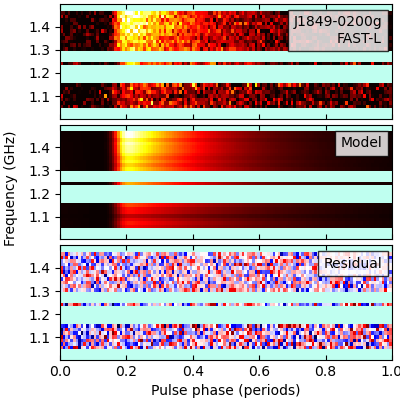}
\includegraphics[width=0.24\textwidth,height=0.24\textwidth]{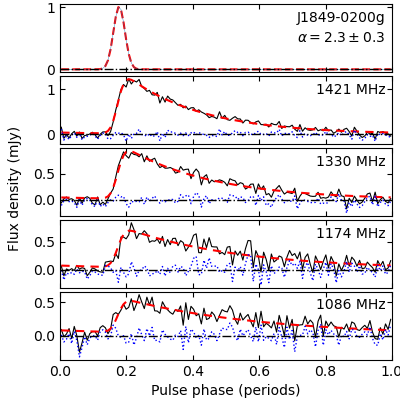}
\caption{--- {\it to be continued.}}
\end{figure*}

\addtocounter{figure}{-1}
\begin{figure*}
\centering
\includegraphics[width=0.24\textwidth,height=0.24\textwidth]{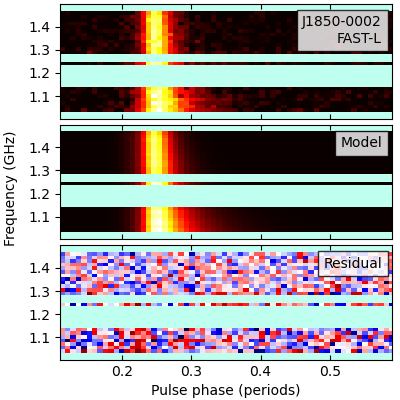}
\includegraphics[width=0.24\textwidth,height=0.24\textwidth]{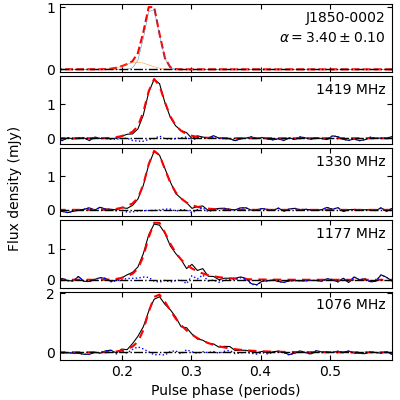}
\includegraphics[width=0.24\textwidth,height=0.24\textwidth]{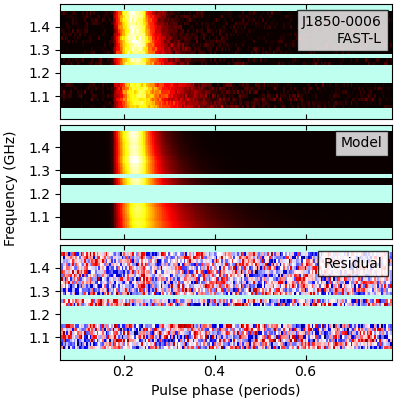}
\includegraphics[width=0.24\textwidth,height=0.24\textwidth]{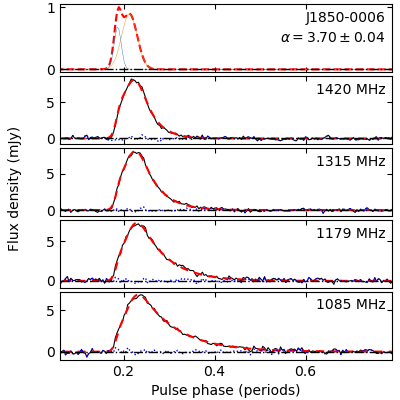}
\includegraphics[width=0.24\textwidth,height=0.24\textwidth]{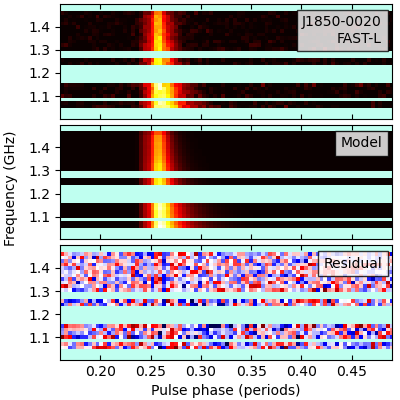}
\includegraphics[width=0.24\textwidth,height=0.24\textwidth]{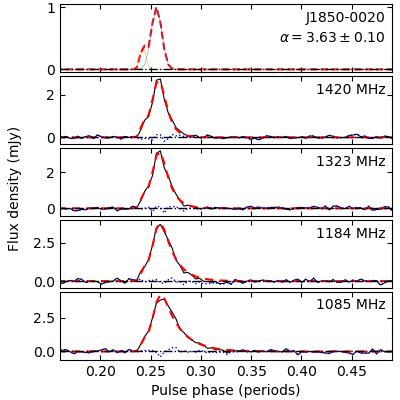}
\includegraphics[width=0.24\textwidth,height=0.24\textwidth]{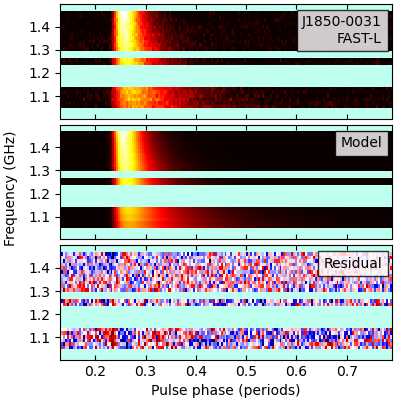}
\includegraphics[width=0.24\textwidth,height=0.24\textwidth]{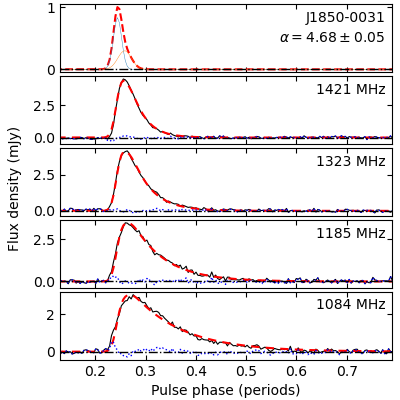}
\includegraphics[width=0.24\textwidth,height=0.24\textwidth]{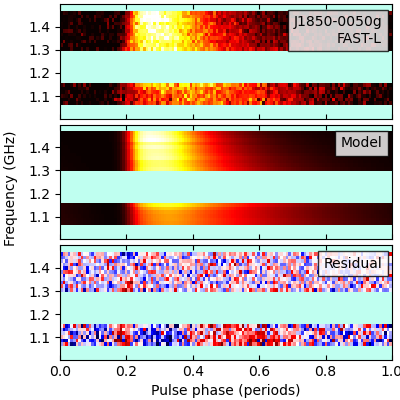}
\includegraphics[width=0.24\textwidth,height=0.24\textwidth]{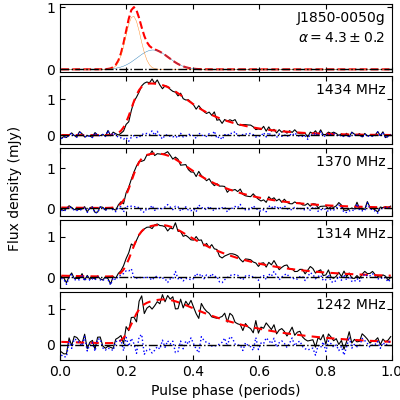}
\includegraphics[width=0.24\textwidth,height=0.24\textwidth]{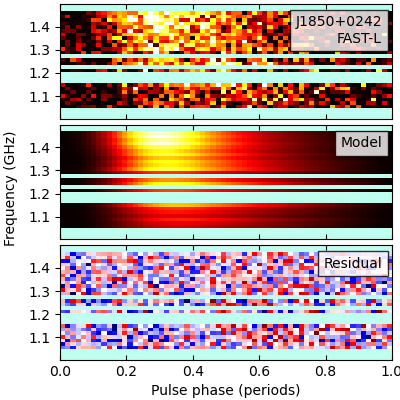}
\includegraphics[width=0.24\textwidth,height=0.24\textwidth]{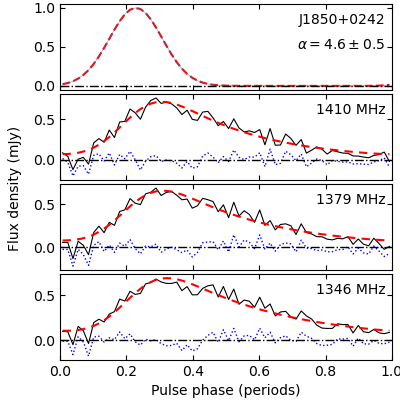}
\includegraphics[width=0.24\textwidth,height=0.24\textwidth]{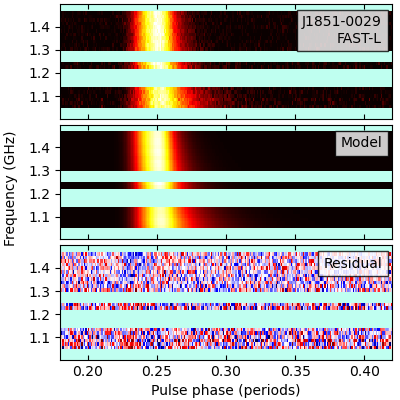}
\includegraphics[width=0.24\textwidth,height=0.24\textwidth]{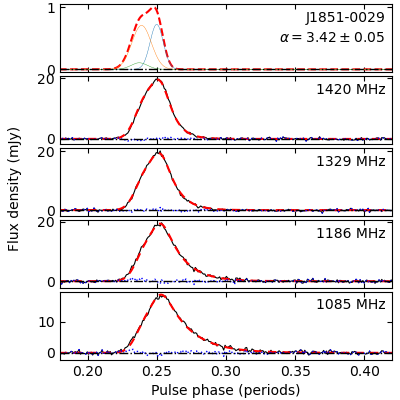}
\includegraphics[width=0.24\textwidth,height=0.24\textwidth]{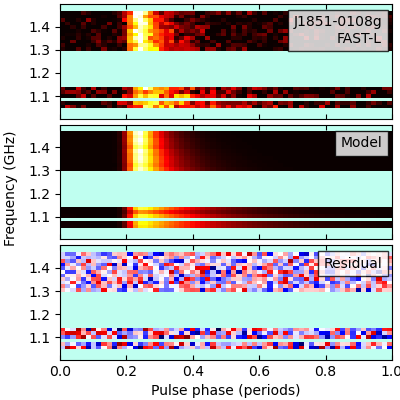}
\includegraphics[width=0.24\textwidth,height=0.24\textwidth]{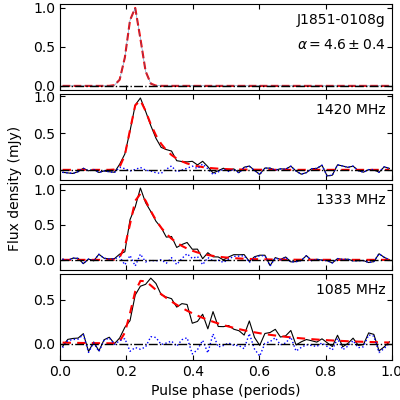}
\includegraphics[width=0.24\textwidth,height=0.24\textwidth]{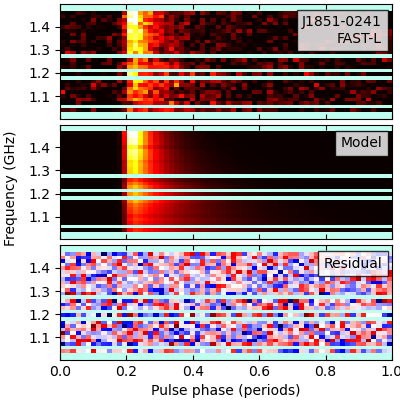}
\includegraphics[width=0.24\textwidth,height=0.24\textwidth]{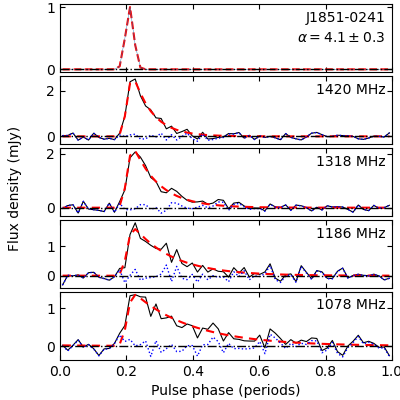}
\includegraphics[width=0.24\textwidth,height=0.24\textwidth]{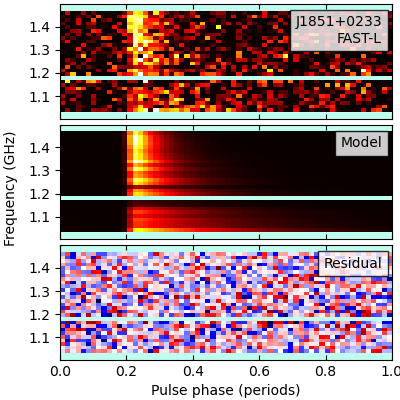}
\includegraphics[width=0.24\textwidth,height=0.24\textwidth]{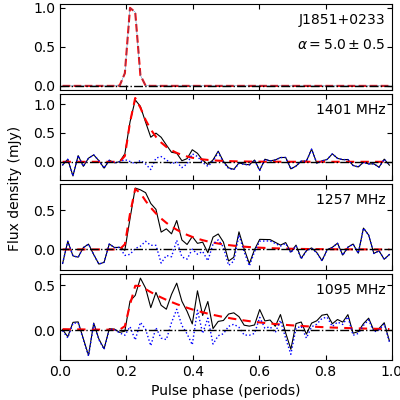}
\caption{--- {\it to be continued.}}
\end{figure*}

\addtocounter{figure}{-1}
\begin{figure*}
\centering
\includegraphics[width=0.24\textwidth,height=0.24\textwidth]{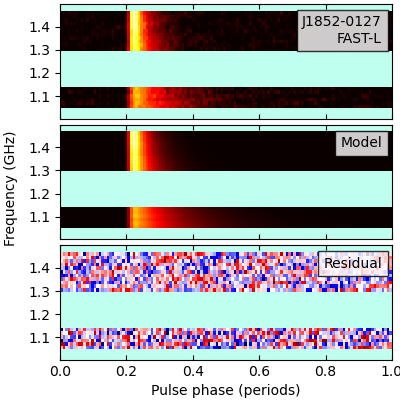}
\includegraphics[width=0.24\textwidth,height=0.24\textwidth]{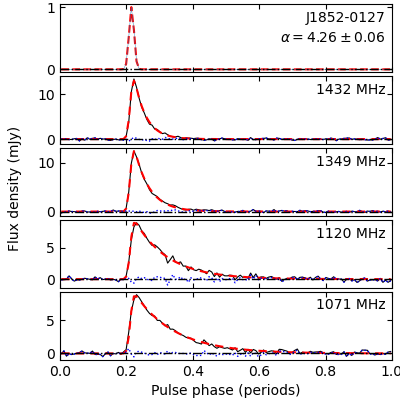}
\includegraphics[width=0.24\textwidth,height=0.24\textwidth]{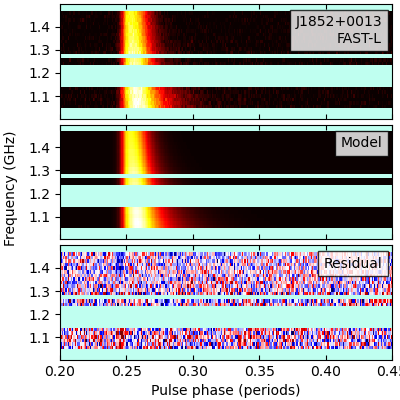}
\includegraphics[width=0.24\textwidth,height=0.24\textwidth]{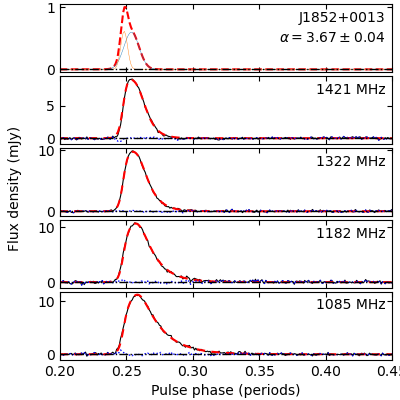}
\includegraphics[width=0.24\textwidth,height=0.24\textwidth]{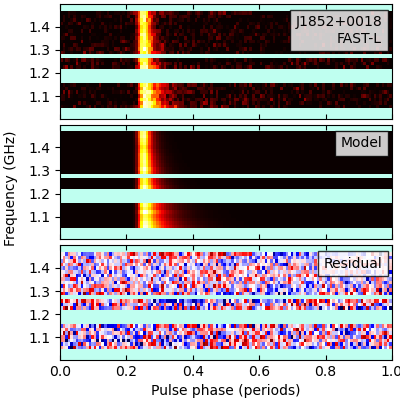}
\includegraphics[width=0.24\textwidth,height=0.24\textwidth]{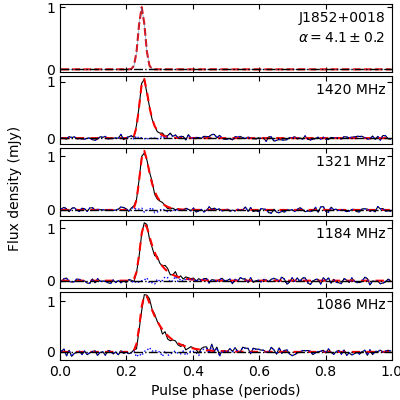}
\includegraphics[width=0.24\textwidth,height=0.24\textwidth]{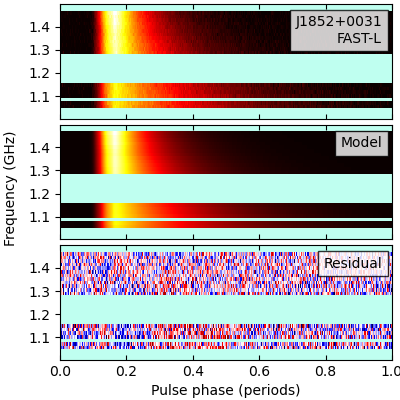}
\includegraphics[width=0.24\textwidth,height=0.24\textwidth]{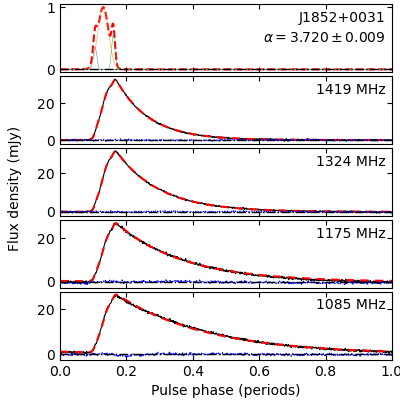}
\includegraphics[width=0.24\textwidth,height=0.24\textwidth]{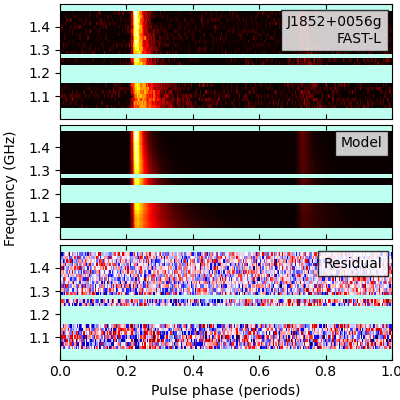}
\includegraphics[width=0.24\textwidth,height=0.24\textwidth]{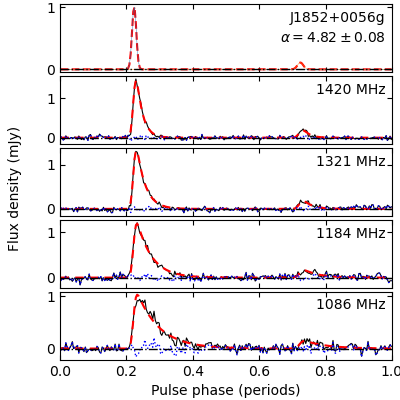}
\includegraphics[width=0.24\textwidth,height=0.24\textwidth]{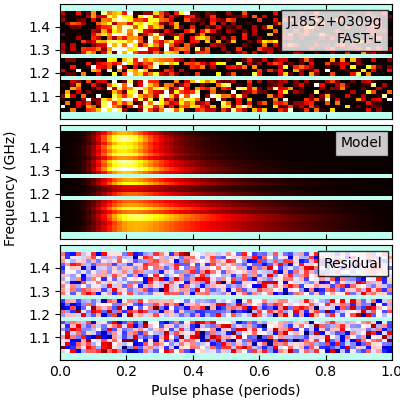}
\includegraphics[width=0.24\textwidth,height=0.24\textwidth]{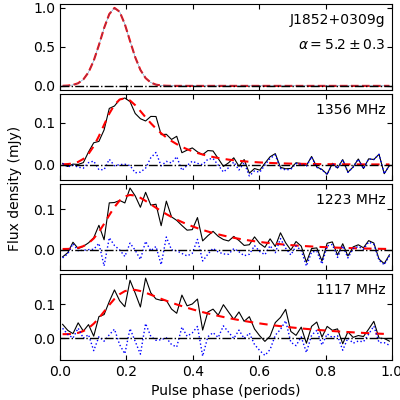}
\includegraphics[width=0.24\textwidth,height=0.24\textwidth]{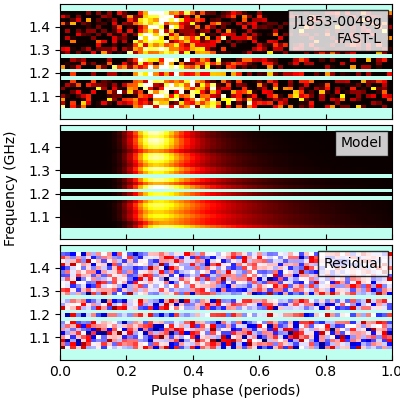}
\includegraphics[width=0.24\textwidth,height=0.24\textwidth]{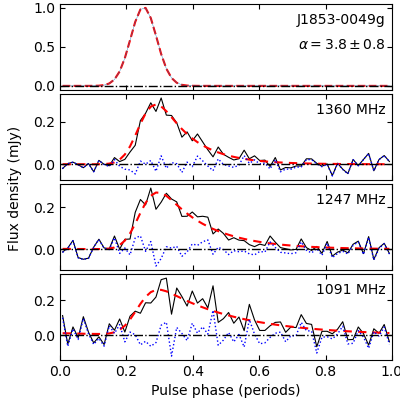}
\includegraphics[width=0.24\textwidth,height=0.24\textwidth]{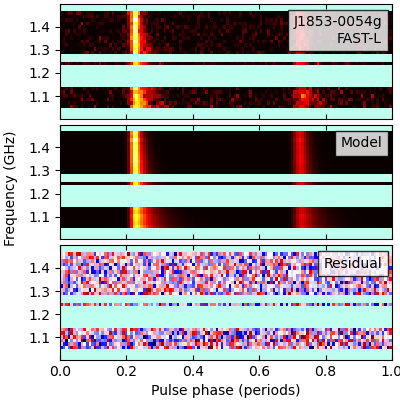}
\includegraphics[width=0.24\textwidth,height=0.24\textwidth]{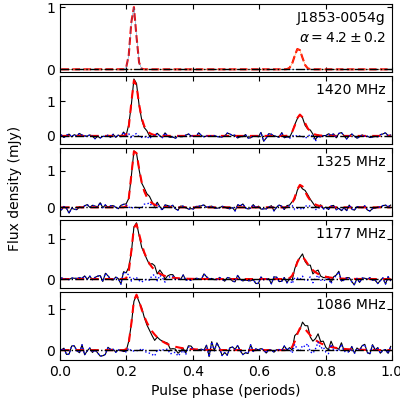}
\includegraphics[width=0.24\textwidth,height=0.24\textwidth]{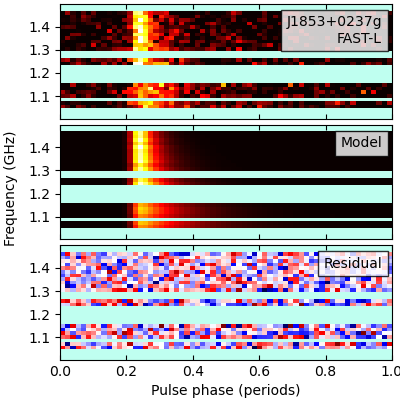}
\includegraphics[width=0.24\textwidth,height=0.24\textwidth]{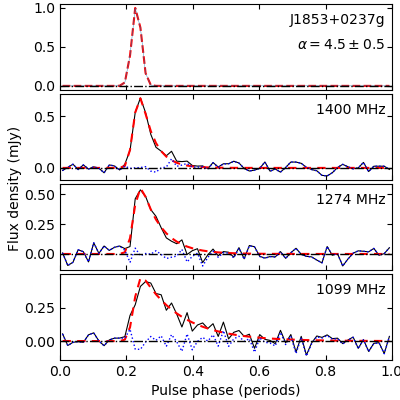}
\includegraphics[width=0.24\textwidth,height=0.24\textwidth]{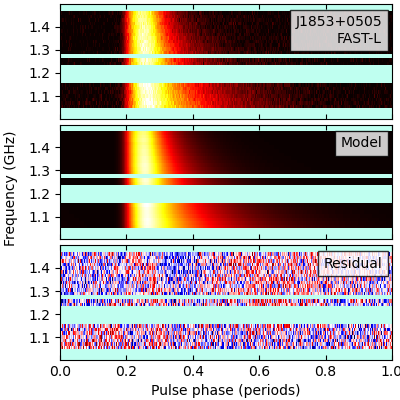}
\includegraphics[width=0.24\textwidth,height=0.24\textwidth]{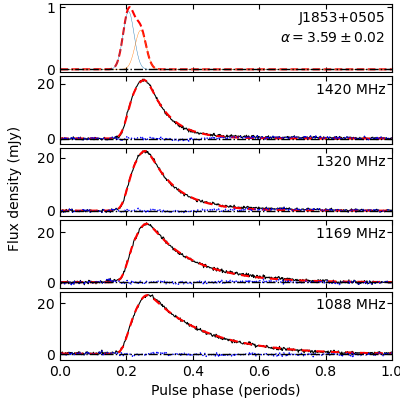}
\caption{--- {\it to be continued.}}
\end{figure*}

\addtocounter{figure}{-1}
\begin{figure*}
\centering
\includegraphics[width=0.24\textwidth,height=0.24\textwidth]{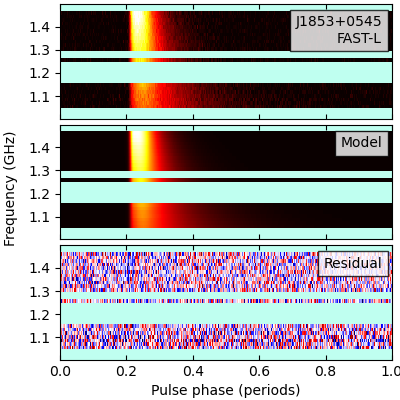}
\includegraphics[width=0.24\textwidth,height=0.24\textwidth]{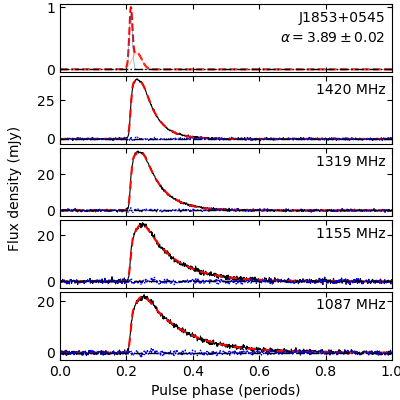}
\includegraphics[width=0.24\textwidth,height=0.24\textwidth]{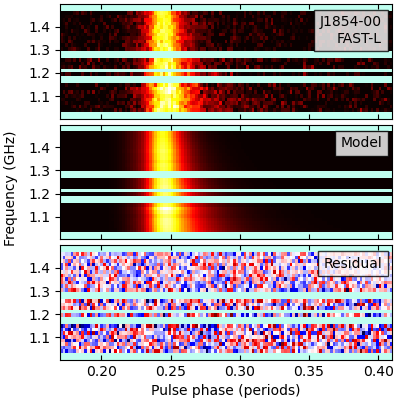}
\includegraphics[width=0.24\textwidth,height=0.24\textwidth]{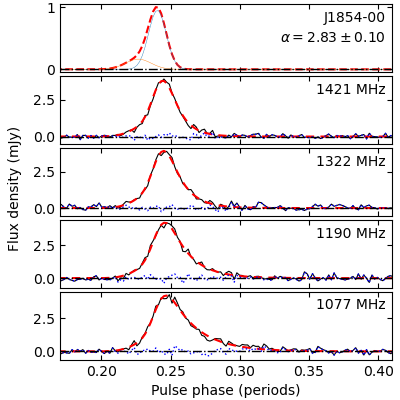}
\includegraphics[width=0.24\textwidth,height=0.24\textwidth]{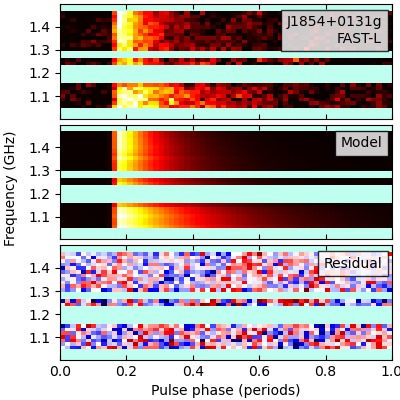}
\includegraphics[width=0.24\textwidth,height=0.24\textwidth]{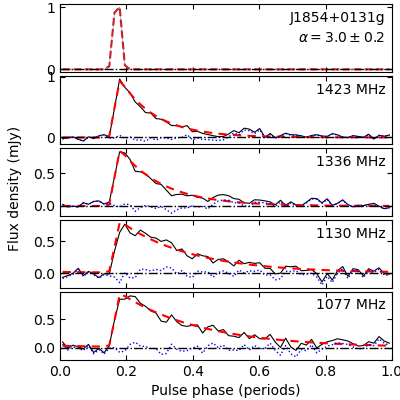}
\includegraphics[width=0.24\textwidth,height=0.24\textwidth]{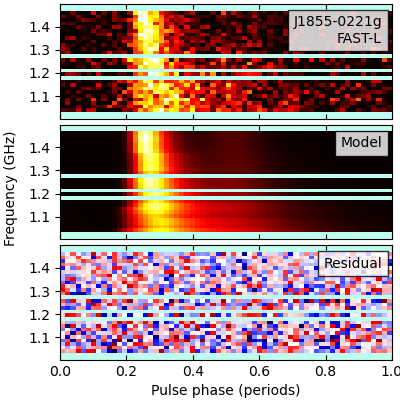}
\includegraphics[width=0.24\textwidth,height=0.24\textwidth]{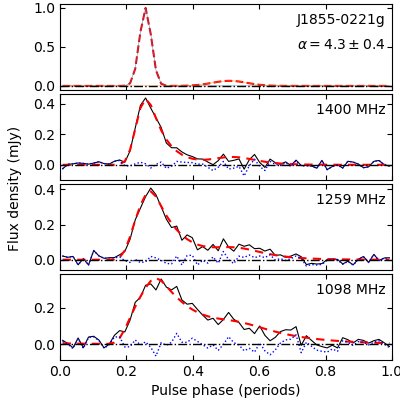}
\includegraphics[width=0.24\textwidth,height=0.24\textwidth]{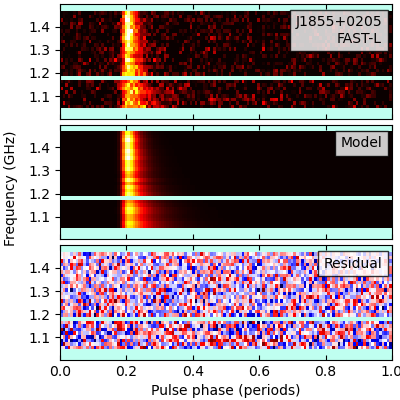}
\includegraphics[width=0.24\textwidth,height=0.24\textwidth]{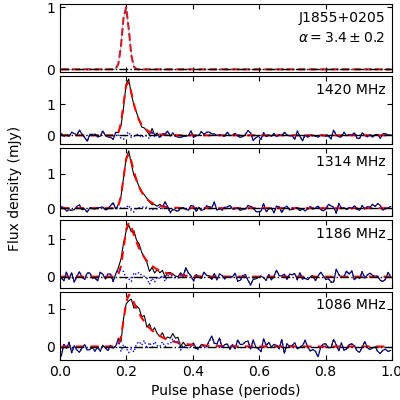}
\includegraphics[width=0.24\textwidth,height=0.24\textwidth]{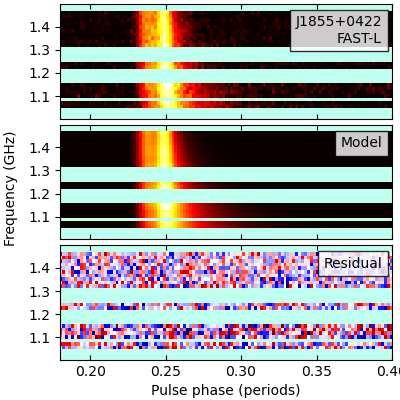}
\includegraphics[width=0.24\textwidth,height=0.24\textwidth]{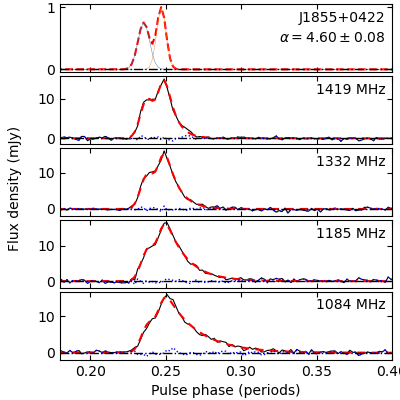}
\includegraphics[width=0.24\textwidth,height=0.24\textwidth]{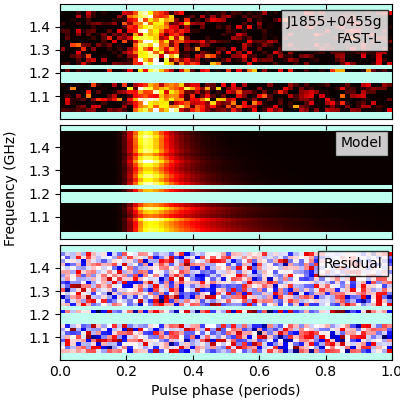}
\includegraphics[width=0.24\textwidth,height=0.24\textwidth]{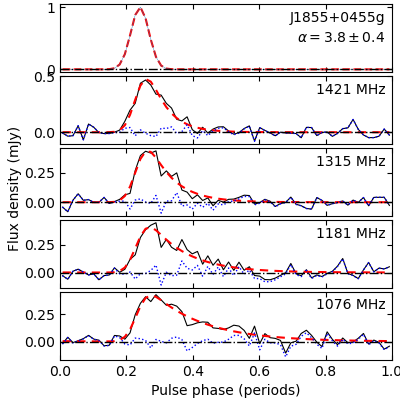}
\includegraphics[width=0.24\textwidth,height=0.24\textwidth]{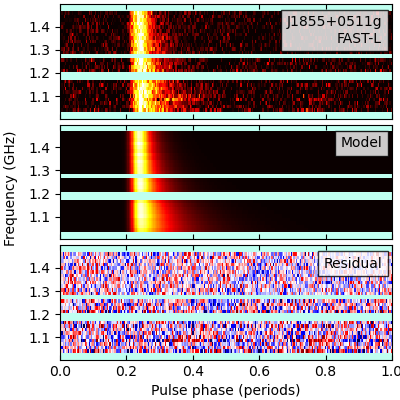}
\includegraphics[width=0.24\textwidth,height=0.24\textwidth]{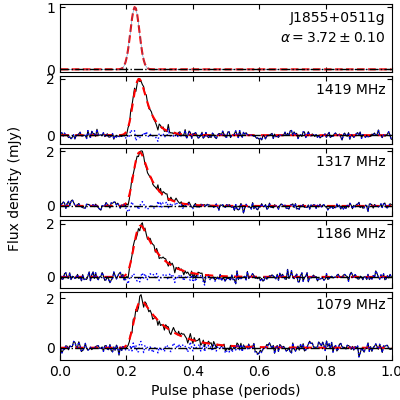}
\includegraphics[width=0.24\textwidth,height=0.24\textwidth]{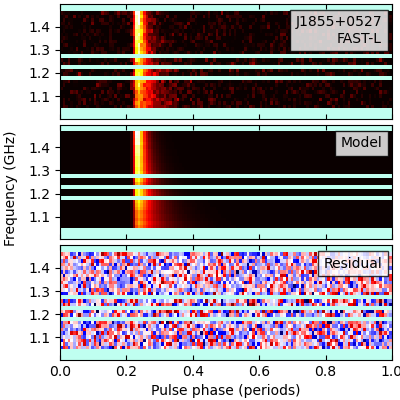}
\includegraphics[width=0.24\textwidth,height=0.24\textwidth]{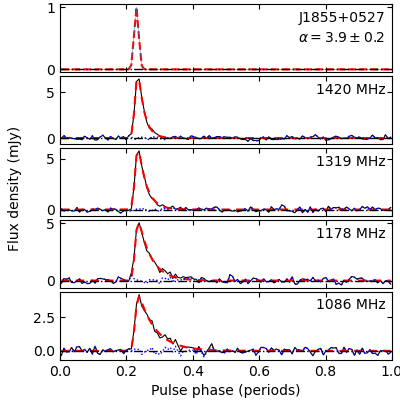}
\includegraphics[width=0.24\textwidth,height=0.24\textwidth]{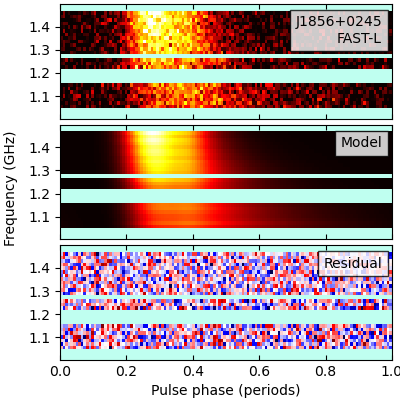}
\includegraphics[width=0.24\textwidth,height=0.24\textwidth]{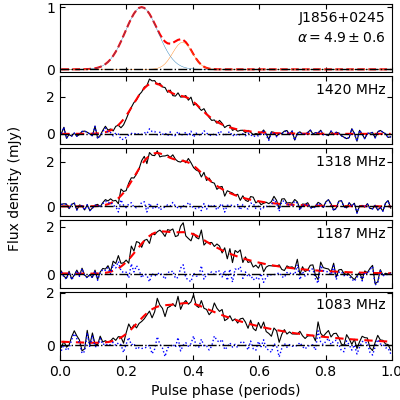}
\caption{--- {\it to be continued.}}
\end{figure*}

\addtocounter{figure}{-1}
\begin{figure*}
\centering
\includegraphics[width=0.24\textwidth,height=0.24\textwidth]{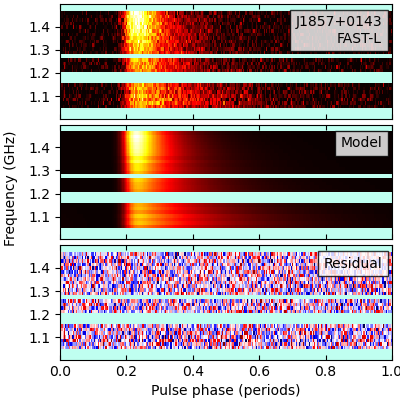}
\includegraphics[width=0.24\textwidth,height=0.24\textwidth]{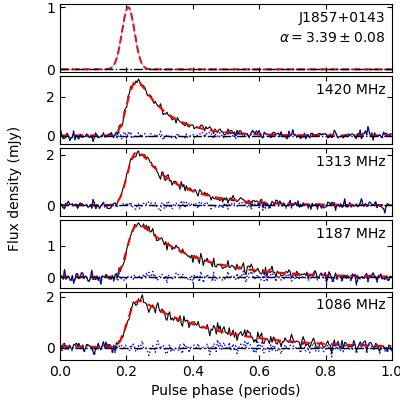}
\includegraphics[width=0.24\textwidth,height=0.24\textwidth]{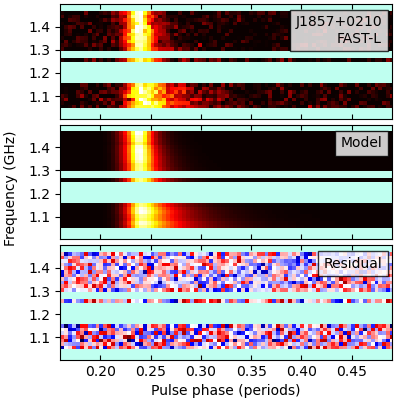}
\includegraphics[width=0.24\textwidth,height=0.24\textwidth]{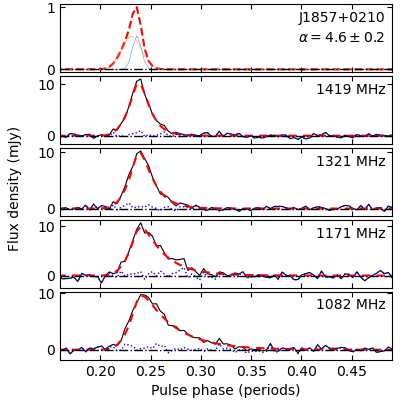}
\includegraphics[width=0.24\textwidth,height=0.24\textwidth]{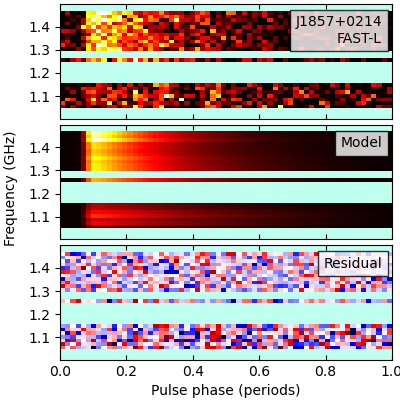}
\includegraphics[width=0.24\textwidth,height=0.24\textwidth]{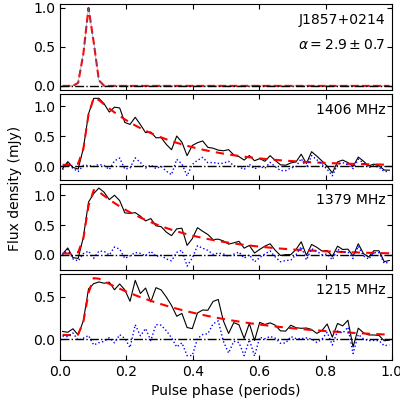}
\includegraphics[width=0.24\textwidth,height=0.24\textwidth]{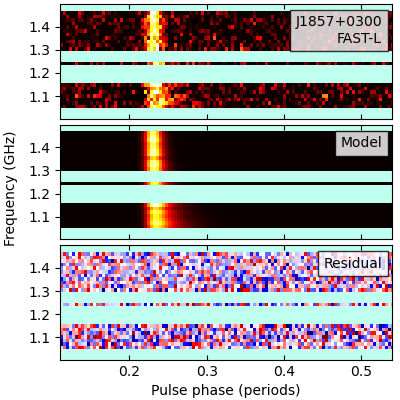}
\includegraphics[width=0.24\textwidth,height=0.24\textwidth]{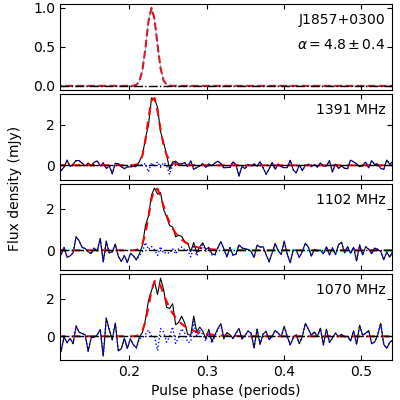}
\includegraphics[width=0.24\textwidth,height=0.24\textwidth]{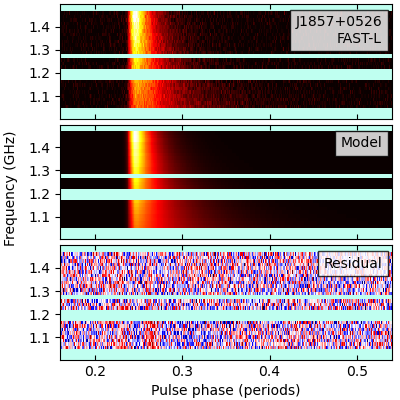}
\includegraphics[width=0.24\textwidth,height=0.24\textwidth]{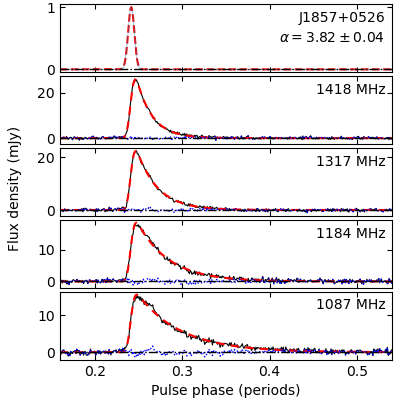}
\includegraphics[width=0.24\textwidth,height=0.24\textwidth]{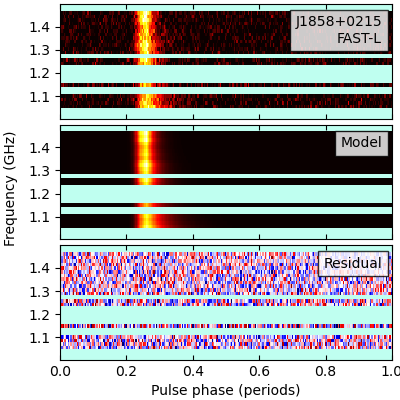}
\includegraphics[width=0.24\textwidth,height=0.24\textwidth]{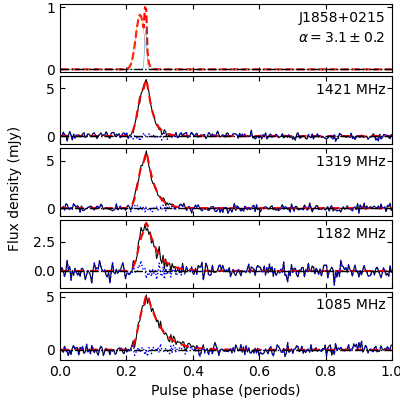}
\includegraphics[width=0.24\textwidth,height=0.24\textwidth]{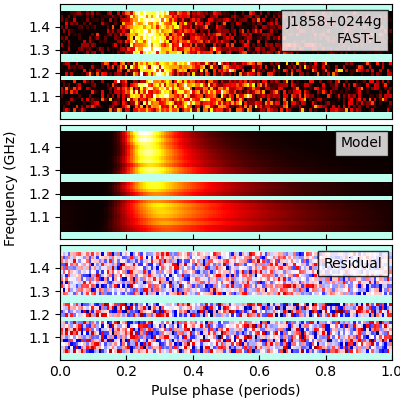}
\includegraphics[width=0.24\textwidth,height=0.24\textwidth]{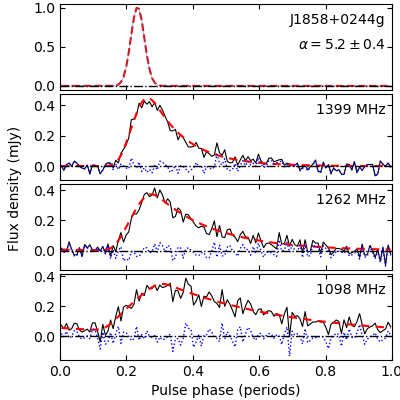}
\includegraphics[width=0.24\textwidth,height=0.24\textwidth]{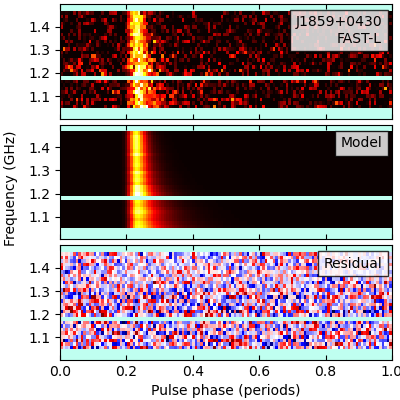}
\includegraphics[width=0.24\textwidth,height=0.24\textwidth]{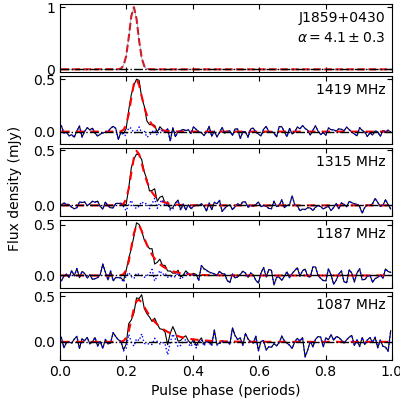}
\includegraphics[width=0.24\textwidth,height=0.24\textwidth]{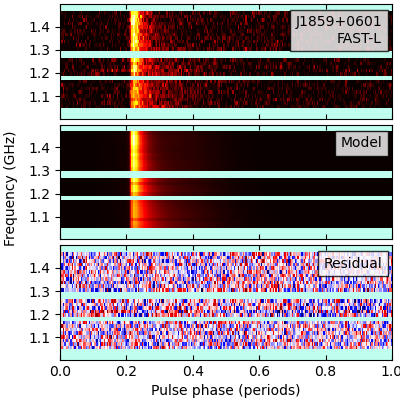}
\includegraphics[width=0.24\textwidth,height=0.24\textwidth]{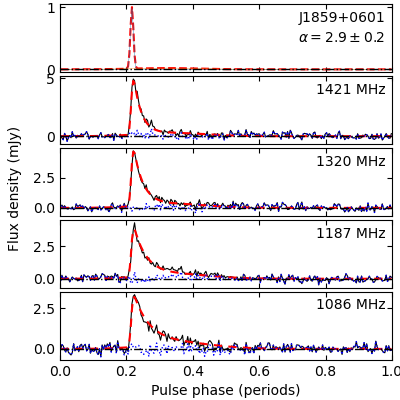}
\includegraphics[width=0.24\textwidth,height=0.24\textwidth]{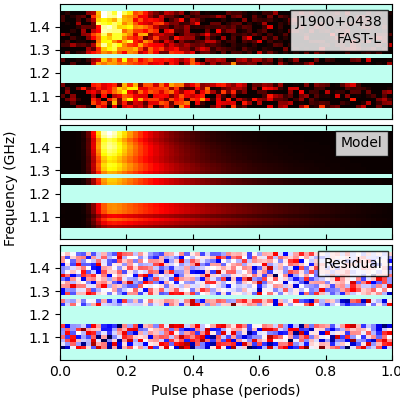}
\includegraphics[width=0.24\textwidth,height=0.24\textwidth]{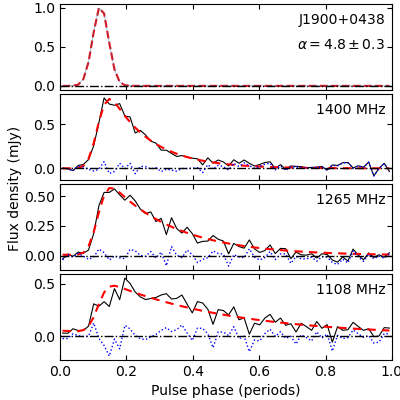}
\caption{--- {\it to be continued.}}
\end{figure*}

\addtocounter{figure}{-1}
\begin{figure*}
\centering
\includegraphics[width=0.24\textwidth,height=0.24\textwidth]{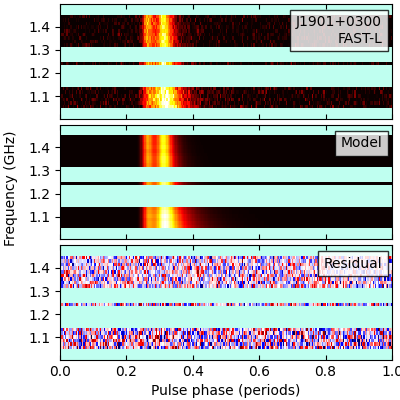}
\includegraphics[width=0.24\textwidth,height=0.24\textwidth]{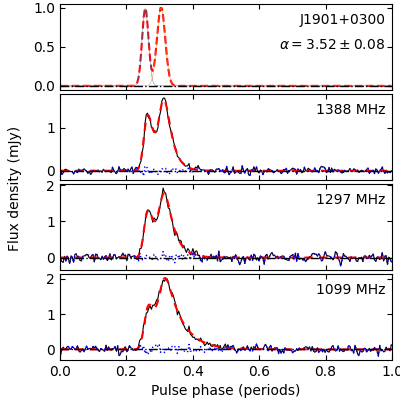}
\includegraphics[width=0.24\textwidth,height=0.24\textwidth]{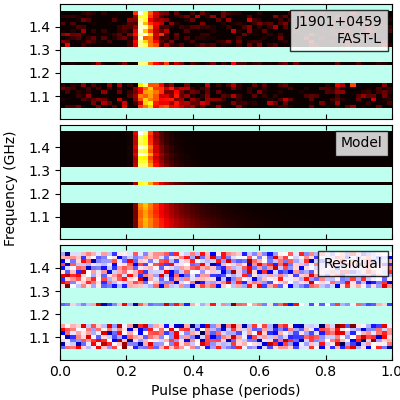}
\includegraphics[width=0.24\textwidth,height=0.24\textwidth]{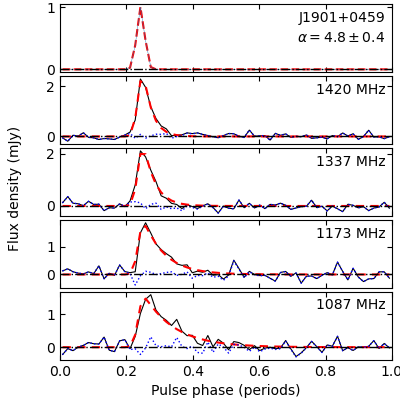}
\includegraphics[width=0.24\textwidth,height=0.24\textwidth]{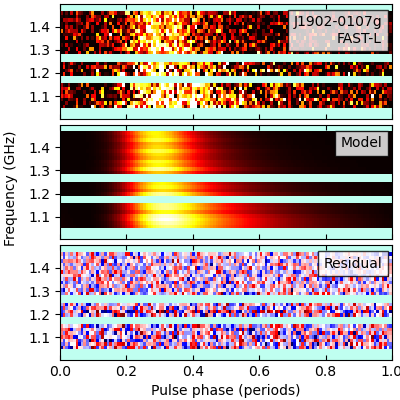}
\includegraphics[width=0.24\textwidth,height=0.24\textwidth]{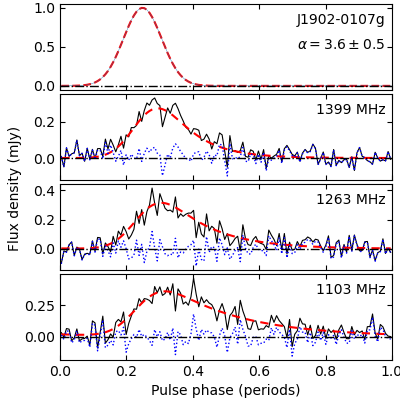}
\includegraphics[width=0.24\textwidth,height=0.24\textwidth]{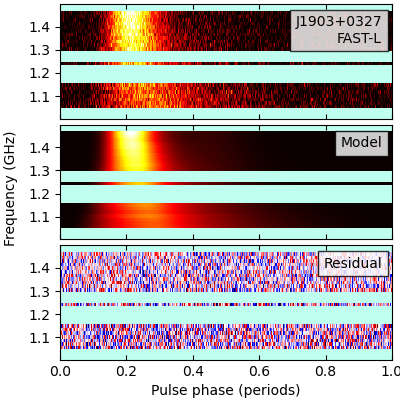}
\includegraphics[width=0.24\textwidth,height=0.24\textwidth]{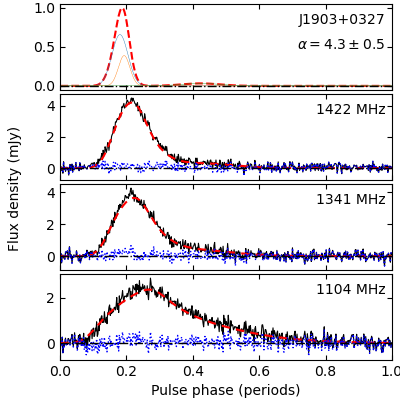}
\includegraphics[width=0.24\textwidth,height=0.24\textwidth]{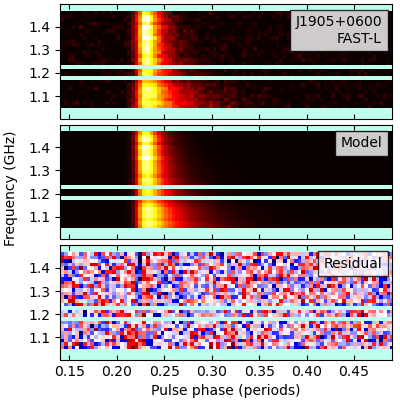}
\includegraphics[width=0.24\textwidth,height=0.24\textwidth]{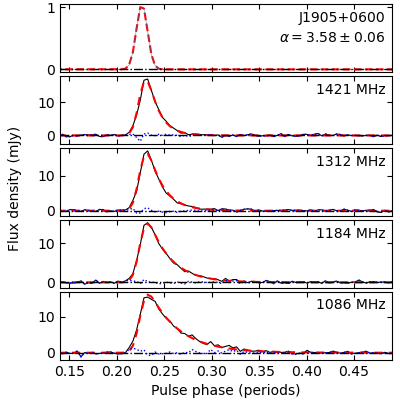}
\includegraphics[width=0.24\textwidth,height=0.24\textwidth]{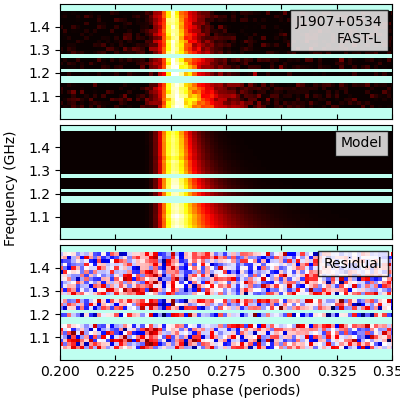}
\includegraphics[width=0.24\textwidth,height=0.24\textwidth]{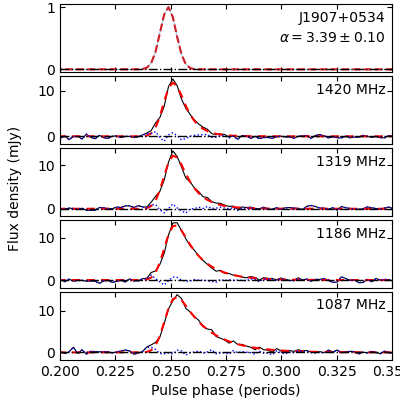}
\includegraphics[width=0.24\textwidth,height=0.24\textwidth]{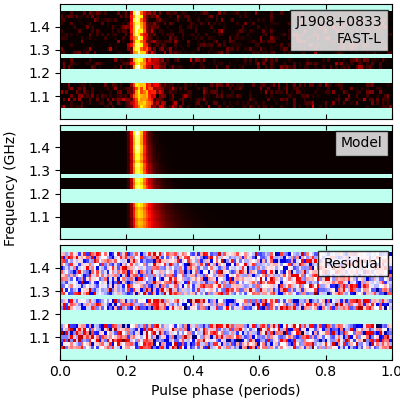}
\includegraphics[width=0.24\textwidth,height=0.24\textwidth]{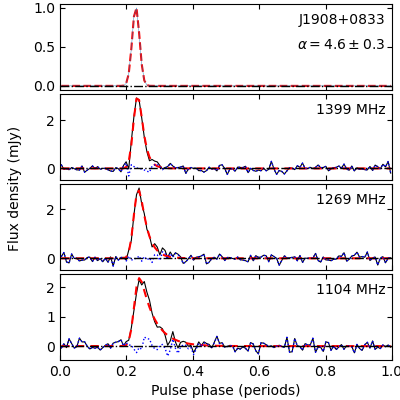}
\includegraphics[width=0.24\textwidth,height=0.24\textwidth]{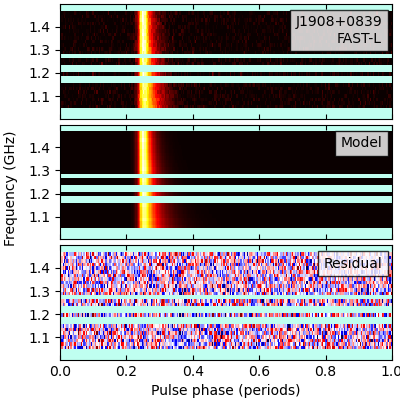}
\includegraphics[width=0.24\textwidth,height=0.24\textwidth]{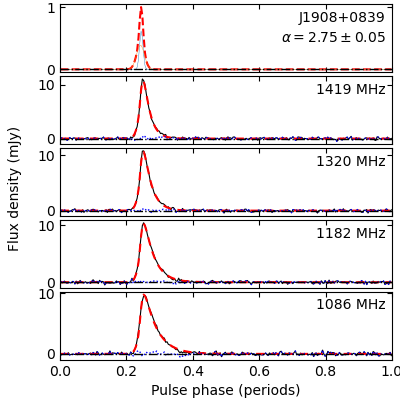}
\includegraphics[width=0.24\textwidth,height=0.24\textwidth]{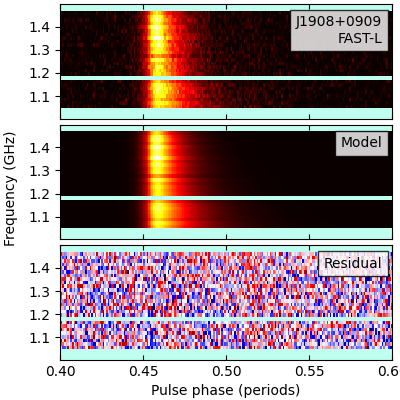}
\includegraphics[width=0.24\textwidth,height=0.24\textwidth]{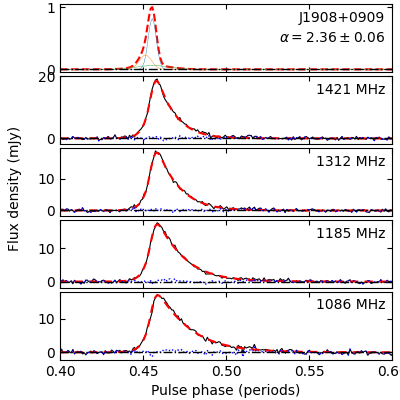}
\includegraphics[width=0.24\textwidth,height=0.24\textwidth]{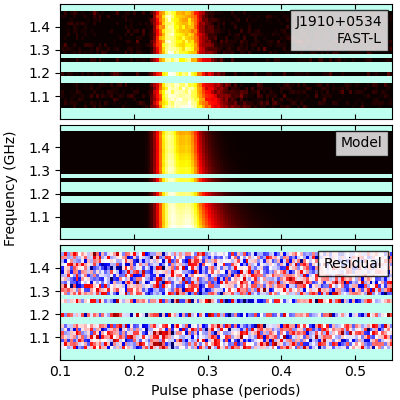}
\includegraphics[width=0.24\textwidth,height=0.24\textwidth]{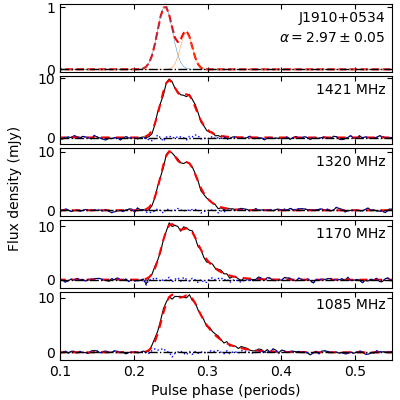}
\caption{--- {\it to be continued.}}
\end{figure*}

\addtocounter{figure}{-1}
\begin{figure*}
\centering
\includegraphics[width=0.24\textwidth,height=0.24\textwidth]{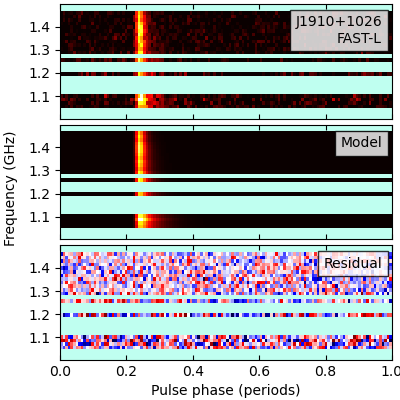}
\includegraphics[width=0.24\textwidth,height=0.24\textwidth]{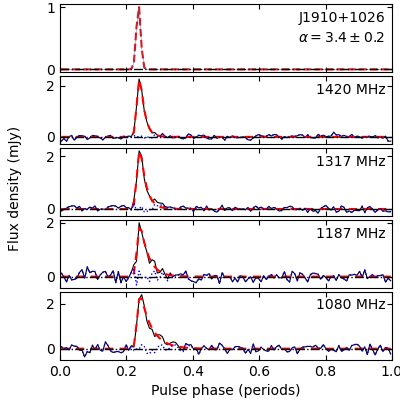}
\includegraphics[width=0.24\textwidth,height=0.24\textwidth]{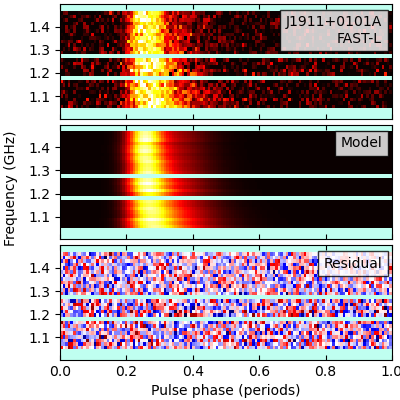}
\includegraphics[width=0.24\textwidth,height=0.24\textwidth]{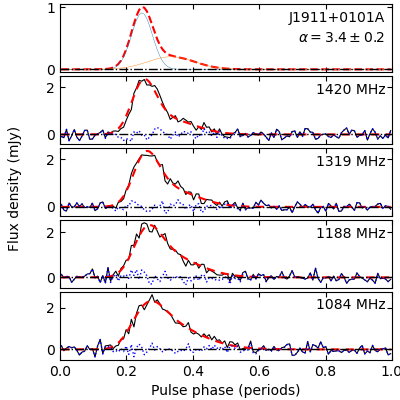}
\includegraphics[width=0.24\textwidth,height=0.24\textwidth]{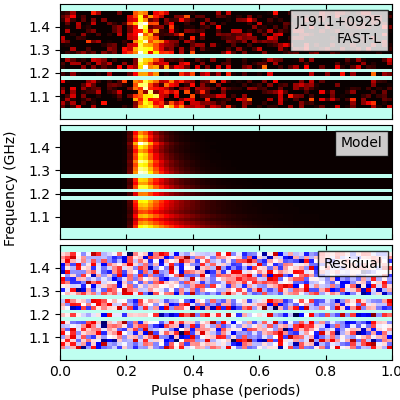}
\includegraphics[width=0.24\textwidth,height=0.24\textwidth]{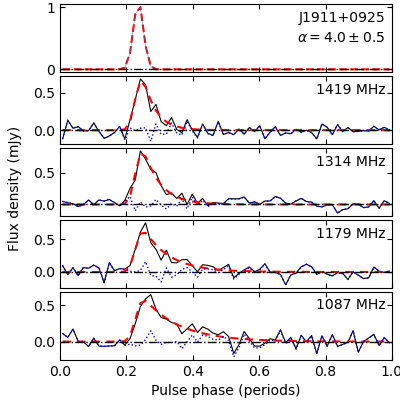}
\includegraphics[width=0.24\textwidth,height=0.24\textwidth]{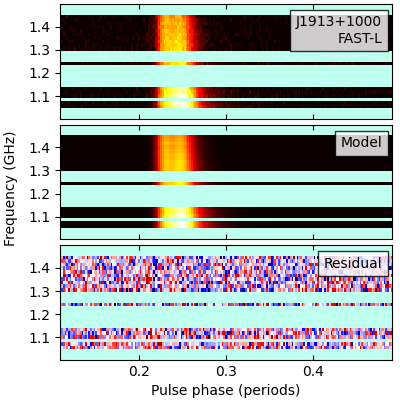}
\includegraphics[width=0.24\textwidth,height=0.24\textwidth]{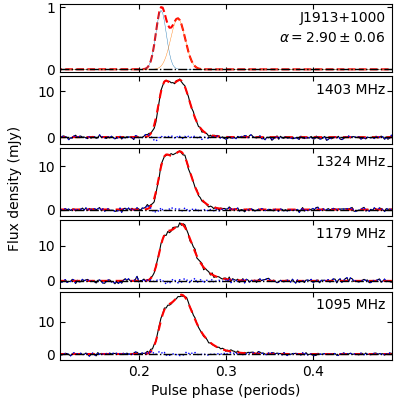}
\includegraphics[width=0.24\textwidth,height=0.24\textwidth]{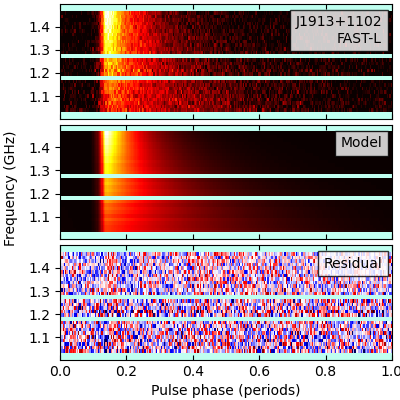}
\includegraphics[width=0.24\textwidth,height=0.24\textwidth]{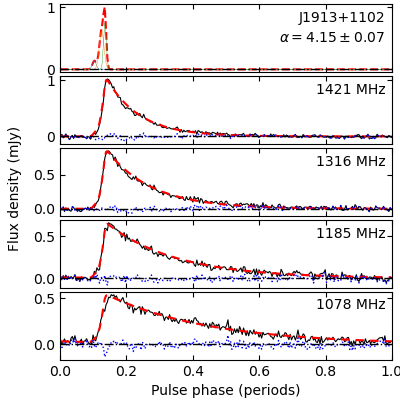}
\includegraphics[width=0.24\textwidth,height=0.24\textwidth]{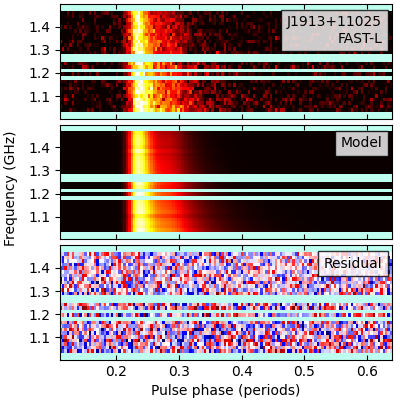}
\includegraphics[width=0.24\textwidth,height=0.24\textwidth]{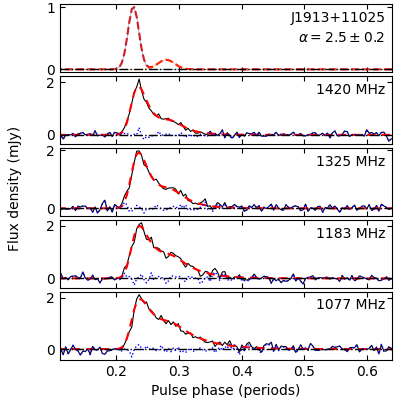}
\includegraphics[width=0.24\textwidth,height=0.24\textwidth]{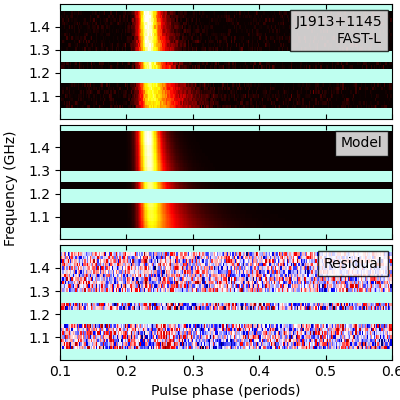}
\includegraphics[width=0.24\textwidth,height=0.24\textwidth]{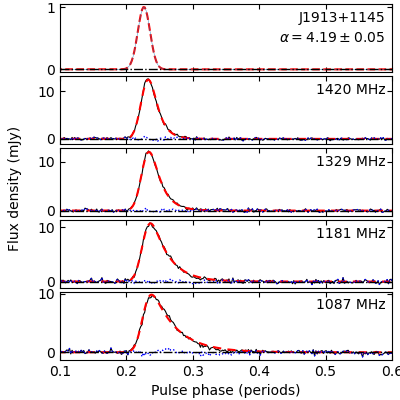}
\includegraphics[width=0.24\textwidth,height=0.24\textwidth]{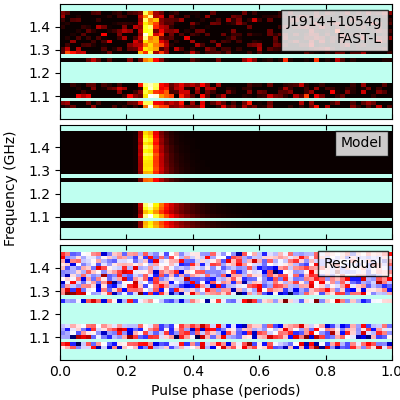}
\includegraphics[width=0.24\textwidth,height=0.24\textwidth]{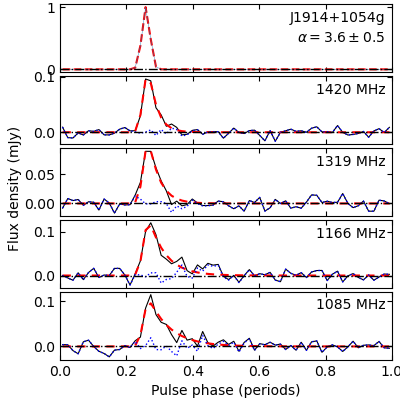}
\includegraphics[width=0.24\textwidth,height=0.24\textwidth]{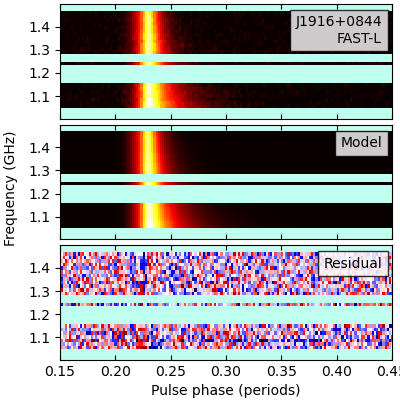}
\includegraphics[width=0.24\textwidth,height=0.24\textwidth]{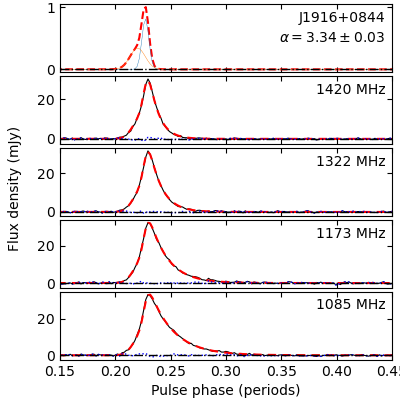}
\includegraphics[width=0.24\textwidth,height=0.24\textwidth]{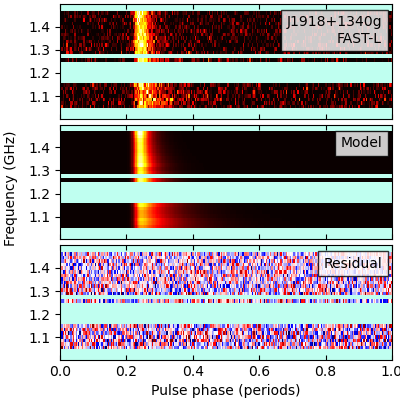}
\includegraphics[width=0.24\textwidth,height=0.24\textwidth]{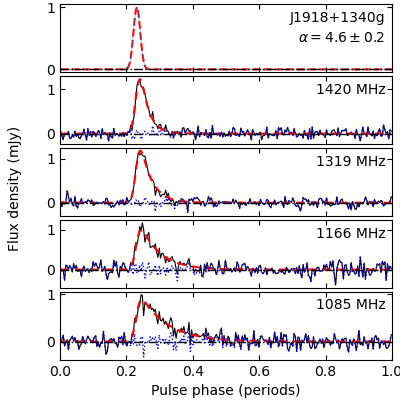}
\caption{--- {\it to be continued.}}
\end{figure*}

\addtocounter{figure}{-1}
\begin{figure*}
\centering
\includegraphics[width=0.24\textwidth,height=0.24\textwidth]{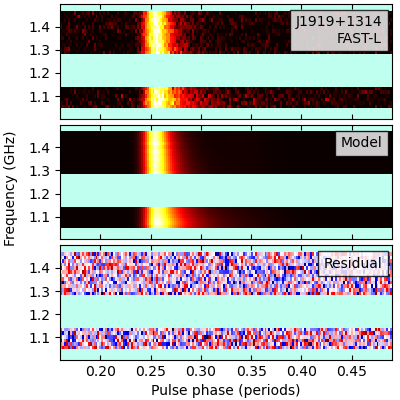}
\includegraphics[width=0.24\textwidth,height=0.24\textwidth]{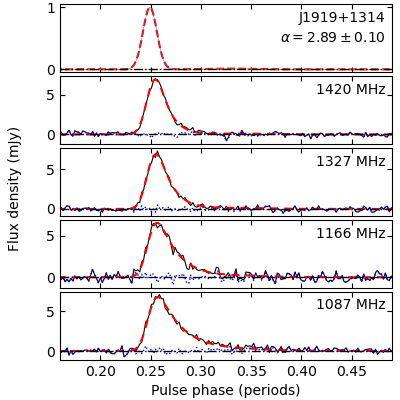}
\includegraphics[width=0.24\textwidth,height=0.24\textwidth]{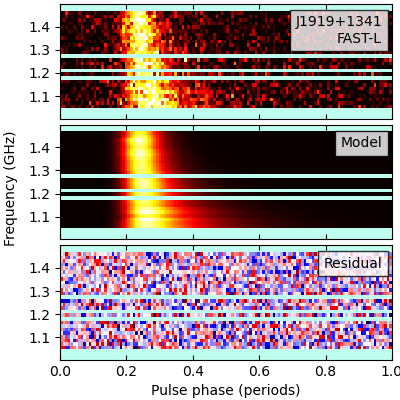}
\includegraphics[width=0.24\textwidth,height=0.24\textwidth]{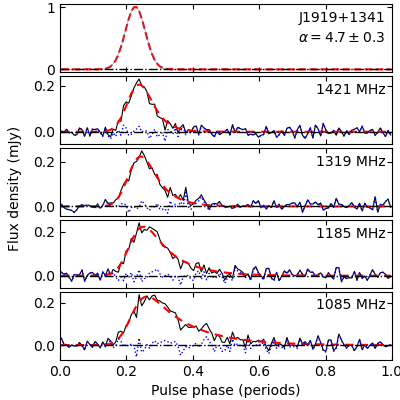}
\includegraphics[width=0.24\textwidth,height=0.24\textwidth]{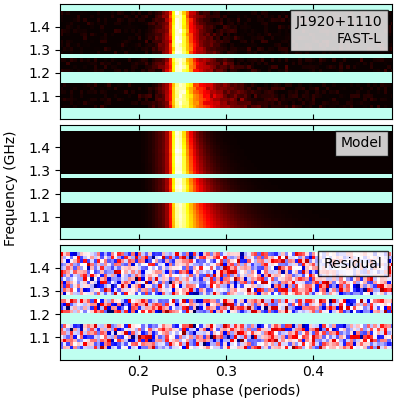}
\includegraphics[width=0.24\textwidth,height=0.24\textwidth]{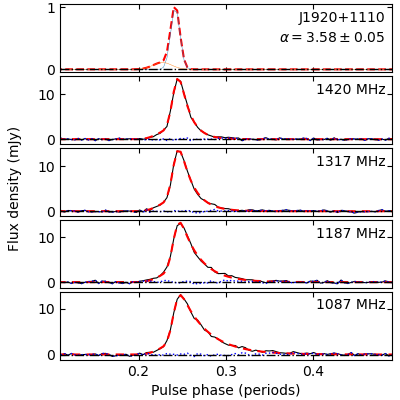}
\includegraphics[width=0.24\textwidth,height=0.24\textwidth]{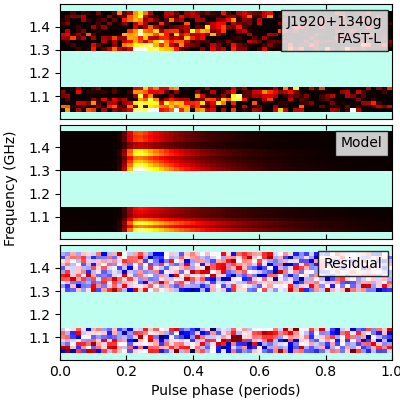}
\includegraphics[width=0.24\textwidth,height=0.24\textwidth]{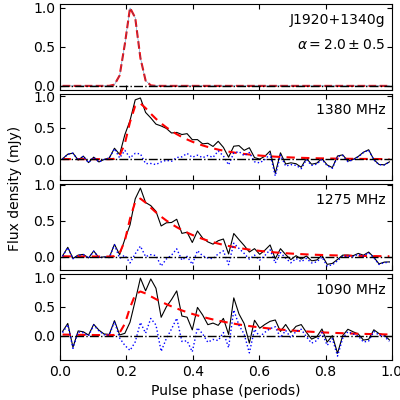}
\includegraphics[width=0.24\textwidth,height=0.24\textwidth]{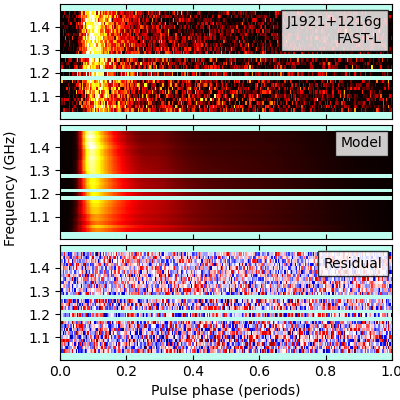}
\includegraphics[width=0.24\textwidth,height=0.24\textwidth]{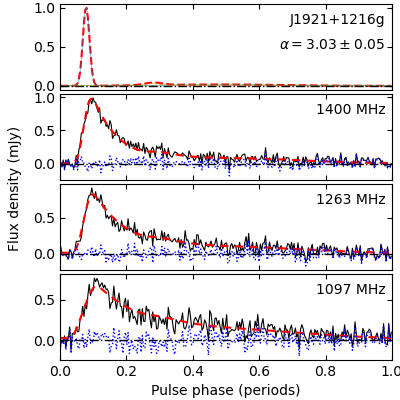}
\includegraphics[width=0.24\textwidth,height=0.24\textwidth]{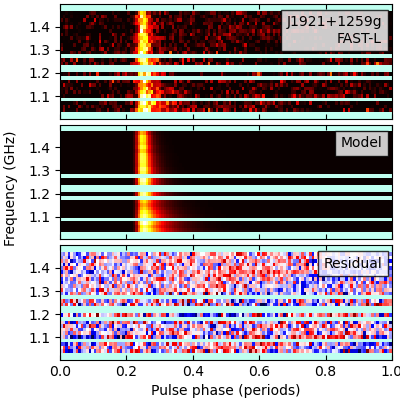}
\includegraphics[width=0.24\textwidth,height=0.24\textwidth]{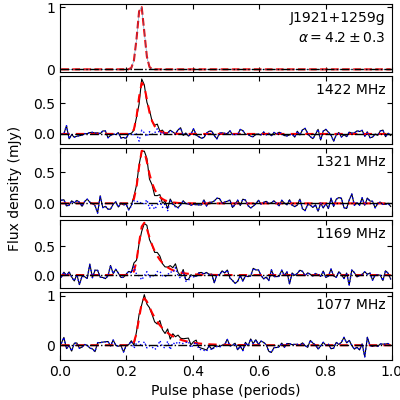}
\includegraphics[width=0.24\textwidth,height=0.24\textwidth]{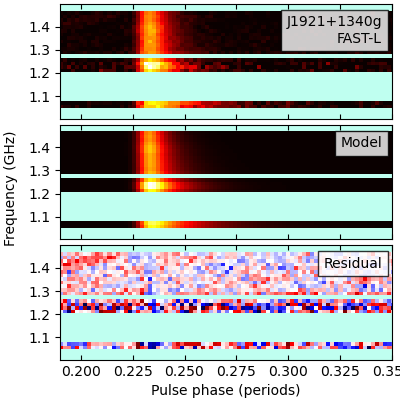}
\includegraphics[width=0.24\textwidth,height=0.24\textwidth]{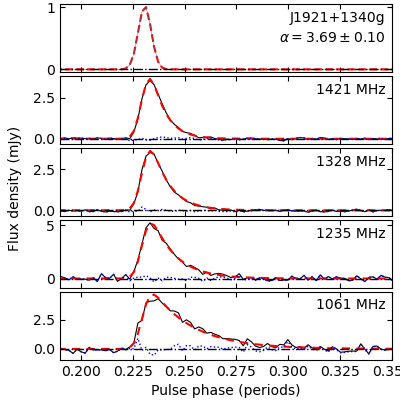}
\includegraphics[width=0.24\textwidth,height=0.24\textwidth]{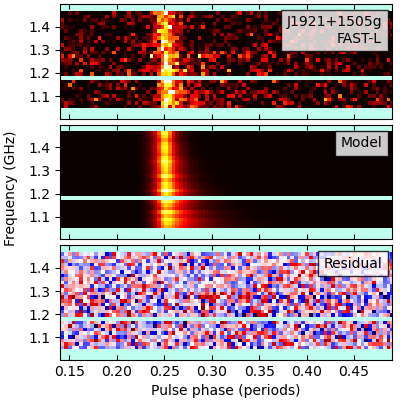}
\includegraphics[width=0.24\textwidth,height=0.24\textwidth]{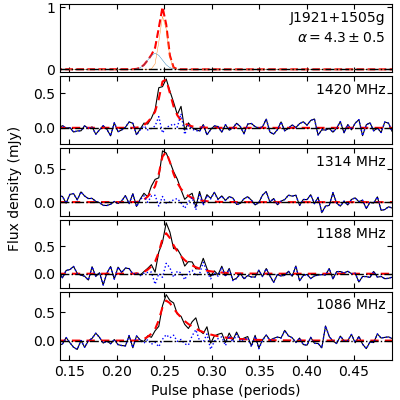}
\includegraphics[width=0.24\textwidth,height=0.24\textwidth]{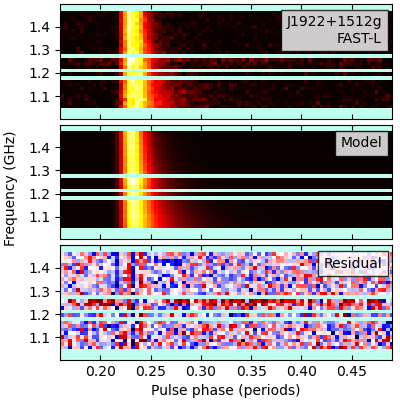}
\includegraphics[width=0.24\textwidth,height=0.24\textwidth]{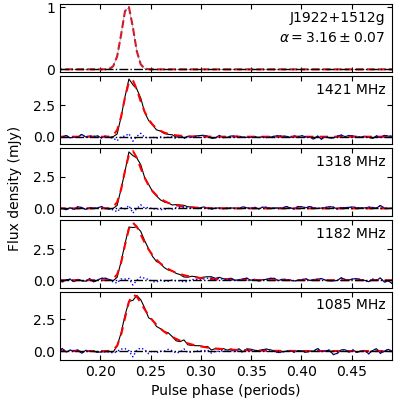}
\includegraphics[width=0.24\textwidth,height=0.24\textwidth]{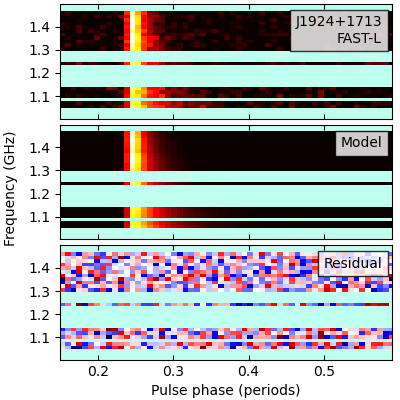}
\includegraphics[width=0.24\textwidth,height=0.24\textwidth]{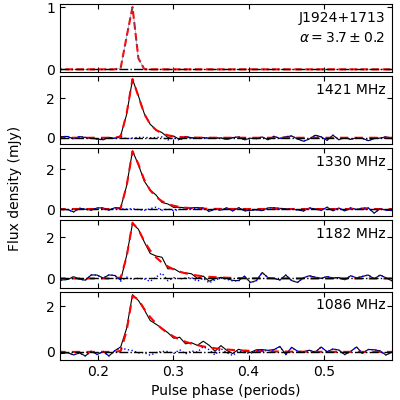}
\caption{--- {\it to be continued.}}
\end{figure*}

\addtocounter{figure}{-1}
\begin{figure*}
\centering
\includegraphics[width=0.24\textwidth,height=0.24\textwidth]{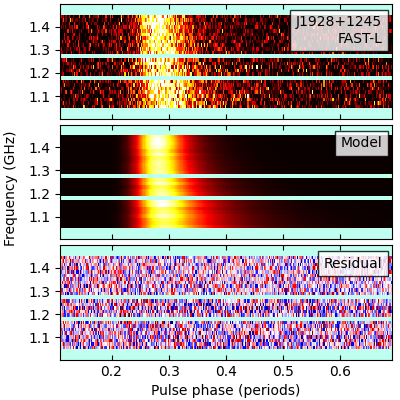}
\includegraphics[width=0.24\textwidth,height=0.24\textwidth]{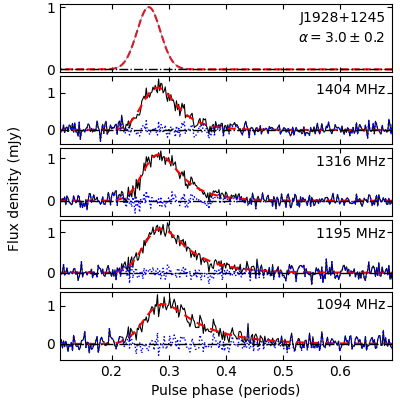}
\includegraphics[width=0.24\textwidth,height=0.24\textwidth]{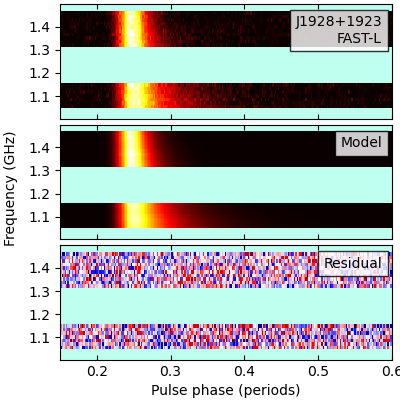}
\includegraphics[width=0.24\textwidth,height=0.24\textwidth]{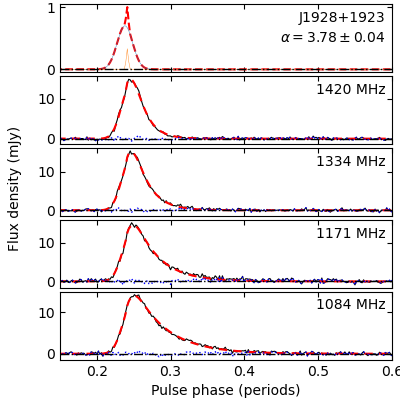}
\includegraphics[width=0.24\textwidth,height=0.24\textwidth]{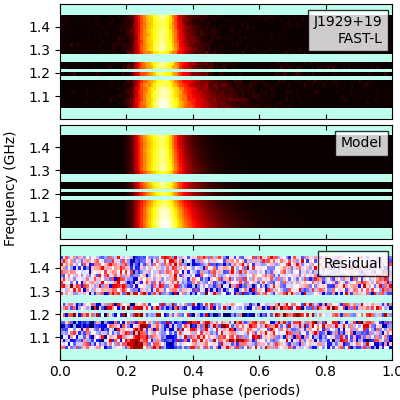}
\includegraphics[width=0.24\textwidth,height=0.24\textwidth]{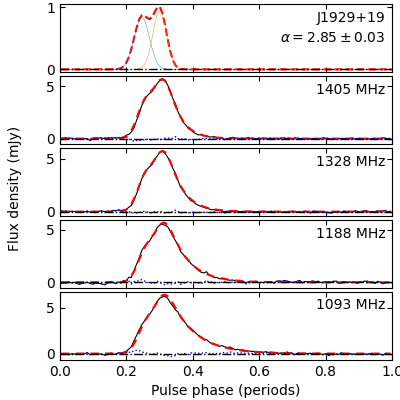}
\includegraphics[width=0.24\textwidth,height=0.24\textwidth]{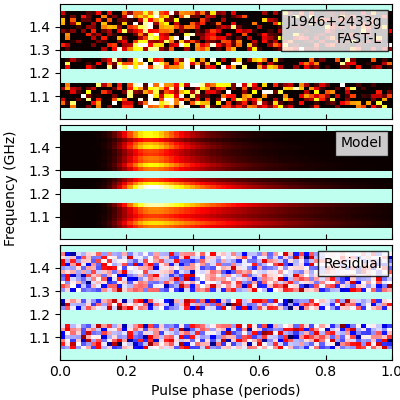}
\includegraphics[width=0.24\textwidth,height=0.24\textwidth]{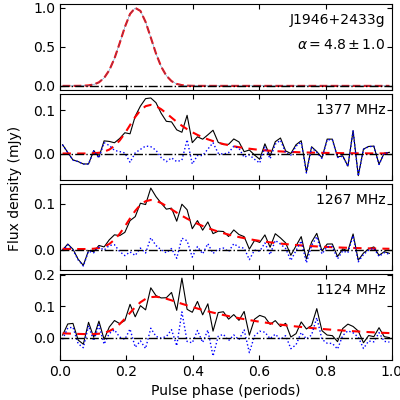}
\includegraphics[width=0.24\textwidth,height=0.24\textwidth]{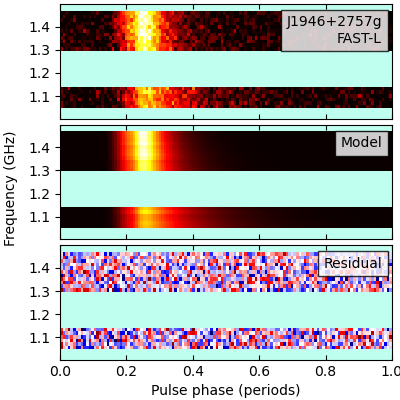}
\includegraphics[width=0.24\textwidth,height=0.24\textwidth]{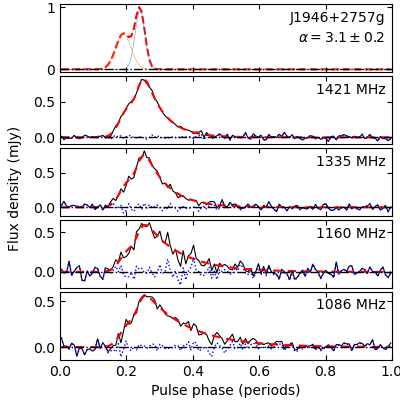}
\includegraphics[width=0.24\textwidth,height=0.24\textwidth]{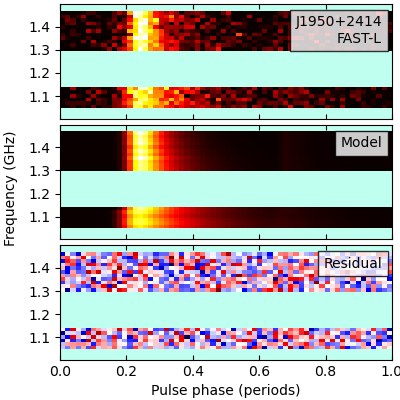}
\includegraphics[width=0.24\textwidth,height=0.24\textwidth]{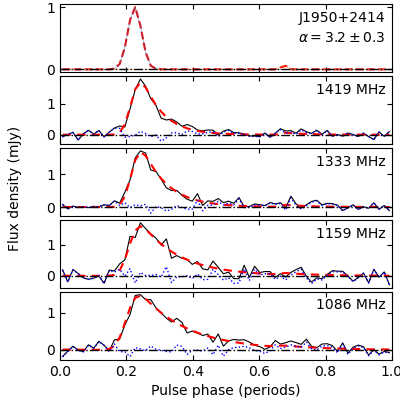}
\includegraphics[width=0.24\textwidth,height=0.24\textwidth]{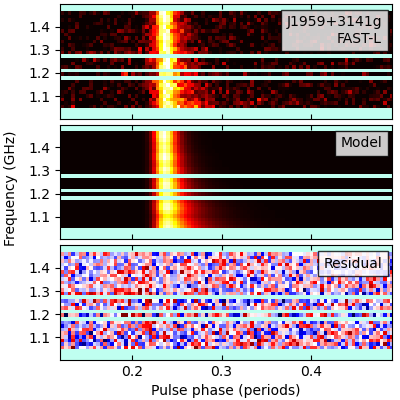}
\includegraphics[width=0.24\textwidth,height=0.24\textwidth]{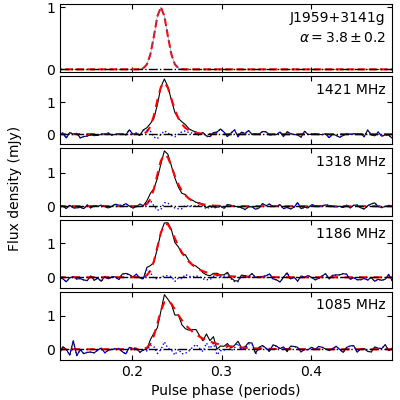}
\includegraphics[width=0.24\textwidth,height=0.24\textwidth]{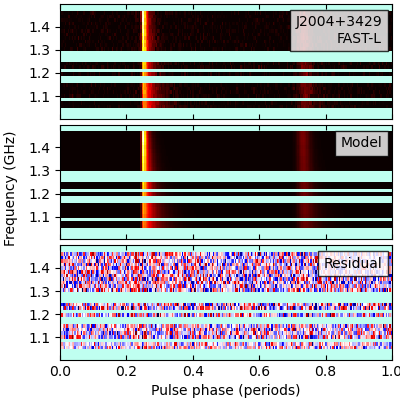}
\includegraphics[width=0.24\textwidth,height=0.24\textwidth]{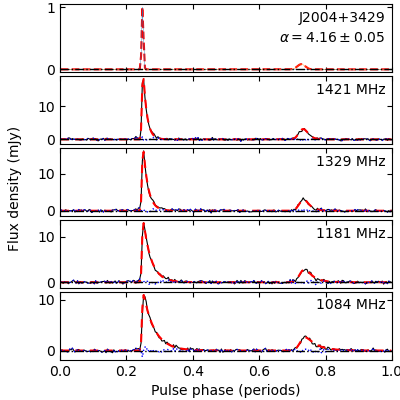}
\includegraphics[width=0.24\textwidth,height=0.24\textwidth]{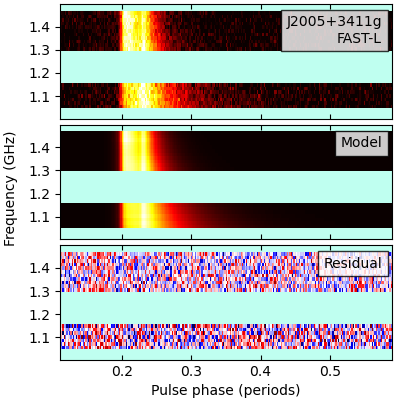}
\includegraphics[width=0.24\textwidth,height=0.24\textwidth]{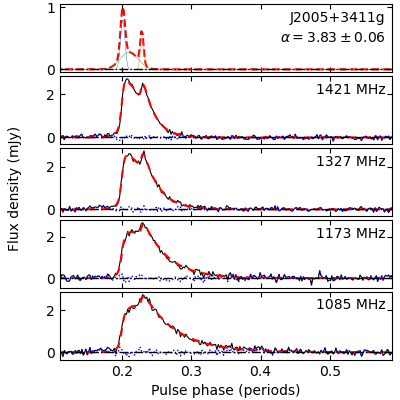}
\includegraphics[width=0.24\textwidth,height=0.24\textwidth]{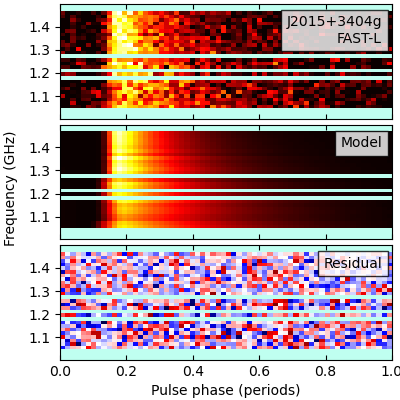}
\includegraphics[width=0.24\textwidth,height=0.24\textwidth]{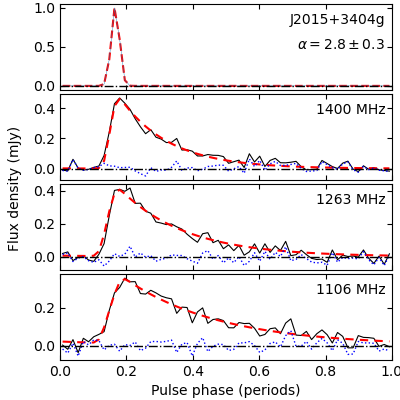}
\caption{--- {\it to be continued.}}
\end{figure*}

\addtocounter{figure}{-1}
\begin{figure*}
\centering
\includegraphics[width=0.24\textwidth,height=0.24\textwidth]{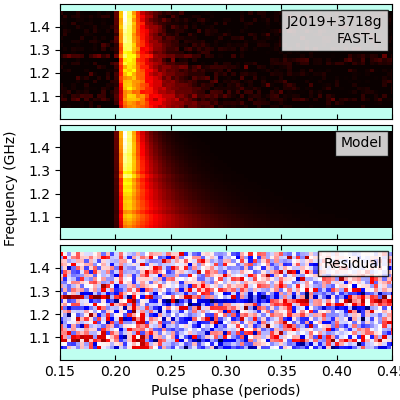}
\includegraphics[width=0.24\textwidth,height=0.24\textwidth]{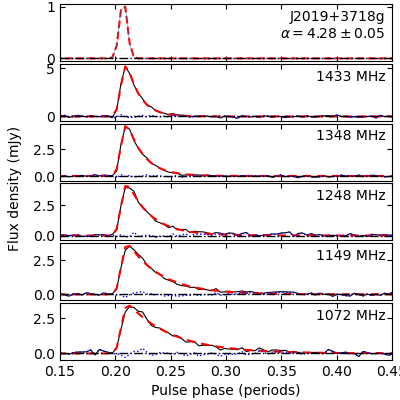}
\includegraphics[width=0.24\textwidth,height=0.24\textwidth]{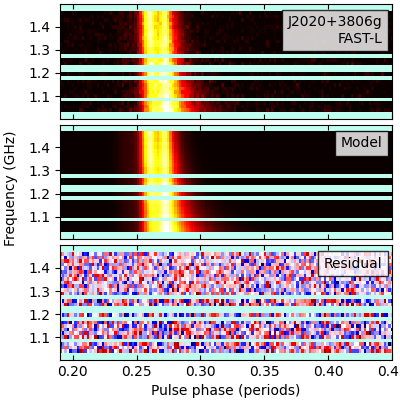}
\includegraphics[width=0.24\textwidth,height=0.24\textwidth]{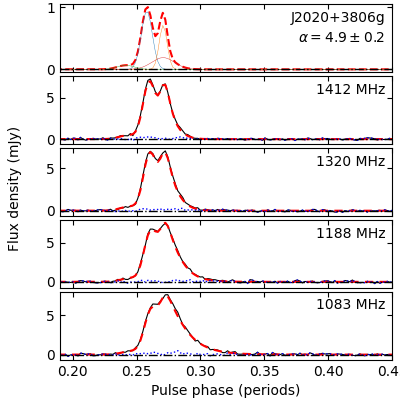}
\includegraphics[width=0.24\textwidth,height=0.24\textwidth]{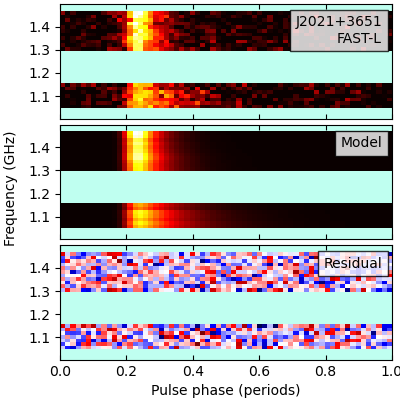}
\includegraphics[width=0.24\textwidth,height=0.24\textwidth]{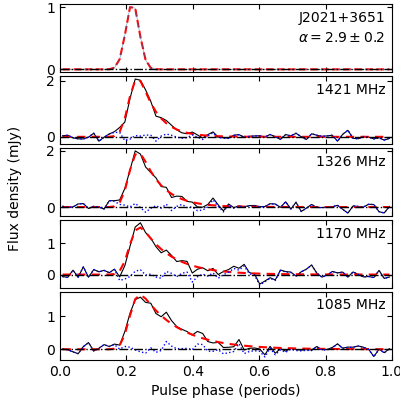}
\includegraphics[width=0.24\textwidth,height=0.24\textwidth]{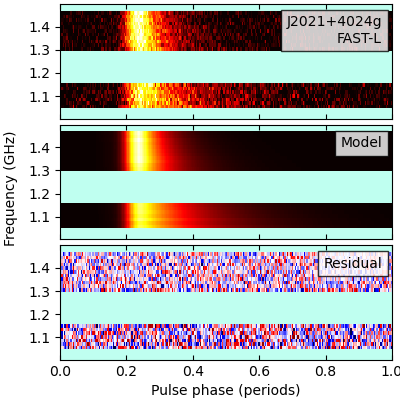}
\includegraphics[width=0.24\textwidth,height=0.24\textwidth]{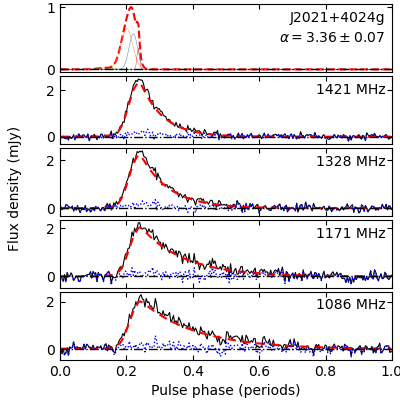}
\includegraphics[width=0.24\textwidth,height=0.24\textwidth]{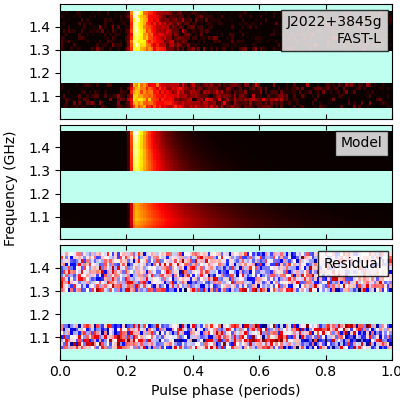}
\includegraphics[width=0.24\textwidth,height=0.24\textwidth]{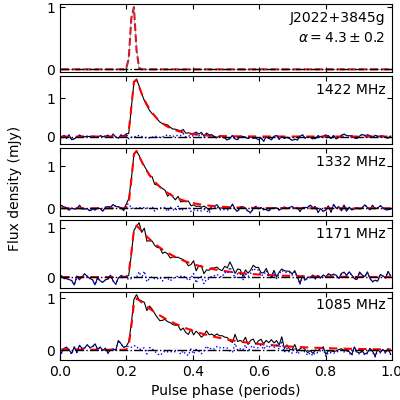}
\includegraphics[width=0.24\textwidth,height=0.24\textwidth]{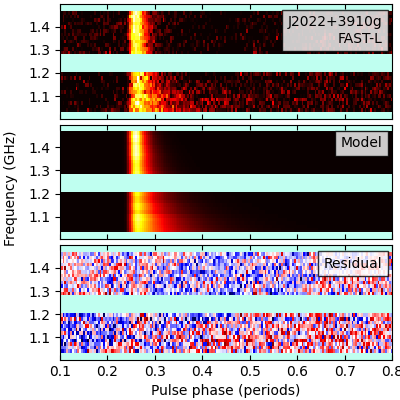}
\includegraphics[width=0.24\textwidth,height=0.24\textwidth]{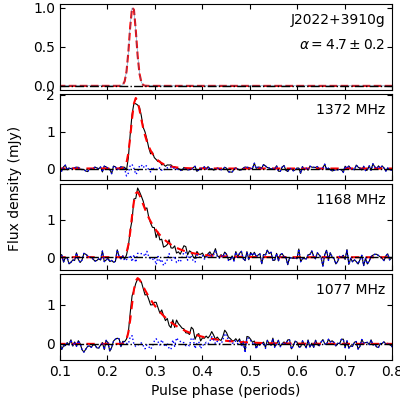}
\includegraphics[width=0.24\textwidth,height=0.24\textwidth]{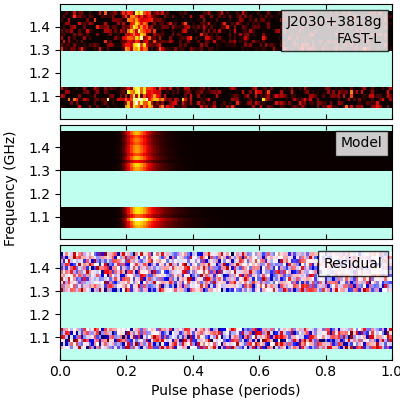}
\includegraphics[width=0.24\textwidth,height=0.24\textwidth]{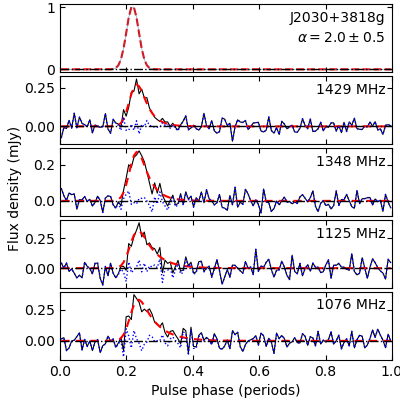}
\includegraphics[width=0.24\textwidth,height=0.24\textwidth]{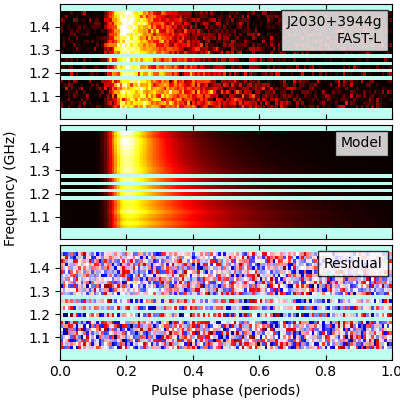}
\includegraphics[width=0.24\textwidth,height=0.24\textwidth]{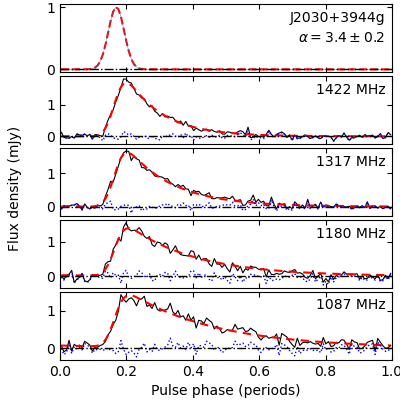}
\includegraphics[width=0.24\textwidth,height=0.24\textwidth]{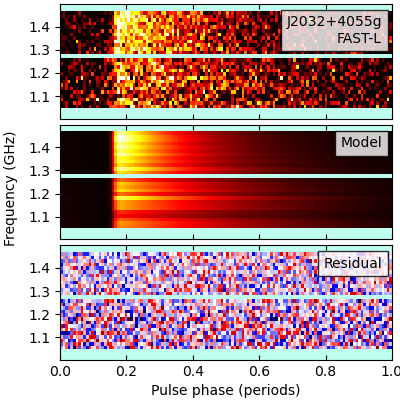}
\includegraphics[width=0.24\textwidth,height=0.24\textwidth]{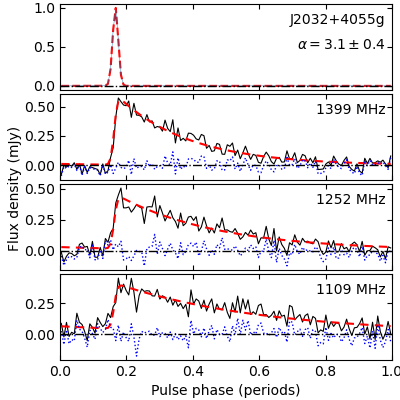}
\includegraphics[width=0.24\textwidth,height=0.24\textwidth]{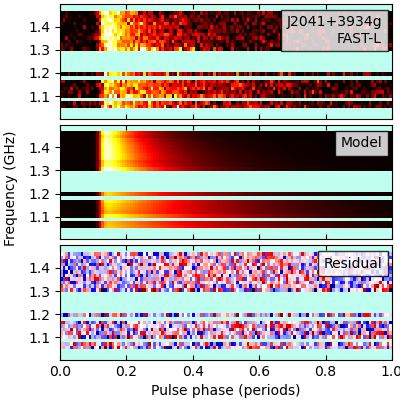}
\includegraphics[width=0.24\textwidth,height=0.24\textwidth]{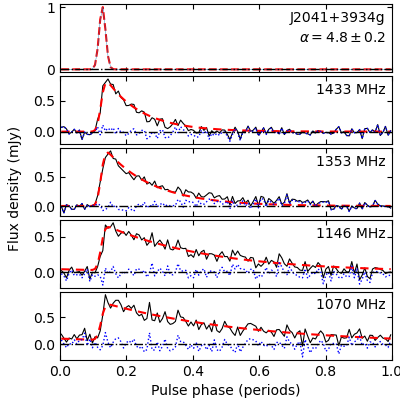}
\caption{--- {\it to be continued.}}
\end{figure*}

\addtocounter{figure}{-1}
\begin{figure*}
\centering
\includegraphics[width=0.24\textwidth,height=0.24\textwidth]{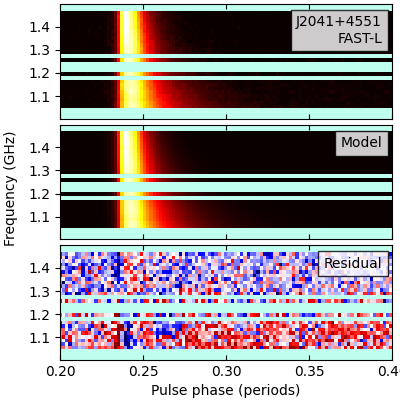}
\includegraphics[width=0.24\textwidth,height=0.24\textwidth]{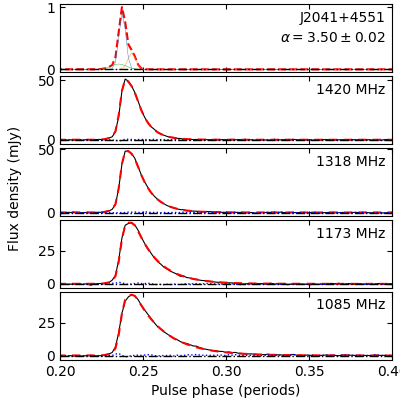}
\includegraphics[width=0.24\textwidth,height=0.24\textwidth]{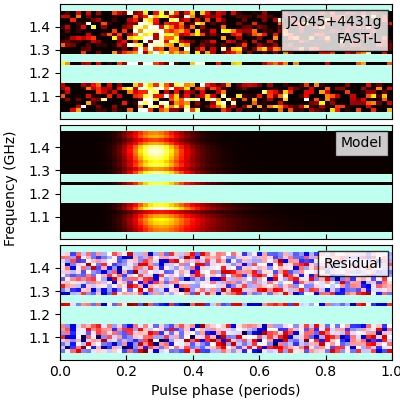}
\includegraphics[width=0.24\textwidth,height=0.24\textwidth]{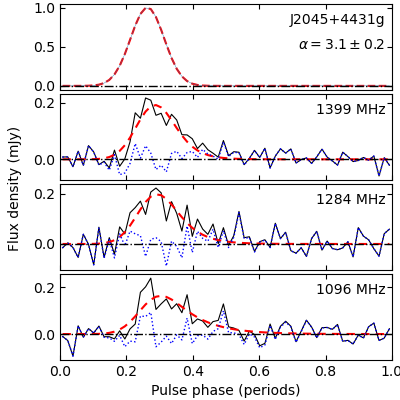}
\includegraphics[width=0.24\textwidth,height=0.24\textwidth]{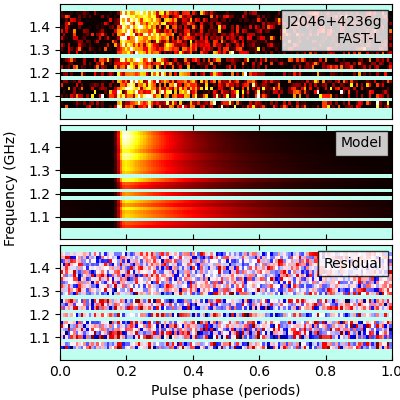}
\includegraphics[width=0.24\textwidth,height=0.24\textwidth]{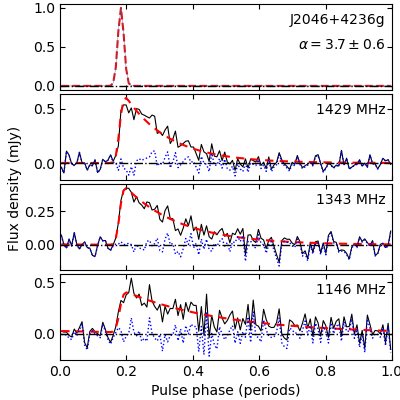}
\includegraphics[width=0.24\textwidth,height=0.24\textwidth]{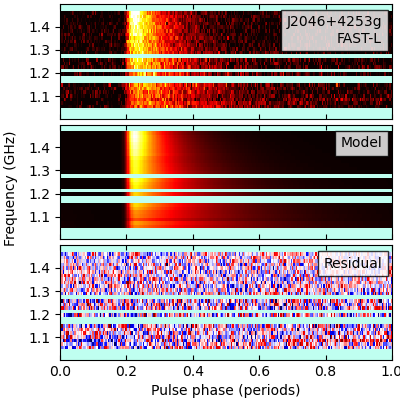}
\includegraphics[width=0.24\textwidth,height=0.24\textwidth]{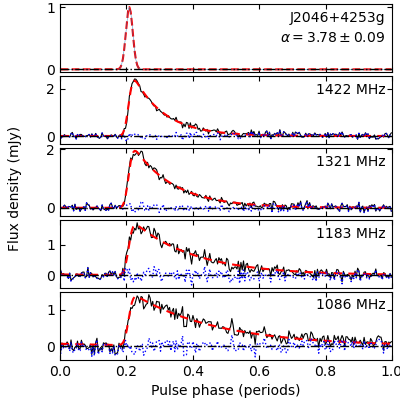}
\includegraphics[width=0.24\textwidth,height=0.24\textwidth]{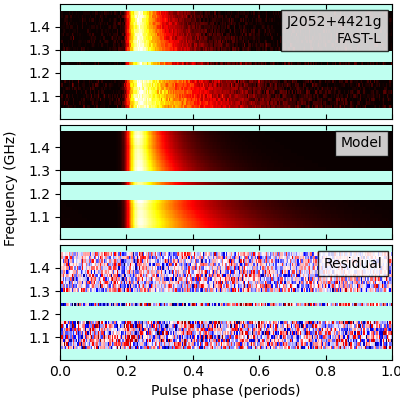}
\includegraphics[width=0.24\textwidth,height=0.24\textwidth]{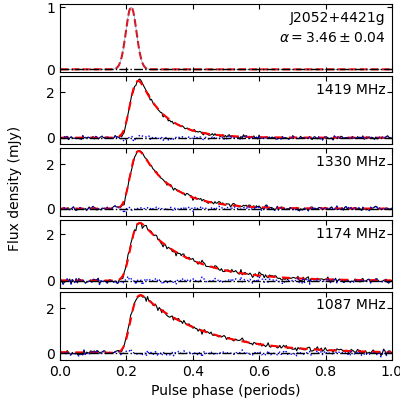}
\includegraphics[width=0.24\textwidth,height=0.24\textwidth]{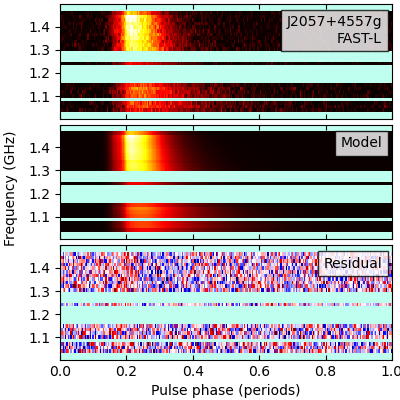}
\includegraphics[width=0.24\textwidth,height=0.24\textwidth]{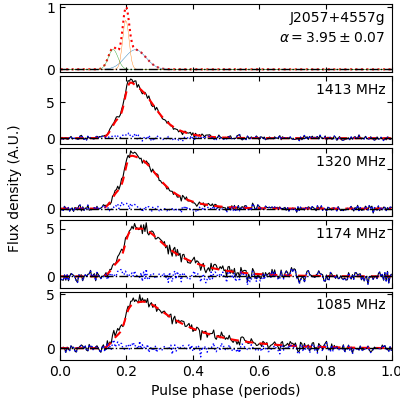}
\includegraphics[width=0.24\textwidth,height=0.24\textwidth]{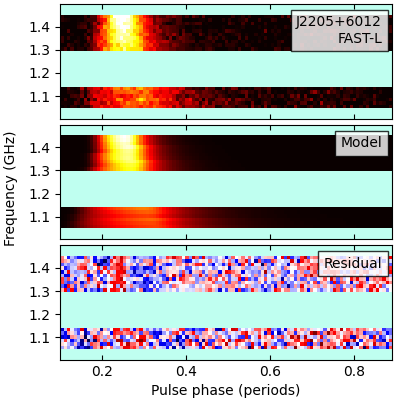}
\includegraphics[width=0.24\textwidth,height=0.24\textwidth]{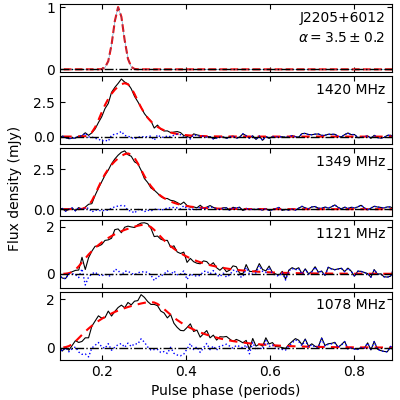}
\hspace{0.24\textwidth}
\hspace{0.24\textwidth}
\caption{--- {\it end.}}
\end{figure*}

\clearpage

\begin{figure*}
\centering
\includegraphics[width=0.24\textwidth,height=0.24\textwidth]{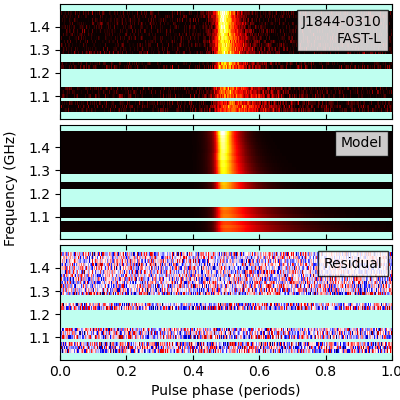}
\includegraphics[width=0.24\textwidth,height=0.24\textwidth]{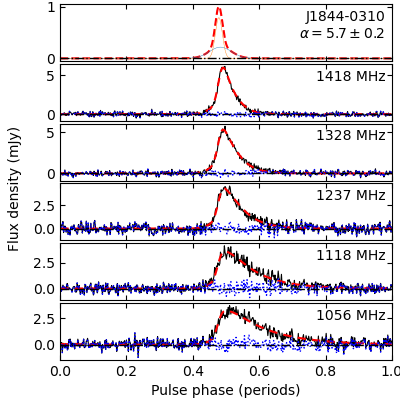}
\includegraphics[width=0.24\textwidth,height=0.24\textwidth]{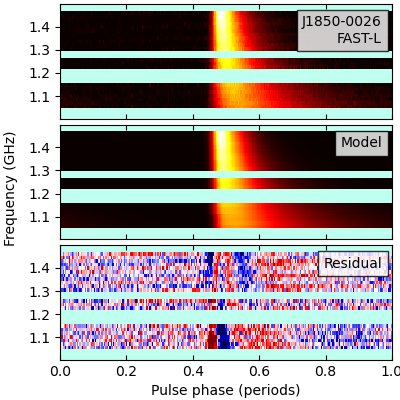}
\includegraphics[width=0.24\textwidth,height=0.24\textwidth]{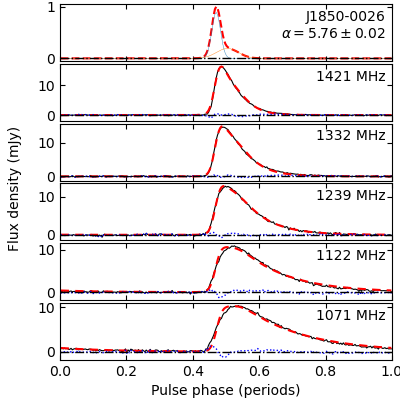}
\caption{The same as Figure~\ref{fig:2d-all} but for two unusual pulsars, PSR J1844-0310 and J1850-0026, which have to include the frequency-evolution for intrinsic profile components for the modeling, otherwise remarkable residuals will be left.}
\label{fig:2psr-bad}
\end{figure*}

\begin{figure*}
\centering
\includegraphics[width=0.24\textwidth,height=0.24\textwidth]{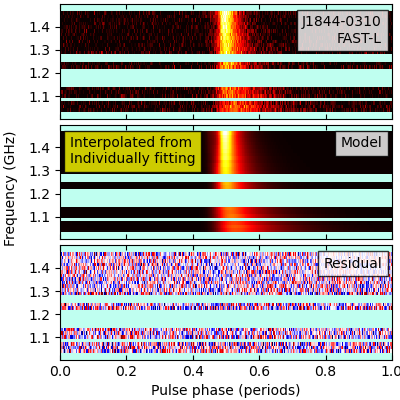}
\includegraphics[width=0.24\textwidth,height=0.24\textwidth]{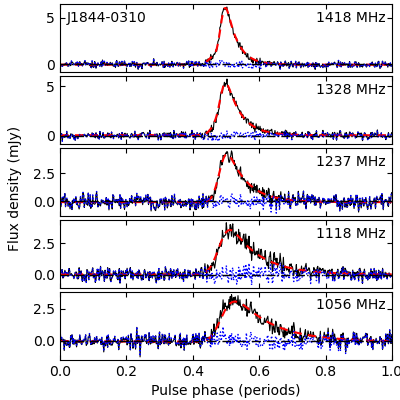}
\includegraphics[width=0.24\textwidth,height=0.24\textwidth]{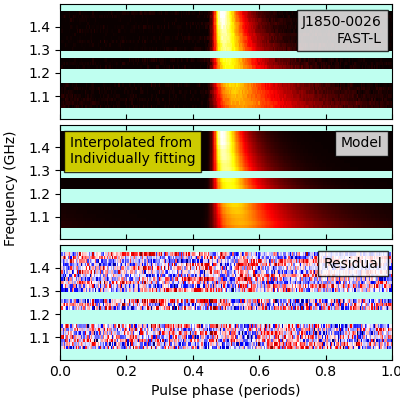}
\includegraphics[width=0.24\textwidth,height=0.24\textwidth]{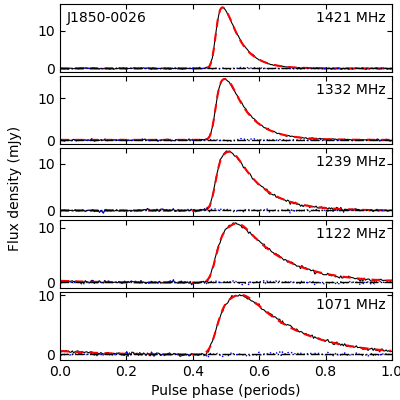}
\caption{The validation plot of two unusual pulsars, PSR J1844-0310 and J1850-0026, with the interpolated intrinsic profile from the individual fitting. Their residuals reach the noise level.}
\end{figure*}

\newpage

\begin{figure*}
  \centering
  \includegraphics[width=0.3\columnwidth]{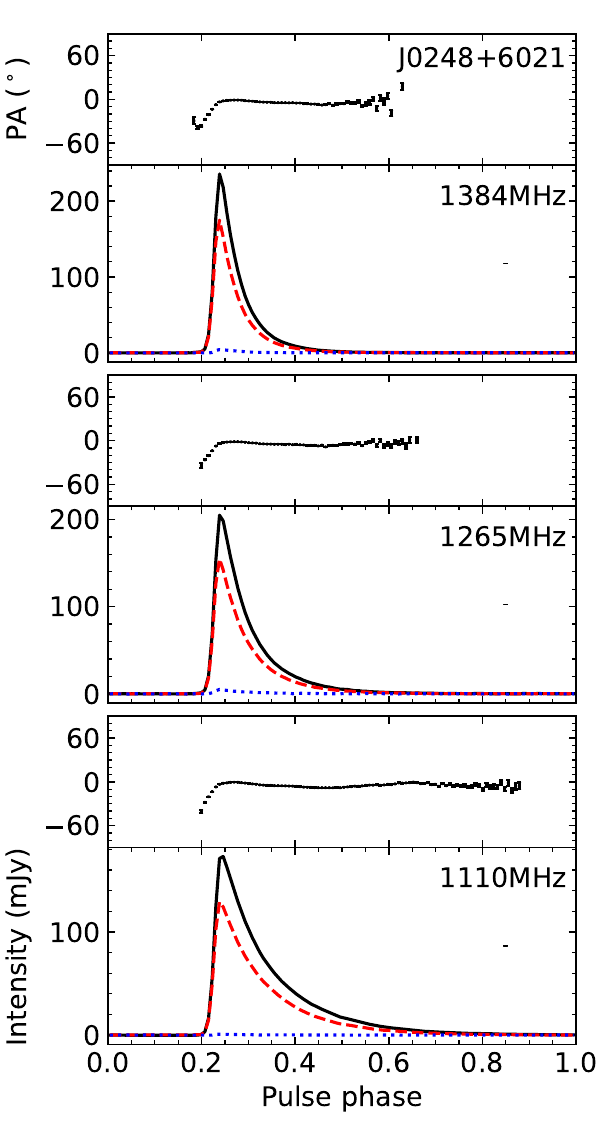}
  \includegraphics[width=0.3\columnwidth]{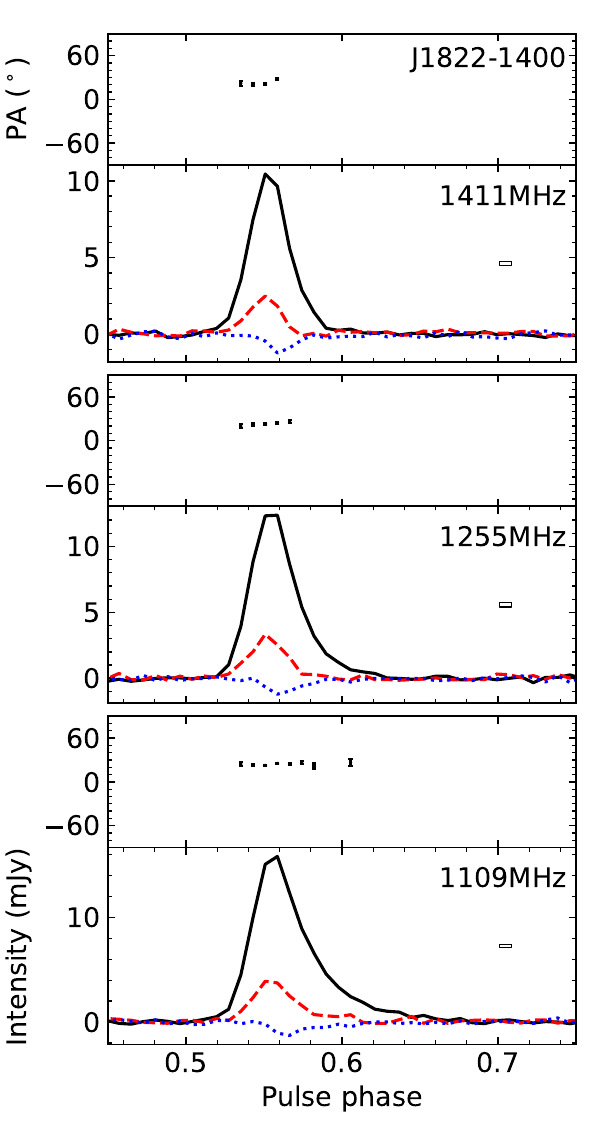}
  \includegraphics[width=0.3\columnwidth]{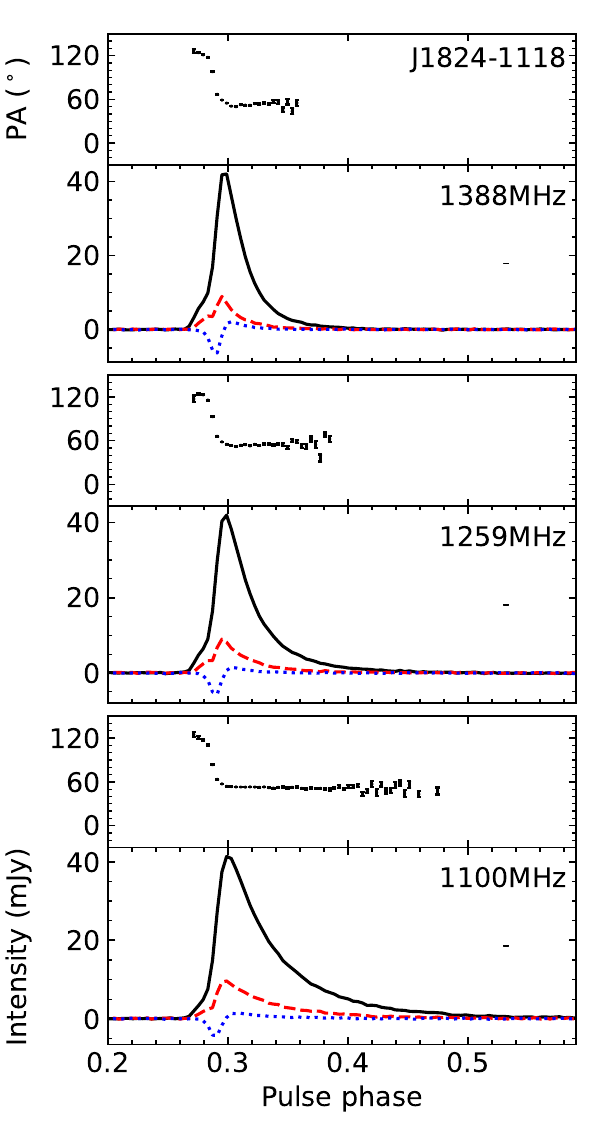}
  \includegraphics[width=0.3\columnwidth]{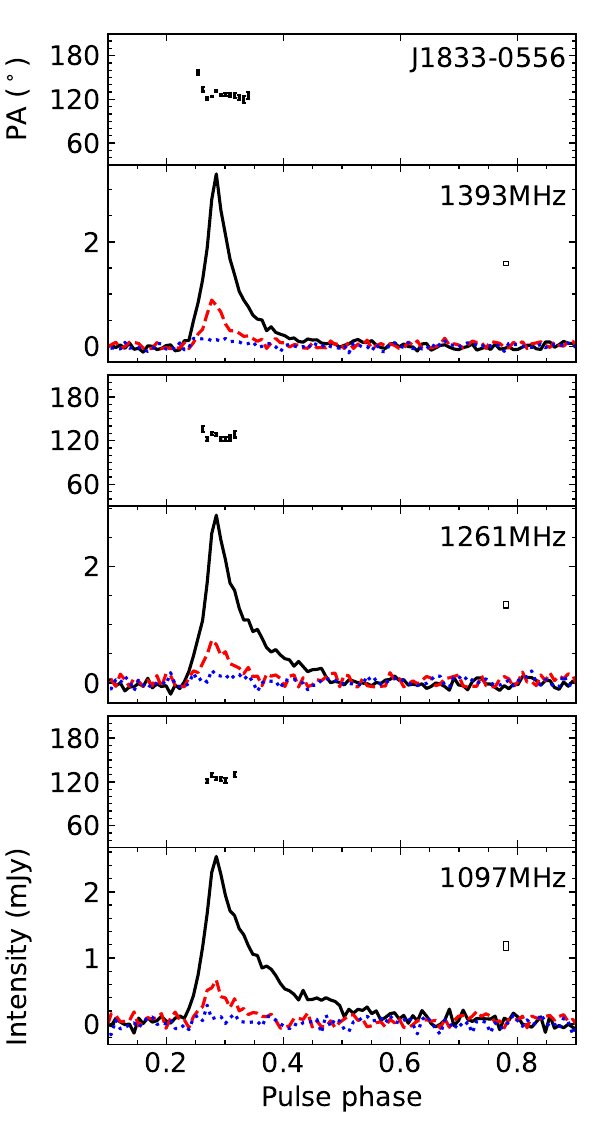}
  \includegraphics[width=0.3\columnwidth]{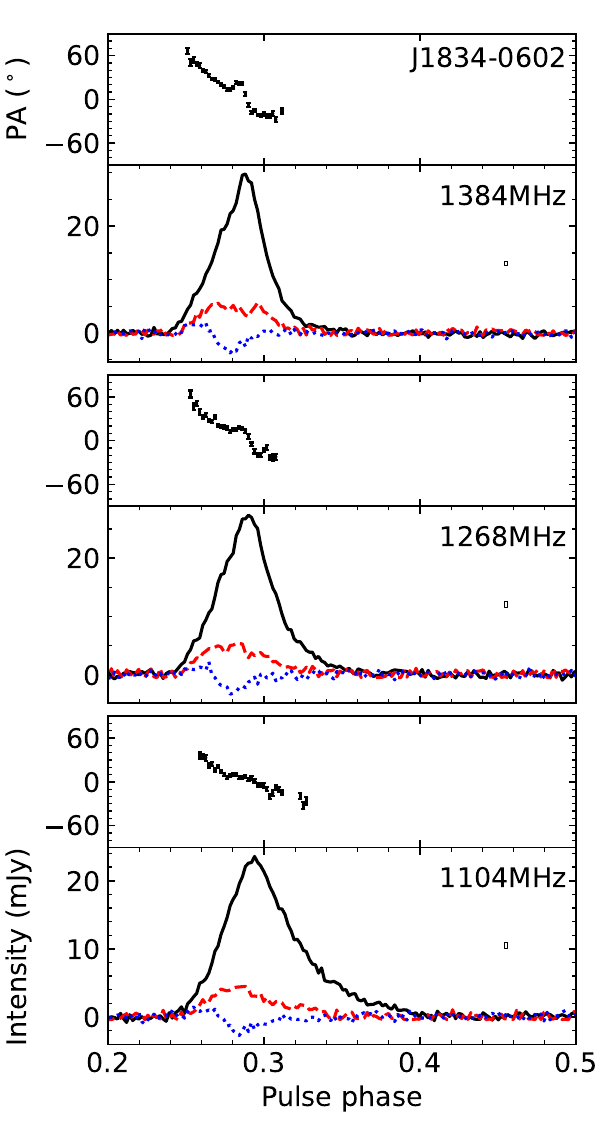}
  \includegraphics[width=0.3\columnwidth]{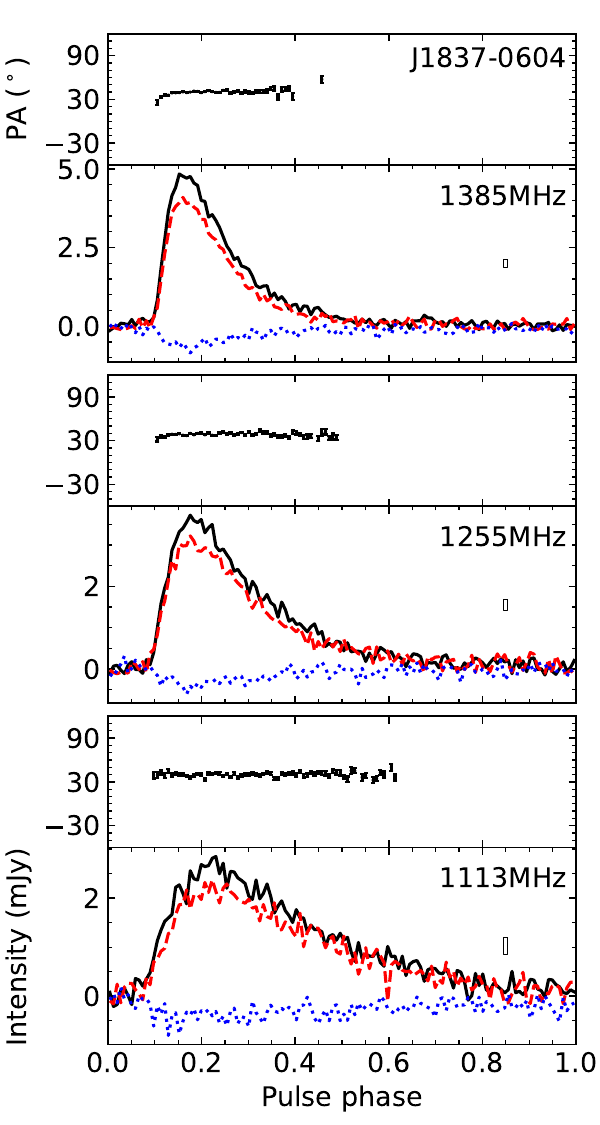}
  \caption{Polarization profiles at 3 subbands of FAST observations for 82 scattered pulsars, including PSR J2052$+$4421 in Figure~\ref{fig:2052profile}. Each profile has the total intensity in the solid line, linear polarization in the dashed line, and circular polarization in the dotted line. The band central frequency and the box for a bin size and $\pm1\sigma$ are marked in the lower sub-panel. The polarization angles are plotted with error-bar in the upper sub-panel. -- {\it to be continued.}}
\label{fig:pol}
\end{figure*}

\addtocounter{figure}{-1}
\begin{figure*}
  \centering
  \includegraphics[width=0.3\columnwidth]{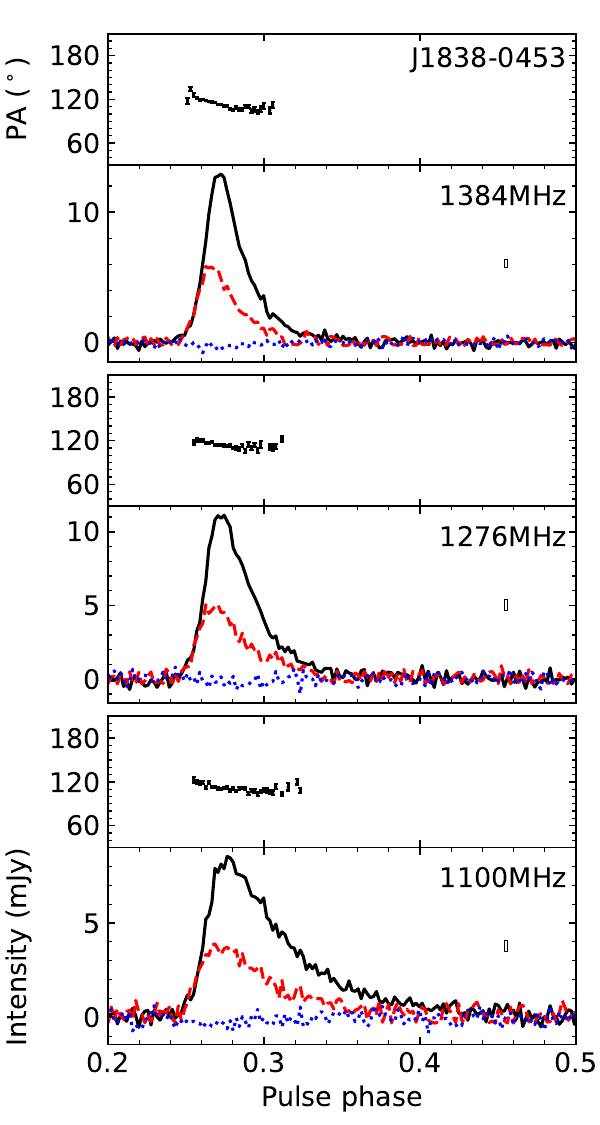}
  \includegraphics[width=0.3\columnwidth]{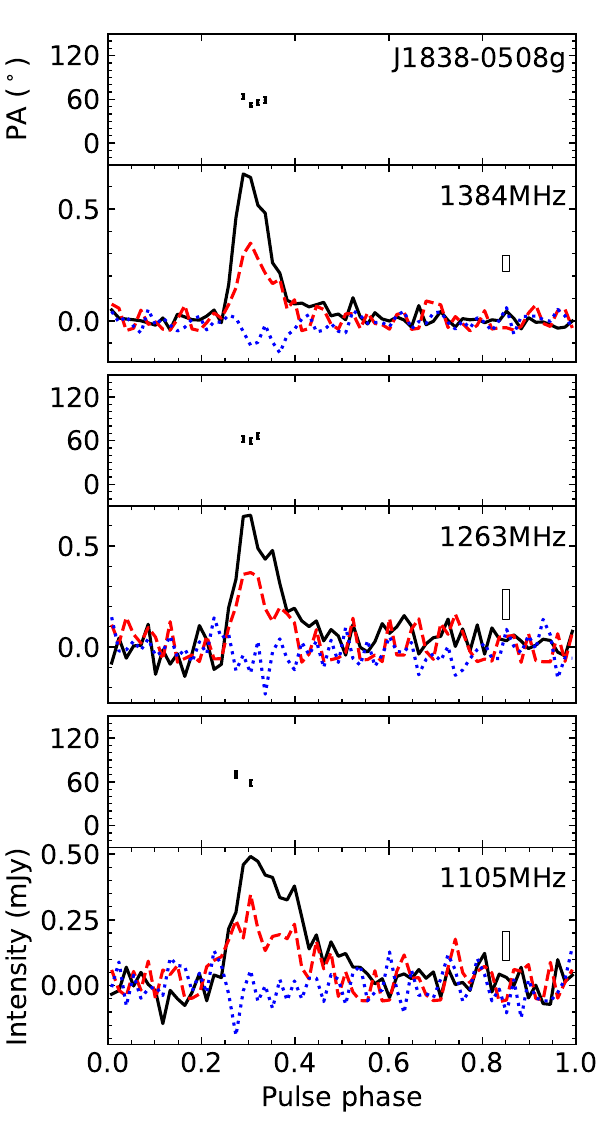}
  \includegraphics[width=0.3\columnwidth]{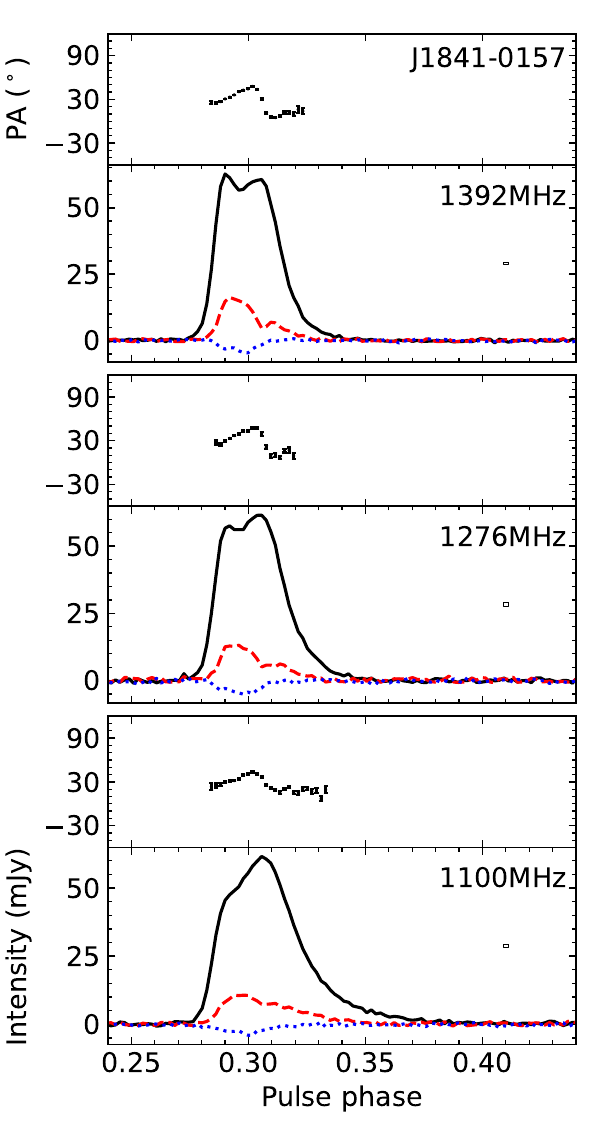}
  \includegraphics[width=0.3\columnwidth]{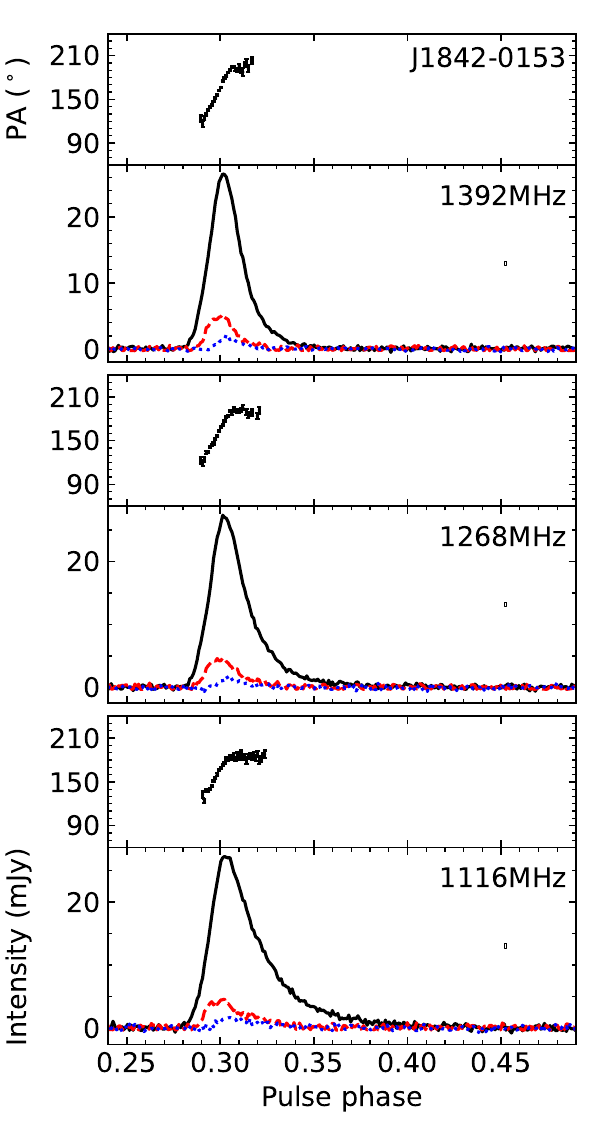}
  \includegraphics[width=0.3\columnwidth]{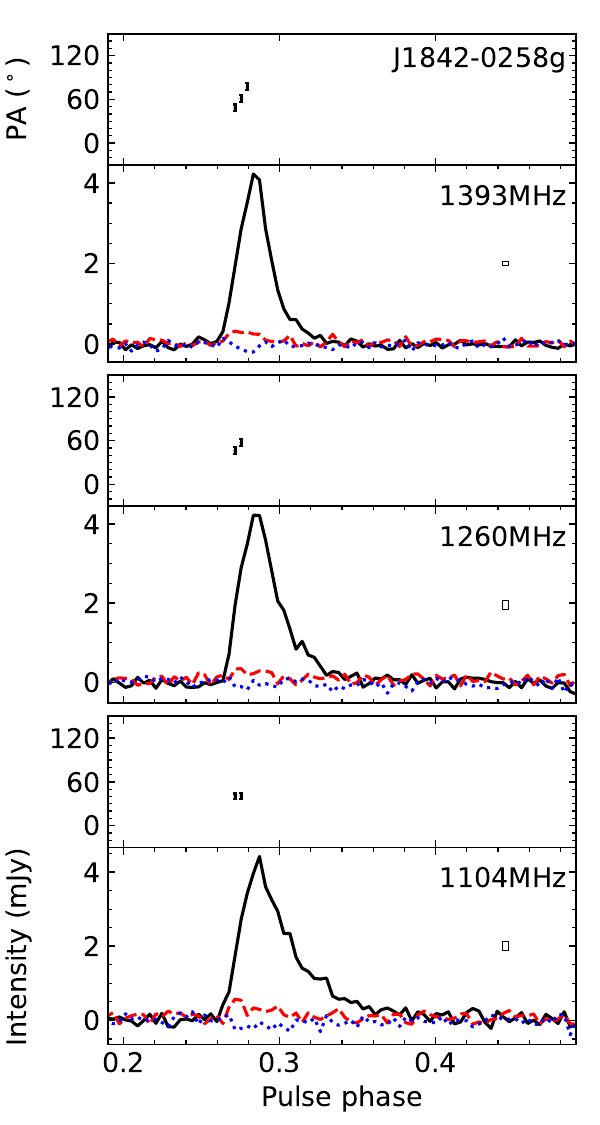}
  \includegraphics[width=0.3\columnwidth]{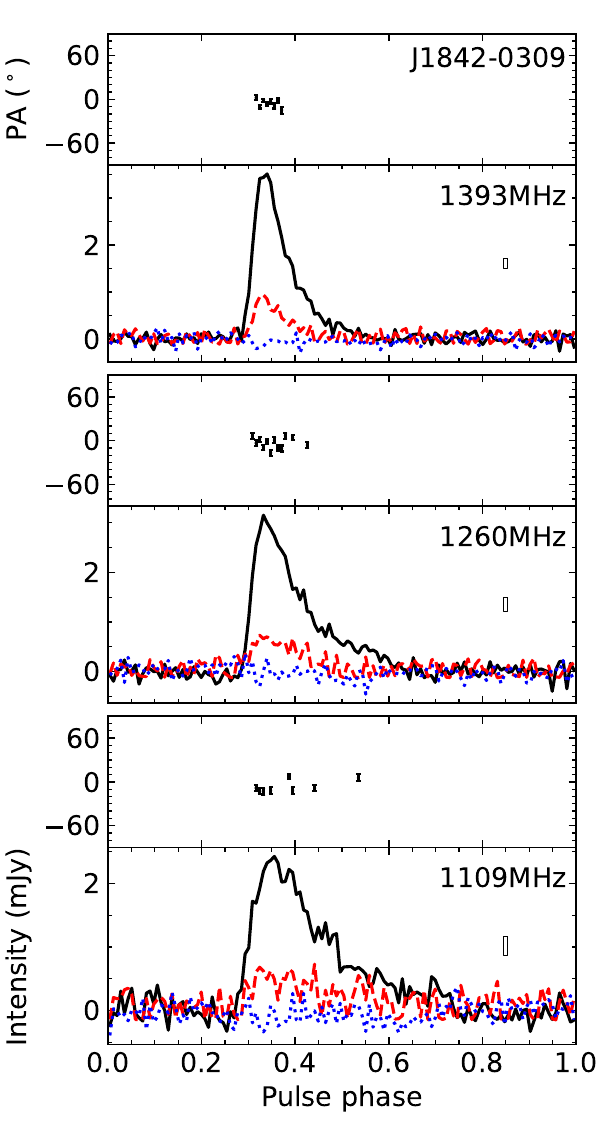}
  \caption{-- {\it continued}}
\end{figure*}

\addtocounter{figure}{-1}
\begin{figure*}
  \centering
  \includegraphics[width=0.3\columnwidth]{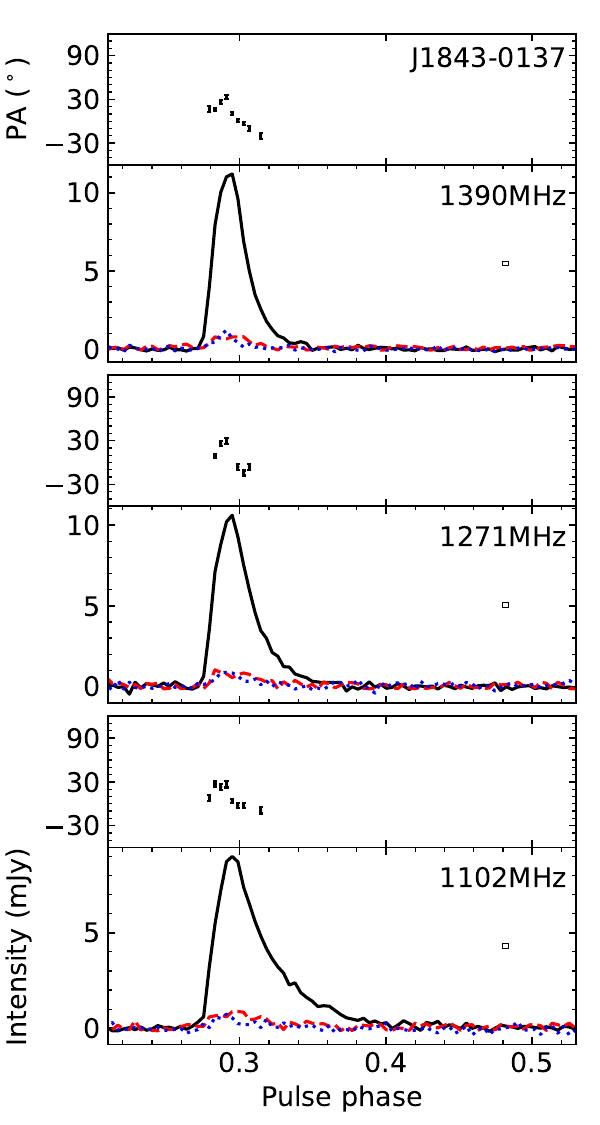}
  \includegraphics[width=0.3\columnwidth]{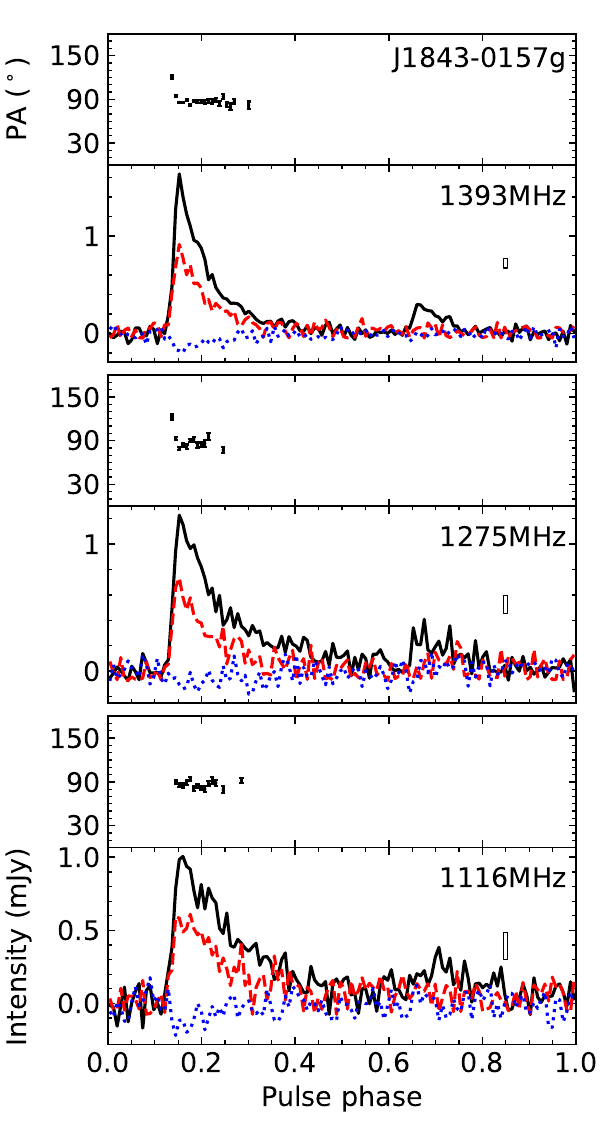}
  \includegraphics[width=0.3\columnwidth]{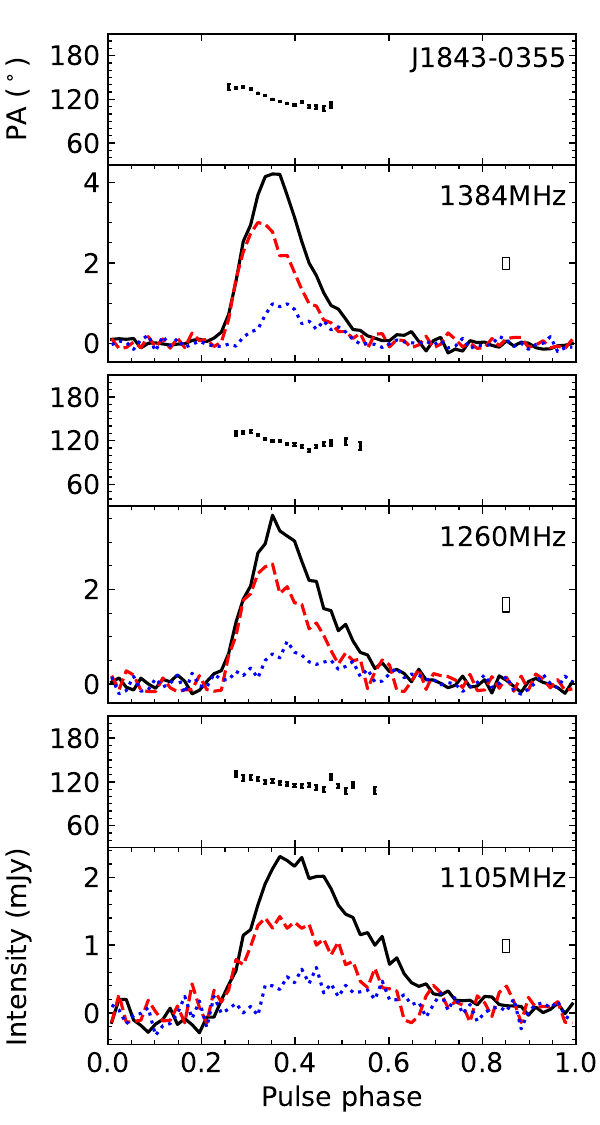}
  \includegraphics[width=0.3\columnwidth]{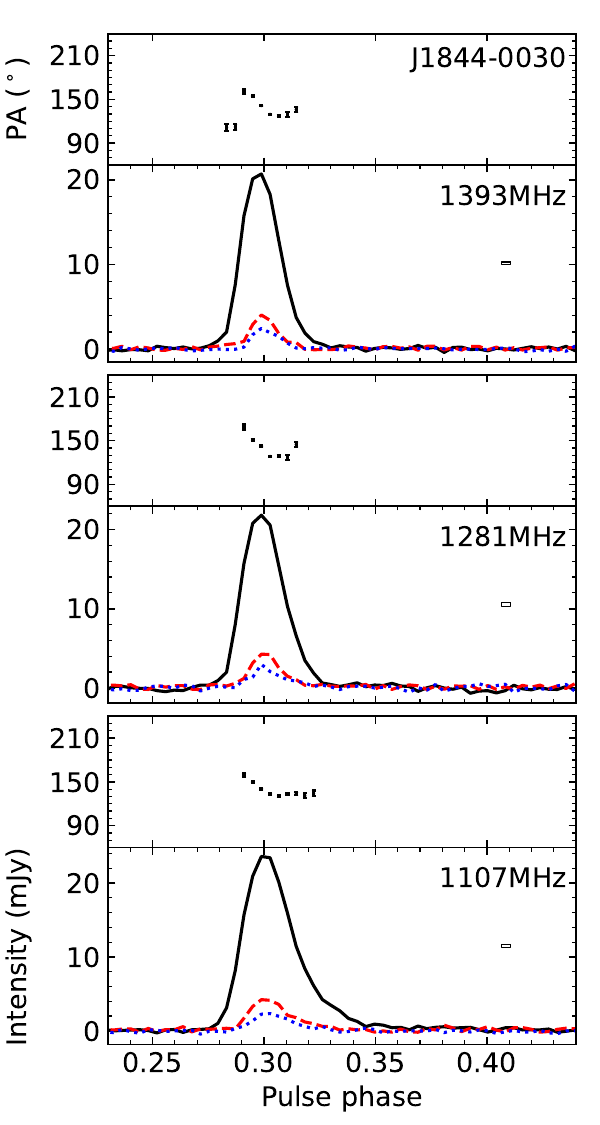}
  \includegraphics[width=0.3\columnwidth]{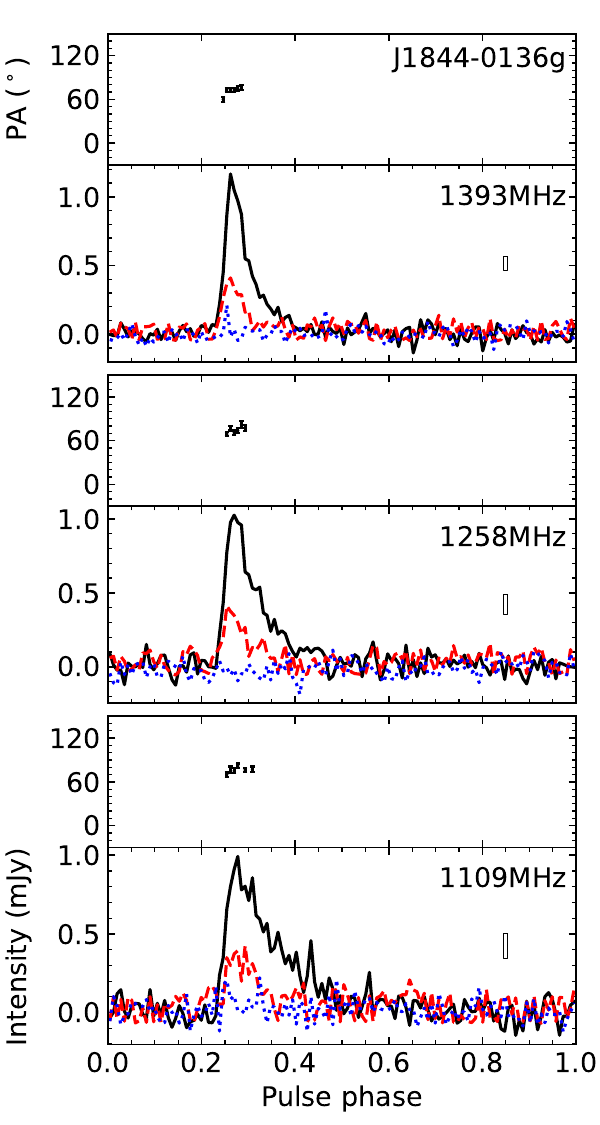}
  \includegraphics[width=0.3\columnwidth]{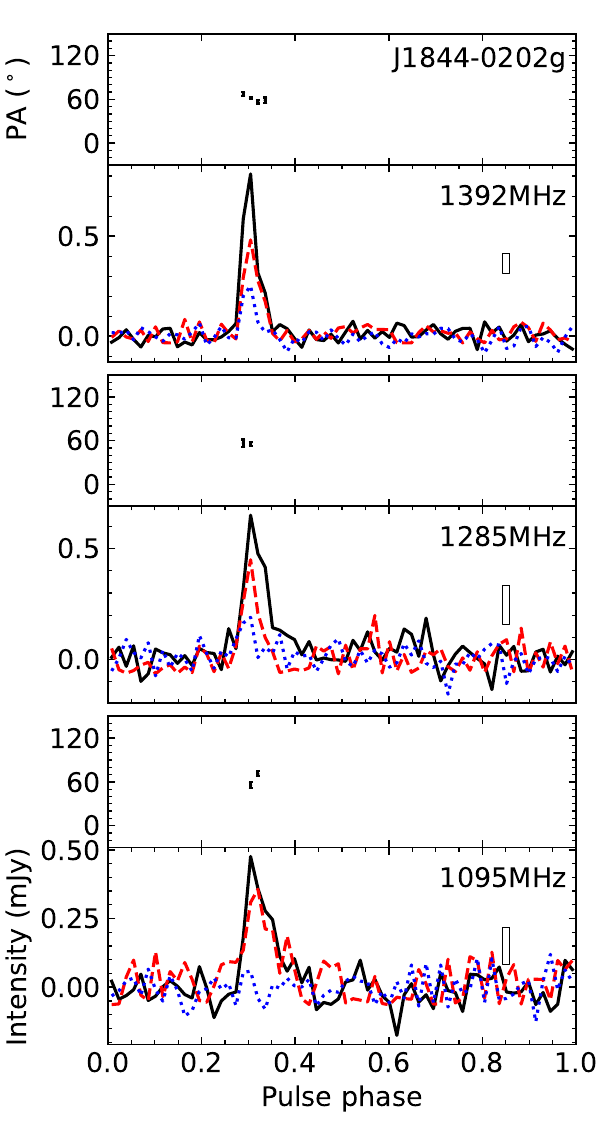}
  \caption{-- {\it continued}}
\end{figure*}

\addtocounter{figure}{-1}
\begin{figure*}
  \centering
  \includegraphics[width=0.3\columnwidth]{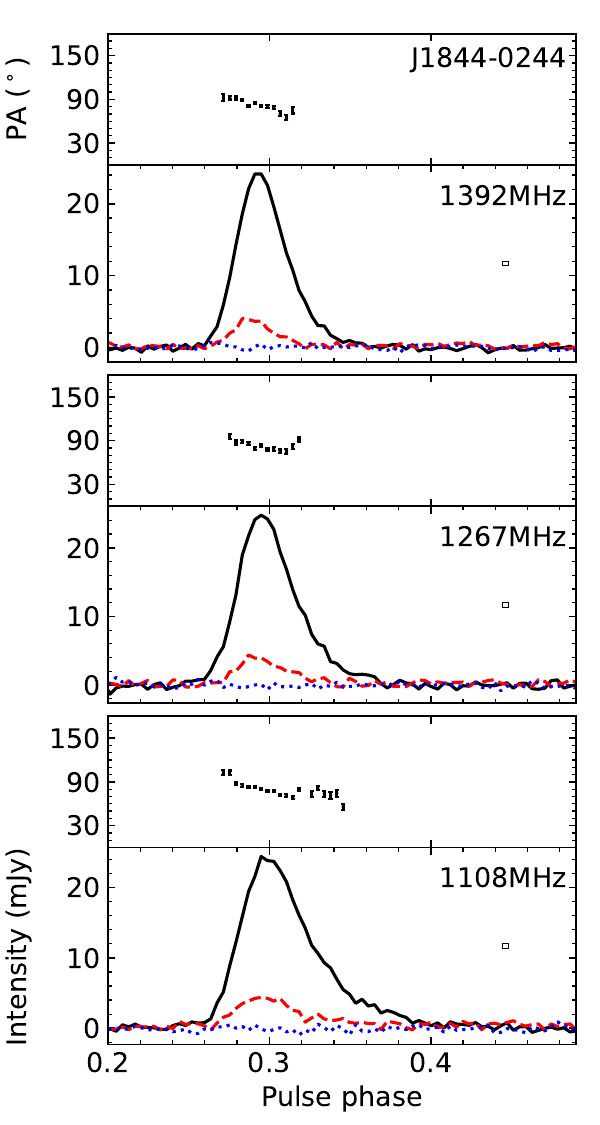}
  \includegraphics[width=0.3\columnwidth]{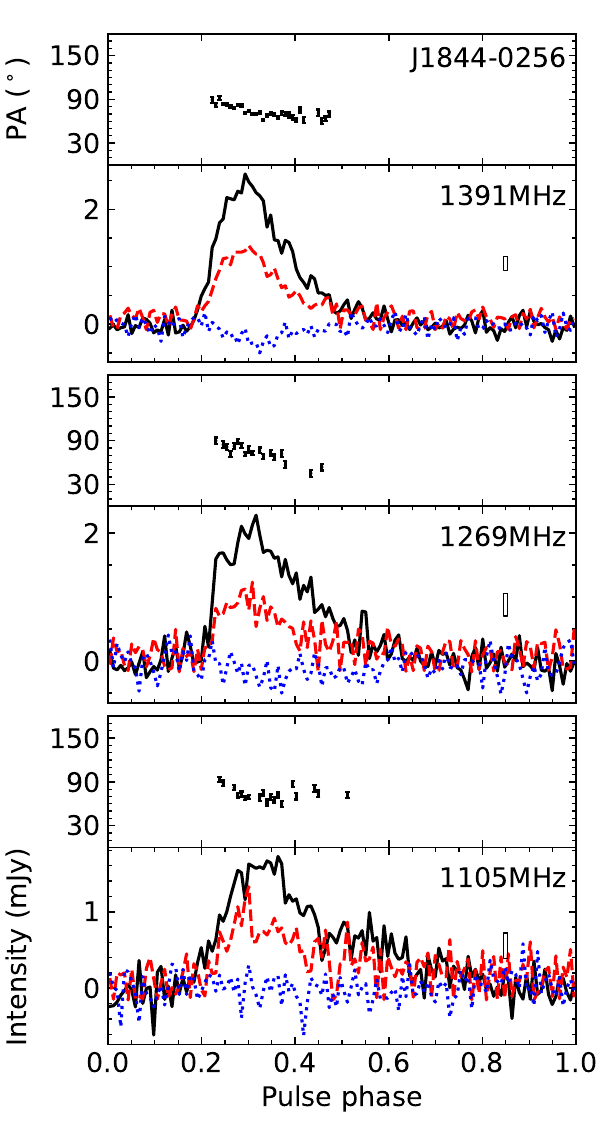}
  \includegraphics[width=0.3\columnwidth]{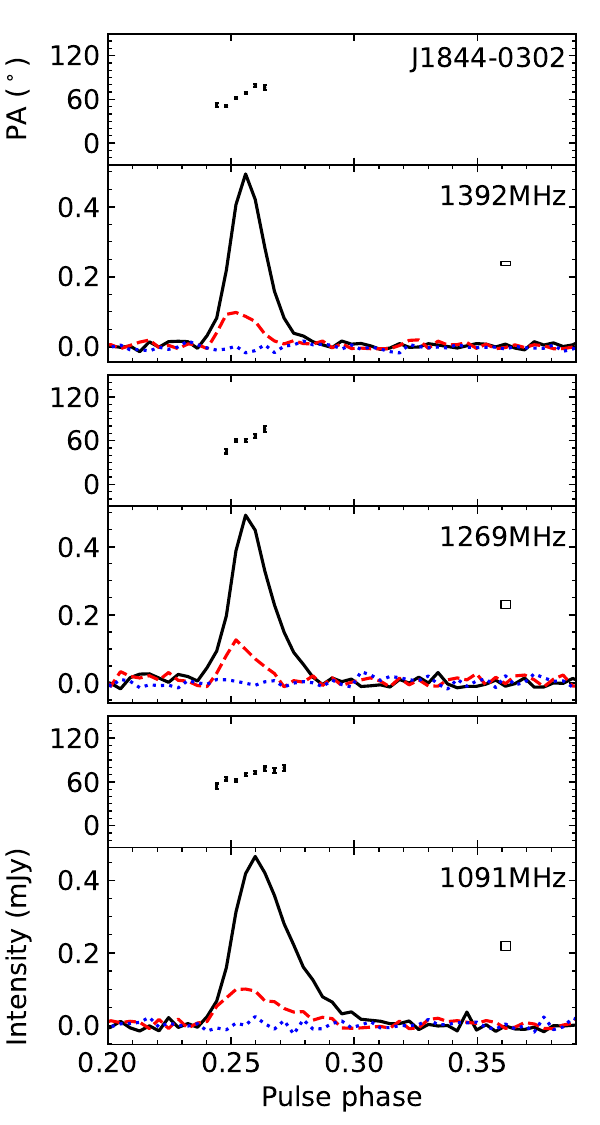}
  \includegraphics[width=0.3\columnwidth]{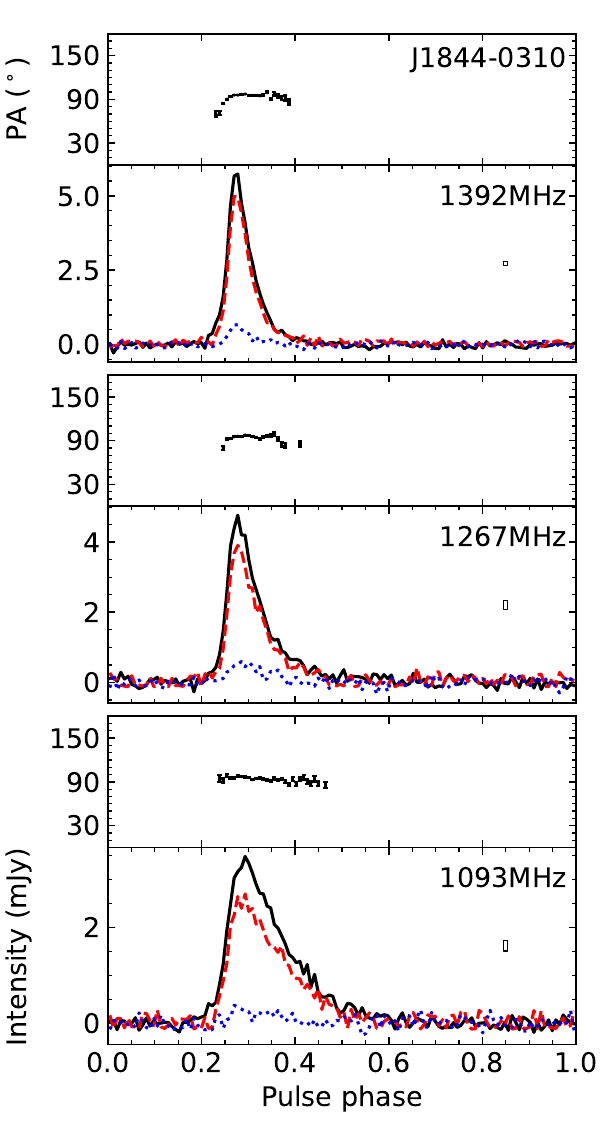}
  \includegraphics[width=0.3\columnwidth]{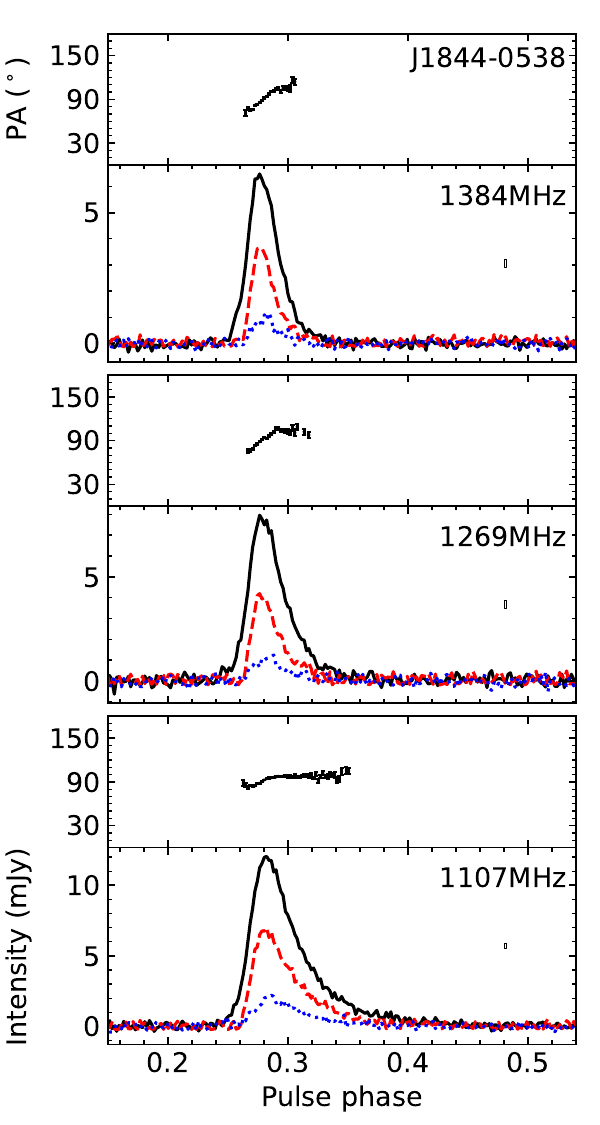}
  \includegraphics[width=0.3\columnwidth]{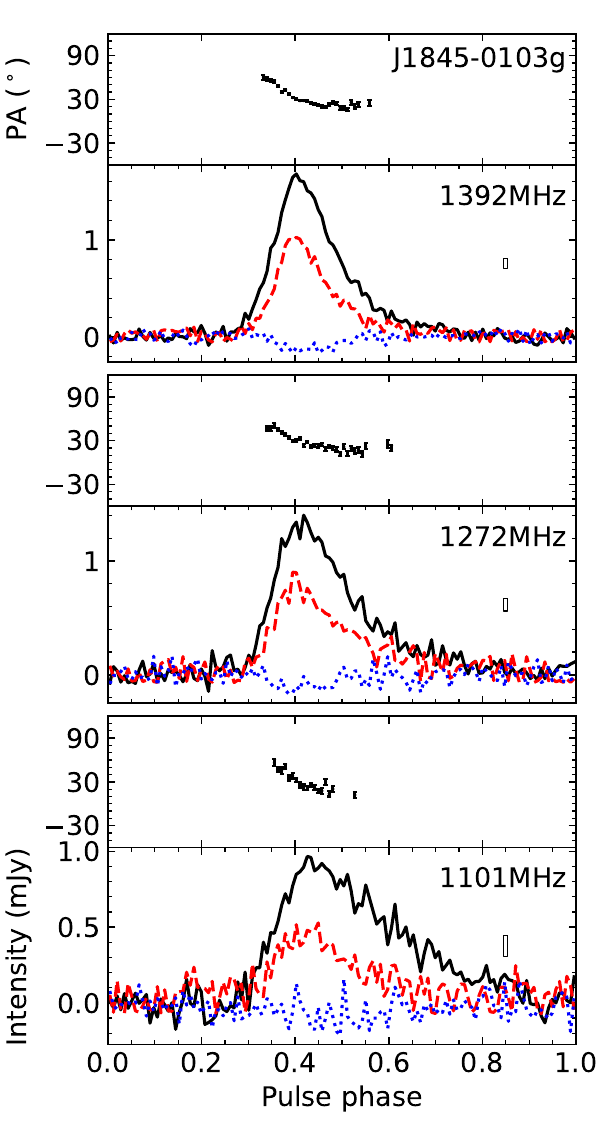}
  \caption{-- {\it continued}}
\end{figure*}

\addtocounter{figure}{-1}
\begin{figure*}
  \centering
  \includegraphics[width=0.3\columnwidth]{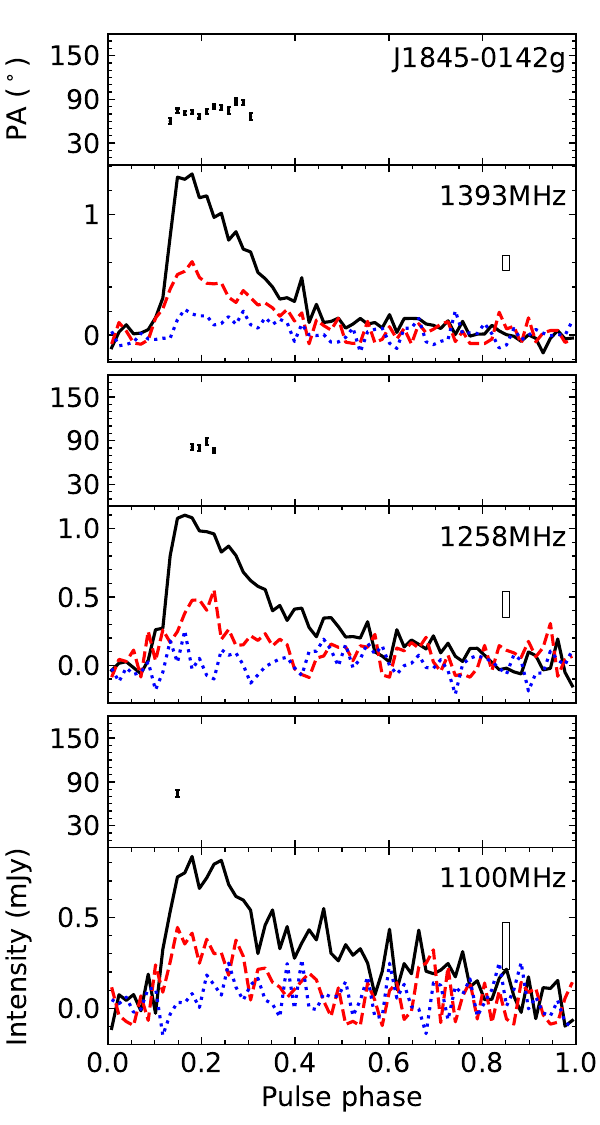}
  \includegraphics[width=0.3\columnwidth]{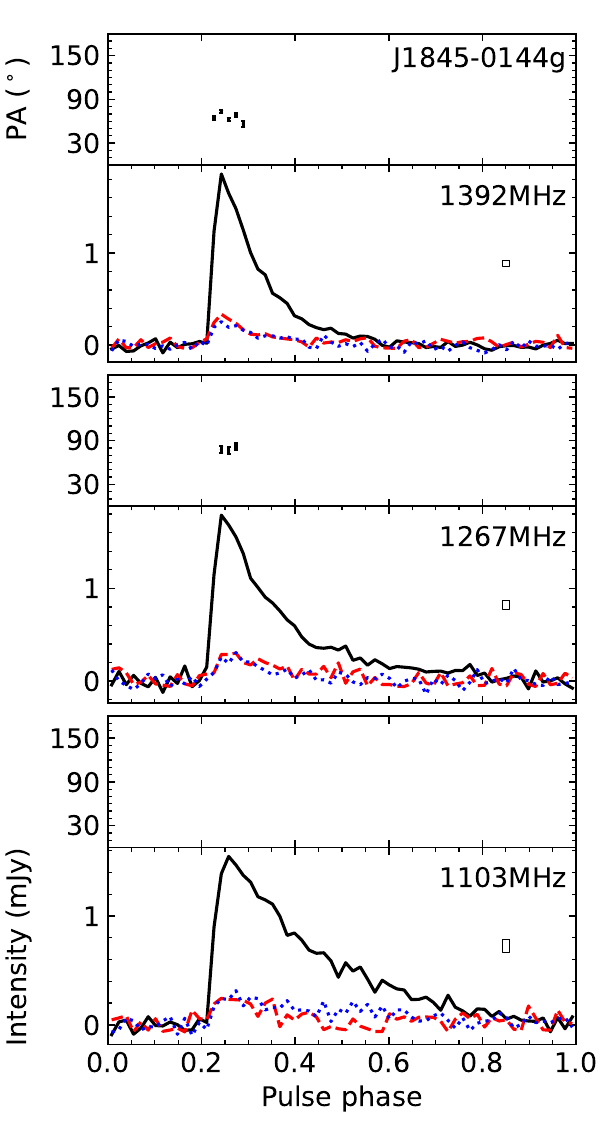}
  \includegraphics[width=0.3\columnwidth]{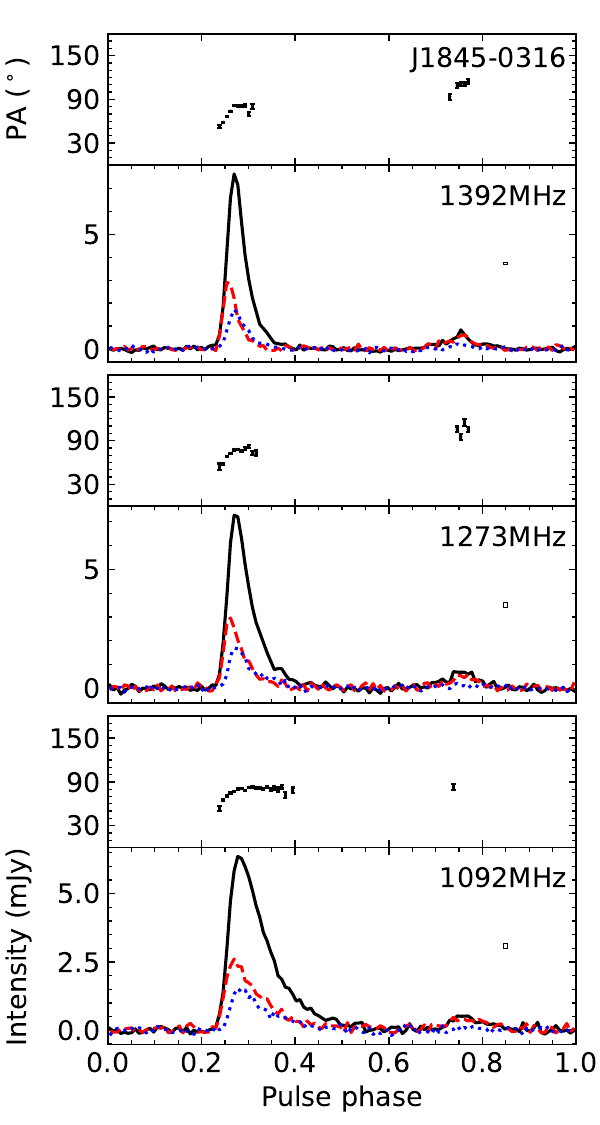}
  \includegraphics[width=0.3\columnwidth]{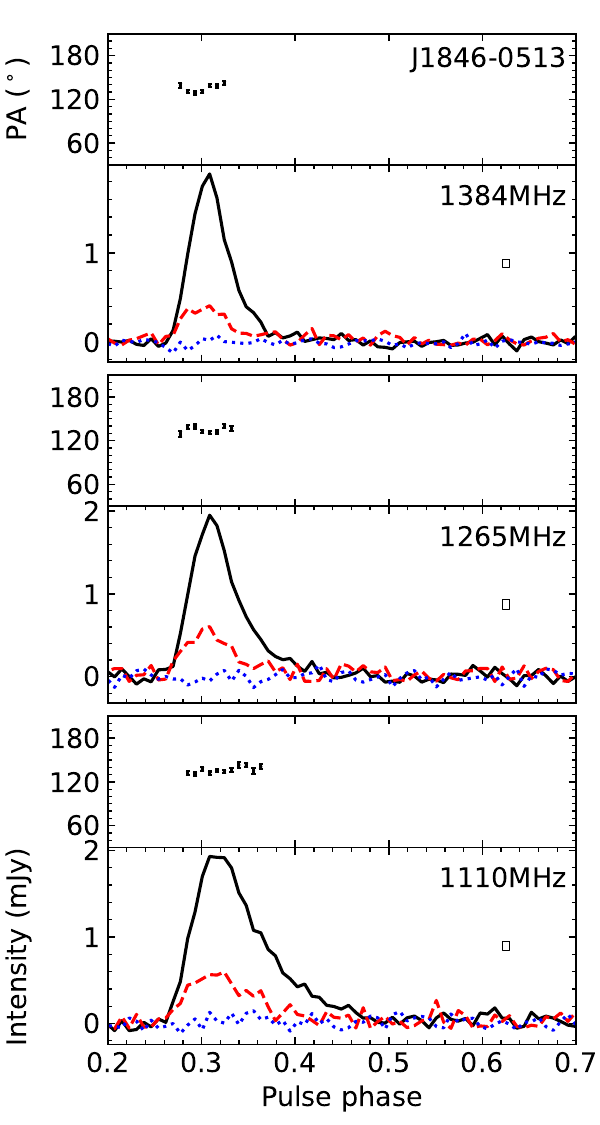}
  \includegraphics[width=0.3\columnwidth]{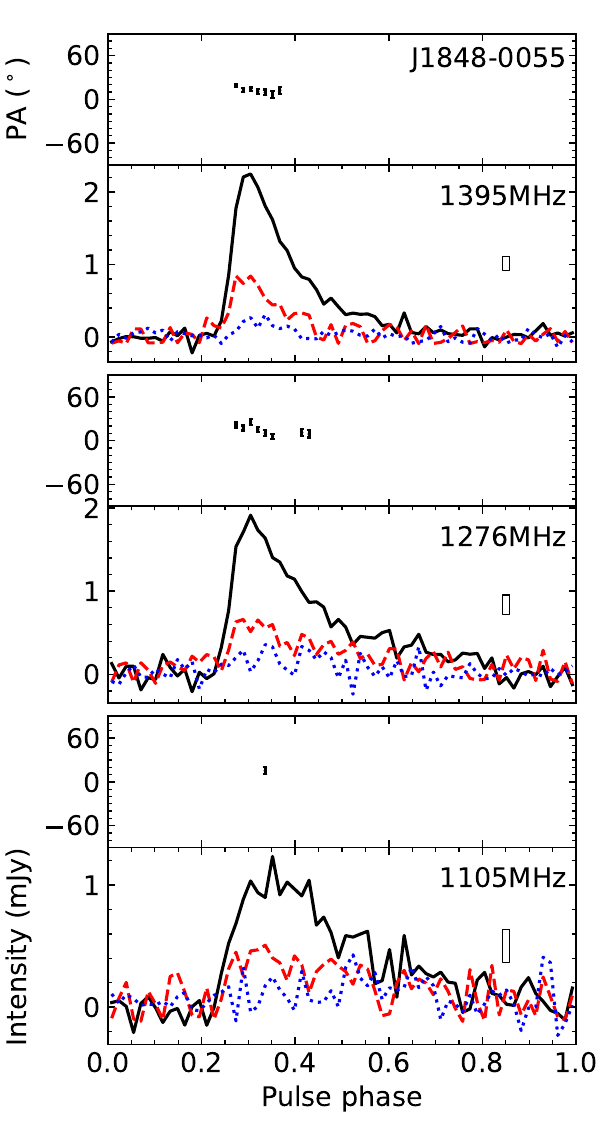}
  \includegraphics[width=0.3\columnwidth]{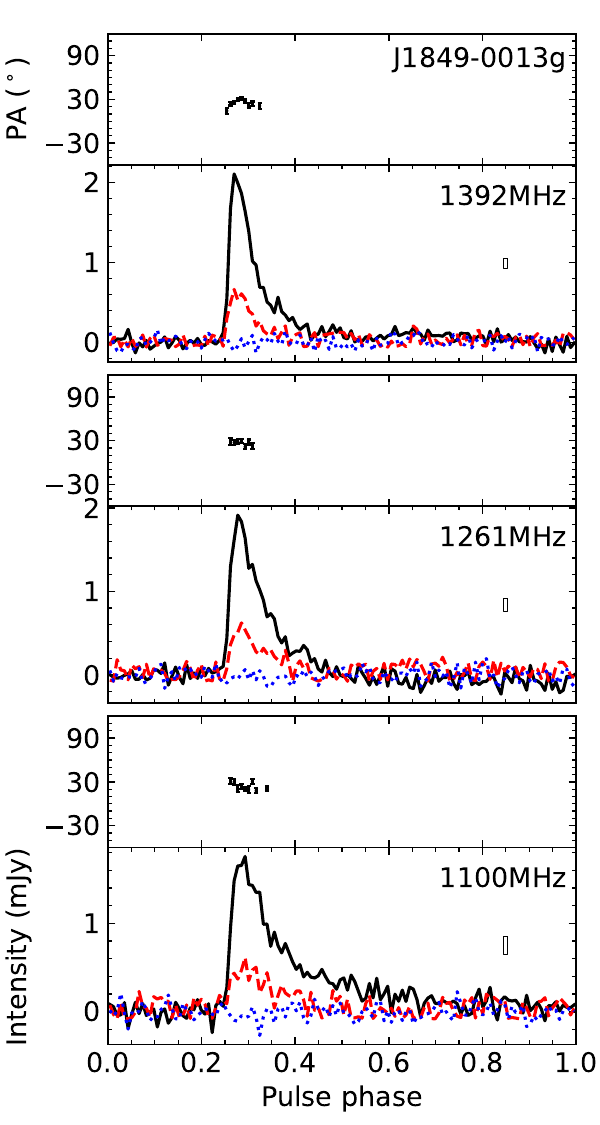}
  \caption{-- {\it continued}}
\end{figure*}

\addtocounter{figure}{-1}
\begin{figure*}
  \centering
  \includegraphics[width=0.3\columnwidth]{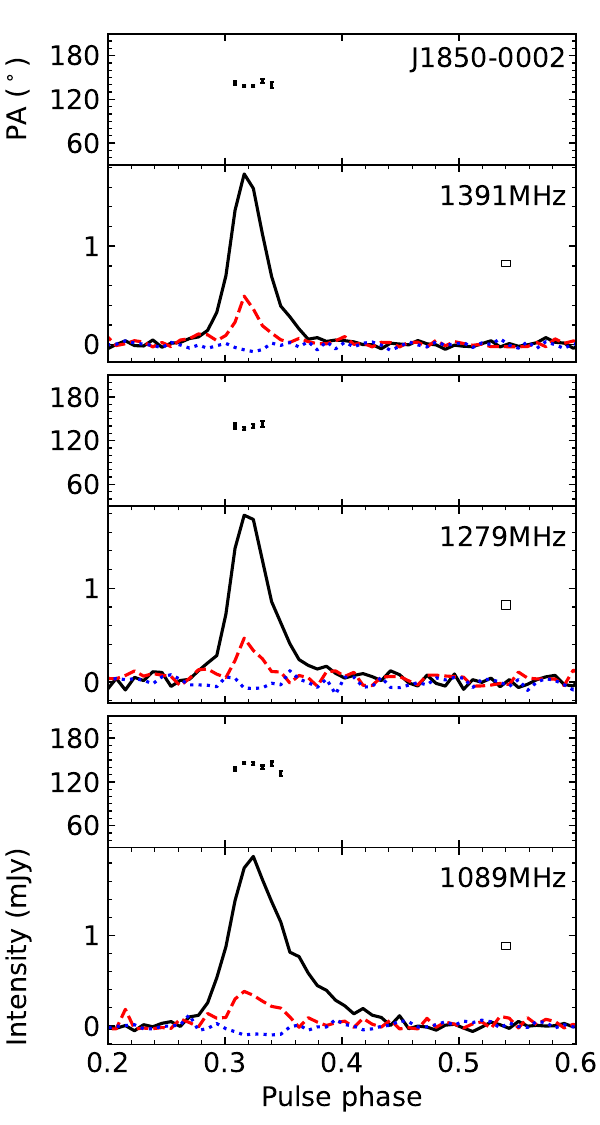}
  \includegraphics[width=0.3\columnwidth]{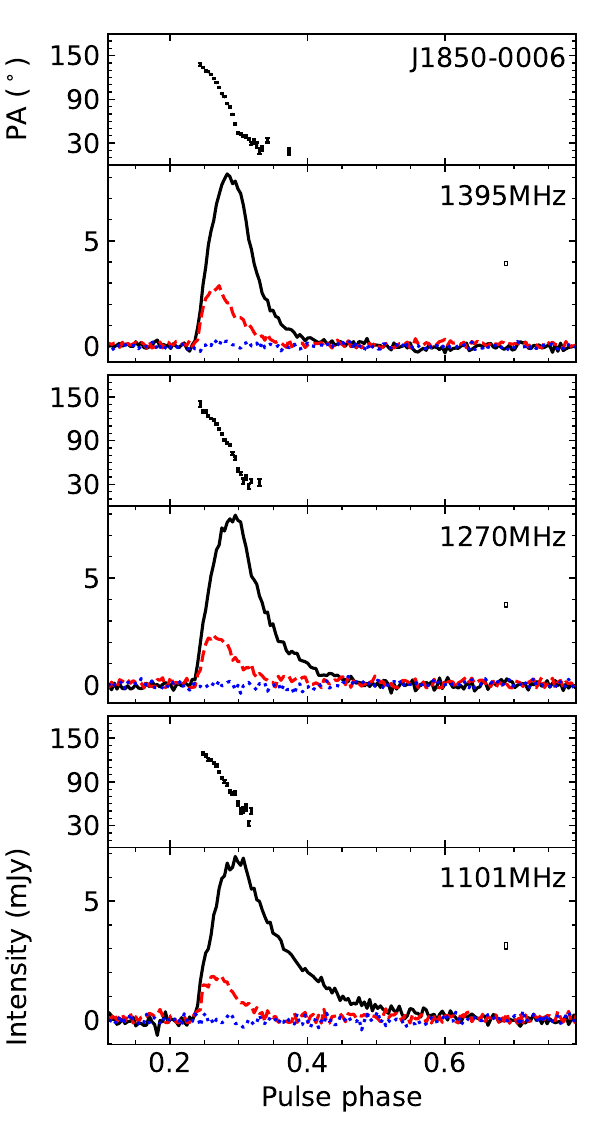}
  \includegraphics[width=0.3\columnwidth]{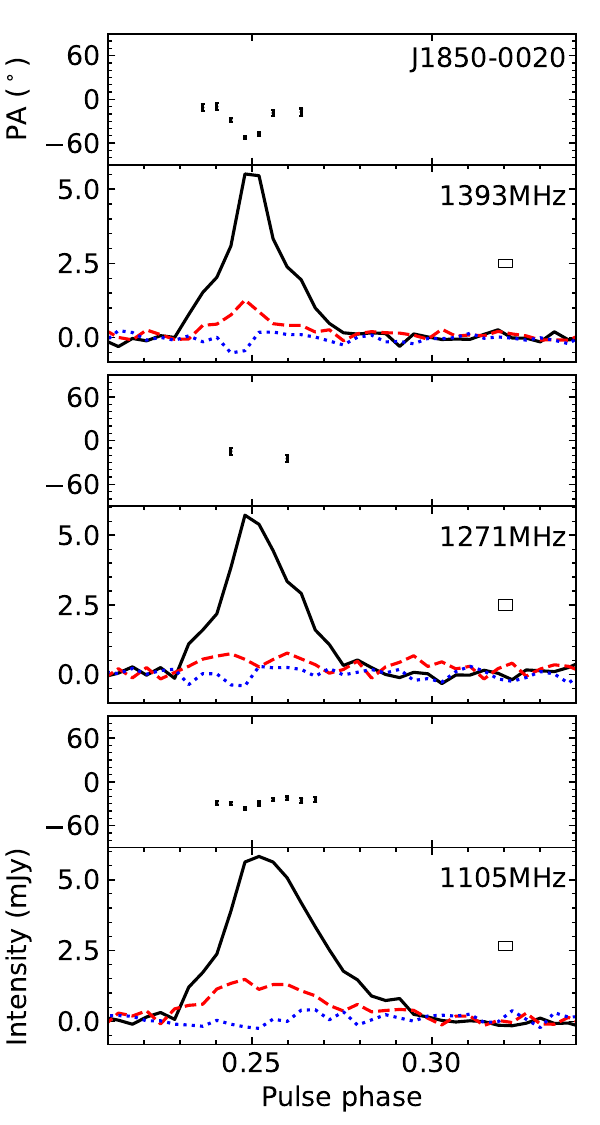}
  \includegraphics[width=0.3\columnwidth]{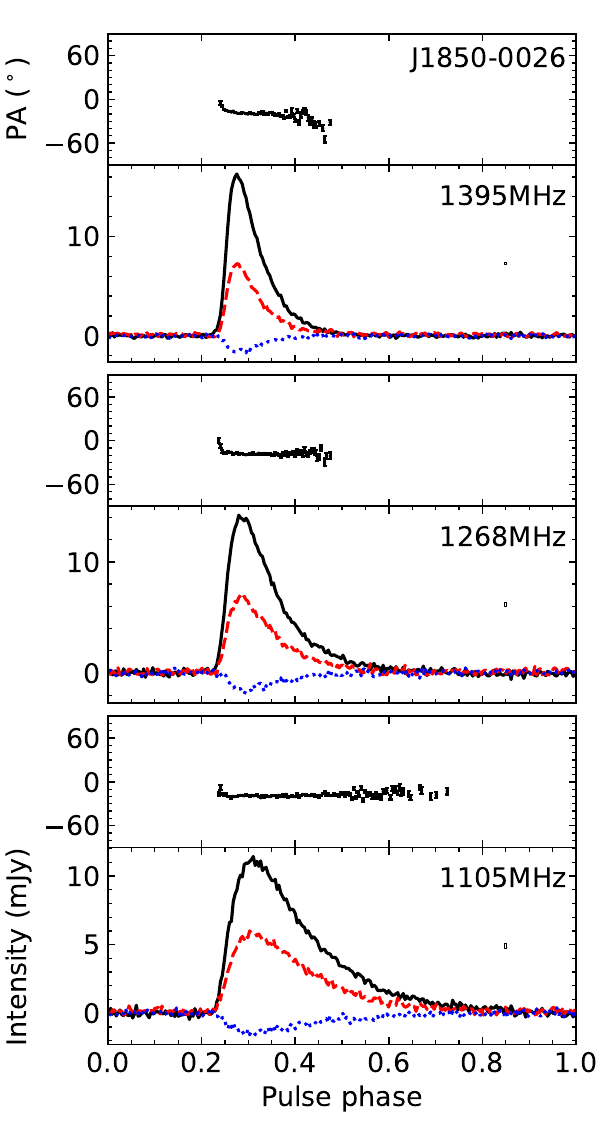}
  \includegraphics[width=0.3\columnwidth]{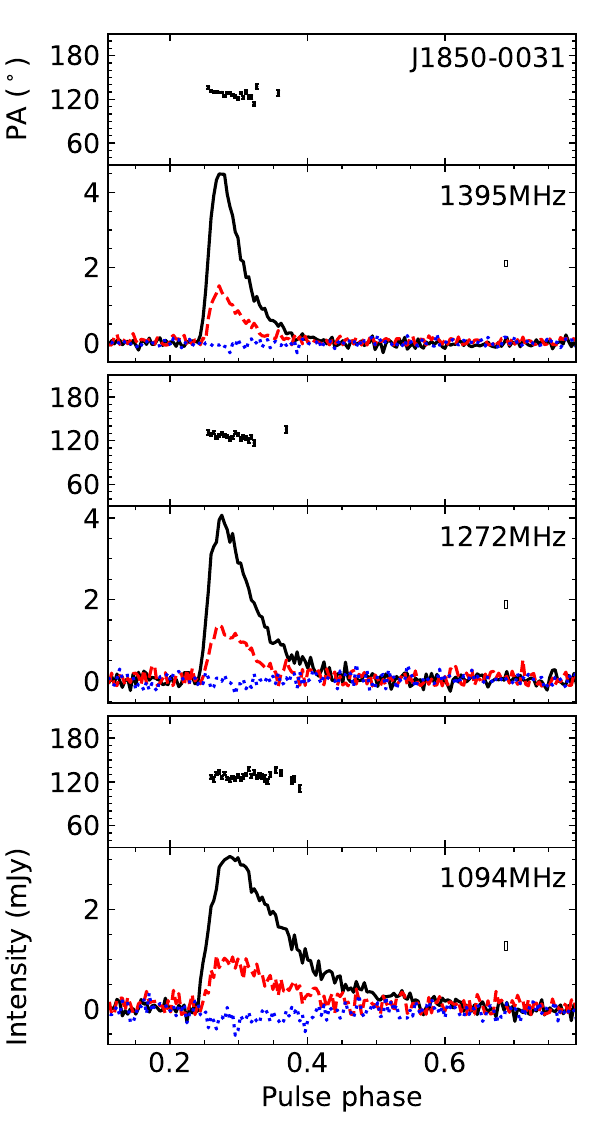}
  \includegraphics[width=0.3\columnwidth]{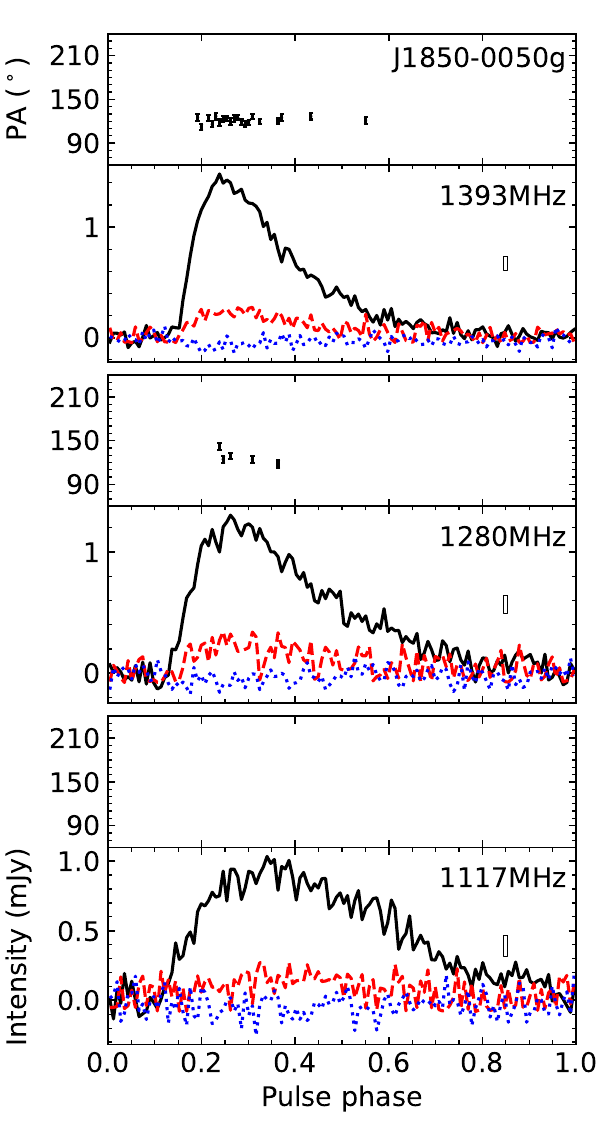}
  \caption{-- {\it continued}}
\end{figure*}

\addtocounter{figure}{-1}
\begin{figure*}
  \centering
  \includegraphics[width=0.3\columnwidth]{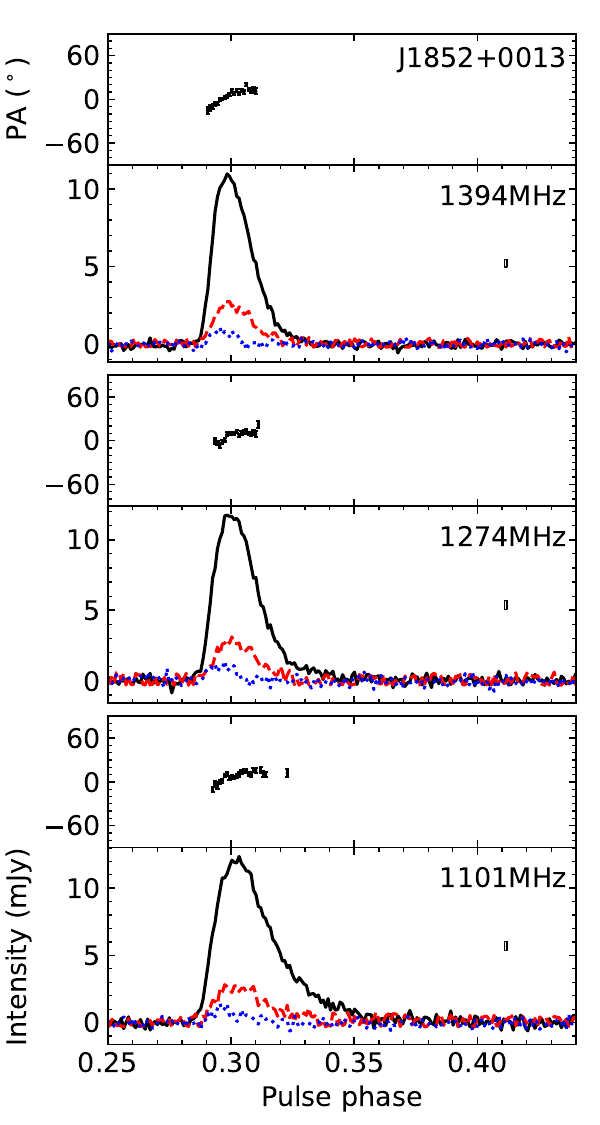}
  \includegraphics[width=0.3\columnwidth]{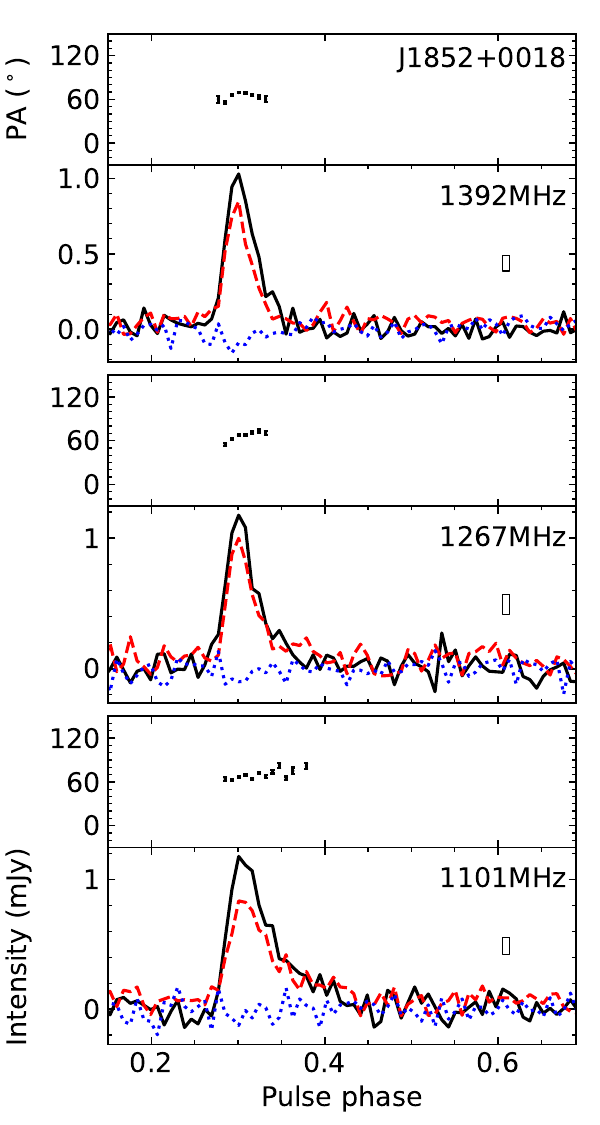}
  \includegraphics[width=0.3\columnwidth]{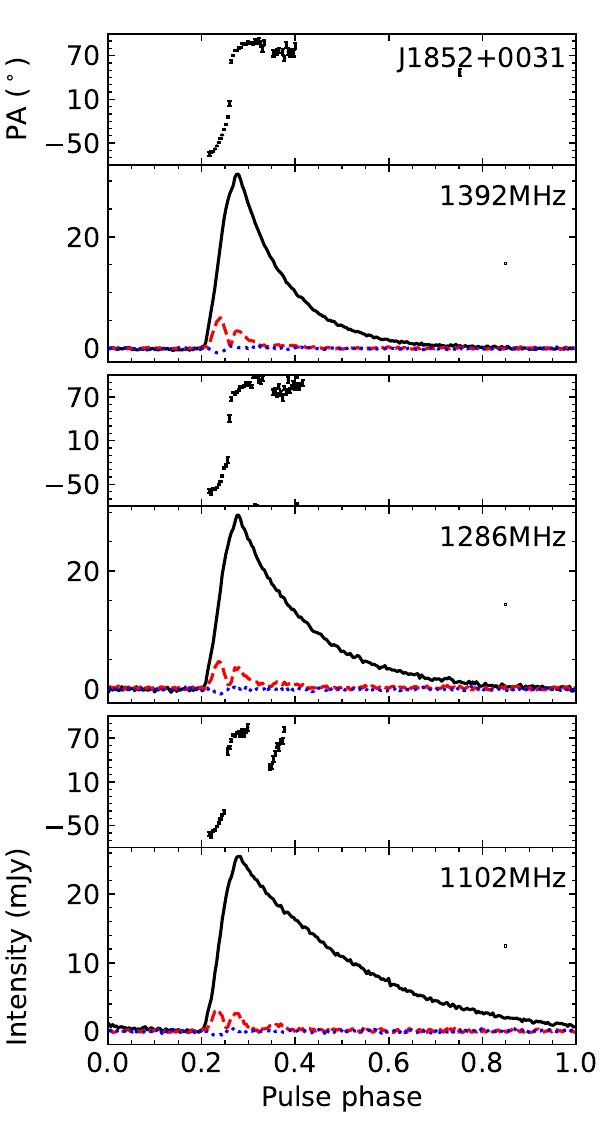}
  \includegraphics[width=0.3\columnwidth]{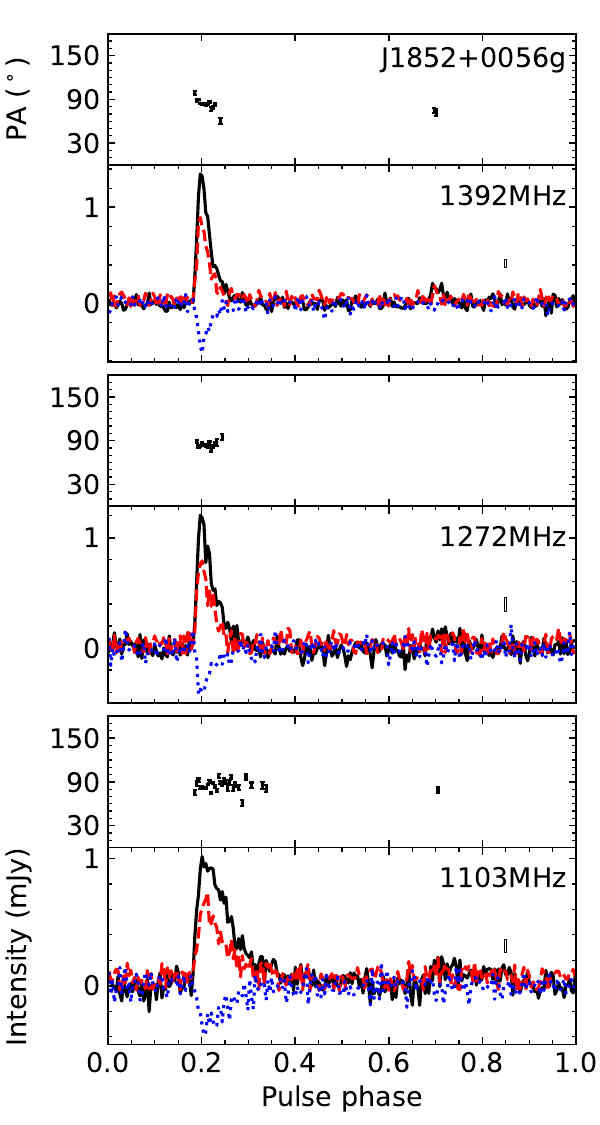}
  \includegraphics[width=0.3\columnwidth]{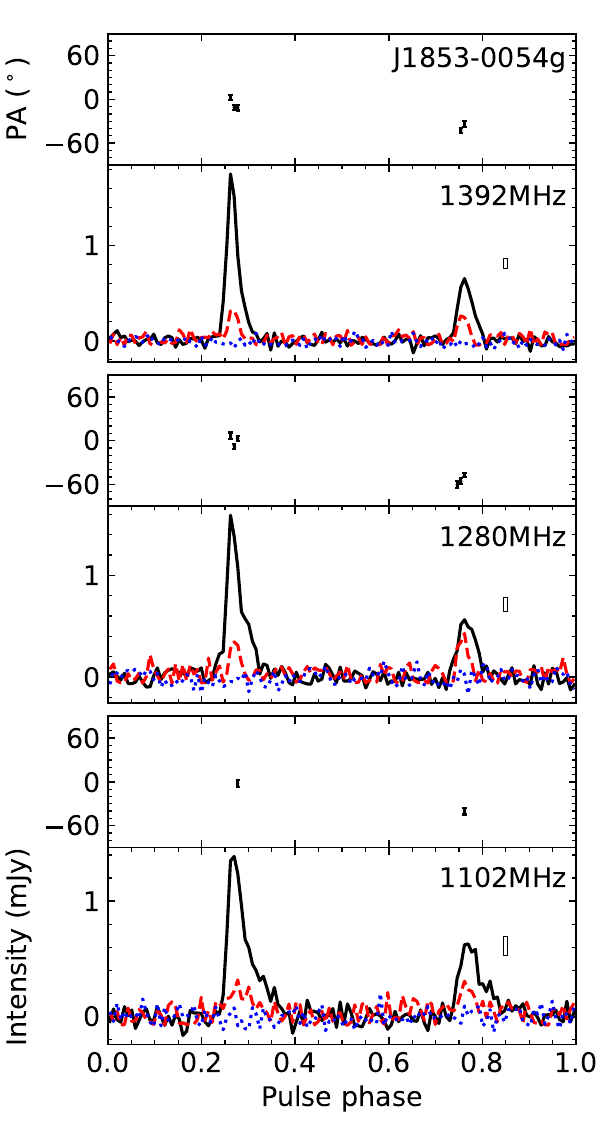}
  \includegraphics[width=0.3\columnwidth]{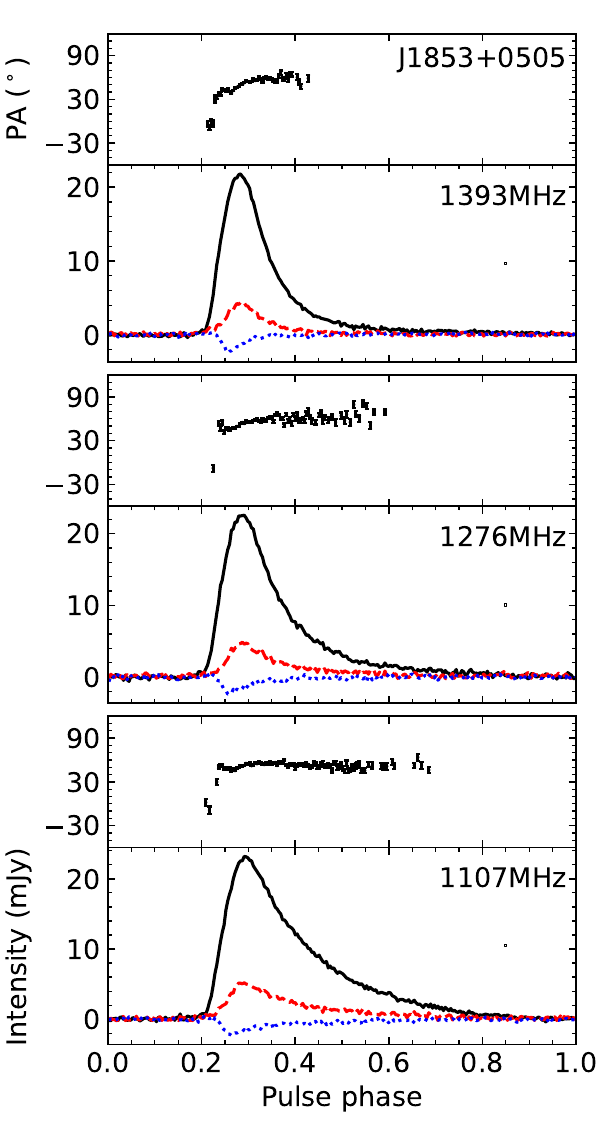}
  \caption{-- {\it continued}}
\end{figure*}

\addtocounter{figure}{-1}
\begin{figure*}
  \centering
  \includegraphics[width=0.3\columnwidth]{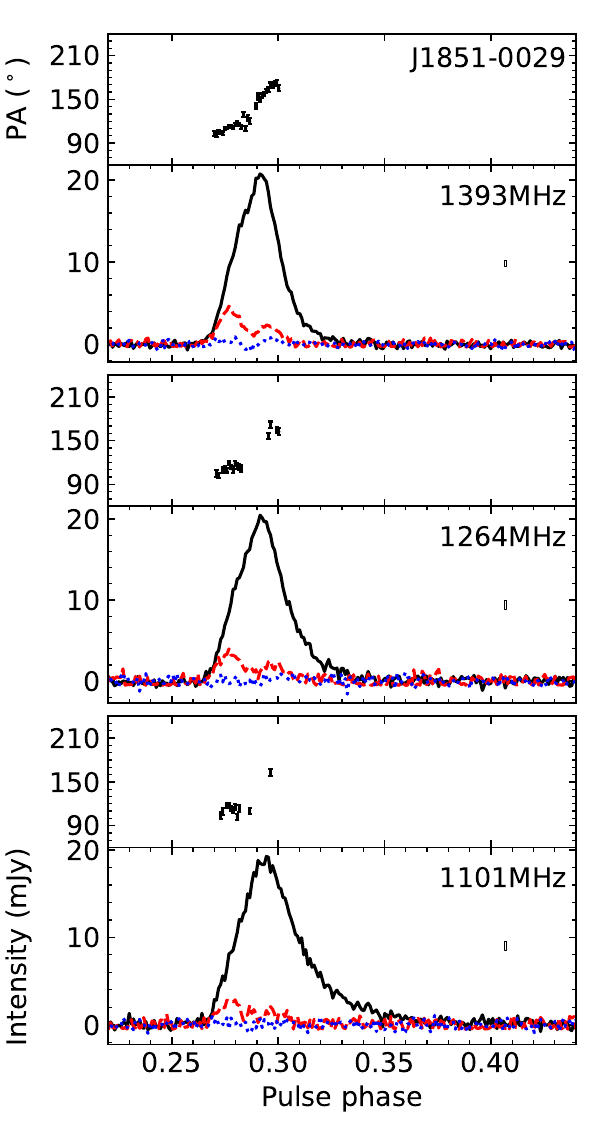}
  \includegraphics[width=0.3\columnwidth]{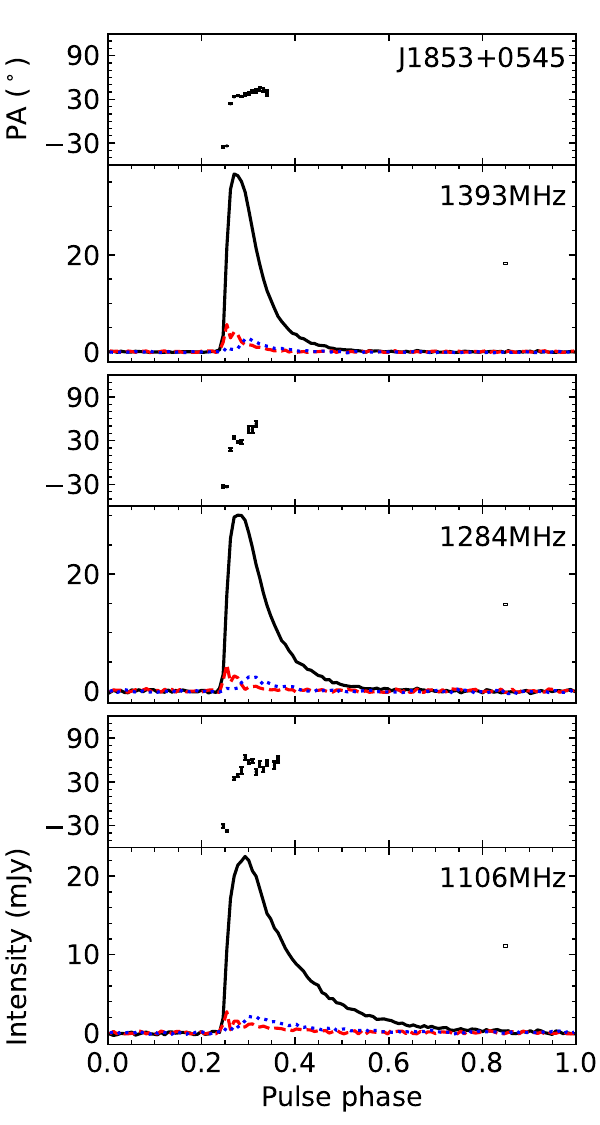}
  \includegraphics[width=0.3\columnwidth]{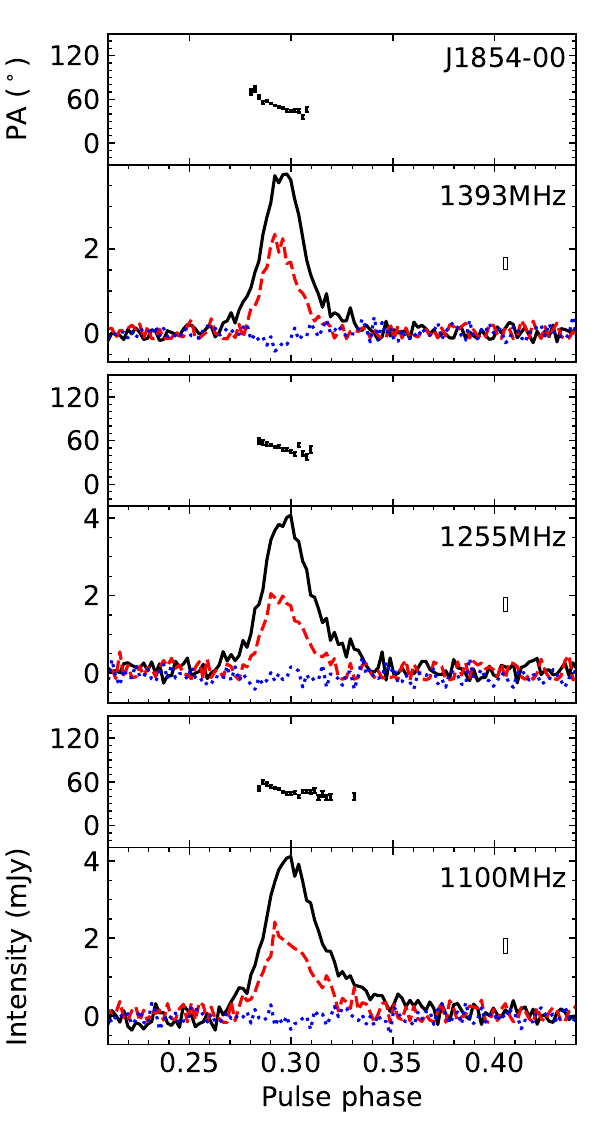}
  \includegraphics[width=0.3\columnwidth]{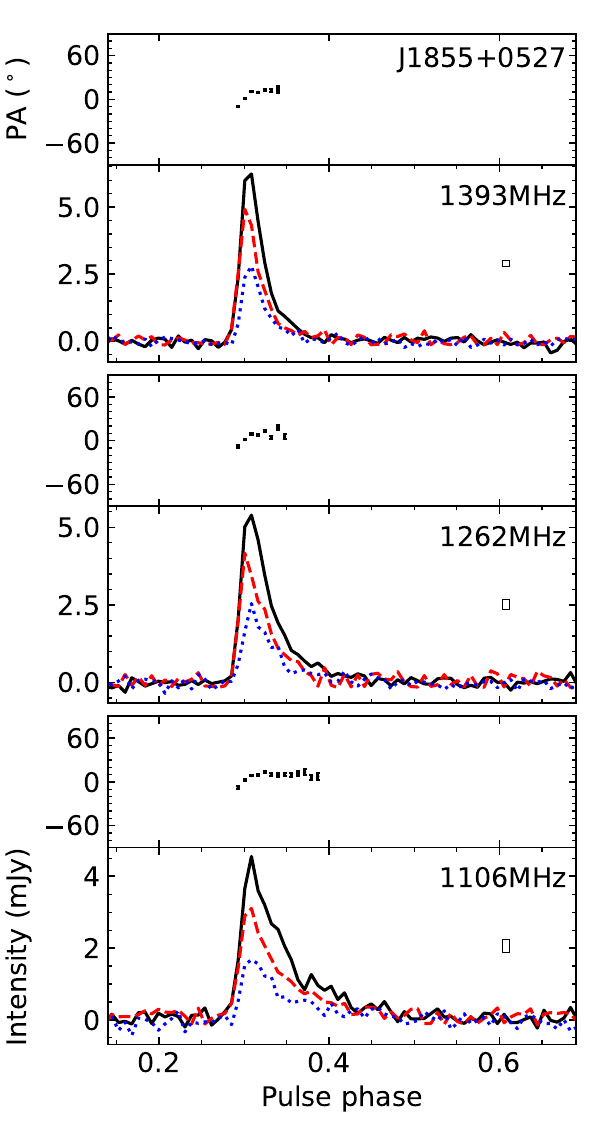}
  \includegraphics[width=0.3\columnwidth]{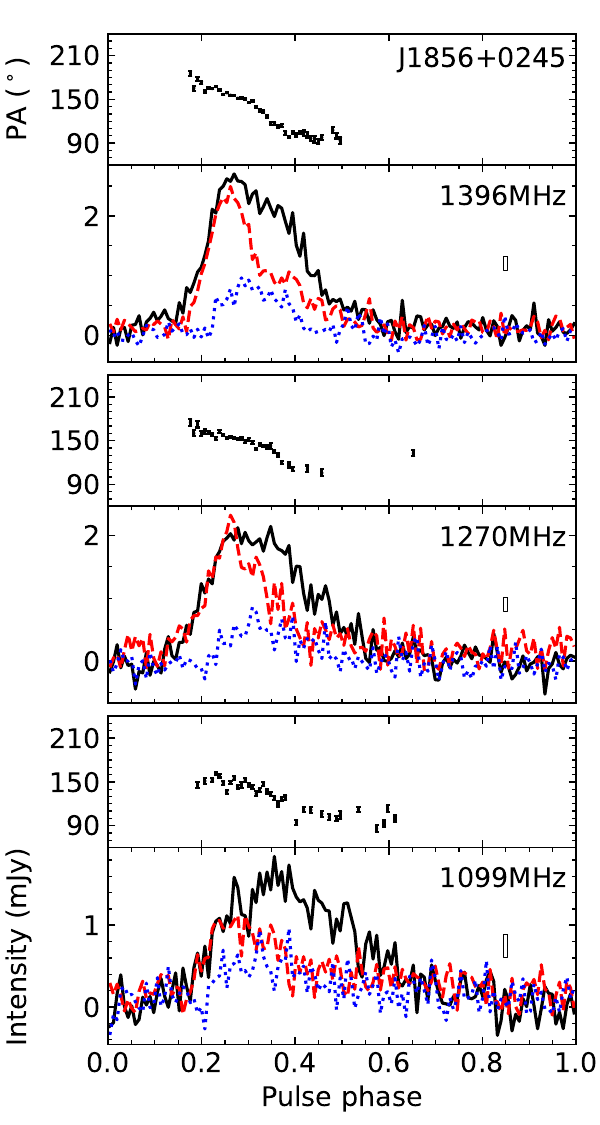}
  \includegraphics[width=0.3\columnwidth]{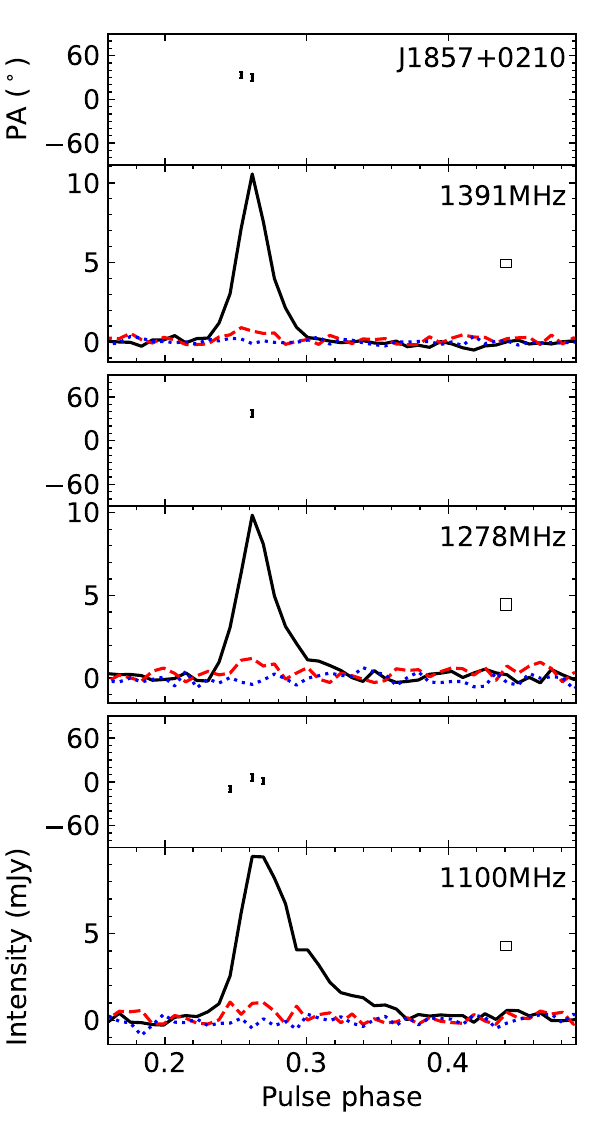}
  \caption{-- {\it continued}}
\end{figure*}

\addtocounter{figure}{-1}
\begin{figure*}
  \centering
  \includegraphics[width=0.3\columnwidth]{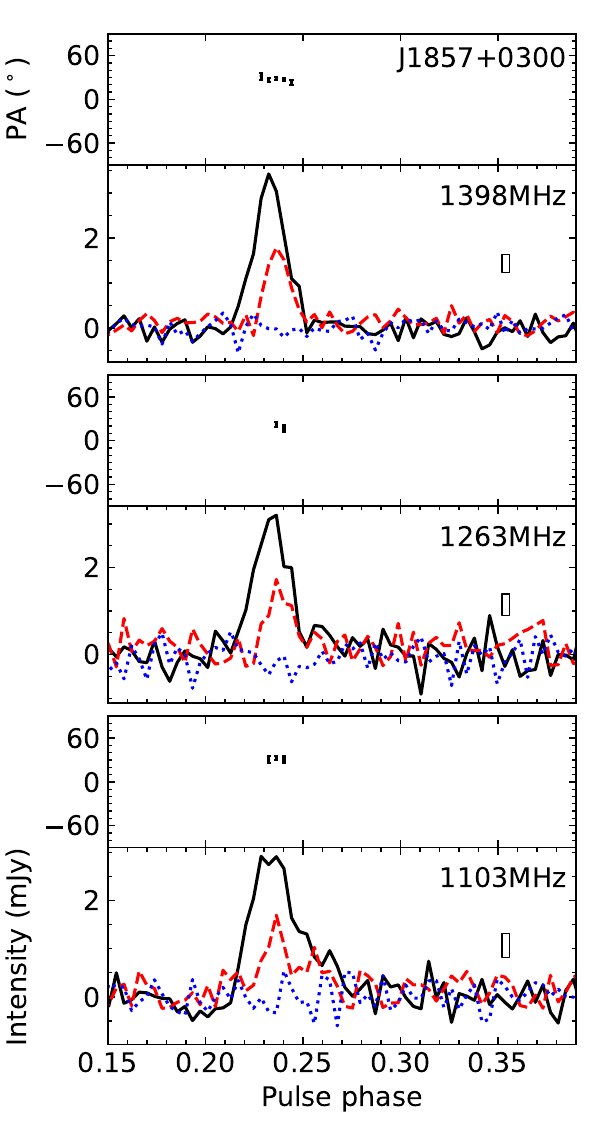}
  \includegraphics[width=0.3\columnwidth]{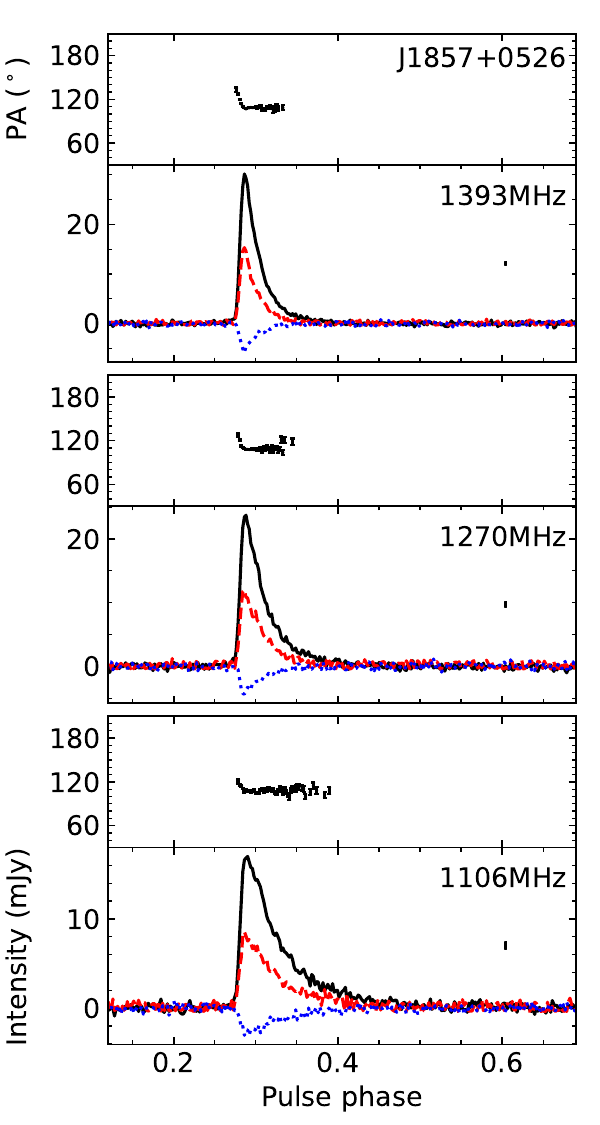}
  \includegraphics[width=0.3\columnwidth]{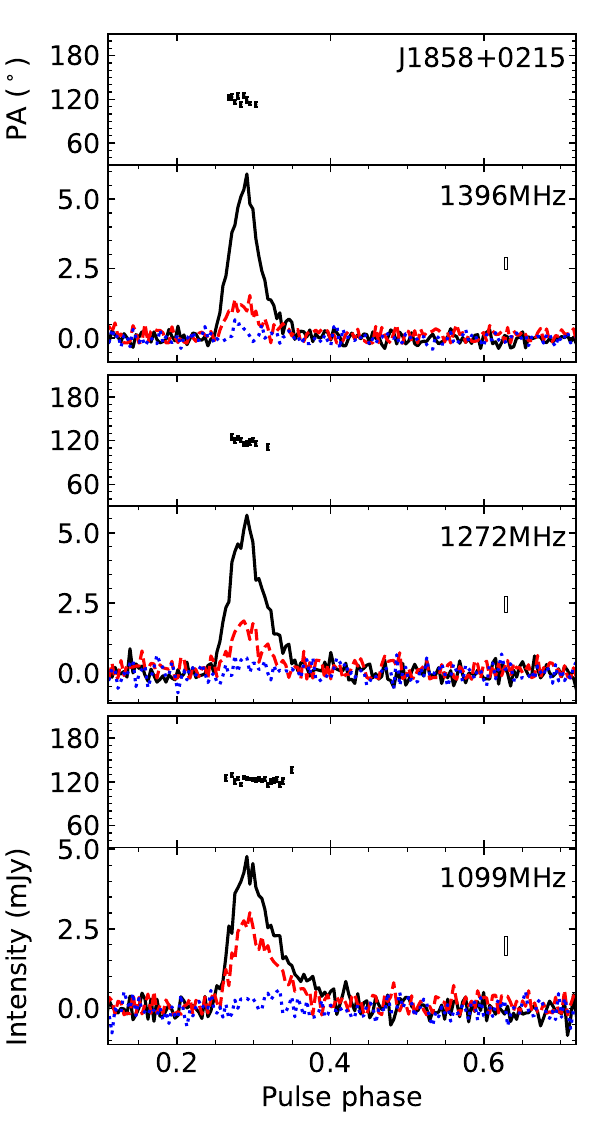}
  \includegraphics[width=0.3\columnwidth]{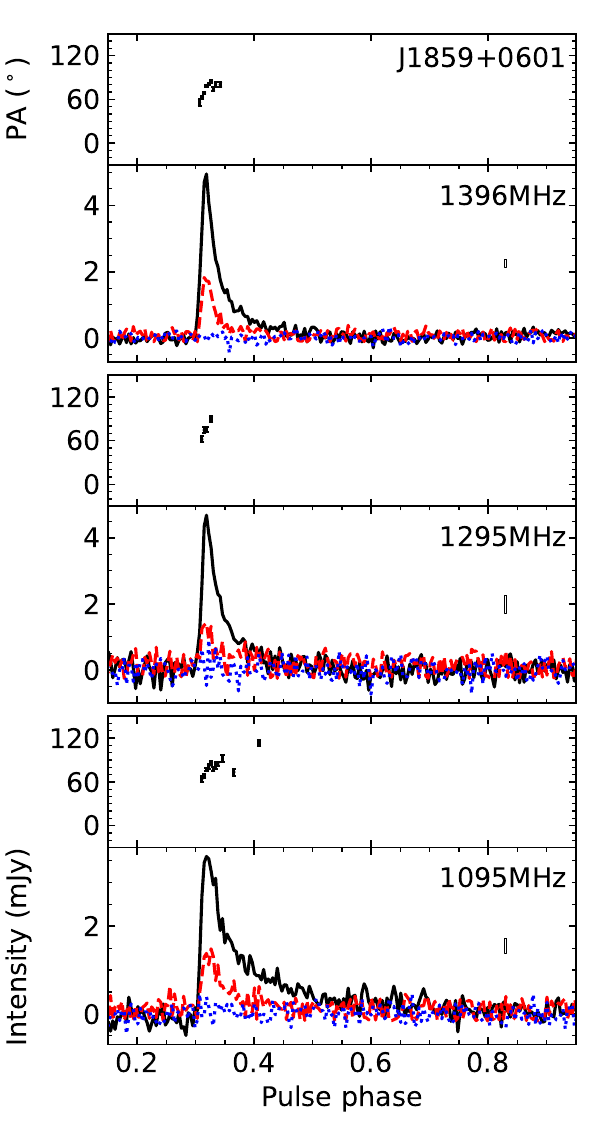}
  \includegraphics[width=0.3\columnwidth]{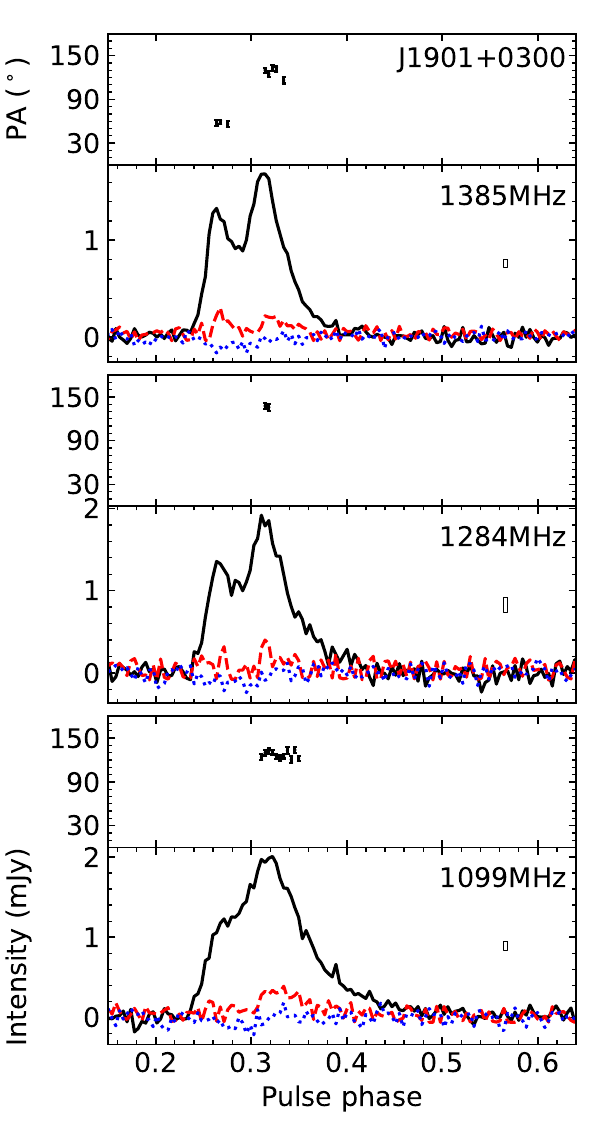}
  \includegraphics[width=0.3\columnwidth]{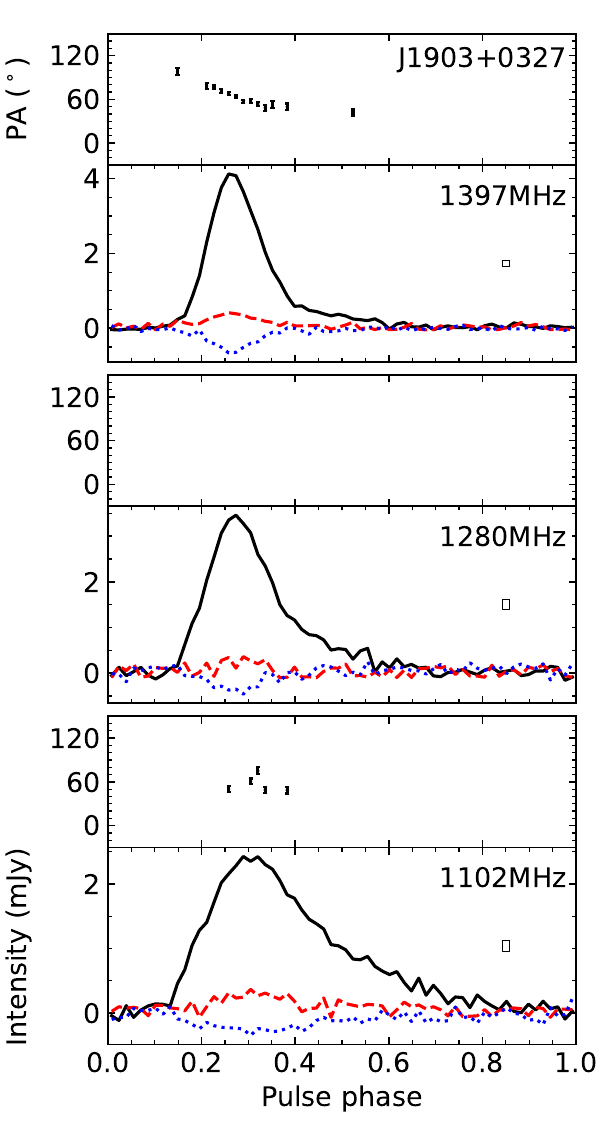}
  \caption{-- {\it continued}}
\end{figure*}

\addtocounter{figure}{-1}
\begin{figure*}
  \centering
  \includegraphics[width=0.3\columnwidth]{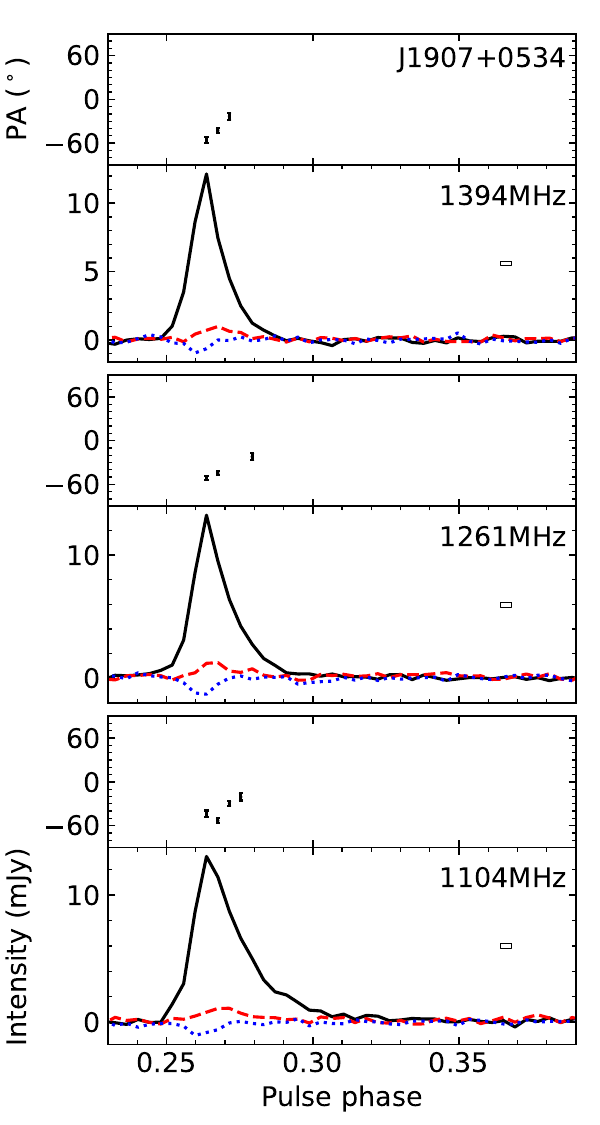}
  \includegraphics[width=0.3\columnwidth]{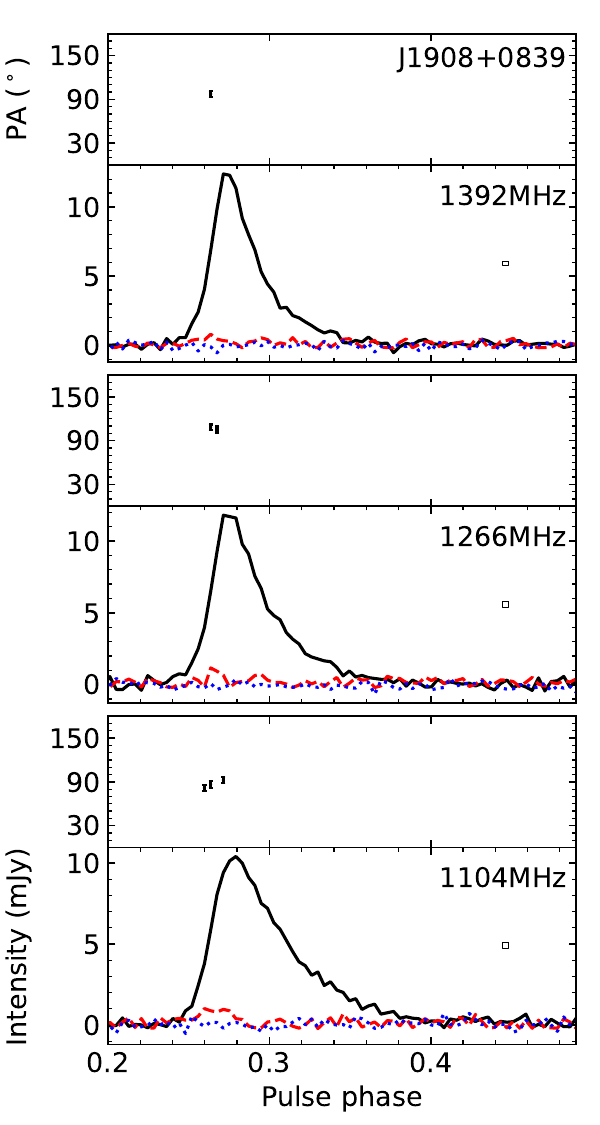}
  \includegraphics[width=0.3\columnwidth]{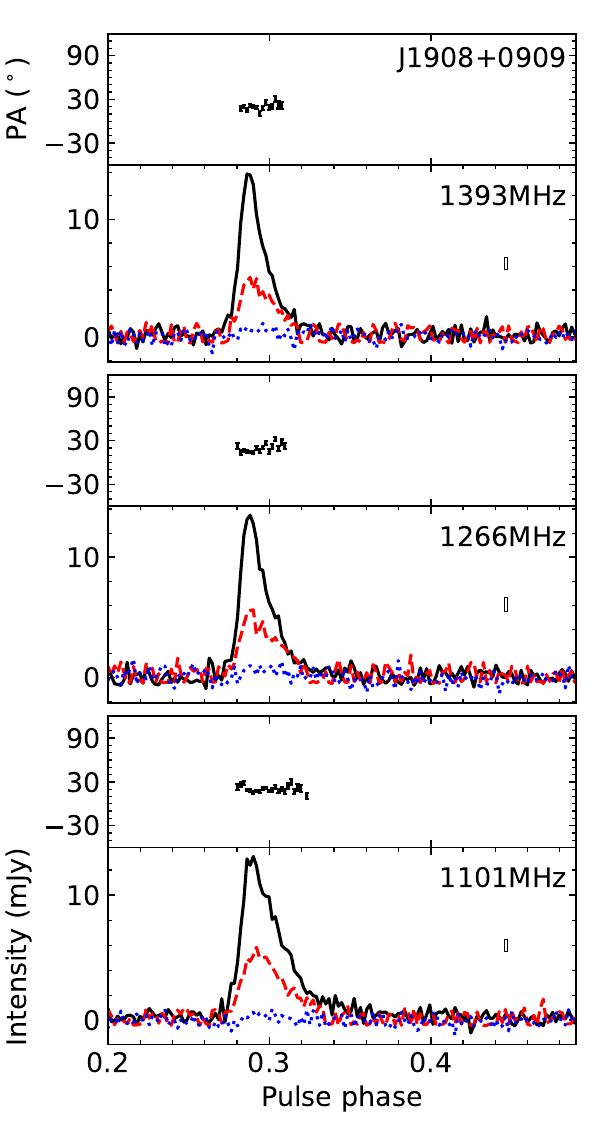}
  \includegraphics[width=0.3\columnwidth]{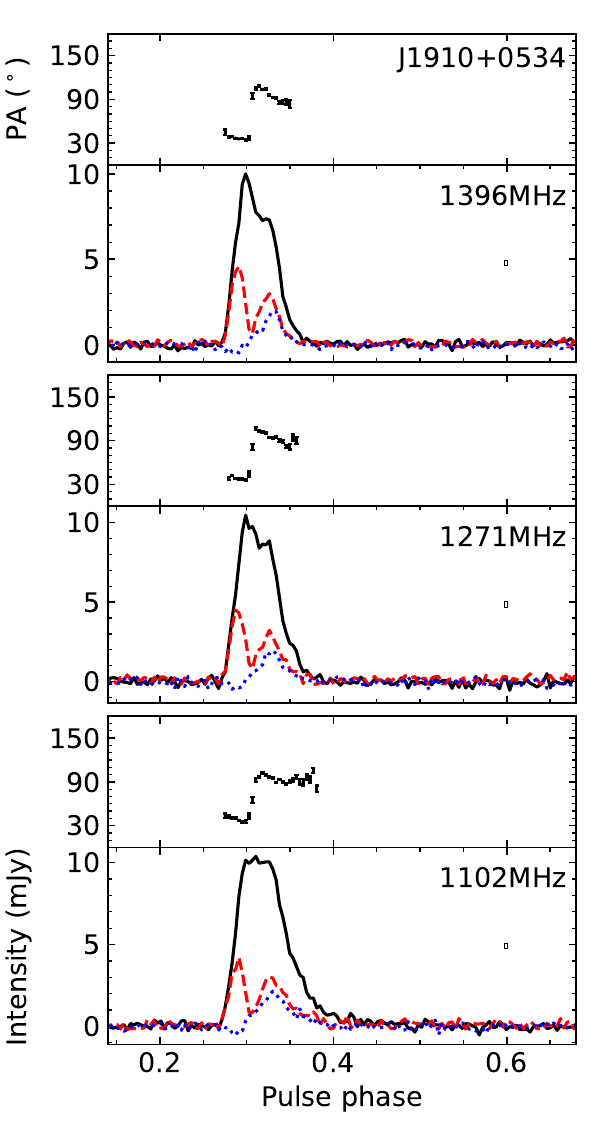}
  \includegraphics[width=0.3\columnwidth]{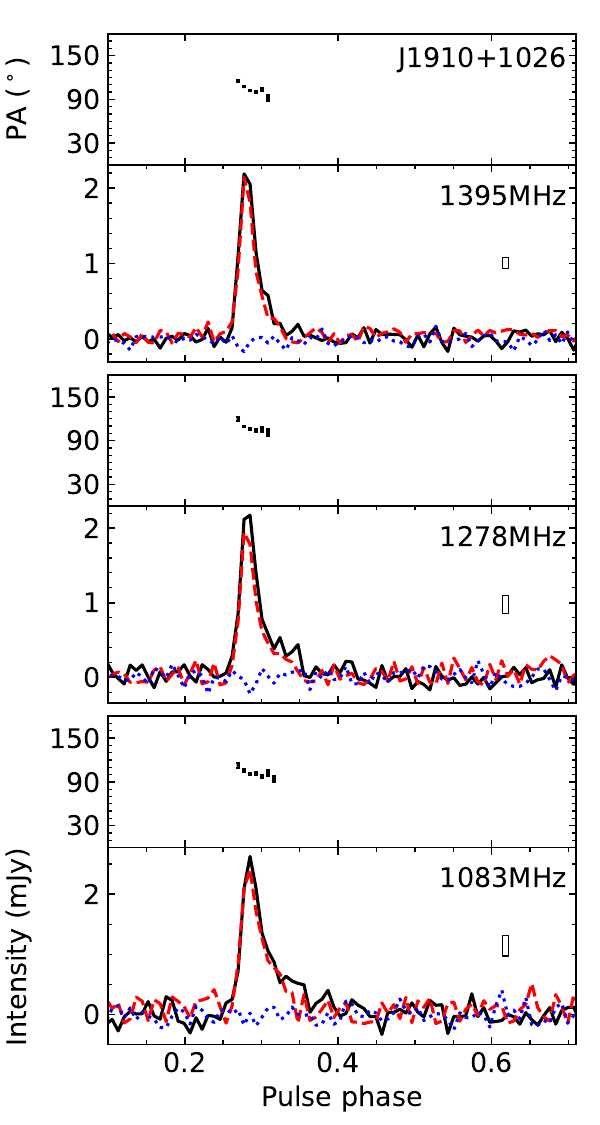}
  \includegraphics[width=0.3\columnwidth]{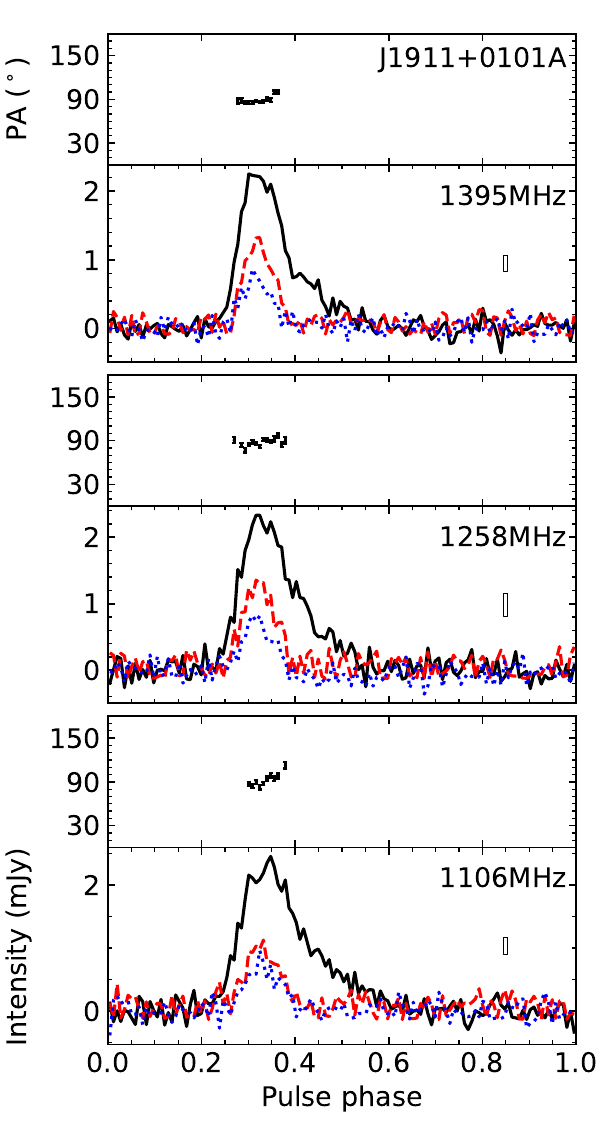}
  \caption{-- {\it continued}}
\end{figure*}

\addtocounter{figure}{-1}
\begin{figure*}
  \centering
  \includegraphics[width=0.3\columnwidth]{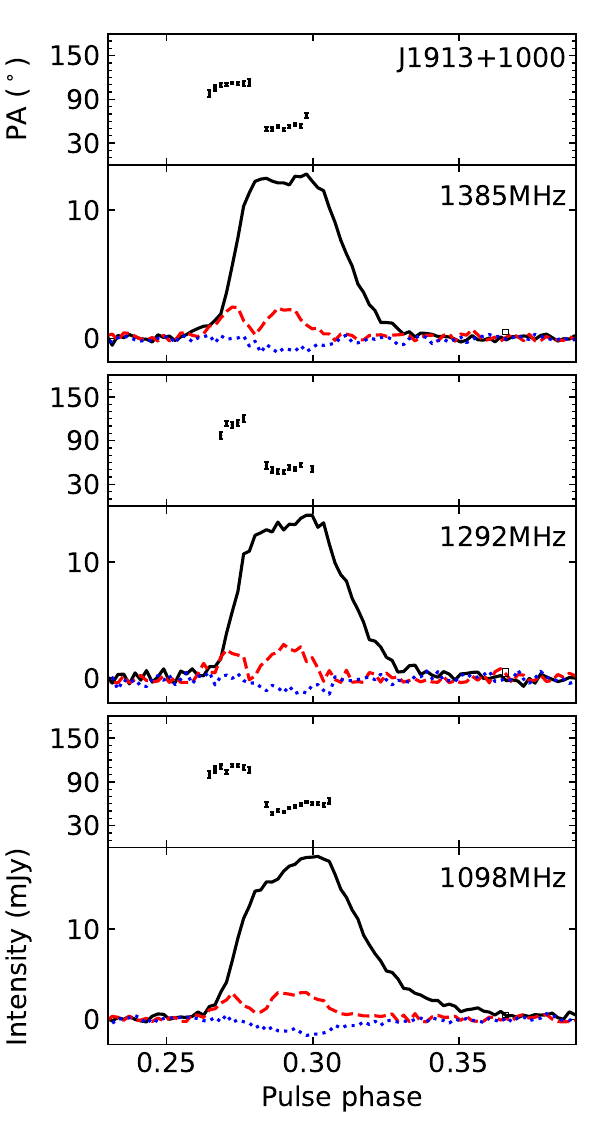}
  \includegraphics[width=0.3\columnwidth]{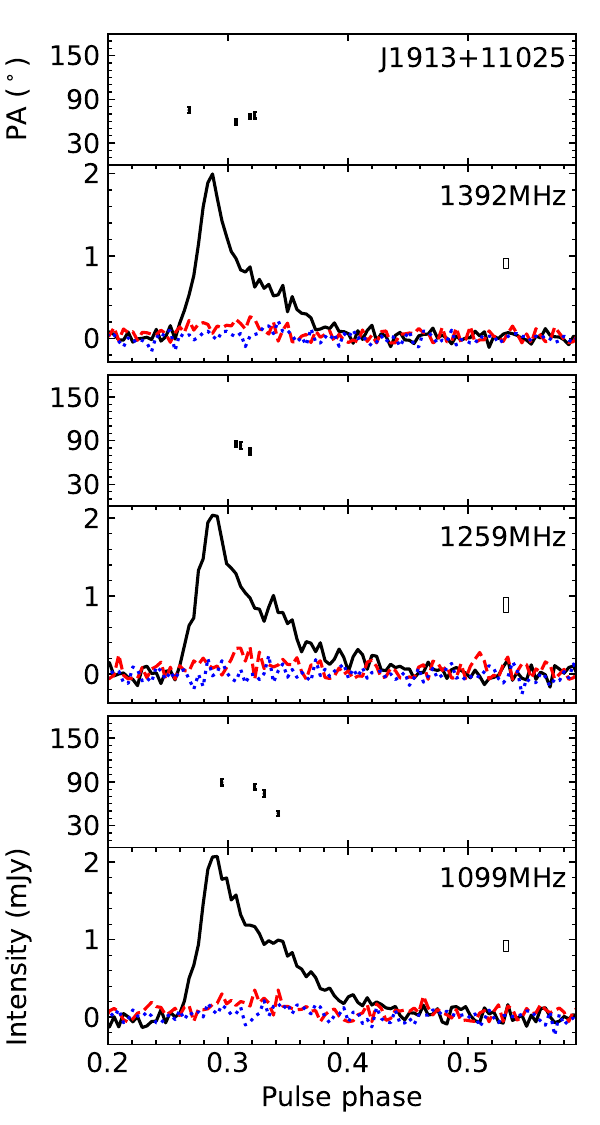}
  \includegraphics[width=0.3\columnwidth]{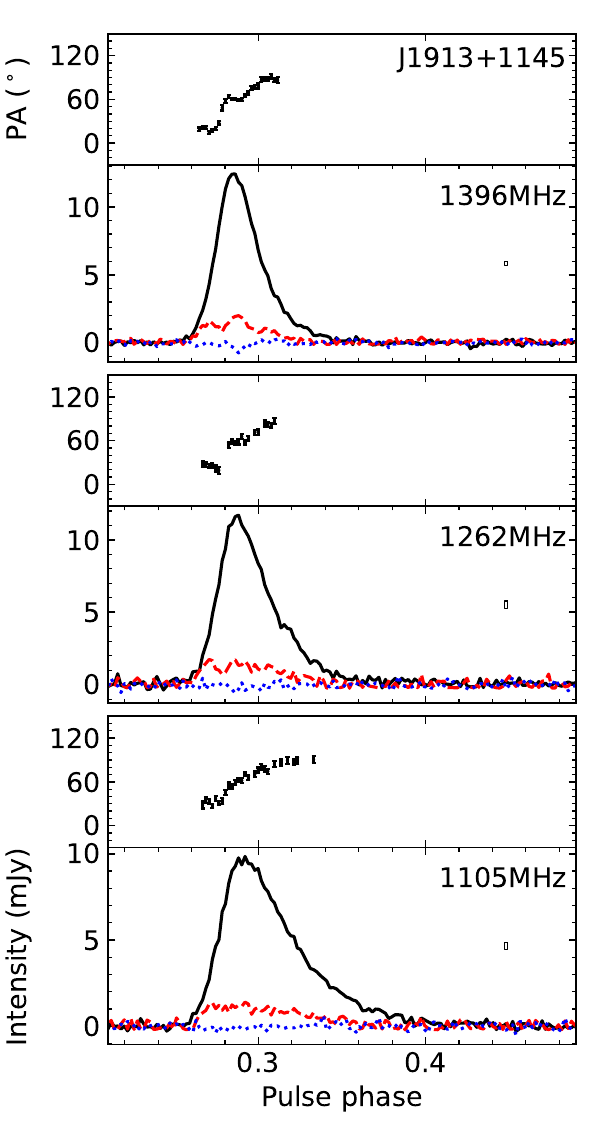}
  \includegraphics[width=0.3\columnwidth]{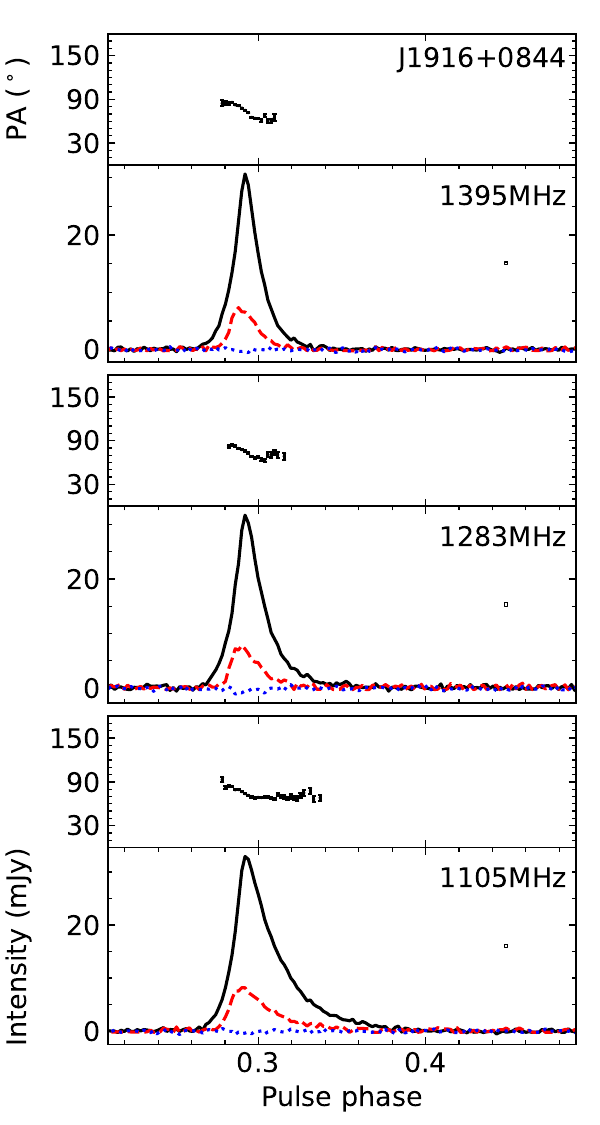}
  \includegraphics[width=0.3\columnwidth]{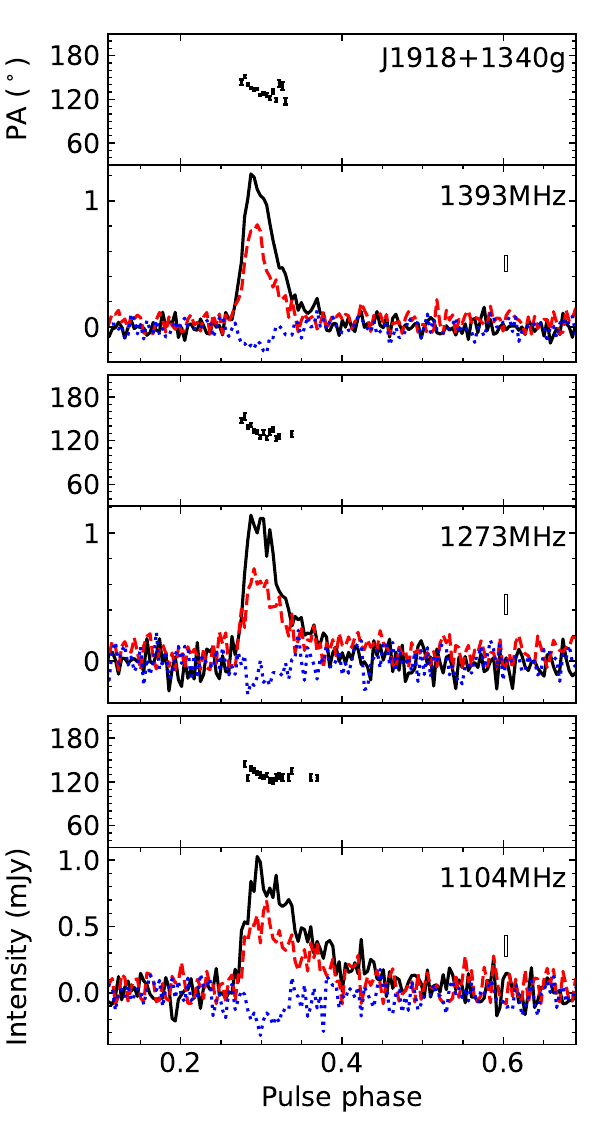}
  \includegraphics[width=0.3\columnwidth]{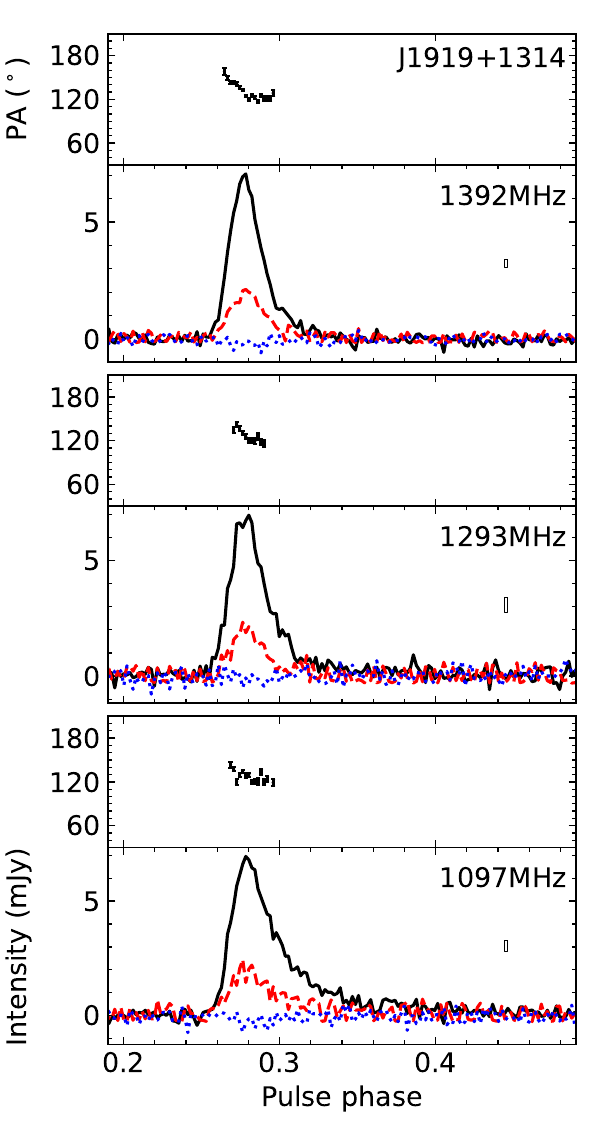}
  \caption{-- {\it continued}}
\end{figure*}

\addtocounter{figure}{-1}
\begin{figure*}
  \centering
  \includegraphics[width=0.3\columnwidth]{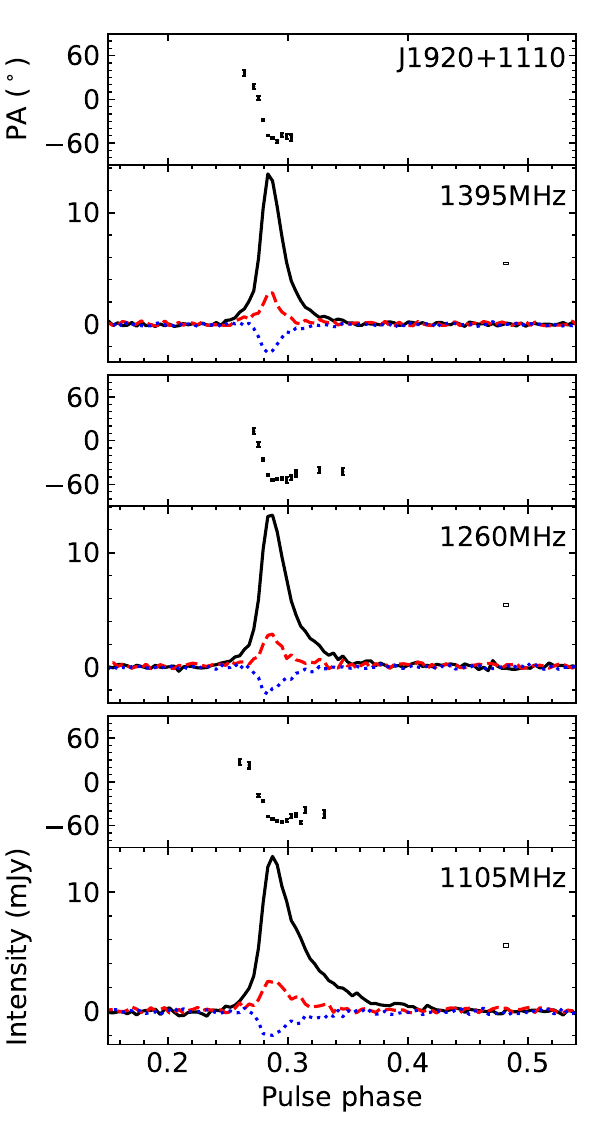}
  \includegraphics[width=0.3\columnwidth]{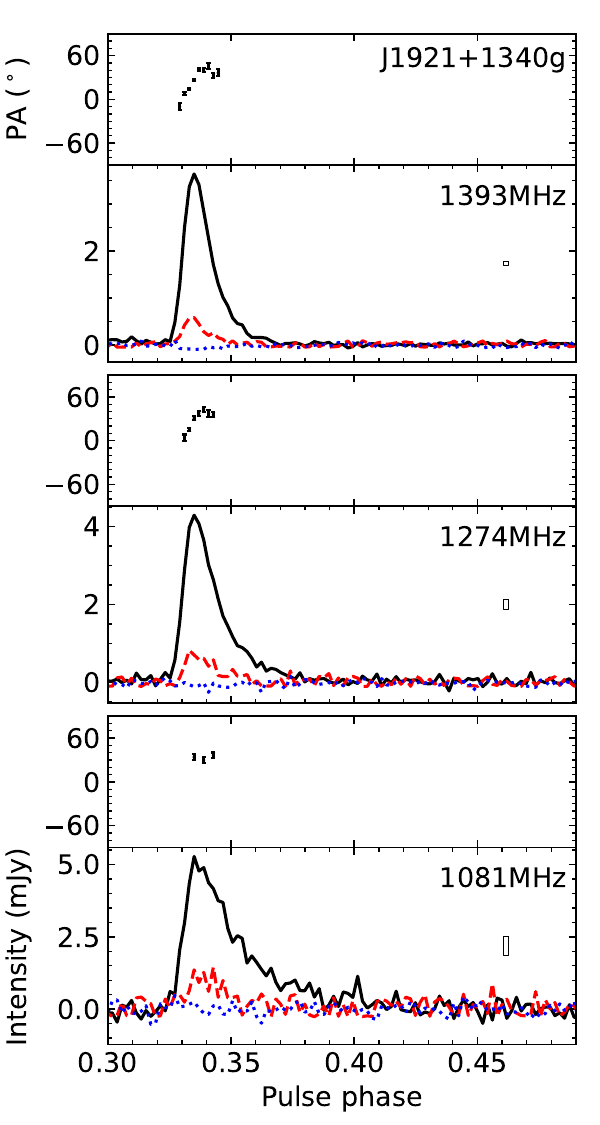}
  \includegraphics[width=0.3\columnwidth]{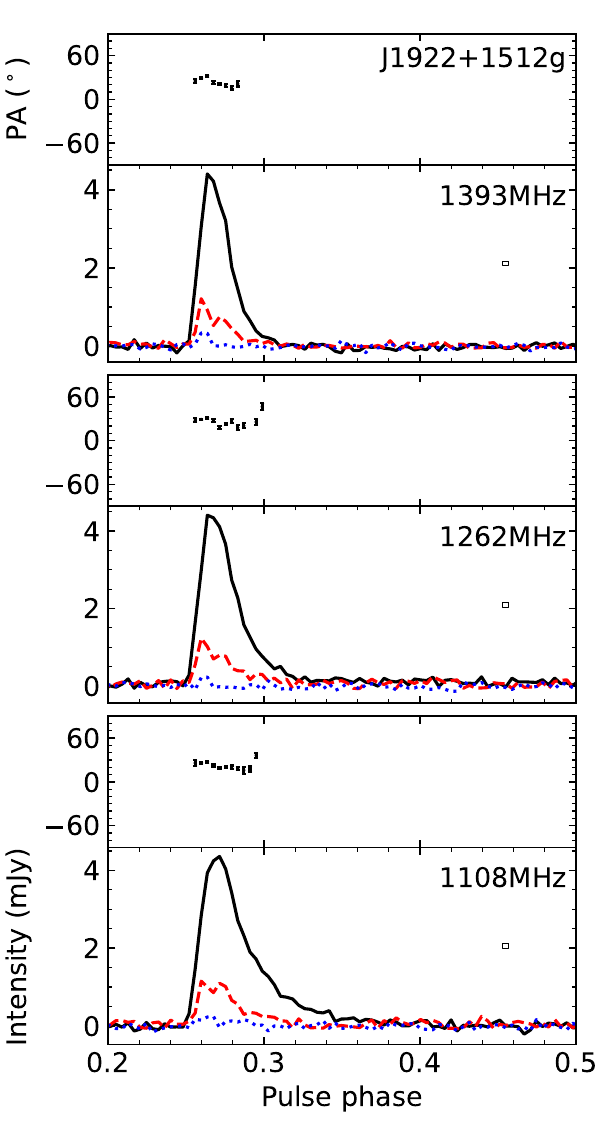}
  \includegraphics[width=0.3\columnwidth]{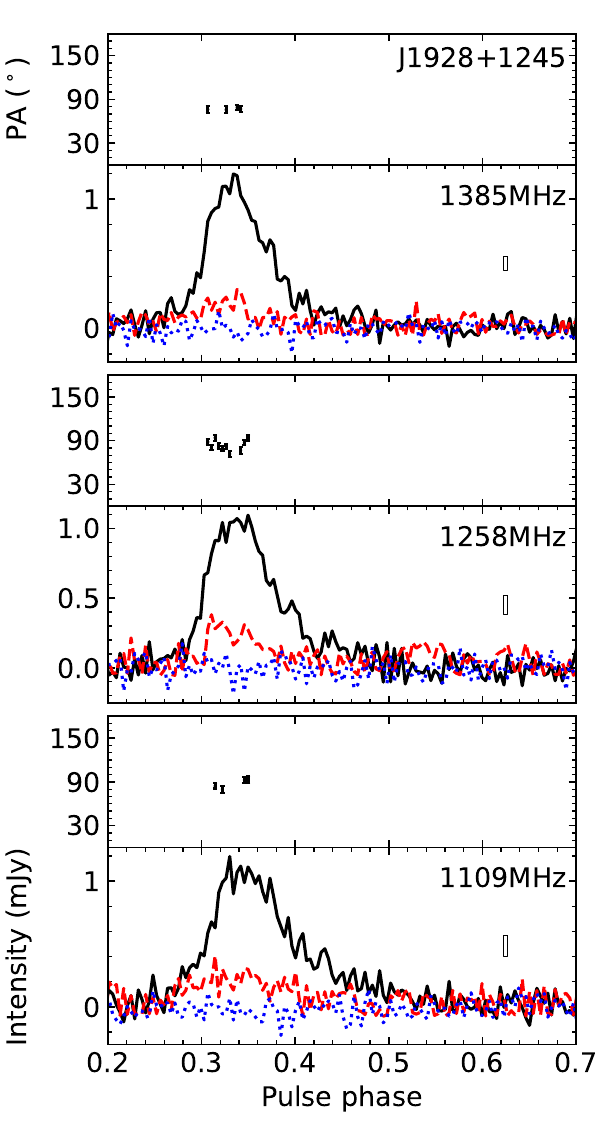}
  \includegraphics[width=0.3\columnwidth]{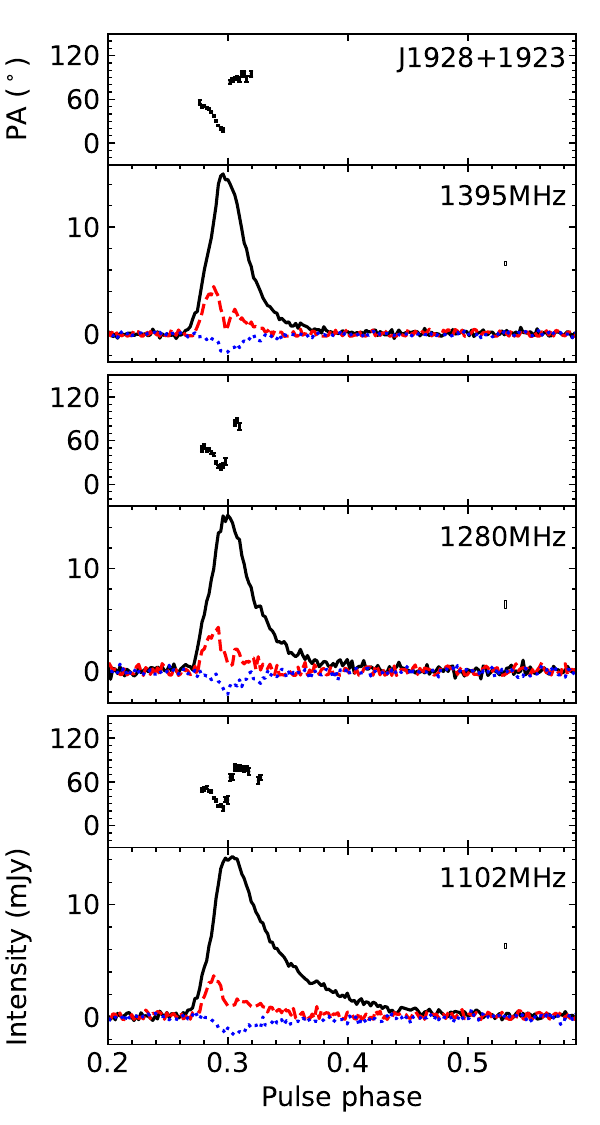}
  \includegraphics[width=0.3\columnwidth]{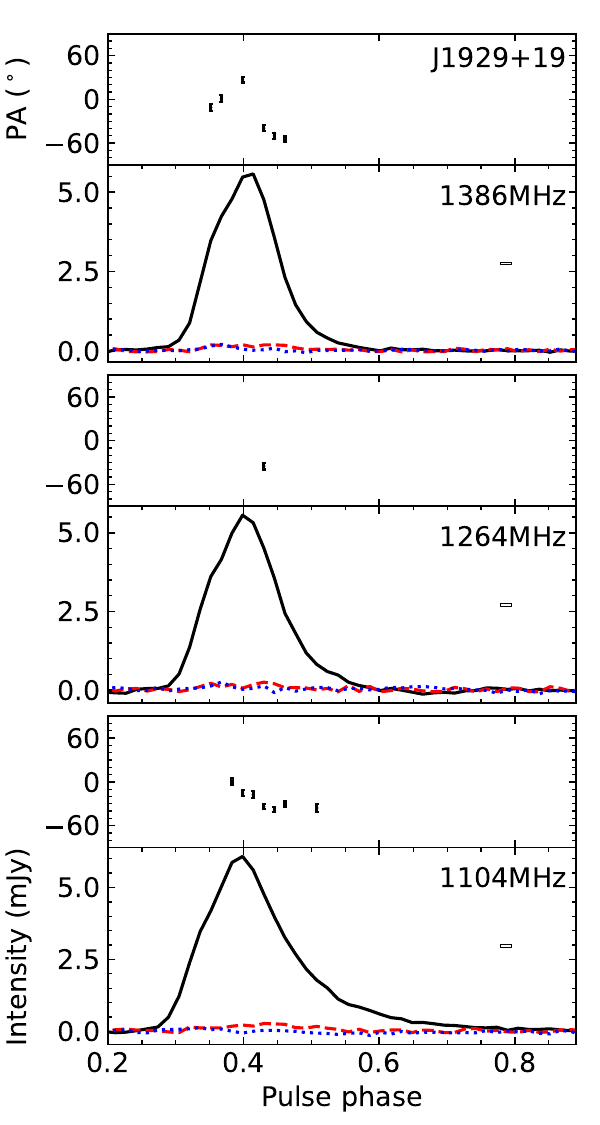}
  \caption{-- {\it continued}}
\end{figure*}

\addtocounter{figure}{-1}
\begin{figure*}
  \centering
  \includegraphics[width=0.3\columnwidth]{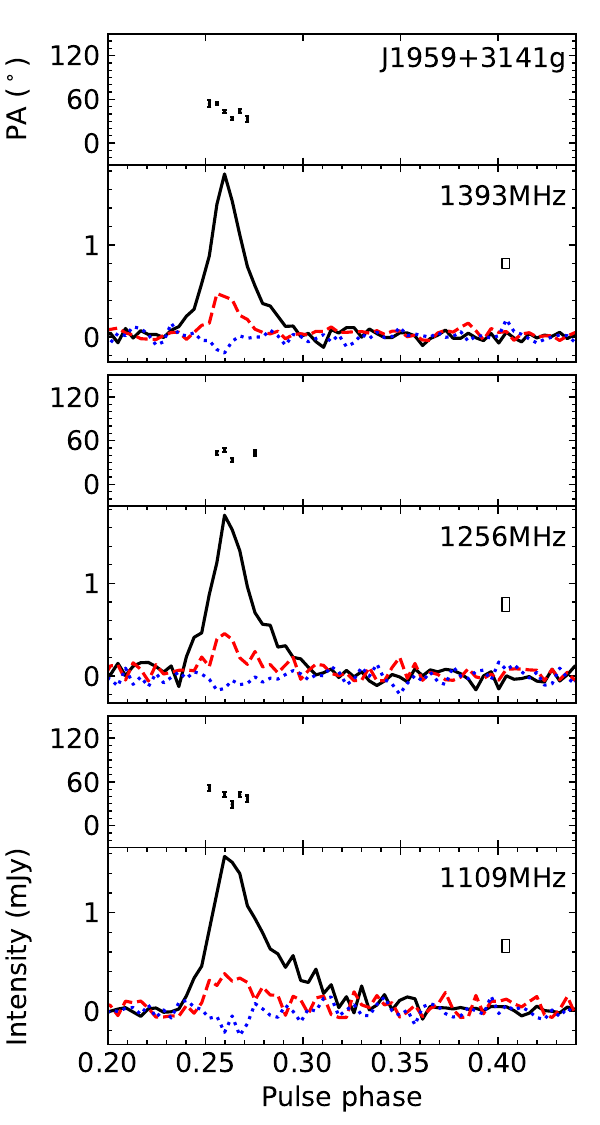}
  \includegraphics[width=0.3\columnwidth]{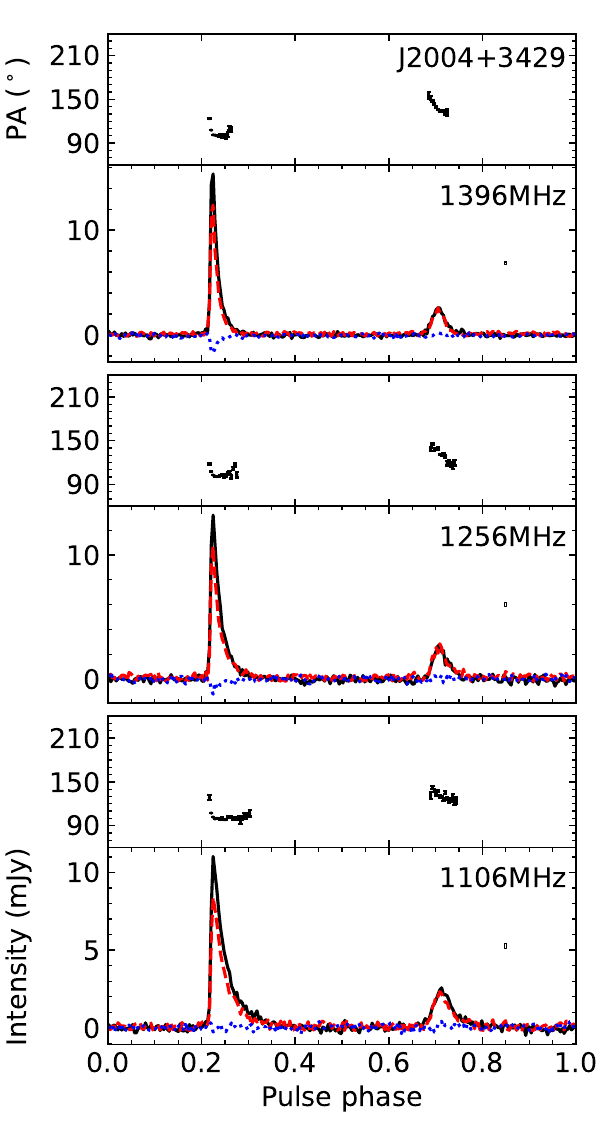}
  \includegraphics[width=0.3\columnwidth]{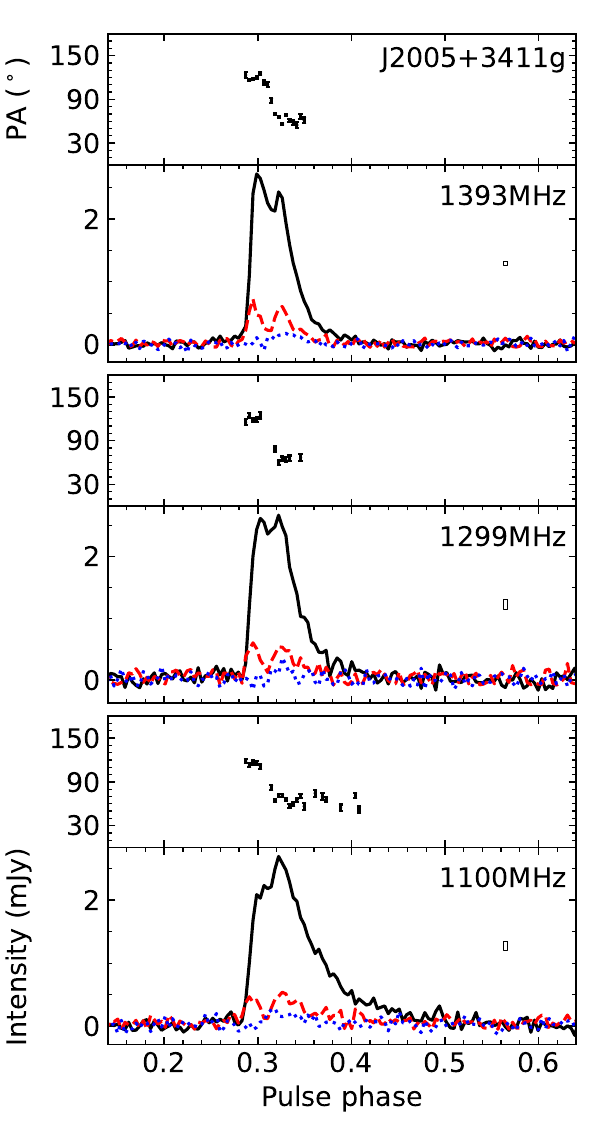}
  \includegraphics[width=0.3\columnwidth]{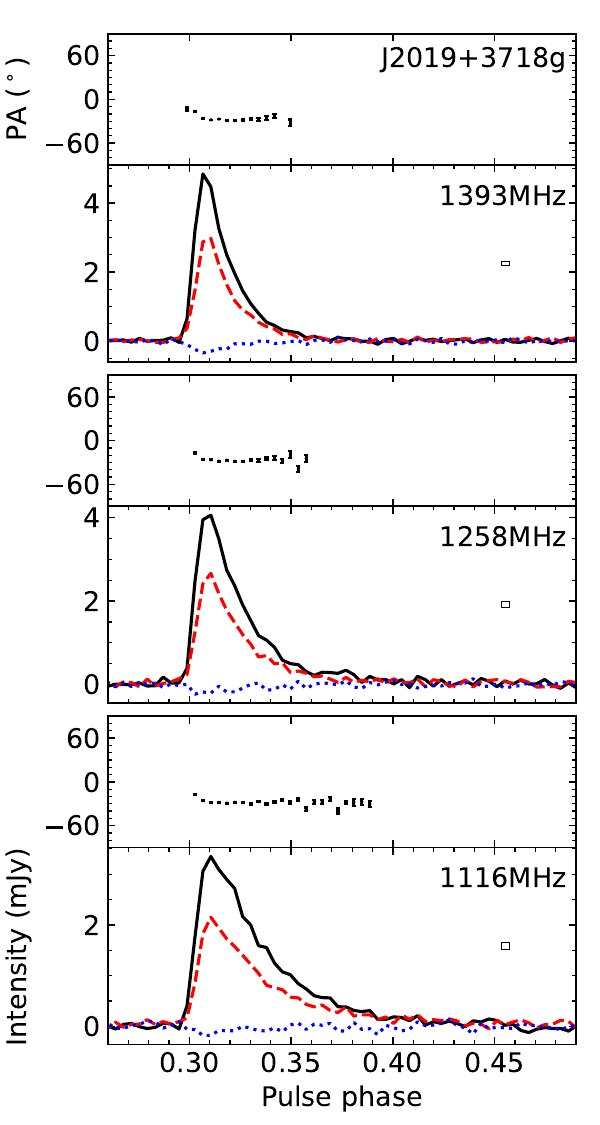}
  \includegraphics[width=0.3\columnwidth]{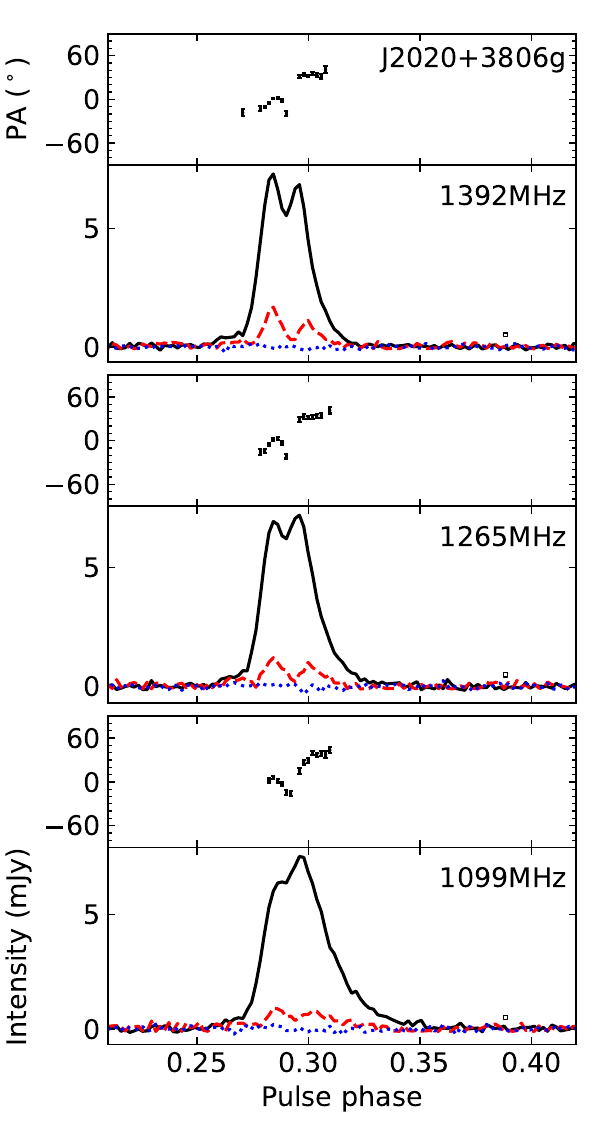}
  \includegraphics[width=0.3\columnwidth]{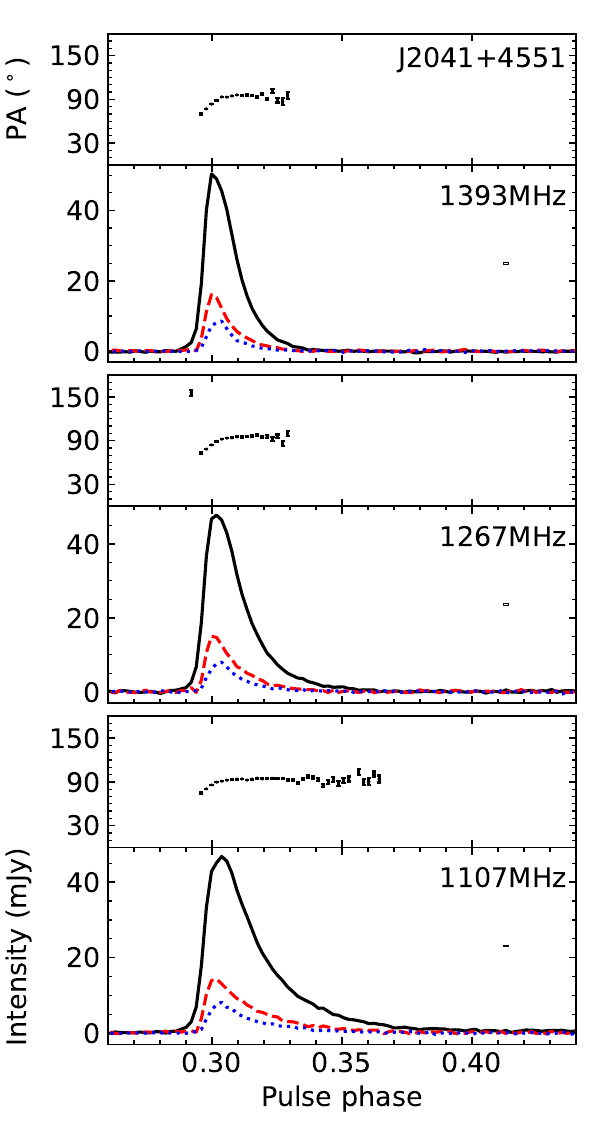}
  \caption{-- {\it continued}}
\end{figure*}

\addtocounter{figure}{-1}
\begin{figure*}
  \centering
  \includegraphics[width=0.3\columnwidth]{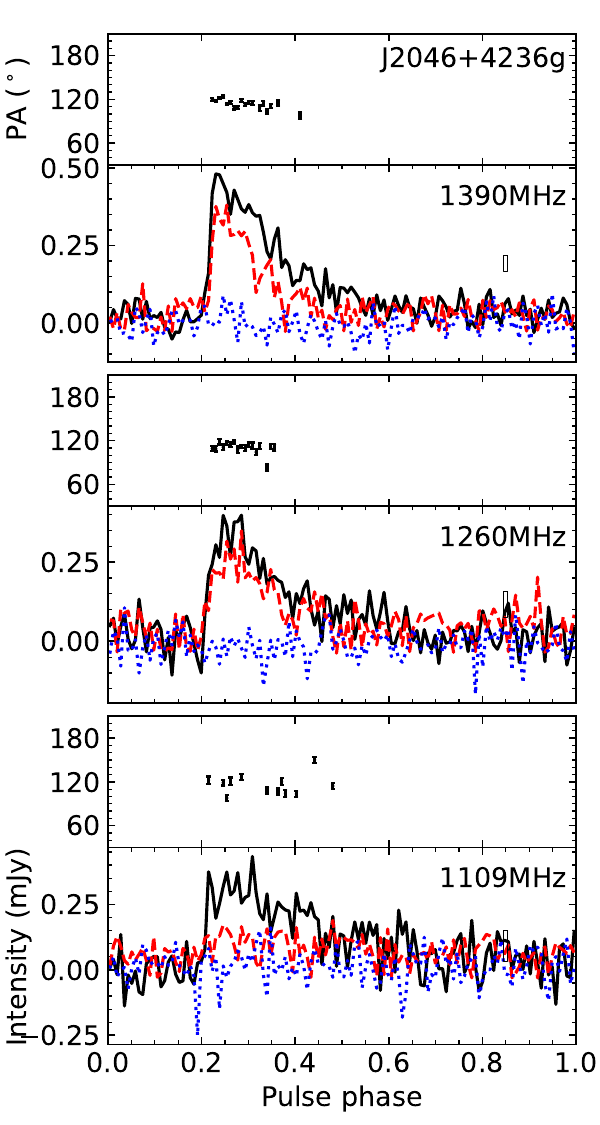}
  \includegraphics[width=0.3\columnwidth]{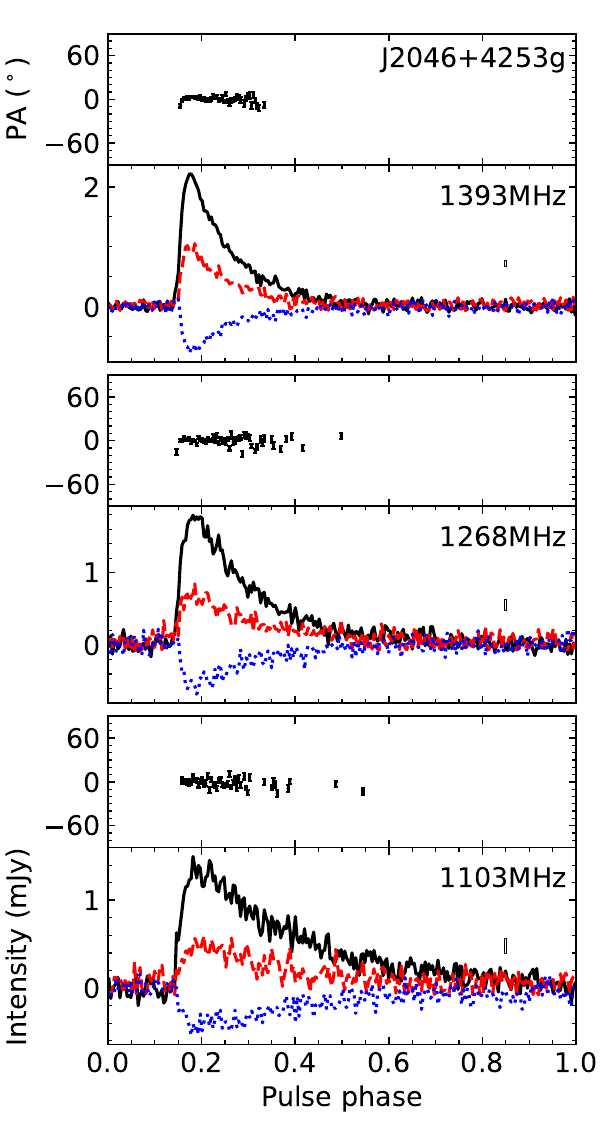}
  \includegraphics[width=0.3\columnwidth]{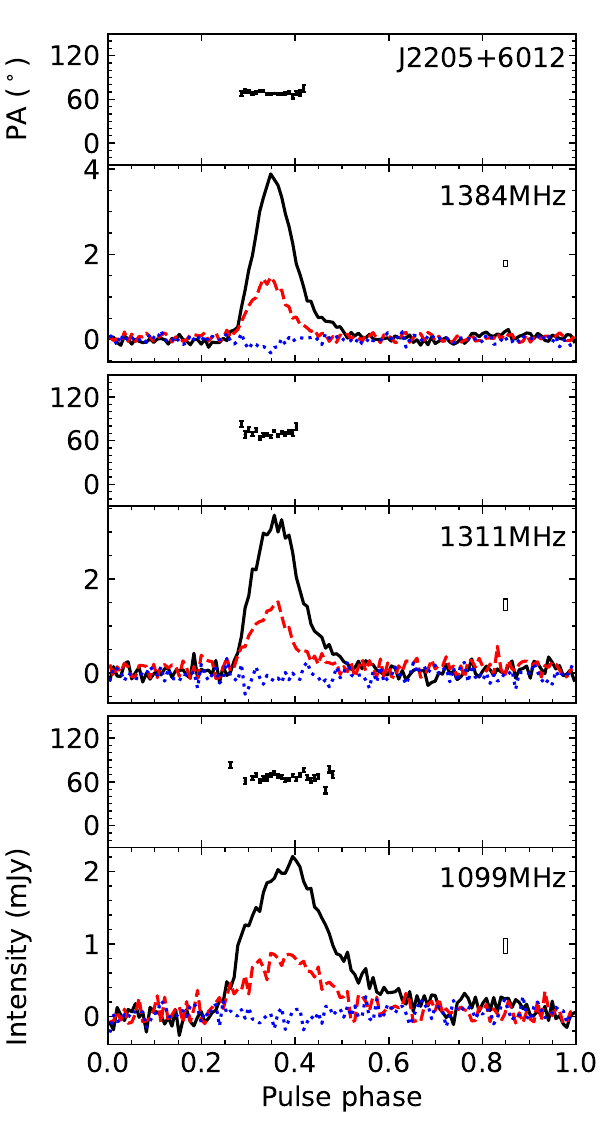}
  \caption{-- {\it  ended}}
\end{figure*}

\begin{table}[hbtp]
    \centering
    \caption{Polarization parameters of subband profiles and the rotation measures for 82 scattered pulsars observed by FAST.}
    \label{tab:pol}
    \footnotesize
    \begin{tabular}{lcrrrrrr}
\hline\noalign{\smallskip}        
\multicolumn{1}{c}{PSR  Name}&\multicolumn{1}{c}{FAST Obs. Date}&\multicolumn{1}{c}{RM}&\multicolumn{1}{c}{$f$}&\multicolumn{1}{c}{$I$}&\multicolumn{1}{c}{$L/I$} &\multicolumn{1}{c}{$V/I$} &\multicolumn{1}{c}{$|V|/I$} \\
\multicolumn{1}{c}{ }&\multicolumn{1}{c}{yyyy-mm-dd}&\multicolumn{1}{c}{($\rm rad/m^2$)}&\multicolumn{1}{c}{(MHz)}&\multicolumn{1}{c}{($\rm \mu Jy$)}&\multicolumn{1}{c}{(\%)}&\multicolumn{1}{c}{(\%)}&\multicolumn{1}{c}{(\%)} \\
\hline\noalign{\smallskip}
J0248+6021  &2023-08-14&$-149.990\pm0.005$&     1110&$ 22263\pm1113$&$  69.4\pm3.0$&$   0.3\pm3.0$&$   0.3\pm3.0$ \\ 
            &&&     1265&$ 17345\pm867$&$  71.9\pm3.0$&$   2.2\pm3.0$&$   2.2\pm3.0$ \\ 
            &&&     1384&$ 15012\pm750$&$  71.6\pm3.0$&$   1.6\pm3.0$&$   1.7\pm3.0$ \\ 
J1822-1400  &2026-04-24&$+895.8\pm0.9$&     1109&$   717\pm 45$&$  19.9\pm3.1$&$  -5.7\pm3.1$&$   5.1\pm3.1$ \\ 
            &&&     1255&$   471\pm 40$&$  19.4\pm3.2$&$  -7.5\pm3.1$&$   5.6\pm3.1$ \\ 
            &&&     1411&$   343\pm 33$&$  18.3\pm3.2$&$  -8.0\pm3.1$&$   7.5\pm3.1$ \\ 
J1824-1118  &2025-08-18&$+184.9\pm0.2$&     1100&$  2527\pm126$&$  22.0\pm3.0$&$   0.7\pm3.0$&$   4.6\pm3.0$ \\ 
            &&&     1259&$  1712\pm 86$&$  19.2\pm3.0$&$  -1.4\pm3.0$&$   5.7\pm3.0$ \\ 
            &&&     1388&$  1387\pm 69$&$  17.6\pm3.0$&$  -1.6\pm3.0$&$   7.8\pm3.0$ \\ 
J1833-0556  &2025-08-07&$+1093.1\pm0.9$&     1097&$   270\pm 15$&$  17.7\pm3.3$&$   6.4\pm3.2$&$   4.0\pm3.2$ \\ 
            &&&     1261&$   225\pm 13$&$  21.4\pm3.3$&$   5.2\pm3.1$&$   5.0\pm3.1$ \\ 
            &&&     1393&$   199\pm 11$&$  23.6\pm3.2$&$   6.0\pm3.1$&$   6.4\pm3.1$ \\ 
J1834-0602  &2025-10-23&$397.7\pm0.6$&     1104&$  1308\pm 73$&$  16.5\pm3.0$&$  -2.3\pm3.0$&$   5.6\pm3.0$ \\ 
            &&&     1268&$  1205\pm 72$&$  21.7\pm3.1$&$  -3.1\pm3.0$&$   6.5\pm3.0$ \\ 
            &&&     1384&$  1127\pm 68$&$  22.9\pm3.1$&$  -2.3\pm3.0$&$   7.1\pm3.0$ \\ 
J1837-0604  &2020-10-10&$324.1\pm0.4$&     1113&$   972\pm 50$&$  82.2\pm3.3$&$ -21.8\pm3.1$&$  20.0\pm3.1$ \\ 
            &&&     1255&$   869\pm 44$&$  84.5\pm3.2$&$ -10.6\pm3.0$&$  10.8\pm3.0$ \\ 
            &&&     1385&$   815\pm 42$&$  79.8\pm3.1$&$ -16.8\pm3.1$&$  15.3\pm3.1$ \\ 
J1838-0453  &2025-09-11&$-44.9\pm0.5$&     1100&$   533\pm 34$&$  41.7\pm3.2$&$  -3.1\pm3.1$&$   1.7\pm3.1$ \\ 
            &&&     1276&$   425\pm 40$&$  45.1\pm3.2$&$  -2.2\pm3.1$&$   0.4\pm3.1$ \\ 
            &&&     1384&$   413\pm 33$&$  43.4\pm3.1$&$  -3.1\pm3.1$&$   1.9\pm3.1$ \\ 
J1838-0508g &2026-01-03&$-172\pm3$&     1105&$    76\pm  9$&$  52.3\pm6.5$&$  -6.1\pm5.8$&$  -6.2\pm5.8$ \\ 
            &&&     1263&$    63\pm 15$&$  52.4\pm7.2$&$ -14.9\pm5.7$&$   8.0\pm5.7$ \\ 
            &&&     1384&$    59\pm  7$&$  49.4\pm5.4$&$ -14.9\pm4.1$&$  13.6\pm4.1$ \\ 
J1841-0157  &2021-09-01&$+100.7\pm0.4$&     1100&$  2384\pm131$&$  15.5\pm3.0$&$  -3.6\pm3.0$&$   3.2\pm3.0$ \\ 
            &&&     1276&$  2055\pm126$&$  15.3\pm3.0$&$  -4.4\pm3.0$&$   3.8\pm3.0$ \\ 
            &&&     1392&$  1953\pm108$&$  17.8\pm3.0$&$  -2.8\pm3.0$&$   3.3\pm3.0$ \\ 
J1842-0153  &2021-09-03&$+139.8\pm0.5$&     1116&$   923\pm 51$&$  12.3\pm3.0$&$   4.5\pm3.0$&$   3.9\pm3.0$ \\ 
            &&&     1268&$   708\pm 43$&$  14.1\pm3.0$&$   3.4\pm3.0$&$   3.0\pm3.0$ \\ 
            &&&     1392&$   604\pm 35$&$  15.3\pm3.0$&$   4.5\pm3.0$&$   3.9\pm3.0$ \\ 
J1842-0258g &2023-10-26&$-356\pm5$&     1104&$   180\pm 16$&$  11.3\pm3.3$&$  -5.9\pm3.1$&$   5.9\pm3.1$ \\ 
            &&&     1260&$   135\pm 17$&$   9.5\pm3.3$&$  -3.4\pm3.2$&$   3.4\pm3.2$ \\ 
            &&&     1393&$   113\pm  8$&$   9.5\pm3.3$&$  -2.0\pm3.2$&$   1.5\pm3.2$ \\ 
J1842-0309  &2023-10-26&$+25\pm2$&     1109&$   442\pm 26$&$  26.3\pm3.5$&$  -7.6\pm3.3$&$   6.9\pm3.3$ \\ 
            &&&     1260&$   390\pm 24$&$  20.7\pm3.4$&$  -5.3\pm3.2$&$   4.4\pm3.2$ \\ 
            &&&     1393&$   319\pm 21$&$  22.5\pm3.3$&$  -4.4\pm3.1$&$   1.3\pm3.1$ \\ 
J1843-0137  &2022-02-02&$+584\pm2$&     1102&$   398\pm 23$&$  10.4\pm3.1$&$   4.8\pm3.1$&$   4.2\pm3.1$ \\ 
            &&&     1271&$   334\pm 26$&$  10.3\pm3.1$&$   6.5\pm3.1$&$   7.1\pm3.1$ \\ 
            &&&     1390&$   312\pm 21$&$   8.9\pm3.0$&$   6.4\pm3.0$&$   6.6\pm3.0$ \\ 
J1843-0157g &2021-09-03&$+230\pm2$&     1116&$   161\pm 12$&$  49.1\pm4.0$&$  -9.1\pm3.4$&$  10.8\pm3.4$ \\ 
            &&&     1275&$   169\pm 10$&$  40.2\pm3.9$&$  -9.9\pm3.4$&$  10.6\pm3.4$ \\ 
            &&&     1393&$   149\pm  8$&$  50.8\pm3.5$&$ -11.5\pm3.1$&$  11.5\pm3.1$ \\ 
J1843-0355  &2021-05-29&$-252.5\pm0.6$&     1105&$   623\pm 33$&$  55.8\pm3.8$&$  20.9\pm3.4$&$  17.9\pm3.4$ \\ 
            &&&     1260&$   634\pm 38$&$  67.0\pm3.7$&$  22.3\pm3.2$&$  20.5\pm3.2$ \\ 
            &&&     1384&$   651\pm 39$&$  65.6\pm3.4$&$  20.0\pm3.1$&$  19.1\pm3.1$ \\ 
J1844-0030  &2020-12-19&$+222.4\pm0.7$&     1107&$   711\pm 41$&$  14.9\pm3.1$&$   6.5\pm3.0$&$   7.4\pm3.0$ \\ 
            &&&     1281&$   508\pm 49$&$  15.1\pm3.1$&$   9.8\pm3.0$&$   9.3\pm3.0$ \\ 
            &&&     1393&$   449\pm 34$&$  13.9\pm3.0$&$   8.0\pm3.0$&$   7.6\pm3.0$ \\ 
J1844-0136g &2024-04-22&$+291\pm3$&     1109&$   110\pm 11$&$  28.5\pm4.0$&$   7.8\pm3.5$&$   8.3\pm3.5$ \\ 
            &&&     1258&$    84\pm 10$&$  33.1\pm4.0$&$  -6.2\pm3.4$&$   4.6\pm3.4$ \\ 
            &&&     1393&$    73\pm  7$&$  32.6\pm3.9$&$   4.7\pm3.4$&$   4.6\pm3.4$ \\ 
\multicolumn{8}{r}{ ... to be continued. } \\
\hline 
    \end{tabular}
\end{table}

\addtocounter{table}{-1}
\begin{table}[hbtp]
    \centering
    \caption{-- {\it continued}}
    \footnotesize
    \begin{tabular}{lcrrrrrr}
\hline\noalign{\smallskip}        
\multicolumn{1}{c}{PSR  Name}&\multicolumn{1}{c}{FAST Obs. Date}&\multicolumn{1}{c}{RM}&\multicolumn{1}{c}{$f$}&\multicolumn{1}{c}{$I$}&\multicolumn{1}{c}{$L/I$} &\multicolumn{1}{c}{$V/I$} &\multicolumn{1}{c}{$|V|/I$} \\
\multicolumn{1}{c}{ }&\multicolumn{1}{c}{yyyy-mm-dd}&\multicolumn{1}{c}{($\rm rad/m^2$)}&\multicolumn{1}{c}{(MHz)}&\multicolumn{1}{c}{($\rm \mu Jy$)}&\multicolumn{1}{c}{(\%)}&\multicolumn{1}{c}{(\%)}&\multicolumn{1}{c}{(\%)} \\
\hline\noalign{\smallskip}
J1844-0202g &2023-02-18&$-585\pm3$&     1095&$    24\pm 20$&$  79.2\pm11.2$&$   0.1\pm6.4$&$   8.1\pm6.4$ \\ 
            &&&     1285&$    32\pm 26$&$  51.6\pm8.3$&$  19.7\pm5.6$&$  14.8\pm5.5$ \\ 
            &&&     1392&$    30\pm 16$&$  63.5\pm5.1$&$  28.1\pm4.4$&$  25.3\pm4.4$ \\ 
J1844-0244  &2022-11-20&$+121.2\pm1.0$&     1108&$  1365\pm 74$&$  19.3\pm3.1$&$  -1.7\pm3.0$&$   2.4\pm3.0$ \\ 
            &&&     1267&$  1103\pm 71$&$  14.8\pm3.1$&$  -0.9\pm3.0$&$   0.6\pm3.0$ \\ 
            &&&     1392&$   934\pm 59$&$  13.6\pm3.0$&$   1.2\pm3.0$&$   0.8\pm3.0$ \\ 
J1844-0256  &2022-11-20&$+251\pm2$&     1105&$   434\pm 25$&$  51.1\pm3.8$&$  -3.1\pm3.5$&$   3.4\pm3.5$ \\ 
            &&&     1269&$   433\pm 27$&$  42.3\pm3.7$&$ -11.2\pm3.2$&$   9.5\pm3.2$ \\ 
            &&&     1391&$   425\pm 24$&$  53.3\pm3.3$&$ -11.3\pm3.1$&$   9.8\pm3.1$ \\ 

J1844-0302  &2021-02-21&$+942.6\pm1.3$&     1091&$    12\pm  1$&$  21.7\pm3.3$&$  -0.1\pm3.1$&$   3.1\pm3.1$ \\ 
            &&&     1269&$    10\pm  1$&$  18.7\pm3.5$&$   0.5\pm3.2$&$   0.4\pm3.2$ \\ 
            &&&     1392&$    10\pm  0$&$  23.0\pm3.3$&$  -2.0\pm3.2$&$   3.9\pm3.2$ \\ 
J1844-0310  &2021-02-21&$+279.6\pm0.3$&     1093&$   488\pm 26$&$  72.1\pm3.3$&$   6.7\pm3.1$&$   7.3\pm3.1$ \\ 
            &&&     1267&$   413\pm 25$&$  79.0\pm3.4$&$  12.0\pm3.1$&$  10.1\pm3.1$ \\ 
            &&&     1392&$   371\pm 20$&$  83.1\pm3.1$&$   9.0\pm3.1$&$   8.5\pm3.1$ \\ 
J1844-0538  &2025-09-26&$+26.1\pm0.4$&     1107&$   622\pm 34$&$  49.7\pm3.1$&$  16.4\pm3.0$&$  16.2\pm3.0$ \\ 
            &&&     1269&$   307\pm 23$&$  38.6\pm3.1$&$  11.9\pm3.1$&$  10.3\pm3.1$ \\ 
            &&&     1384&$   209\pm 17$&$  43.1\pm3.1$&$  11.5\pm3.1$&$  10.6\pm3.1$ \\ 
J1845-0103g &2021-11-11&$-1032.8\pm0.7$&     1101&$   275\pm 14$&$  38.7\pm3.5$&$ -10.1\pm3.3$&$   8.4\pm3.3$ \\ 
            &&&     1272&$   282\pm 14$&$  52.8\pm3.4$&$  -5.9\pm3.1$&$   7.4\pm3.1$ \\ 
            &&&     1392&$   281\pm 14$&$  53.5\pm3.2$&$  -5.1\pm3.0$&$   7.1\pm3.0$ \\ 
J1845-0142g &2024-02-03&$-397\pm2$&     1100&$   239\pm 19$&$  34.5\pm5.1$&$  14.8\pm4.2$&$   4.4\pm4.2$ \\ 
            &&&     1258&$   270\pm 17$&$  32.9\pm4.2$&$   5.6\pm3.6$&$   6.5\pm3.6$ \\ 
            &&&     1393&$   256\pm 15$&$  43.9\pm3.9$&$  12.5\pm3.4$&$  13.8\pm3.4$ \\ 
J1845-0144g &2022-01-13&$562\pm3$&     1103&$   367\pm 19$&$  11.2\pm3.4$&$  20.4\pm3.2$&$  17.9\pm3.2$ \\ 
            &&&     1267&$   299\pm 15$&$  16.8\pm3.6$&$  10.7\pm3.3$&$   8.9\pm3.3$ \\ 
            &&&     1392&$   210\pm 11$&$  17.5\pm3.2$&$  13.2\pm3.3$&$  11.9\pm3.3$ \\ 
J1845-0316  &2024-01-10&$-72.7\pm0.4$&     1092&$   697\pm 35$&$  40.1\pm3.1$&$  21.0\pm3.0$&$  20.7\pm3.0$ \\ 
            &&&     1273&$   529\pm 28$&$  36.6\pm3.2$&$  23.0\pm3.1$&$  21.4\pm3.1$ \\ 
            &&&     1392&$   435\pm 22$&$  35.6\pm3.1$&$  21.5\pm3.0$&$  20.7\pm3.0$ \\ 
J1846-0513  &2025-09-21&$232\pm2$&     1110&$   183\pm 11$&$  27.5\pm3.4$&$   1.0\pm3.3$&$   1.0\pm3.3$ \\ 
            &&&     1265&$   120\pm 11$&$  30.3\pm3.5$&$  -1.6\pm3.3$&$   0.8\pm3.3$ \\ 
            &&&     1384&$    94\pm  8$&$  24.8\pm3.3$&$  -2.1\pm3.1$&$   2.0\pm3.1$ \\ 
J1848-0055  &2021-10-24&$-429\pm3$&     1105&$   278\pm 23$&$  41.6\pm4.6$&$  17.4\pm4.0$&$  12.3\pm4.0$ \\ 
            &&&     1276&$   401\pm 24$&$  36.9\pm4.1$&$  13.3\pm3.8$&$  10.3\pm3.8$ \\ 
            &&&     1395&$   340\pm 22$&$  29.9\pm3.7$&$   8.6\pm3.2$&$   6.7\pm3.2$ \\ 
J1849-0013g &2023-09-16&$-103\pm2$&     1100&$   230\pm 15$&$  25.3\pm3.5$&$  -2.6\pm3.3$&$   2.3\pm3.3$ \\ 
            &&&     1261&$   155\pm 13$&$  30.4\pm3.6$&$  -4.3\pm3.2$&$   1.6\pm3.2$ \\ 
            &&&     1392&$   164\pm 10$&$  30.1\pm3.5$&$   0.8\pm3.3$&$   1.2\pm3.3$ \\ 
J1850-0002  &     1089&$   120\pm  7$&$  16.6\pm3.2$&$  -2.5\pm3.1$&$   6.4\pm3.1$ \\ 
            &&&     1279&$    82\pm  8$&$  21.8\pm3.6$&$  -1.5\pm3.3$&$   3.1\pm3.3$ \\ 
            &&&     1391&$    68\pm  6$&$  21.4\pm3.2$&$  -3.1\pm3.1$&$   2.5\pm3.1$ \\ 
J1850-0006  &2022-11-06&$152.3\pm1.4$&     1101&$   831\pm 42$&$  16.0\pm3.1$&$  -1.6\pm3.0$&$   1.7\pm3.0$ \\ 
            &&&     1270&$   720\pm 37$&$  21.1\pm3.1$&$  -0.7\pm3.0$&$   1.8\pm3.0$ \\ 
            &&&     1395&$   619\pm 32$&$  27.7\pm3.0$&$   1.3\pm3.0$&$   2.2\pm3.0$ \\ 
J1850-0020  &2021-12-26&$+97\pm3$&     1105&$   202\pm 24$&$  28.6\pm3.4$&$   0.5\pm3.2$&$   5.1\pm3.2$ \\ 
            &&&     1271&$   141\pm 33$&$  16.8\pm3.7$&$   0.8\pm3.3$&$   5.1\pm3.3$ \\ 
            &&&     1393&$   121\pm 18$&$  22.2\pm3.5$&$  -1.4\pm3.3$&$   6.2\pm3.3$ \\ 
J1850-0026  &2020-08-19&$663.8\pm0.2$&     1105&$  2356\pm118$&$  51.9\pm3.0$&$ -12.7\pm3.0$&$  12.5\pm3.0$ \\ 
            &&&     1268&$  1852\pm 93$&$  45.8\pm3.0$&$ -10.1\pm3.0$&$  10.5\pm3.0$ \\ 
            &&&     1395&$  1462\pm 73$&$  44.1\pm3.0$&$  -9.9\pm3.0$&$   9.7\pm3.0$ \\ 
\multicolumn{8}{r}{ ... to be continued. } \\
\hline 
    \end{tabular}
\end{table}

\addtocounter{table}{-1}
\begin{table}[hbtp]
    \centering
    \caption{-- {\it continued}}
    \footnotesize
    \begin{tabular}{lcrrrrrr}
\hline\noalign{\smallskip}        
\multicolumn{1}{c}{PSR  Name}&\multicolumn{1}{c}{FAST Obs. Date}&\multicolumn{1}{c}{RM}&\multicolumn{1}{c}{$f$}&\multicolumn{1}{c}{$I$}&\multicolumn{1}{c}{$L/I$} &\multicolumn{1}{c}{$V/I$} &\multicolumn{1}{c}{$|V|/I$} \\
\multicolumn{1}{c}{ }&\multicolumn{1}{c}{yyyy-mm-dd}&\multicolumn{1}{c}{($\rm rad/m^2$)}&\multicolumn{1}{c}{(MHz)}&\multicolumn{1}{c}{($\rm \mu Jy$)}&\multicolumn{1}{c}{(\%)}&\multicolumn{1}{c}{(\%)}&\multicolumn{1}{c}{(\%)} \\
\hline\noalign{\smallskip}
J1850-0031  &2020-04-02&$-206.0\pm0.8$&     1094&$   406\pm 21$&$  31.2\pm3.2$&$  -7.6\pm3.1$&$   7.2\pm3.1$ \\ 
            &&&     1272&$   321\pm 18$&$  30.7\pm3.2$&$   1.1\pm3.1$&$   3.2\pm3.1$ \\ 
            &&&     1395&$   262\pm 14$&$  29.5\pm3.1$&$  -1.3\pm3.0$&$   2.0\pm3.0$ \\ 
J1850-0050g &2021-09-03&$596\pm5$&     1117&$   435\pm 22$&$  14.8\pm3.3$&$  -6.8\pm3.2$&$   3.9\pm3.2$ \\ 
            &&&     1280&$   406\pm 21$&$  18.3\pm3.3$&$  -5.1\pm3.1$&$   4.6\pm3.1$ \\ 
            &&&     1393&$   347\pm 18$&$  19.8\pm3.1$&$  -5.9\pm3.0$&$   5.5\pm3.0$ \\ 
J1851-0029  &2021-03-06&$648.6\pm0.9$&     1101&$   684\pm 48$&$   9.7\pm3.1$&$   0.9\pm3.0$&$   0.6\pm3.0$ \\ 
            &&&     1264&$   570\pm 53$&$  14.5\pm3.1$&$   1.2\pm3.0$&$   0.9\pm3.0$ \\ 
            &&&     1393&$   516\pm 40$&$  16.8\pm3.0$&$   2.0\pm3.0$&$   1.5\pm3.0$ \\ 
J1852+0013  &2020-08-26&$462.7\pm0.7$&     1101&$   350\pm 31$&$  19.6\pm3.1$&$   3.9\pm3.0$&$   3.3\pm3.0$ \\ 
            &&&     1274&$   272\pm 27$&$  21.0\pm3.2$&$   4.1\pm3.1$&$   3.8\pm3.1$ \\ 
            &&&     1394&$   205\pm 25$&$  23.3\pm3.1$&$   4.5\pm3.0$&$   4.6\pm3.0$ \\ 
J1852+0018  &2020-09-15&$75\pm2$&     1101&$    80\pm 10$&$  71.6\pm4.2$&$  -0.8\pm4.1$&$   5.1\pm4.1$ \\ 
            &&&     1267&$    56\pm 13$&$  72.4\pm4.7$&$  -5.9\pm3.6$&$   9.3\pm3.6$ \\ 
            &&&     1392&$    47\pm  8$&$  68.2\pm4.2$&$  -8.9\pm3.6$&$   6.9\pm3.6$ \\      
J1852+0031  &2021-11-14&$315.8\pm0.4$&     1102&$  7059\pm353$&$   4.1\pm3.0$&$   0.5\pm3.0$&$   0.8\pm3.0$ \\ 
            &&&     1286&$  5518\pm276$&$   8.2\pm3.0$&$  -0.4\pm3.0$&$   1.2\pm3.0$ \\ 
            &&&     1392&$  4682\pm234$&$   7.9\pm3.0$&$   0.6\pm3.0$&$   1.5\pm3.0$ \\ 
J1852+0056g &2021-02-07&$-90.0\pm0.9$&     1103&$   103\pm  6$&$  59.7\pm3.4$&$ -22.2\pm3.2$&$  25.5\pm3.3$ \\ 
            &&&     1272&$    53\pm  8$&$  58.0\pm3.8$&$ -30.5\pm3.4$&$  28.9\pm3.3$ \\ 
            &&&     1392&$    53\pm  4$&$  57.3\pm3.4$&$ -26.7\pm3.2$&$  26.8\pm3.2$ \\ 
J1853-0054g &2022-01-13&$879\pm3$&     1102&$   105\pm 12$&$  25.8\pm4.1$&$  -1.5\pm3.4$&$   5.9\pm3.4$ \\ 
            &&&     1280&$    84\pm 11$&$  24.4\pm3.9$&$  -4.5\pm3.5$&$   3.7\pm3.5$ \\ 
            &&&     1392&$    75\pm  9$&$  19.4\pm3.6$&$  -4.8\pm3.2$&$   1.2\pm3.2$ \\ 
J1853+0505  &2021-04-22&$10.8\pm0.3$&     1107&$  4838\pm242$&$  19.9\pm3.0$&$  -6.1\pm3.0$&$   5.8\pm3.0$ \\ 
            &&&     1276&$  3658\pm183$&$  19.6\pm3.0$&$  -4.0\pm3.0$&$   5.0\pm3.0$ \\ 
            &&&     1393&$  2926\pm146$&$  17.3\pm3.0$&$  -2.3\pm3.0$&$   5.0\pm3.0$ \\ 
J1853+0545  &2020-08-01&$77.8\pm0.6$&     1106&$  3440\pm172$&$   5.4\pm3.0$&$   9.1\pm3.0$&$   8.8\pm3.0$ \\ 
            &&&     1284&$  3133\pm157$&$   5.8\pm3.0$&$   6.5\pm3.0$&$   6.3\pm3.0$ \\ 
            &&&     1393&$  3067\pm153$&$   8.0\pm3.0$&$   6.2\pm3.0$&$   6.1\pm3.0$ \\ 
J1854-00    &2021-04-15&$567.6\pm0.7$&     1100&$   146\pm 19$&$  43.7\pm3.4$&$  -2.7\pm3.2$&$   1.4\pm3.2$ \\ 
            &&&     1255&$   128\pm 19$&$  36.3\pm3.5$&$  -6.1\pm3.3$&$   3.9\pm3.3$ \\ 
            &&&     1393&$   103\pm 15$&$  43.8\pm3.4$&$  -0.9\pm3.2$&$   3.4\pm3.2$ \\ 
J1855+0527  &2020-04-19&$+124.6\pm0.6$&     1106&$   273\pm 27$&$  66.6\pm3.7$&$  35.6\pm3.5$&$  36.0\pm3.5$ \\ 
            &&&     1262&$   251\pm 26$&$  62.1\pm3.9$&$  38.5\pm3.3$&$  40.3\pm3.3$ \\ 
            &&&     1393&$   222\pm 24$&$  70.6\pm3.6$&$  39.4\pm3.3$&$  41.5\pm3.3$ \\ 
J1856+0245  &2020-11-04&$+267.7\pm0.9$&     1097&$   499\pm 28$&$  62.1\pm4.3$&$  31.1\pm3.3$&$  31.0\pm3.3$ \\ 
            &&&     1267&$   531\pm 32$&$  67.0\pm4.3$&$  18.2\pm3.5$&$  17.9\pm3.5$ \\ 
            &&&     1399&$   631\pm 33$&$  63.3\pm3.6$&$  20.4\pm3.1$&$  21.9\pm3.1$ \\
J1857+0210  &2021-01-26&$+136\pm6$&     1100&$   526\pm 45$&$  10.3\pm3.4$&$  -1.8\pm3.4$&$  -0.7\pm3.4$ \\ 
            &&&     1278&$   324\pm 71$&$  12.1\pm3.8$&$  -1.5\pm3.5$&$   0.7\pm3.5$ \\ 
            &&&     1391&$   289\pm 56$&$   9.7\pm3.3$&$   1.0\pm3.2$&$  -1.1\pm3.2$ \\ 
J1857+0300  &2020-05-20&$+884\pm3$&     1103&$   107\pm 35$&$  38.8\pm5.2$&$  -3.0\pm4.2$&$  14.8\pm4.2$ \\ 
            &&&     1263&$   104\pm 33$&$  30.2\pm6.2$&$  -5.0\pm4.7$&$   2.4\pm4.7$ \\ 
            &&&     1398&$    71\pm 37$&$  39.0\pm4.6$&$  -2.8\pm3.8$&$   1.1\pm3.8$ \\ 
J1857+0526  &2020-05-14&$959.5\pm0.3$&     1106&$   960\pm 57$&$  45.3\pm3.1$&$ -16.2\pm3.0$&$  14.9\pm3.0$ \\ 
            &&&     1270&$   849\pm 57$&$  45.1\pm3.1$&$ -15.0\pm3.1$&$  13.6\pm3.1$ \\ 
            &&&     1393&$   843\pm 51$&$  44.1\pm3.1$&$ -15.1\pm3.0$&$  15.4\pm3.0$ \\ 
J1858+0215  &2020-04-18&$-1.5\pm1.3$&     1099&$   306\pm 37$&$  57.8\pm3.5$&$   4.4\pm3.3$&$   5.1\pm3.3$ \\ 
            &&&     1272&$   287\pm 38$&$  28.6\pm3.5$&$   3.7\pm3.3$&$   3.8\pm3.3$ \\ 
            &&&     1396&$   256\pm 29$&$  24.2\pm3.3$&$   5.6\pm3.1$&$   3.3\pm3.1$ \\ 
\multicolumn{8}{r}{ ... to be continued. } \\
\hline 
    \end{tabular}
\end{table}

\addtocounter{table}{-1}
\begin{table}[hbtp]
    \centering
    \caption{-- {\it continued}}
    \footnotesize
    \begin{tabular}{lcrrrrrr}
\hline\noalign{\smallskip}        
\multicolumn{1}{c}{PSR  Name}&\multicolumn{1}{c}{FAST Obs. Date}&\multicolumn{1}{c}{RM}&\multicolumn{1}{c}{$f$}&\multicolumn{1}{c}{$I$}&\multicolumn{1}{c}{$L/I$} &\multicolumn{1}{c}{$V/I$} &\multicolumn{1}{c}{$|V|/I$} \\
\multicolumn{1}{c}{ }&\multicolumn{1}{c}{yyyy-mm-dd}&\multicolumn{1}{c}{($\rm rad/m^2$)}&\multicolumn{1}{c}{(MHz)}&\multicolumn{1}{c}{($\rm \mu Jy$)}&\multicolumn{1}{c}{(\%)}&\multicolumn{1}{c}{(\%)}&\multicolumn{1}{c}{(\%)} \\
\hline\noalign{\smallskip}
J1859+0601  &2020-08-21&$76\pm2$&     1095&$   306\pm 19$&$  29.9\pm3.5$&$   3.6\pm3.3$&$   4.9\pm3.3$ \\ 
            &&&     1295&$   192\pm 33$&$  32.4\pm4.3$&$   1.9\pm3.5$&$   5.8\pm3.5$ \\ 
            &&&     1396&$   206\pm 16$&$  32.2\pm3.4$&$   0.9\pm3.2$&$   2.6\pm3.2$ \\ 
J1901+0300  &2020-10-28&$-34\pm3$&     1099&$   203\pm 11$&$  14.9\pm3.1$&$  -1.0\pm3.1$&$   3.5\pm3.1$ \\ 
            &&&     1284&$   147\pm 11$&$  10.0\pm3.4$&$  -4.5\pm3.1$&$   5.9\pm3.1$ \\ 
            &&&     1385&$   129\pm  7$&$  12.0\pm3.1$&$  -5.8\pm3.1$&$   5.2\pm3.1$ \\ 
J1903+0327  &2020-05-23&$239\pm10$&     1102&$   752\pm 38$&$  11.5\pm3.1$&$ -12.3\pm3.1$&$  12.4\pm3.1$ \\ 
            &&&     1280&$   649\pm 35$&$   5.0\pm3.3$&$  -5.3\pm3.1$&$   6.4\pm3.1$ \\ 
            &&&     1397&$   628\pm 32$&$  10.9\pm3.1$&$ -13.1\pm3.1$&$  12.1\pm3.1$ \\ 
J1907+0534  &2020-05-14&$607\pm4$&     1104&$   290\pm 28$&$  11.3\pm3.2$&$  -4.6\pm3.1$&$   4.2\pm3.1$ \\ 
            &&&     1261&$   216\pm 32$&$   9.6\pm3.3$&$  -4.7\pm3.2$&$   5.8\pm3.2$ \\ 
            &&&     1394&$   172\pm 24$&$   9.7\pm3.3$&$  -3.4\pm3.2$&$   3.9\pm3.2$ \\ 
J1908+0839  &2020-10-26&$821\pm8$&     1104&$   622\pm 35$&$   6.6\pm3.2$&$   2.3\pm3.1$&$   2.4\pm3.1$ \\ 
            &&&     1266&$   533\pm 35$&$   6.4\pm3.2$&$  -1.6\pm3.1$&$   0.9\pm3.1$ \\ 
            &&&     1392&$   484\pm 29$&$   7.5\pm3.1$&$   1.1\pm3.1$&$   2.7\pm3.1$ \\ 
J1908+0909  &2022-11-07&$370.8\pm1.0$&     1101&$   409\pm 51$&$  42.0\pm3.3$&$   0.5\pm3.1$&$   4.0\pm3.1$ \\ 
            &&&     1266&$   301\pm 70$&$  48.3\pm3.6$&$   5.2\pm3.3$&$   5.5\pm3.3$ \\ 
            &&&     1393&$   311\pm 59$&$  38.4\pm3.7$&$   6.7\pm3.3$&$   3.7\pm3.3$ \\ 
J1910+0534  &2021-06-21&$349.7\pm0.4$&     1102&$   738\pm 38$&$  29.0\pm3.1$&$  11.7\pm3.0$&$  13.1\pm3.0$ \\ 
            &&&     1271&$   547\pm 35$&$  34.5\pm3.1$&$   9.6\pm3.0$&$  12.3\pm3.0$ \\ 
            &&&     1396&$   483\pm 30$&$  35.6\pm3.1$&$   9.2\pm3.0$&$  13.1\pm3.0$ \\ 
J1910+1026  &2020-11-23&$207.3\pm0.9$&     1083&$   115\pm 25$&$  74.9\pm6.3$&$  -4.2\pm4.5$&$  -4.2\pm4.5$ \\ 
            &&&     1278&$    79\pm 27$&$  75.1\pm5.2$&$   0.0\pm3.7$&$   0.5\pm3.7$ \\ 
            &&&     1395&$    65\pm 19$&$  85.3\pm4.5$&$  -6.8\pm3.7$&$   4.1\pm3.7$ \\ 
J1911+0101A &2020-12-02&$124.8\pm1.3$&     1106&$   391\pm 23$&$  27.0\pm3.5$&$  17.1\pm3.2$&$  16.3\pm3.2$ \\ 
            &&&     1258&$   323\pm 25$&$  36.3\pm3.5$&$  13.0\pm3.2$&$  20.6\pm3.2$ \\ 
            &&&     1395&$   294\pm 19$&$  37.3\pm3.5$&$  21.8\pm3.3$&$  18.9\pm3.3$ \\ 
J1913+1000  &2021-01-13&$499.3\pm0.9$&     1098&$   843\pm 48$&$  12.5\pm3.0$&$  -5.6\pm3.0$&$   5.3\pm3.0$ \\ 
            &&&     1292&$   573\pm 47$&$  11.4\pm3.1$&$  -5.5\pm3.1$&$   4.0\pm3.1$ \\ 
            &&&     1385&$   515\pm 34$&$  12.9\pm3.0$&$  -4.3\pm3.0$&$   4.6\pm3.0$ \\ 
J1913+11025 &2025-08-23&$1373\pm7$&     1099&$   141\pm  9$&$  15.6\pm3.2$&$   6.9\pm3.1$&$   7.4\pm3.1$ \\ 
            &&&     1259&$   121\pm 11$&$  13.4\pm3.4$&$   0.9\pm3.1$&$   5.0\pm3.1$ \\ 
            &&&     1392&$    98\pm  7$&$  14.5\pm3.3$&$   6.7\pm3.1$&$   6.1\pm3.1$ \\ 
J1913+1145  &2020-05-23&$921.8\pm0.8$&     1105&$   527\pm 30$&$  15.5\pm3.1$&$  -0.1\pm3.0$&$   0.5\pm3.0$ \\ 
            &&&     1262&$   455\pm 32$&$  16.8\pm3.1$&$  -2.1\pm3.1$&$   0.1\pm3.1$ \\ 
            &&&     1396&$   406\pm 25$&$  16.5\pm3.0$&$  -1.5\pm3.0$&$   2.2\pm3.0$ \\ 
J1916+0844  &2020-12-15&$601.0\pm0.4$&     1105&$  1049\pm 56$&$  24.4\pm3.0$&$  -0.6\pm3.0$&$   0.4\pm3.0$ \\ 
            &&&     1283&$   729\pm 53$&$  23.0\pm3.0$&$  -1.8\pm3.0$&$   1.7\pm3.0$ \\ 
            &&&     1395&$   603\pm 41$&$  24.0\pm3.0$&$  -0.4\pm3.0$&$   0.4\pm3.0$ \\ 
J1918+1340g &2020-08-07&$740\pm2$&     1104&$    82\pm  8$&$  55.2\pm4.2$&$ -18.2\pm3.5$&$  18.1\pm3.5$ \\ 
            &&&     1273&$    68\pm  9$&$  58.2\pm4.0$&$  -7.4\pm3.6$&$   9.4\pm3.6$ \\ 
            &&&     1393&$    56\pm  9$&$  57.3\pm3.7$&$ -12.0\pm3.3$&$  12.2\pm3.3$ \\ 
J1919+1314  &2021-10-12&$1981\pm1$&     1097&$   283\pm 24$&$  28.5\pm3.4$&$  -3.3\pm3.2$&$   1.1\pm3.2$ \\ 
            &&&     1293&$   211\pm 36$&$  25.3\pm3.5$&$  -2.4\pm3.2$&$   1.8\pm3.2$ \\ 
            &&&     1392&$   182\pm 23$&$  30.1\pm3.2$&$  -4.6\pm3.1$&$   3.3\pm3.1$ \\ 
J1920+1110  &2020-05-09&$654.0\pm0.9$&     1105&$   563\pm 32$&$  17.6\pm3.1$&$ -12.4\pm3.0$&$  12.0\pm3.0$ \\ 
            &&&     1260&$   464\pm 26$&$  20.3\pm3.1$&$ -12.0\pm3.1$&$  13.0\pm3.1$ \\ 
            &&&     1395&$   362\pm 23$&$  19.0\pm3.1$&$ -14.9\pm3.1$&$  15.7\pm3.1$ \\ 
J1921+1340g &2023-02-14&$3270\pm3$&     1081&$   131\pm 42$&$  16.5\pm3.7$&$  -0.7\pm3.3$&$  -1.1\pm3.3$ \\ 
            &&&     1274&$    84\pm 13$&$  13.5\pm3.3$&$  -3.5\pm3.1$&$   2.1\pm3.1$ \\ 
            &&&     1393&$    68\pm  4$&$  11.6\pm3.2$&$  -1.6\pm3.1$&$   1.6\pm3.1$ \\ 
\multicolumn{8}{r}{ ... to be continued. } \\
\hline 
    \end{tabular}
\end{table}

\addtocounter{table}{-1}
\begin{table}[hbtp]
    \centering
    \caption{-- {\it continued}}
    \footnotesize
    \begin{tabular}{lcrrrrrr}
\hline\noalign{\smallskip}        
\multicolumn{1}{c}{PSR  Name}&\multicolumn{1}{c}{FAST Obs. Date}&\multicolumn{1}{c}{RM}&\multicolumn{1}{c}{$f$}&\multicolumn{1}{c}{$I$}&\multicolumn{1}{c}{$L/I$} &\multicolumn{1}{c}{$V/I$} &\multicolumn{1}{c}{$|V|/I$} \\
\multicolumn{1}{c}{ }&\multicolumn{1}{c}{yyyy-mm-dd}&\multicolumn{1}{c}{($\rm rad/m^2$)}&\multicolumn{1}{c}{(MHz)}&\multicolumn{1}{c}{($\rm \mu Jy$)}&\multicolumn{1}{c}{(\%)}&\multicolumn{1}{c}{(\%)}&\multicolumn{1}{c}{(\%)} \\
\hline\noalign{\smallskip}
J1922+1512g &2022-05-13&$-242.9\pm1.0$&     1108&$   176\pm 11$&$  20.8\pm3.2$&$   2.0\pm3.1$&$   4.2\pm3.1$ \\ 
            &&&     1262&$   152\pm  9$&$  25.4\pm3.3$&$  -1.8\pm3.1$&$   6.0\pm3.1$ \\ 
            &&&     1393&$   109\pm  7$&$  23.2\pm3.2$&$   2.7\pm3.1$&$   2.1\pm3.1$ \\ 
J1928+1245  &2020-09-29&$+53\pm5$&     1109&$   123\pm  9$&$  25.3\pm3.5$&$  -3.1\pm3.2$&$   3.1\pm3.2$ \\ 
            &&&     1258&$    98\pm  8$&$  21.6\pm3.4$&$  -1.7\pm3.1$&$   8.1\pm3.1$ \\ 
            &&&     1385&$    98\pm  6$&$  20.5\pm3.5$&$  -0.6\pm3.2$&$   1.8\pm3.2$ \\ 
J1928+1923  &2020-08-04&$346.5\pm0.6$&     1102&$   908\pm 48$&$  13.3\pm3.0$&$  -9.1\pm3.0$&$   8.4\pm3.0$ \\ 
            &&&     1280&$   680\pm 47$&$  15.3\pm3.1$&$  -8.9\pm3.1$&$   7.2\pm3.1$ \\ 
            &&&     1395&$   585\pm 33$&$  17.4\pm3.1$&$  -8.3\pm3.0$&$   6.7\pm3.0$ \\ 
J1929+19    &2020-11-21&$+414\pm3$&     1104&$   967\pm 48$&$   4.5\pm3.0$&$  -0.7\pm3.0$&$   2.1\pm3.0$ \\ 
            &&&     1264&$   688\pm 35$&$   3.8\pm3.1$&$   2.6\pm3.0$&$   2.3\pm3.0$ \\ 
            &&&     1386&$   652\pm 32$&$   4.3\pm3.0$&$   1.9\pm3.0$&$   1.4\pm3.0$ \\ 
J1959+3141g &2023-08-26&$192\pm3$&     1109&$    61\pm  8$&$  20.3\pm3.9$&$  -1.0\pm3.3$&$   8.5\pm3.3$ \\ 
            &&&     1256&$    47\pm 11$&$  24.1\pm3.6$&$  -5.8\pm3.2$&$   4.2\pm3.2$ \\ 
            &&&     1393&$    42\pm  7$&$  23.1\pm3.5$&$  -3.7\pm3.4$&$   3.5\pm3.4$ \\ 
J2004+3429  &2020-04-19&$158.7\pm0.3$&     1106&$   522\pm 29$&$  75.6\pm3.2$&$   3.3\pm3.1$&$   3.1\pm3.1$ \\ 
            &&&     1256&$   436\pm 26$&$  80.6\pm3.2$&$  -3.8\pm3.1$&$   5.9\pm3.1$ \\ 
            &&&     1396&$   389\pm 24$&$  76.5\pm3.1$&$  -8.7\pm3.0$&$  10.3\pm3.0$ \\ 
J2005+3411g &2021-11-14&$208.3\pm1.2$&     1100&$   204\pm 11$&$  20.1\pm3.1$&$   7.9\pm3.1$&$   6.3\pm3.1$ \\ 
            &&&     1299&$   153\pm 11$&$  21.7\pm3.3$&$   3.8\pm3.2$&$   4.6\pm3.2$ \\ 
            &&&     1393&$   138\pm  7$&$  22.9\pm3.1$&$   5.2\pm3.0$&$   5.5\pm3.0$ \\ 
J2019+3718g &2023-10-17&$59.2\pm0.4$&     1116&$   133\pm 10$&$  61.4\pm3.1$&$  -4.3\pm3.1$&$   4.8\pm3.1$ \\ 
            &&&     1258&$   119\pm 10$&$  60.2\pm3.3$&$  -5.2\pm3.1$&$   4.8\pm3.1$ \\ 
            &&&     1393&$   102\pm 10$&$  62.4\pm3.1$&$  -6.8\pm3.0$&$   6.0\pm3.0$ \\ 
J2020+3806g &2025-02-28&$+1108.5\pm0.9$&     1099&$   251\pm 15$&$  11.5\pm3.1$&$  -0.1\pm3.0$&$   1.0\pm3.0$ \\ 
            &&&     1265&$   208\pm 14$&$  12.8\pm3.1$&$   0.3\pm3.0$&$   0.4\pm3.0$ \\ 
            &&&     1392&$   185\pm 13$&$  16.4\pm3.0$&$  -0.5\pm3.0$&$   1.1\pm3.0$ \\ 
J2041+4551  &2022-05-02&$-235.9\pm0.2$&     1107&$  1354\pm 69$&$  25.6\pm3.0$&$  13.0\pm3.0$&$  12.5\pm3.0$ \\ 
            &&&     1267&$  1009\pm 52$&$  23.7\pm3.0$&$  11.6\pm3.0$&$  12.5\pm3.0$ \\ 
            &&&     1393&$   824\pm 43$&$  24.9\pm3.0$&$  13.0\pm3.0$&$  13.3\pm3.0$ \\ 
J2046+4236g &2023-10-15&$978\pm3$&     1109&$    89\pm  6$&$  38.9\pm3.7$&$   4.0\pm3.8$&$  10.5\pm3.8$ \\ 
            &&&     1260&$    69\pm  5$&$  64.4\pm4.4$&$  -8.3\pm3.3$&$  14.6\pm3.3$ \\ 
            &&&     1390&$    94\pm  5$&$  53.1\pm3.6$&$  -1.2\pm3.4$&$  10.2\pm3.4$ \\ 
J2046+4253g &2022-09-07&$-92.6\pm0.8$&     1103&$   354\pm 18$&$  33.6\pm3.2$&$ -33.3\pm3.1$&$  31.2\pm3.1$ \\ 
            &&&     1268&$   302\pm 15$&$  42.3\pm3.2$&$ -27.5\pm3.1$&$  25.5\pm3.1$ \\ 
            &&&     1393&$   267\pm 13$&$  43.3\pm3.1$&$ -30.7\pm3.1$&$  30.3\pm3.1$ \\ 
J2052+4421g &2023-11-23&$-274.2\pm0.2$&     1100&$   664\pm 33$&$  65.6\pm3.1$&$  12.1\pm3.0$&$  11.6\pm3.0$ \\ 
            &&&     1290&$   454\pm 23$&$  64.1\pm3.1$&$  21.1\pm3.1$&$  19.3\pm3.1$ \\ 
            &&&     1393&$   381\pm 19$&$  63.0\pm3.1$&$  11.8\pm3.0$&$  12.2\pm3.0$ \\ 
J2205+6012  &2023-10-18&$-87.1\pm0.8$&     1099&$   466\pm 24$&$  38.6\pm3.3$&$   4.5\pm3.2$&$   2.1\pm3.2$ \\ 
            &&&     1311&$   415\pm 24$&$  39.0\pm3.3$&$  -4.8\pm3.2$&$   4.5\pm3.2$ \\ 
            &&&     1384&$   408\pm 21$&$  34.9\pm3.1$&$  -2.2\pm3.1$&$   3.7\pm3.1$ \\ 
            \multicolumn{8}{r}{ ... ended. } \\
\hline 
\hline
    \end{tabular}
\end{table}

\newpage

\end{document}